\documentclass[10pt]{iopart}

\usepackage{graphicx}
\usepackage{subcaption}
\usepackage{fancyhdr}
\usepackage{hyperref}
\usepackage{comment}
\usepackage{tikz}
\usepackage{xcolor}
\usepackage{iopams} 
\usepackage{ulem}

\usepackage{cuted}

\newcommand{\eqref}[1]{(\ref{#1})}

\expandafter\let\csname equation*\endcsname\relax
\expandafter\let\csname endequation*\endcsname\relax
\usepackage{amsmath}

\def\be{\begin{equation}}
\def\ee{\end{equation}}
\def\ber{\begin{eqnarray}}
\def\eer{\end{eqnarray}}

\newcommand{\ie}{{\it i.e.~}} 	
\newcommand{\eg}{{\it e.g.~}} 	
\newcommand\commentout[1]{}

\def\br{{\bf r}}

\def\bq{{\bf q}}
\def\hb{{\hat b}}

\newcommand{\BLG}{bilayer graphene}
\newcommand{\QPC}{quantum point contact}
\newcommand{\SGM}{SGM}

\newcommand{\fig}{figure~}
\newcommand{\eq}{equation~}
\newcommand{\tab}{table~}
\newcommand{\sectionname}{section~}
\newcommand{\reference}{reference~}

\begin{document}

\topical{Electron quantum optics in graphene}

\author{Himadri Chakraborti$^1$, Cosimo Gorini$^1$, Angelika Knothe$^{2, \footnotemark}$\footnotetext[1]{angelika.knothe@physik.uni-regensburg.de}, Ming-Hao Liu$^3$, P\'eter Makk$^{4,5, \footnotemark}$ \footnotetext[3]{makk.peter@ttk.bme.hu},  Fran\c{c}ois D.\ Parmentier$^1$, David Perconte$^6$, Klaus Richter$^2$, Preden Roulleau$^1$, Benjamin Sac\'ep\'e$^6$, Christian Sch\"onenberger$^7$, Wenmin Yang$^6$}


\address{$^1$Universit\'e Paris-Saclay, CEA, CNRS, SPEC, 91191, Gif-sur-Yvette, France}

\address{$^2$Institut f\"ur Theoretische Physik, Universität Regensburg, D-93040 Regensburg, Germany}

\address{$^3$Department of Physics and Center for Quantum Frontiers of Research and Technology (QFort), National Cheng Kung University, Tainan 70101, Taiwan}

\address{$^4$Department of Physics, Institute of Physics, Budapest University of Technology and Economics, M\"{u}egyetem rkp. 3., H-1111 Budapest, Hungary}

\address{$^5$MTA-BME Correlated van der Waals Structures Momentum Research Group, M\"uegyetem rkp. 3., H-1111 Budapest, Hungary}
\address{$^6$Universit\'e Grenoble Alpes, CNRS, Grenoble INP, Institut N\'eel, 38000 Grenoble, France}

\address{$^7$Nanoelectronics Group, Department of Physics, University of Basel, Basel, Switzerland}



\begin{abstract}
In the last decade, graphene has become an exciting platform for electron optical experiments, in many aspects superior to conventional two-dimensional electron gases (2DEGs). A major advantage, besides the ultra-large mobilities, is the fine control over the electrostatics,
which gives the possibility of realising gap-less and compact p-n interfaces with high precision. The latter host non-trivial states, \eg, snake states in moderate magnetic fields, and serve as building blocks of complex electron interferometers. Thanks to the Dirac spectrum and its non-trivial Berry phase, the internal (valley and sublattice) degrees of freedom, and the possibility to tailor the band structure using proximity effects, such interferometers open up a completely new playground based on novel device architectures. In this review, we introduce the theoretical background of graphene electron optics, fabrication methods used to realise electron-optical devices, and techniques for corresponding numerical simulations. Based on this, we give a comprehensive review of ballistic transport experiments and simple building blocks of electron optical devices both in single and bilayer graphene, highlighting the novel physics that is brought in compared to conventional 2DEGs. After describing the different magnetic field regimes in graphene p-n junctions and nanostructures, we conclude by discussing the state of the art in graphene-based Mach-Zender and Fabry-Perot interferometers.

\end{abstract}

\newpage

\tableofcontents

%
%
\submitto{\JPCM}
%
\maketitle
%
\ioptwocol
%
%

\section{Preface}

\subsection{Electron quantum optics; why graphene?}

Since its discovery~\cite{Novoselov2004}, graphene has emerged as a wonder material and as a playground of different fascinating condensed matter physics ideas. Due to its high electrical and thermal conductivity that comes together with an almost $100$\,\% optical transparency for the visible optical spectrum, monolayer and bilayer graphene, as well as the recently discovered twisted bilayer and multilayer graphene~\cite{Cao2018}, are promising materials for various application areas including advanced electronics and novel solutions for sustainability. Advancement of nanofabrication techniques over the past twenty years led to the realization of several interesting concepts stemming from the linear energy-momentum relation near the Fermi energy of undoped graphene, where the two-dimensional Fermi surface shrinks to two Fermi points, known Dirac points. The two separate Fermi surfaces are termed valleys (the K and K' points) and for low-energy, each valley hosts mass-less quasi-particles that are alike relativistic Dirac/Weyl fermions.

Electron quantum optics is a field of research where the focus is on the wave nature of electrons in solids. The goal is to explore fundamental properties of fermionic quasi-particles in solids by conducting prototypic experiments known from optics. A typical traditional free space quantum optics experiment is an interference experiment that converts the quantum nature into a measurable intensity pattern as a function of a control parameter. Examples are the text-book two-slit interference experiment, Fabry-P{\'e}rot, Mach-Zehnder and Michelson interferometers, and intensity-correlation experiments along the line of Hanbury-Brown and Twist~\cite{Hanbury-Brown1954,Hanbury-Brown1956}. These experiments make use of the large velocity of photons given by the speed of light which transforms even a rather short temporal coherence of a light source of $1$\,ns into an appreciable coherence length of $0.3$\,m. Taking a laser source, some light apertures and mirrors, it is straightforward to construct such interference experiments on an optical table.

In the solid state, electrons move at much lower speeds. The velocity is an interesting parameter as it can be dressed by all interactions: it is usually enhanced by strong electron-electron repulsion, but it can also be lowered through strong coupling with lattice degrees of freedom, such as phonons. For a typical doping, the group velocity of quasi-particles in semiconductors is only of order $10^5$\,m/s. Noting that relaxation times through electron-phonon interaction can easily be as short as $1$\,ps, which yields an electron mean-free path $l_{mfp}$ of only $\approx 100$\,nm. Hence, it seems impossible to perform free electron propagation (ballistic) electron optics experiments. First, one has to cool the solid state material to low enough temperatures that relaxation through the lattice is sufficiently suppressed. In the second place, one has to work with materials of very high quality, since also inherent disorder, such as atomic defects and grain boundaries, limit the mobility, and, hence, $l_{mfp}$. Electron optics experiments become feasible when the ballistic mean-free path $l_{mfp}$ is larger or comparable to system sizes that can be patterned with current state-of-the-art micro and nano-fabrication technology. So samples are of size $L > 0.1$\,$\mu$m, and typically in the range of $L\simeq 1-10$\,$\mu$m.

A second requirement is that the de Broglie wavelength $\lambda$ of the electrons (or holes) in the solid-state material should be $\lambda \ll l_{mfp}$. This is the limit of quasi-classical quantum optics, where propagation along optical trajectories can be defined and engineered through smooth electrostatic profiles.
Moreover, beyond conventional optical properties of light rays, trajectories of Dirac charge carriers can be additionally bent and controlled through magnetic fields, a hallmark and unique stronghold of Dirac electron quantum optics.
We note here that the notion of quasi-classical electron propagation looses its meaning in the opposite limit $\lambda \simeq L$. Here, localized confined states appear, which are realized in quantum dots and cavities. Such structures will also be covered in this review.

The field for electron quantum optics widened to a great extent when graphene devices could be fabricated with high enough purity yielding large enough ballistic mean-free paths $l_{mfp}$. The large Fermi velocity of graphene of $v\simeq 10^6$\,m/s helped to move on from disordered graphene ribbons to graphene that is ballistic over macroscopic distances. Moreover the electric field controllable ambipolar nature of graphene~\cite{Novoselov2004} provides a pristine way to control the global carrier density over the ion-implantation method of doping in conventional semiconductors. Realization of junctions of opposite carrier densities (electron and hole-type) in a single graphene sheet through the local control of carrier density using electrostatic gates is a milestone in electron quantum optics in general~\cite{huard2007transport, lemme2007graphene, williams2007quantum, abanin2007quantized}. The ability of fabricating p-n junctions gave rise to `relativistic' condensed matter physics where relativistic quantum phenomena like Klein tunneling~\cite{Katsnelson2006}, Veselago lensing~\cite{Cheianov2007}, particle collimation~\cite{Cheianov2006}, and quasibound states~\cite{silvestrov2007quantum} were demonstrated on chip-size devices. 

\subsection{Traditional quantum optics in a 2DEG}

Two-dimensional electron gases (2DEGs) confined in a GaAs quantum well in hetero\-structures of AlGaAs/GaAs-based semiconductors~\cite{Dingle1978} have been the historically dominant material in which electron quantum optics experiments could be realized. This is due to various factors: among the technical ones, the use of microelectronics industry growth and processing techniques has allowed obtaining high quality 2DEGs on wafer scale, that can be electrically connected using highly transparent ohmic contacts~\cite{Ketterson1985}. Due to the low mass of the electrons in the conduction band, very high mobilities have been achieved~\cite{Chung2022}.
Record electron mobilities are $\mu > 50 \cdot 10^6$\,cm$^2$/Vs yielding macroscopic mean-free paths $l_{mfp}> 250$\,$\mu$m.
Perhaps even more importantly, the semiconducting nature of AlGaAs/GaAs hetero\-structures permits the use of electrostatic depletion gates with which one can shape the potential landscape in the 2DEG. Thereby, one can gain control over electron wave trajectories. The most commonly used gating structure in electron quantum optics experiments with GaAs quantum wells is a split gate realising a quantum point contact (QPC)~\cite{VanWees1988}: a saddle point-like constriction in the 2DEG through which the electrons are channeled.

\begin{figure*}[t]
	\begin{center}
	  \includegraphics[width=\textwidth]{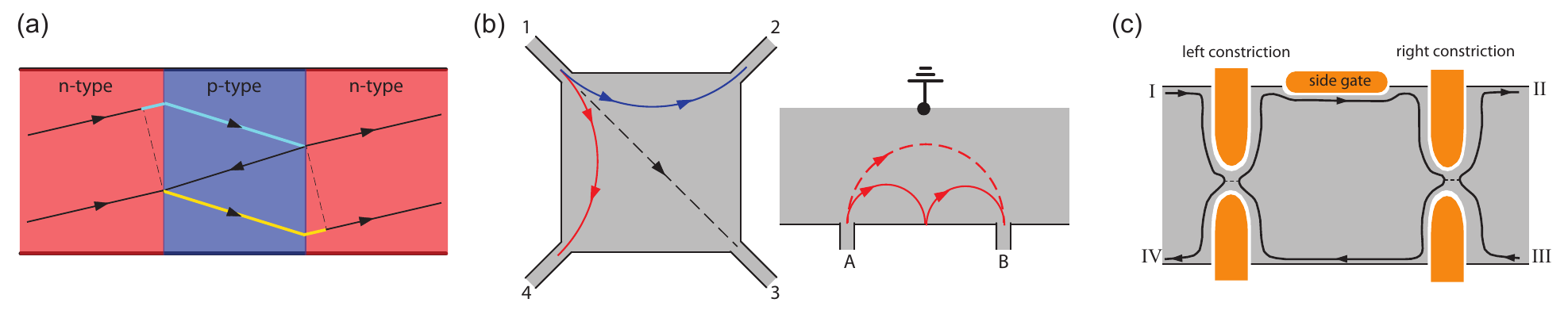}
    \end{center}
	\caption{\small {\bf Examples of electron optical devices}. (a) shows a {\it Fabry-P{\'e}rot interferometer} where the ``mirrors'' are defined by the boundaries between two regions with different electron concentrations. Specific to graphene, the two regions can also have opposite doping, indicated here in red as n-type and in blue as p-type. In this bipolar case, electron refraction is negative at the interface, which is something quite peculiar. Constructive interference occurs when the phase acquired along the blue or yellow path is a multiple of $2\pi$. The phase can be controlled through gates that tune the electron concentrations in, for example, the middle region. (b) shows another interesting device concept which is based on {\it transverse magnetic focusing}. The Lorentz force due to a magnetic field applied perpendicular to the graphene plane deflects the electron trajectories. In the red case, the electrons are deflected from the source contact 1 into drain contact 2. Changing either the direction of the magnetic field or the sign of the charge carrier type, flips the orientation of the circular motion from the red to the blue case. The right figure in (b) shows transverse magnetic focusing along one sample edge. The resonance condition for scattering from source contact A into drain contact B can be obtained for different strengths of magnetic fields. Finally, (c) shows a {\it quantum Hall interferometer} that makes use of edge states (dark black lines with arrows indicating propagation direction). One can measure, for example, the transmission probability from contact {\rm I} to contact {\rm II} or the reflection probability to {\rm IV}. The edge state can be seen as a one-dimensional propagating electron wave that is scattered with some probabilities at the left and right constrictions. With the aid of the side gate in the interior region the path length can be adjusted resulting in Aharonov-Bohm oscillation in the measurements.}
	\label{fig:CS0-Introduction}
\end{figure*}

For electron-transport experiments at low temperature, which are at the focus in this review, only electrons (or holes) at low energies are considered. According to the de Broglie relation and parabolic energy dispersion of quasi-free electrons with mass $m_e$, the Fermi wavelength $\lambda_F=h/\sqrt{2m_e E_F}$, where $E_F$ is the Fermi energy and $h$ the Planck constant. For a low effective mass and Fermi energy, the Fermi wavelength can become large, reaching values $\lambda_F > 100$\,nm. Since structure on the $100$\,nm scale can easily be fabricated today, QPCs can be designed that can electrostatically be tuned to pass none, only one, or a few (transverse) channels through the constriction. This has led to the seminal work of quantized conductance in QPCs, first conducted by B. van Wees {\it et al.}~\cite{VanWees1988}.

The most elementary electron optics experiment could start with a QPC that is tuned to transmit exactly one single channel. If the channel opens adiabatically on the exit side, a spherical electron wave would emerge. One could then place a gate electrode in the shape of a half circle some distance behind the QPC and applying a strongly negative voltage so that the gate acts as a mirror for electrons. If the mirror is smooth enough, it would reflect the electrons back to the QPC. Due to the wave nature, this arrangement will give rise to so-called Fabry-P{\'e}rot interferences, which, depending on the wavelength and the distance from the QPC to the mirror, can be constructive or destructive. This interference pattern can be made visible in three ways: (i) by changing the Fermi energy of the 2DEG with e.g. a global gate, or (ii) by tuning the voltage that is applied over the QPC. 
In case (i), one would measure the linear-response electrical conductance at small bias voltage as a function of gate voltage, and in case (ii), one would measure the differential conductance as a function of bias voltage. The (iii) way is by applying a tunable magnetic field $B$, which is also a very important tuning knob in quantum-interference experiment, and it will very often show up in this review.

\subsection{Graphene for quantum electron optics}

In recent years graphene devices with amazing qualities could be obtained, either by current-annealing suspended graphene \cite{Bolotin2008,Du2008} or by encapsulation into single-crystalline hexagonal boron nitride (h-BN)~\cite{Dean2010,Wang2012,Wang2013} complemented with one-dimensional (1D) edge contacts~\cite{Wang2013} which improved the electronic performance of graphene-based devices drastically, as explained in detail in \sectionname\ref{sec:Fabrication}. All these developments led to two-dimensional (2D) gate-tunable electrical conductors with mobilities that typically reach values $\mu_e > 10^5$\,cm$^2$/Vs, in some cases even above $10^6$\,cm$^2$/Vs. Correspondingly, large mean-free paths $l_{mfp}$ exceeding $1$~$\mu$m could be established and ballistic transport became possible in samples of “mesoscopic”, and even macroscopic size~\cite{Du2008,Bolotin2008,Young2009,Rickhaus2013,Banszerus2016}. If in addition the coherence length is sufficiently long, which requires low bias and low temperature experiments, quantum interference starts to play a decisive role. In graphene devices, all kind of quantum coherent interference effects have been observed in a surprisingly clean fashion.

Three generic electron-optical device concepts that very often appear in the literature are introduced in \fig\ref{fig:CS0-Introduction}.

In a perpendicular magnetic field $B$ the quantum effects are very different for zero, small, intermediate and large fields. For very small fields, the magnetic field adds a weak Lorentz force which bends semiclassical trajectories slightly~\cite{Shytov2008}. This adds a small phase term in quantum interference as a correction to the zero-field case. Nonetheless, this has interesting consequences. At intermediate fields the bending can become significant, leading to the formation of new bounds states within a finite size graphene device~\cite{Williams2017}. If the magnetic field increases beyond a critical value $B_c$ for which the cyclotron radius due to the ballistic motion shrinks below the sample size $L$, edges-states form at the boundary of the samples~\cite{Rickhaus2015}. They can often still be treated in a semiclassical manner as so-called skipping orbits and they are crucial to understand transport in devices where p-n junctions are realized. But if the magnetic field is so large that Landau quantization becomes dominant, this is when the bulk of the sample becomes gapped, one is entering the quantum Hall regime~\cite{Novoselov2005,Zhang2005}. While the interior of the sample is gapped in the quantum Hall state, compressible conducting channels form along the edges of the crystal. These channels are known as edges states. In the integer quantum Hall regime the edge channels are chiral and the number of channels depends on the filling factor $\nu=nh/eB$ which depends on carrier density $n$ and magnetic field strength $B$.

The edge channels of the quantum Hall state form ideal channels, since backscattering is absent in wide enough samples. The transport channels can be seen as analogs to single- or few-mode optical fibres. They are ballistic over lengths reaching the millimeter scale. In GaAs quantum wells they became the tool to explore a very large number of electron quantum optics experiments, such as electronic quantum interferometers in Fabry-P{\'e}rot~\cite{Zhang2009} and Mach-Zehnder~\cite{Ji2003} geometries, the realization of on-demand single electron sources~\cite{Feve2007,Dubois2013}, Hanbury-Brown and Twiss correlations measurements in continuous~\cite{Henny1999,Oliver1999} and single-excitations~\cite{Bocquillon2012} electron beams, or Hong-Ou Mandel two-particle interferences~\cite{Bocquillon2013}. Lately, the ability to combine these electron quantum optics schemes with the fractionally charged anyonic excitations of the fractional quantum Hall effect has further expanded the field~\cite{Nakamura2020,Bartolomei2020} with the perspective of developing yet another quantum information processing platform relying on non-abelian statistics~\cite{Nayak2008}. A quantum Hall edge channel can be selectively transmitted with a probability between zero and unity, fully gate-tunable. As the edge channels form the electronic analogue of fiber optics, quantum point contacts are the electronic equivalent of tunable beam-splitters, and are therefore ubiquitous in electron quantum optics experiments realized in AlGaAs/GaAs 2DEGs.

Transferring electron quantum optics experiments from GaAs 2DEGs to graphene has become an important task in the field in the past decade, as the quality and mobility of the available graphene samples increased dramatically. Indeed, the honeycomb lattice of graphene and its semi-metallic band structure give rise to extremely rich quantum Hall effects~\cite{Goerbig2011}, which can be explored through electron quantum optics experiments. In particular, the ambipolarity of graphene, as well as the strong role of electronic interactions in the emergence of a quantum Hall ferromagnetism where both spin and valley symmetries are broken~\cite{Young2012, Kharitonov2012ML, Kharitonov2012BL, Knothe2016}, greatly expands the playground for electron quantum optics. Furthermore, the structural differences between AlGaAs/GaAs 2DEGs and graphene, particularly with respect to the electronic confinement at the edge of the sample, allows testing the hypotheses upon which our current understanding of edge channel transport is based.

Similar to monolayer graphene, electron quantum optics experiments were early on also conducted with bilayer graphene. In contrast to monolayer graphene, bilayer graphene has additionally to the valley degree also a layer degree of freedom. The layer degree can directly be accessed through the charge density in the two layers using a double gated stack, which today typically starts with a graphite bottom gate, followed by a h-BN gate dielectric, bilayer graphene, followed by the top h-BN gate dielectrics, and ending by a top graphite gate. A symmetric gate voltage will add the same amount of charge to the two layers. However, if gating is asymmetric one can induced opposite charge in the two layers. This gives rise to a so-called displacement field, which -- crucially -- opens a gap in bilayer graphene. In gapped bilayer graphene one can realize the exact analogue of the QPC that was/is used in GaAs quantum well structures. Gating allows to fully deplete a region. This has also allowed to define gate controlled channels where the channel width is not determined by the natural graphene edges or by etched edges, but rather by a smooth bounding potential defined by the global gate structure~\cite{Allen2012, Droscher2012, Goossens2012, Overweg2018}. This has been an important milestone, since it allowed to engineer ballistic few-mode channels without a magnetic field, simply due to the presence of the gap in bilayer graphene with non-zero displacement field. In earlier etched graphene ribbons the edge roughness was too large and strongly limited to scattering mean-free path.

\section{Introduction to graphene}
\subsection{Band structure}

Graphene has a hexagonal lattice with a of two-atom basis as depicted in \fig\ref{fig graphene lattice}, where $a\approx 0.142$~nm is the carbon-carbon bond length, $\mathbf{a}_1=a_c(1/2,\sqrt{3}/2)$ and $\mathbf{a}_2=a_c(-1/2,\sqrt{3}/2)$ are the primitive vectors, $a_c=\sqrt{3}a$ is the hexagonal lattice constant, and the basis vectors of the two atoms are $\mathbf{d}_A=(0,0),\mathbf{d}_B=(0,a)$. The corresponding reciprocal primitive vectors can be chosen as $\mathbf{b}_1 = 2\pi/\sqrt{3}a_c(\sqrt{3},1)$ and $\mathbf{b}_2 = 2\pi/\sqrt{3}a_c(-\sqrt{3},1)$ as shown in \fig\ref{fig graphene BZ}, where the empty dots are part of the hexagonal reciprocal lattice, and the yellow area bounded by a dashed hexagon is the (first) Brillouin zone (BZ), some symmetry points of which are marked.

\subsubsection{Band structure of single layer graphene.}
\label{sec BS of graphene}

\begin{figure*}
\begin{center}
\subcaptionbox{\label{fig graphene lattice}}{
\begin{tikzpicture}[scale=2]
\def\Rdot{2pt}
\foreach \n/\m in {0/0,1/0,0/1,1/-1,-1/1,-1/0,0/-1}{
\fill [shading=ball,ball color=white] ({(\n-\m)*(1/2)},{(\n+\m)*sqrt(3)/2}) circle (\Rdot);
}
\foreach \n/\m in {0/0,-1/0,0/-1,1/-1,-1/1,1/0,0/1}{
\fill [shading=ball,ball color=pink] ({(\n-\m)*(1/2)},{(\n+\m)*sqrt(3)/2+1/sqrt(3)}) circle (\Rdot);
}
\draw [very thick,-latex] (0,0) -- (1/2,{sqrt(3)/2}) node [midway,right] {$\mathbf{a}_1$};
\draw [very thick,-latex] (0,0) -- (-1/2,{sqrt(3)/2}) node [midway,left] {$\mathbf{a}_2$};
\draw [very thick,-latex,densely dashed] (0,0) node [below] {$\mathbf{d}_A$} -- (0,{sqrt(3)/3}) node [above=1mm] {$\mathbf{d}_B$};
\draw [|<->|] (1.2,0) -- ++(0,{(1/sqrt(3)}) node [midway,fill=white] {$a$};
\end{tikzpicture}
}
\hfill
\subcaptionbox{\label{fig graphene BZ}}{
\begin{tikzpicture}[scale=2]
\fill [yellow,draw=black,densely dashed] (0:{1/sqrt(3)})
\foreach \ang in {60,120,...,300}{-- (\ang:{1/sqrt(3)})}
-- cycle;
\def\Rdot{2pt}
\foreach \n/\m in {0/0,1/0,0/1,1/1,-1/-1,-1/0,0/-1}{
\draw ({(\n-\m)*(sqrt(3)/2)},{(\n+\m)*1/2}) circle (\Rdot);
}
\draw [very thick,latex-latex] ({sqrt(3)/2},1/2) -- (0,0) node [pos=0.4,above] {$\mathbf{b}_1$} -- ({-sqrt(3)/2},1/2) node [pos=0.6,above] {$\mathbf{b}_2$};
\foreach \ang in {0,120,240}{
\fill (\ang:{1/sqrt(3)}) circle (1pt);
\fill [red] ({60+\ang}:{1/sqrt(3)}) circle (1pt);
}
\fill [cyan] (-30:{1/2}) circle (1pt) node [right] {$M$};

\node at (0:{1/sqrt(3)}) [right] {$K$};
\node at (-60:{1/sqrt(3)}) [below,red] {$K'$};
\node at (0,0) [below left] {$\Gamma$};
\end{tikzpicture}
}
\hfill
\subcaptionbox{\label{fig graphene BS curves}}{
\includegraphics[height=5.5cm]{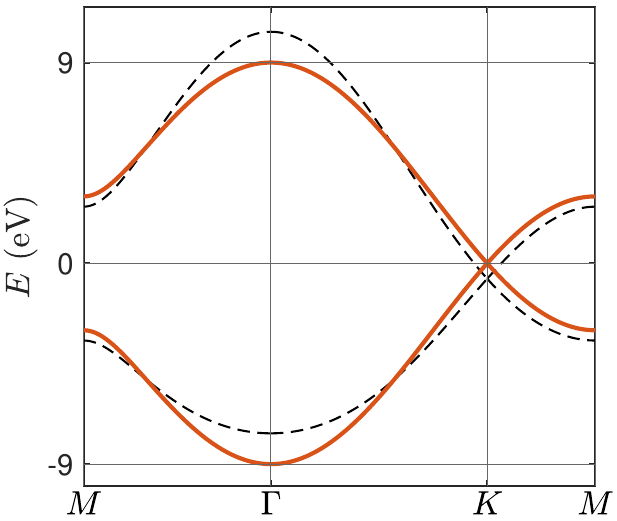}
}

\subcaptionbox{\label{fig graphene BS surface}}{
\includegraphics[height=5.5cm]{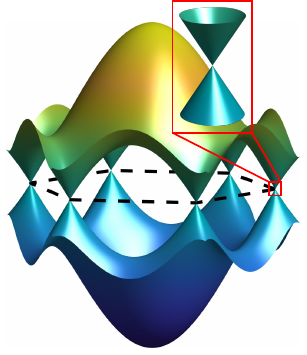}
}
\hfill
\subcaptionbox{\label{fig Dirac cone}}{
\includegraphics[scale=0.6]{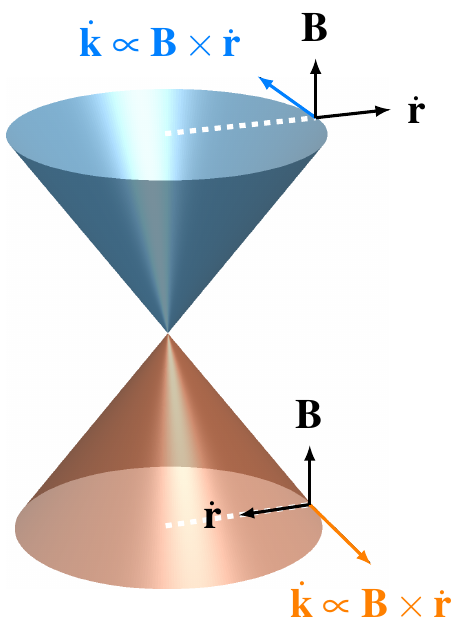}
}
\hfill
\subcaptionbox{\label{fig Dirac cones pn}}{
\includegraphics[scale=0.6]{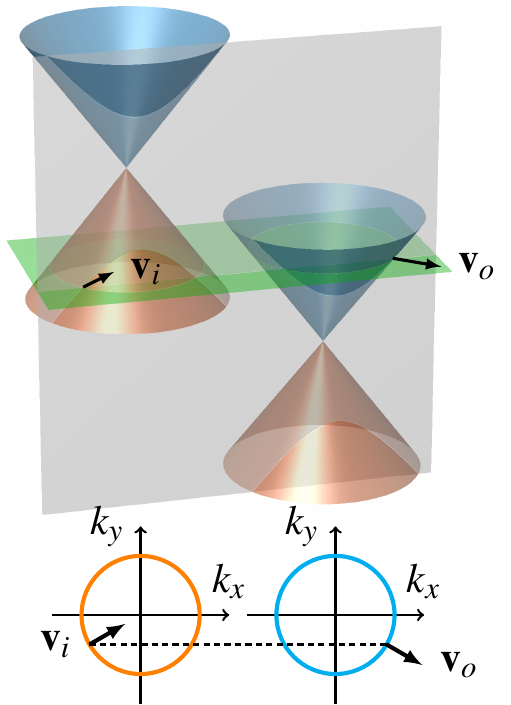}
}
\end{center}
\caption{(a) Lattice structure of graphene which is composed of two hexagonal sublattices (gray and pink balls). (b) The corresponding reciprocal lattice (open circles) and the first Brillouin zone (yellow hexagon). (c) Band structures along $k$-path of $M\Gamma KM$ with $(t,t')=(-3,0.23)$~eV (black dashed) and $(t,t')=(-3,0)$~eV (red solid). (d) Surface plot of the band structure for the case of $t'=0$. (e) The Dirac cone of graphene, i.e., its low energy band structure. (f) Schematics of shifted Dirac cones of a graphene p-n junction, assuming a Bloch electron incident from the p-region at velocity $\mathbf{v}_i$ and transmitted into the n-region at velocity $\mathbf{v}_o$.}
\end{figure*}
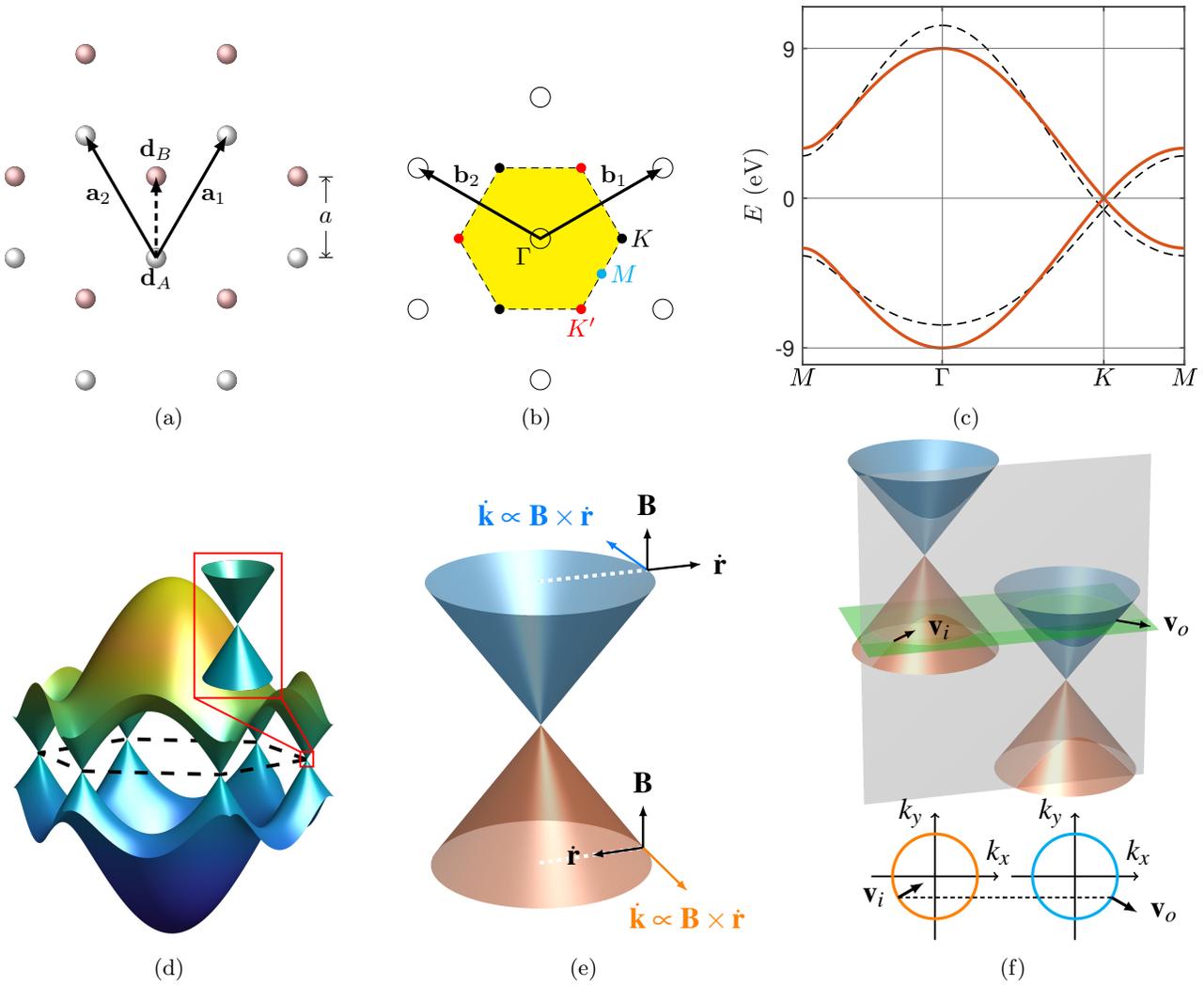

Using a two-basis tight-binding model considering only $p_z$-orbitals up to second nearest neighbors, the energy bands of graphene within the entire Brillouin zone can be written as
\begin{equation}
E(\mathbf{k}) = \epsilon_p+t'F(\mathbf{k}) \pm t\sqrt{3+F(\mathbf{k})}
\label{eq E(k) graphene full}
\end{equation}
where $\epsilon_p$ is the $p_z$-orbital energy (often set to zero), $t$ and $t'$ are the nearest and second-nearest neighbor hopping energy (equivalent to Slater-Koster parameter $V_{pp\pi}$ and $V_{pp\pi}'$), respectively, and the function $F(\mathbf{k})$ is defined by 
\begin{equation}
F(\mathbf{k}) = 2\left(\cos k_x a + 2\cos\frac{k_x a}{2}\cos\frac{\sqrt{3}k_y a}{2}\right)\ .
\label{eq F(k) for E(k) graphene}
\end{equation}
The graphene band structure based on \eq\ref{eq E(k) graphene full}--\eq\ref{eq F(k) for E(k) graphene} along the $k$-path $M\Gamma K M$ points (marked on the Brillouin zone shown in \fig\ref{fig graphene BZ}) is shown in \fig\ref{fig graphene BS curves}, considering $t'\neq 0$ and $t'=0$ cases. The surface plot of the band structure for the $t'=0$ case is shown in \fig\ref{fig graphene BS surface}, where the conic structure centered at $K$ can be clearly seen. By Taylor expansion of \eq\ref{eq F(k) for E(k) graphene} at $\mathbf{k}=K+\mathbf{q}$ with $|\mathbf{q}| a \ll 1$, it can be shown that up to terms quadratic in $q$, $F\approx-3+9q^2a^2/4$. Substituted into \eq\ref{eq E(k) graphene full}, we have $E(K_x+q_x,K_y+q_y) \approx \epsilon_p-3t'\pm (3a t/2)q$, which can be briefly written as
\begin{equation}
E_\sigma(k) \approx -3t'+\sigma\hbar v_F k\,
\label{eq E(k) graphene low E}
\end{equation}
where $k$ is relative to $K$, $\sigma=\pm$ is the band index ($\sigma = +$ for the electron branch and $\sigma=-$ for the hole branch), $\epsilon_p=0$ is chosen, and the Fermi velocity of graphene, $v_F$, is defined via
\begin{equation}
\hbar v_F \equiv \frac{3}{2}|t|a\,
\label{eq hbarvF}
\end{equation}
which is about $0.639$~eV~nm when using the commonly used approximate value of $t=-3$~eV. From this the value of $v_F$ is about one nanometer per femtosecond, or $v_F=10^6$~m/s, which is $1/300$ of speed of light. Note that $t'$ in \eq\ref{eq E(k) graphene low E} appears to be just a trivial band offset, but in the scope of strained graphene with nonuniform hopping, $t'$ may play an interesting role of pseudoscalar potential \cite{Guinea2010,Choi2010,Grassano2020,Wang2021}.\\

\subsubsection{Graphene Landau levels.}
\label{Sec:QHE}

\begin{figure}
\includegraphics[width=\columnwidth]{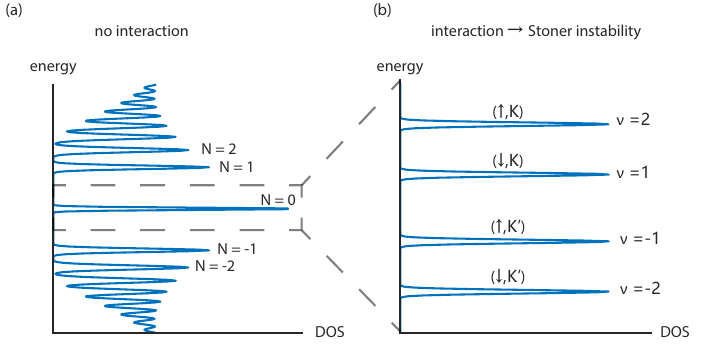}
\caption{(a) In the absence of interaction, each graphene Landau level (denoted by the index N) is four time degenerate and each level energy is given by $E_N = \hbar \omega_C \sqrt{N+1/2}$. (b) Electron interactions lift up the spin and valley degeneracy through a Stoner instability, leading to broken symmetry states Landau level classified by the index $\nu$. }
\label{fig graphene Landau level}
\end{figure}

 Under high perpendicular magnetic field $B$, the band structure of graphene leads to a peculiar quantum Hall effect \cite{Goerbig2011}, characterized by a electron-hole symmetric spectrum of four-fold (spin and valley) degenerate Landau levels with energies $E_N=\mathrm{sign}(N)\sqrt{2e\hbar v_\mathrm{F}^2 N B}$, with $-e$ the electron charge, $\hbar=h/2\pi$ the reduced Planck constant, $N$ the Landau level index (positive for electron, negative for holes), and $v_\mathrm{F}\approx 10^6$m$/$s is the Fermi velocity. For large magnetic fields, electron-electron interactions lift the spin and valley symmetries such that each Landau level splits into four sublevels that are fully spin and valley polarized \cite{Goerbig2011,Young2012}. The $0$-th Landau level, pinned at zero energy, also splits into four sub-levels that reflect its half electron, half hole nature. If the Zeeman energy can be neglected with respect to interaction-driven spin and valley gaps (which generally occurs unless a strong in-plane magnetic field is applied \cite{Young2014}, or a high-constant dielectric is used to screen interactions \cite{Veyrat2019}), the $0$-th Landau level splits into two electron-type sublevels with equal valley polarization (\textit{e.g.} $K$) and opposite spins polarizations, and two hole-type sublevels with equal valley polarization (but opposite to that of the electron-type sublevels, \textit{e.g.} $K'$), and opposite spin polarizations. Thus, at quarter filling, that is at filling factor $\nu=n_{\mathrm{e}/\mathrm{h}}h/eB=\pm1$ ($n_{\mathrm{e}/\mathrm{h}}$ is the carrier density with respect to charge neutrality), both bulk and edge become fully spin and valley polarized \cite{Goerbig2011,Young2012, Knothe2015}. At filling factor $\nu=0$, the spin and valley symmetry breakings lead to a fully insulating state with no edge channels, the gap of which is about 150 K at $B=10~$T, ten times larger than the Zeeman gap \cite{Young2012}. This insulating $\nu=0$ state is crucial in many electron quantum optics experiments, as it allows to locally deplete the electron gas using electrostatic gates, in a manner similar to experiments realized in AlGaAs/GaAs heterostructures. Thus, the experiments described in this section mostly rely on locally changing the filling factor using gates to control the trajectories of edge channels, as well as their coupling.

\subsubsection{Band structure of bilayer graphene.}
\label{sec:BandStuctureBLG}

\begin{figure}[htbp]
	\centering
	\includegraphics[width=\columnwidth]{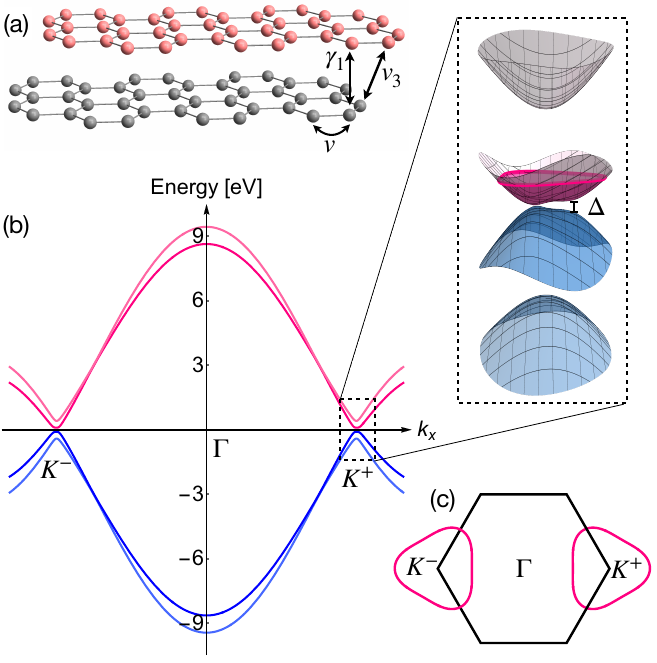}
	\caption{{\bf Lattice and electronic structure of bilayer graphene.} (a) Lattice structure of Bernal stacked bilayer graphene. We indicate nearest neighbour intra-layer hopping ($v$), vertical inter-layer coupling ($\gamma_1$), and skew inter-layer coupling ($\gamma_3$). (b) Cut through the bilayer graphene band structure tracing the corners $K^{\pm}$ and the centre $\Gamma$ of the Brillouin zone. There are four bands, two valence bands, and two conduction bands. Near the $K$ points, one conduction band and one valence band are split from zero by an energy of the order of $\gamma_1$, while the remaining two bands constitute the low-energy bands. These low-energy bands touch at the K-points for zero inter-layer asymmetry (with an approximately parabolic dispersion), while finite inter-layer asymmetry opens a gap $\Delta$. The inset shows the bands around the Brillouin zone corners as described by \eq\ref{eqn:BLG_disp}. (c) The skew hopping breaks rotational symmetry, leading to trigonally warped bands where the Fermi lines are of valley-dependent, triangular shape.}
	\label{fig:BLG}
\end{figure}

Similar to the tight-binding description of monolayer graphene, cf.~\sectionname\ref{sec BS of graphene}, one can obtain the band structure of Bernal stacked bilayer graphene \cite{Mccann2006,Mccann2007,Mccann2013}, taking into account intra-layer and inter-layer hoppings and inter-layer asymmetry, see \fig\ref{fig:BLG}.  One finds four bands, two valence bands, and two conduction bands, see \fig\ref{fig:BLG}b. In the low-energy expansion around the K-points (cf.~the discussion around \eq\ref{eq E(k) graphene low E}), bilayer graphene's  bands can be described by 
\newpage
\begin{strip}
\begin{equation}
 E^{2} = \frac{\gamma_1^2}{2}+\frac{\Delta^2}{4}+\left(v^2+\frac{v_3^2}{2}\right)k^2 + (-1)^{\alpha}\sqrt{\frac{(\gamma_1^2-v_3^2k^2)^2}{4}+v^2 k^2\left(\gamma_1^2+\Delta^2+v_3^2k^2\right)+2\xi\gamma_1v_3v^2k^3 \cos{3\varphi}}, \label{eqn:BLG_disp}
\end{equation}
\end{strip}
where $\alpha=1$ ($\alpha=2$) yields the low-energy (split) bands, $\xi=\pm1$ indexes the two valleys $K^{\pm}$, $\varphi=\arctan{[k_x/k_y]}$ is the polar angle of the momentum, and $\Delta$ is the interlayer asymmetry gap. Equation \ref{eqn:BLG_disp} captures vertical inter-layer coupling of the dimer sites ($\gamma_1$),  in-plane nearest-neighbour intra-layer hopping ($v$), and skew inter-layer coupling between non-dimer orbitals ($v_3$). 

The latter skew hopping parameter breaks rotational symmetry and induces trigonal warping to the parabolic bands, leading to triangularly deformed Fermi surfaces with opposing orientation in the  $K^{\pm}$ valleys, see \fig\ref{fig:BLG}c. 

\subsubsection{Density of states.}\label{sec dos}

Given the energy band of a certain material, the corresponding density of states $\rho(E)$, i.e., the density of the number of states at energy $E$, is generally given by the sum of contributions from all bands: $\rho(E)=\sum_\sigma \rho_n(E)$ where 
\begin{equation}
\rho_\sigma(E) = \frac{gA}{(2\pi)^2}\oint_{E_\sigma(\mathbf{k})=E}\frac{d\ell_k}{|\nabla_k E_\sigma(\mathbf{k})|} 
\label{eq dos 2d general}
\end{equation}
is the density of states of the $\sigma$th band. Here, $g$ is the degeneracy factor and $A$ is the area of the 2D material. The above closed contour integral generally needs to be done numerically, but can be greatly simplified when the energy band is isotropic: $E_\sigma(\mathbf{k}) = E_\sigma(k)$, such as graphene at low energy. For bilayer graphene, on the other hand, the low-energy bands given by \eq\ref{eqn:BLG_disp} are isotropic only when the skew interlayer hopping $v_3=0$.


For graphene, we need to sum up all contributions from the six Dirac cones within the first BZ, with each cone described by \eq\ref{eq E(k) graphene low E} and shared by three BZs, a factor of $6/3=2$, also called the valley degeneracy $g_v=2$, should be taken into account. Together with the spin degeneracy $g_s=2$, the total degeneracy factor in \eq\ref{eq dos 2d general} for low-energy graphene should be set to $g=g_s g_v=4$. Since there is no overlap of energy bands, the contribution to the density of states is from either the electron branch ($\sigma=+$) or the hole branch ($\sigma=-$). Using $|\nabla_k E_\sigma(\mathbf{k})|=\hbar v_F$ from \eq\ref{eq E(k) graphene low E}, the density of states per unit area, $D(E)=\rho(E)/A$, is given by
\begin{equation}
D(E) = \frac{2|E|}{\pi(\hbar v_F)^2}
\label{eq dos graphene}
\end{equation}
for the case of $t'=0$ with the Dirac point at $E=0$. For the case of $t'\neq 0$ with the Dirac point shifted to $-3t'$, as indicated by \eq\ref{eq E(k) graphene low E} and seen in \fig\ref{fig graphene BS curves}, the density of states \eq\ref{eq dos graphene} should be modified with $|E|\rightarrow|E+3t'|$. 
For BLG the quasi-quadratic dispersion relation leads to enhanced DOS close to CNP, leading to better screening of disorder and also to the enhancement of correlation effects.

\subsubsection{Energy carrier density relation.}\label{sec E vs n}

At zero temperature, the carrier density $n$ as a function of energy $E$ can be obtained via
\begin{equation}
n(E) = \int_{E_0}^E D(E')dE'
\label{eq n(E) general}
\end{equation}
where $E_0$ is the charge neutrality energy which is zero for graphene with $t'=0$. By inverting the relation, the Fermi energy as a function carrier density, $E(n)$, can be obtained. When energy bands are analytically available and are isotropic in $\mathbf{k}$, i.e., $E(\mathbf{k})=E(k)$, the result from the above described approach is equivalent to replacing $k$ in the energy dispersion with 
\begin{equation}
n=g\frac{k^2}{4\pi}\ .
\label{eq n(k) general}
\end{equation}
For graphene, we have $g=4$ and may replace $k$ in \eq\ref{eq E(k) graphene low E} with $\sqrt{\pi|n|}$ (for the case of $t'=0$) to obtain
\begin{equation}
E(n) = \mathrm{sgn}(n)\hbar v_F \sqrt{\pi|n|}\ ,
\label{eq E(n) graphene}
\end{equation}
where $n>0$ and $n<0$ correspond to n- and p-type graphene, respectively. 

\subsection{Semiclassical description for motion of carriers}
\label{semiclassic}

In solids, the semiclassical dynamics of Bloch electrons (without the correction from the Berry curvature \cite{Chang1995}) is governed by \cite{Ashcroft1976}
\begin{equation}
\mathbf{\dot{r}} = \frac{1}{\hbar }\nabla_k E_\sigma(\mathbf{k}) \label{eq r dot general}
\end{equation}
\begin{equation}
\mathbf{\dot{k}} = -\frac{e}{\hbar}(\mathbf{E}_e+\mathbf{\dot{r}\times B})
\label{eq k dot general}
\end{equation}
where $\nabla_k$ is the gradient operator with respect to $\mathbf{k}$, $E_\sigma(\mathbf{k})$ is the energy band, $\sigma$ is the band index, $\hbar$ is the reduced Planck constant, $-e$ is the electron charge, $\mathbf{E}_e$ is the electric field, and $\mathbf{B}$ is the magnetic field. For two-dimensional materials arranged in the $x$-$y$ plane, the position vector is $\mathbf{r}=(x,y)$, and the wave vector is also two-dimensional, $\mathbf{k}=(k_x,k_y)$. Given an energy band $E_\sigma(\mathbf{k})$, \eq\ref{eq r dot general} stands for two first-order ordinary differential equations (ODEs) for $\dot{x}$ and $\dot{y}$. Together with \eq\ref{eq k dot general} that describes another two first-order  ODEs, one for $\dot{k}_x$ and the other for $\dot{k}_y$, \eq\ref{eq r dot general} and \eq\ref{eq k dot general} represent four first-order coupled ODEs that can be numerically solved to describe the semiclassical dynamics of Bloch electrons in 2D. Without having to bother with such numerics, however, the following section provides a simple understanding of the basic properties of Dirac electrons in graphene based on \eq\ref{eq r dot general}--\eq\ref{eq k dot general}.

\subsubsection{Cyclotron motion.}
\label{Sec:cyclotron}

To describe the semiclassical motion of the Bloch electron in the present focus of graphene, it is sufficient to adopt the low-energy dispersion \eq\ref{eq E(k) graphene low E}, which exhibits a conic band structure, known as the Dirac cone, already shown in the inset of \fig\ref{fig graphene BS surface} and now elaborated in \fig\ref{fig Dirac cone}, where carriers occupying the upper branch ($\sigma=+$) behave like negatively charged electrons and those occupying the lower branch ($\sigma=-$) behave like positively charged holes. The behaviors of electron-like and hole-like carriers can be understood by considering Bloch electrons in graphene applied with only $\mathbf{B}=(0,0,B)=B\hat{\mathbf{e}}_z$ where $B$ is constant and $\hat{\mathbf{e}}_z$ is the unit vector along the $z$ axis. The absence of $\mathbf{E}_e$ simplifies \eq\ref{eq k dot general} to
\begin{equation} 
\mathbf{\dot{k}} =-\frac{e}{\hbar}\mathbf{\dot{r}}\times \mathbf{B}\ .
\label{k dot no E}
\end{equation}
By a bit of mathematical processing \cite{Ashcroft1976}, the above \eq\ref{k dot no E} leads to the following geometric relation:
\begin{equation}
\mathbf{r}(t) - \mathbf{r}(0) = \frac{\hbar}{eB}\left[\mathbf{k}(t)-\mathbf{k}(0)\right]\times \hat{\mathbf{e}}_z\ .
\label{geometric relation}
\end{equation}
which indicates that the real-space trajectory of a Bloch electron is just its reciprocal-space trajectory scaled by $\hbar/eB$ and rotated about the $z$-axis by 90 degrees clockwise. 

Since the magnetic force is perpendicular to the group velocity $\dot{\mathbf{r}}$ and does not alter the kinetic energy, the motion of a Bloch electron in a constant magnetic field is a constant-energy motion. Put in another way, its reciprocal-space trajectory is a constant-energy contour, i.e., the Fermi contour for Bloch electrons at the Fermi energy. At sufficiently low energies, the Fermi contour is a circle of radius $k$, with which \eq\ref{geometric relation} indicates that the corresponding real-space trajectory is a circle of radius
\begin{equation}
r_c = \frac{\hbar k}{eB}\ ,
\label{eq cyclotron radius}
\end{equation}
called cyclotron radius. Whereas the above discussion is valid independent of the band structure, the cyclotron radius formula \eq\ref{eq cyclotron radius} is applicable also for other 2D materials with isotropic dispersion relation. 

\subsubsection{Electron and hole orbits.} 

Equation \ref{k dot no E}, together with the group velocity $\dot{\mathbf{r}}$ given by \eq\ref{eq r dot general}, allows us to distinguish between electron and hole orbits. Since the group velocity $\dot{\mathbf{r}}$ is directed along the energy gradient, which is radially outward for the electron branch and inward for the hole branch (see \fig\ref{fig Dirac cone}), how $\dot{\mathbf{k}}$ evolves with time follows the direction of $\mathbf{B}\times\dot{\mathbf{r}}$, leading to counterclockwise and clockwise orbits for electrons and holes, respectively; see \fig\ref{fig Dirac cone}. 

Note that the above argument is valid not only for graphene. What makes graphene different from its linear energy dispersion \eq\ref{eq E(k) graphene low E} is that the group velocity \eq\ref{eq r dot general} explicitly reads
\begin{equation}
\mathbf{v} = \sigma v_F \frac{\mathbf{k}}{|\mathbf{k}|}\ ,
\label{eq v graphene}
\end{equation}
whose magnitude is always $v_F$, independent of energy. 

\subsubsection{Cyclotron frequency.}

The energy-independent magnitude of the group velocity \eq\ref{eq v graphene} leads to distinct behaviors of electrons in graphene compared to non-Dirac materials where the energy dispersion is not linear in $k$. Take the cyclotron motion for example. Because of the constant $|\mathbf{v}|=v_F$, the smaller the cyclotron radius, the shorter the time needed for the electron to complete a cycle, which is in sharp contrast with electrons in usual two-dimensional electron gas (2DEG) where $E(k) = \hbar^2 k^2/2m^\ast$, the magnitude of group velocity then clearly depends on energy: $|\mathbf{v}|=\hbar k/m^\ast=\sqrt{2E/m^\ast}$, leading to energy-independent cyclotron frequency $|\mathbf{v}|/r_c=eB/m^\ast$. Using the effective mass of GaAs, $m^\ast = 0.067m_e$, $m_e$ the bare electron mass, the cyclotron frequency in 2DEG confined in GaAs is about $2.63$~THz under $B=1$~T. On the other hand, the cyclotron frequency in graphene is energy-dependent: $|\mathbf{v}|/r_c=eBv_F^2/E$, which ranges between a few THz ($E>100$~meV) to about one hundred THz ($E\lesssim 10$~meV) under the same magnetic field of $B=1$~T.

\subsubsection{Negative refraction.}\label{sec negative refraction}

Although the problem of transmission across a graphene p-n junction will be elaborated in more details in \sectionname\ref{sec graphene interface}, the above semiclassical description allows us to easily understand the origin of negative refraction in graphene.

Suppose a graphene sheet is subject to an on-site energy band offset arranged in a way that the Dirac cone in the left (right) region is shifted upward (downward) in energy, forming a graphene p-n junction; see \fig\ref{fig Dirac cones pn}, where the horizontal green plane is the global Fermi energy, and the circles in the bottom part of the figure are the Fermi circles in the corresponding regions. Because the left region is p-type, the group velocity points radially inward as explained above, and a possible incoming state with positive $x$-component of the group velocity vector $\mathbf{v}_i$ is shown in \fig\ref{fig Dirac cones pn}. After transmission across the p-n junction with the $k_y$ component conserved, the Bloch electron occupies a state in the n-region at the Fermi energy with the resulting group velocity $\mathbf{v}_o$ which points radially outward of the Dirac cone (see \fig\ref{fig Dirac cones pn}), leading naturally to the negative refraction of the Bloch electron because $\mathbf{v}_i$ and $\mathbf{v}_o$ lie on opposite side of the incidence normal (dashed line in \fig\ref{fig Dirac cones pn}).

The above explanation considers the special case of a symmetric p-n junction where the Fermi energies in the p and n regions are equal in magnitude and opposite in signs. Whereas more general cases will be elaborated in \sectionname\ref{sec snell}, from the simple picture based on the semiclassical description here, it can be seen that the origin of the negative refraction for electrons traversing p-n junctions in graphene arises from the opposite energy gradient of the conductance and valence bands. 

\subsection{Graphene junctions}\label{sec graphene interface}

If one uses two gates, for example, one bottom and one top-gate, one can define regions of different doping and gate-controlled p-n devices. If the doping has the same sign on the two sides (n-n' or p-p') it is called a unipolar junction, if it changes sign it is a bipolar (p-n or n-p) junction. The sharpness of the change in potential profile can also vary: the carrier density can switch sharply or smoothly, depending on the geometrical parameters and on screening properties in general. If the gate electrodes are very close to the graphene layer, the potential step is more abrupt as compared to the case when thicker insulating barrier layers are employed. The sharpness of the potential affects how electrons traverse the p-n region. In the ``abrupt'' case, the p-n junction can be seen as a ``thin'' scattering region. 

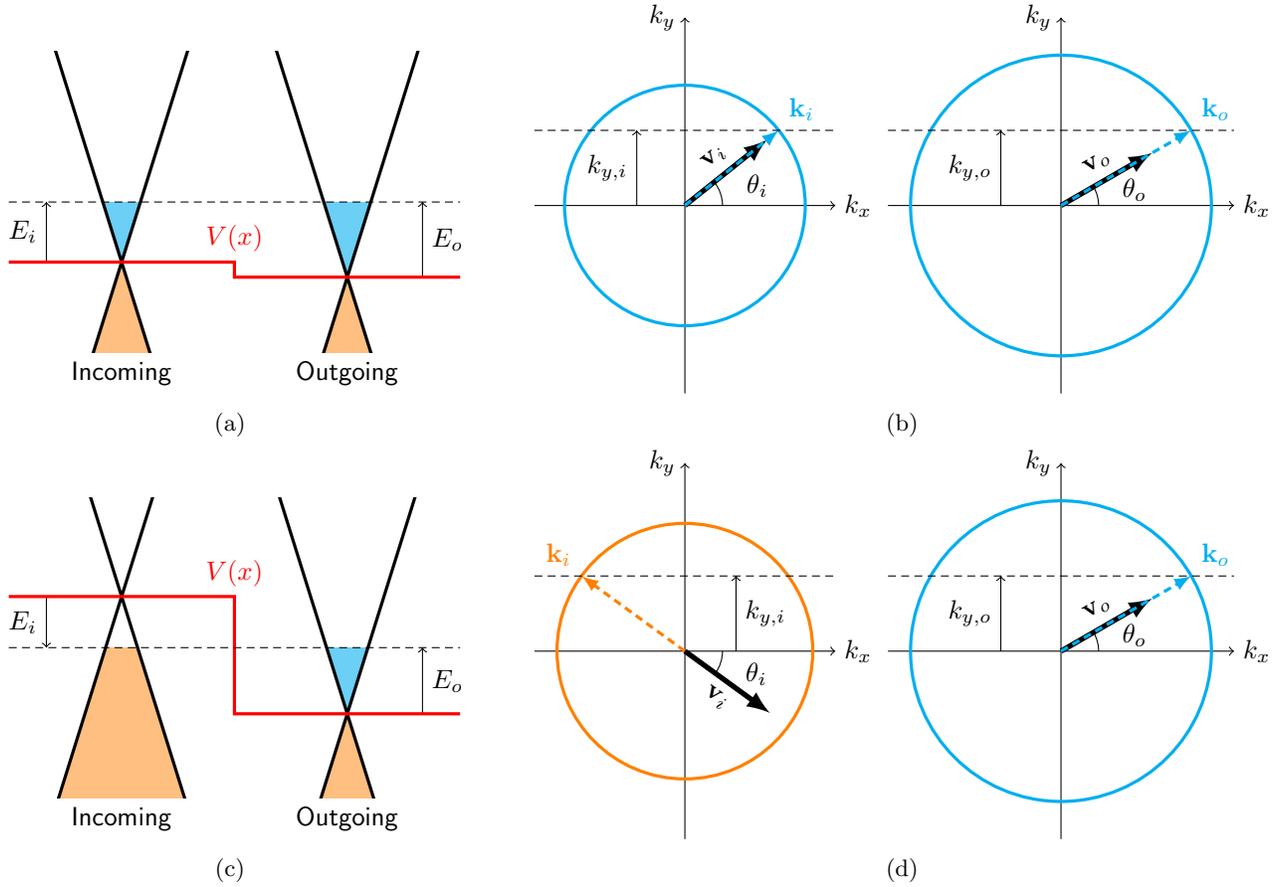
\begin{figure*}
\centering
\subcaptionbox{\label{fig band offset profile:nn}}{
\begin{tikzpicture}[yscale=0.4]
\def\Ei{-2cm}
\def\Eo{-2.5cm}
\def\hbarvf{8}
\begin{scope}
\clip (-3,-5) rectangle (3,5);
\begin{scope}[xshift=-1.5cm,yshift=\Ei]
\draw [->] (-1,0) -- ++(0,-\Ei) node [midway,left] {$E_i$};
\begin{scope}
\clip (-1,-\Ei) rectangle (1,{-6cm-\Ei});
\fill [cyan!50] (-1,{\hbarvf*1}) -- (0,0) -- (1,{\hbarvf*1});
\fill [orange!50] (-1,{\hbarvf*-1}) -- (0,0) -- (1,{\hbarvf*-1});
\end{scope}
\draw [very thick] (-1,{\hbarvf*-1}) -- (1,{\hbarvf*1}) -- (-1,{\hbarvf*1}) -- (1,{\hbarvf*-1});
\end{scope}
\begin{scope}[xshift=1.5cm,yshift=\Eo]
\draw [->] (1,0) -- ++(0,-\Eo) node [midway,right] {$E_o$};
\begin{scope}
\clip (-1,-\Eo) rectangle (1,{-6cm-\Eo});
\fill [cyan!50] (-1,{\hbarvf*1}) -- (0,0) -- (1,{\hbarvf*1});
\fill [orange!50] (-1,{\hbarvf*-1}) -- (0,0) -- (1,{\hbarvf*-1});
\end{scope}
\draw [very thick] (-1,{\hbarvf*-1}) -- (1,{\hbarvf*1}) -- (-1,{\hbarvf*1}) -- (1,{\hbarvf*-1});
\end{scope}
\end{scope}
\draw [very thick,red] (-3,\Ei) -- (0,\Ei) node [above] {$V(x)$} |- (3,\Eo);
\draw [densely dashed] (-3,0) -- (3,0);
\node at (-1.5,-5) [below] {\textsf{Incoming}};
\node at (1.5,-5) [below] {\textsf{Outgoing}};
\end{tikzpicture}
}
\hfill
\subcaptionbox{\label{fig Fermi circles for Snell:nn}}{
\begin{tikzpicture}
\def\ky{1cm}
\def\vf{1.4cm}
\begin{scope}
\def\kxmax{2cm}
\draw [->] (-\kxmax,0) -- (\kxmax,0) node [right] {$k_x$};
\draw [->] (0,-2.5) -- (0,2.5) node [left] {$k_y$};
\def\kf{1.6cm}
\draw [very thick,cyan] circle (\kf);
\draw [line width=2pt,-latex] (0,0) -- ({asin(\ky/\kf)}:\vf) node [midway,sloped,above] {$\mathbf{v}_i$};
\draw [very thick,-latex,densely dashed,cyan] (0,0) -- ({sqrt(\kf^2-\ky^2)},{\ky}) node [above right] {$\mathbf{k}_i$};
\draw (0:0.5) arc (0:{asin(\ky/\kf)}:0.5);
\node at ({asin(\ky/\kf)/2.5}:1) {$\theta_i$};
\draw [densely dashed] (-\kxmax,\ky) -- (\kxmax,\ky);
\draw [->] ({-\kf/2.5},0) -- ++(0,\ky) node [midway,left] {$k_{y,i}$};
\end{scope}
\begin{scope}[xshift=5cm]
\def\kxmax{2.3cm}
\draw [->] (-\kxmax,0) -- (\kxmax,0) node [right] {$k_x$};
\draw [->] (0,-2.5) -- (0,2.5) node [left] {$k_y$};
\def\kf{2cm}
\draw [very thick,cyan] circle (\kf);
\draw [line width=2pt,-latex] (0,0) -- ({asin(\ky/\kf)}:\vf) node [midway,sloped,above] {$\mathbf{v}_o$};
\draw [very thick,-latex,densely dashed,cyan] (0,0) -- ({sqrt(\kf^2-\ky^2)},{\ky}) node [above right] {$\mathbf{k}_o$};
\draw (0:0.5) arc (0:{asin(\ky/\kf)}:0.5);
\node at ({asin(\ky/\kf)/2.5}:1) {$\theta_o$};
\draw [densely dashed] (-\kxmax,\ky) -- (\kxmax,\ky);
\draw [->] ({-\kf/2.5},0) -- ++(0,\ky) node [midway,left] {$k_{y,o}$};
\end{scope}
\end{tikzpicture}
}

\subcaptionbox{\label{fig band offset profile:pn}}{
\begin{tikzpicture}[yscale=0.4]
\def\Ei{1.7cm}
\def\Eo{-2.2cm}
\def\hbarvf{8}
\begin{scope}
\clip (-3,-5) rectangle (3,5);
\begin{scope}[xshift=-1.5cm,yshift=\Ei]
\draw [->] (-1,0) -- ++(0,-\Ei) node [midway,left] {$E_i$};
\begin{scope}
\clip (-1,-\Ei) rectangle (1,{-6cm-\Ei});
\fill [orange!50] (-1,{\hbarvf*-1}) -- (1,{\hbarvf*1}) -- (-1,{\hbarvf*1}) -- (1,{\hbarvf*-1});
\end{scope}
\draw [very thick] (-1,{\hbarvf*-1}) -- (1,{\hbarvf*1}) -- (-1,{\hbarvf*1}) -- (1,{\hbarvf*-1});
\end{scope}
\begin{scope}[xshift=1.5cm,yshift=\Eo]
\draw [->] (1,0) -- ++(0,-\Eo) node [midway,right] {$E_o$};
\begin{scope}
\clip (-1,-\Eo) rectangle (1,{-6cm-\Eo});
\fill [cyan!50] (-1,{\hbarvf*1}) -- (0,0) -- (1,{\hbarvf*1});
\fill [orange!50] (-1,{\hbarvf*-1}) -- (0,0) -- (1,{\hbarvf*-1});
\end{scope}
\draw [very thick] (-1,{\hbarvf*-1}) -- (1,{\hbarvf*1}) -- (-1,{\hbarvf*1}) -- (1,{\hbarvf*-1});
\end{scope}
\end{scope}
\draw [very thick,red] (-3,\Ei) -- (0,\Ei) node [above] {$V(x)$} |- (3,\Eo);
\draw [densely dashed] (-3,0) -- (3,0);
\node at (-1.5,-5) [below] {\textsf{Incoming}};
\node at (1.5,-5) [below] {\textsf{Outgoing}};
\end{tikzpicture}
}
\hfill
\subcaptionbox{\label{fig Fermi circles for Snell:pn}}{
\begin{tikzpicture}
\def\ky{1cm}
\def\vf{1.4cm}
\begin{scope}
\def\kxmax{2cm}
\draw [->] (-\kxmax,0) -- (\kxmax,0) node [right] {$k_x$};
\draw [->] (0,-2.5) -- (0,2.5) node [left] {$k_y$};
\def\kf{1.7cm}
\draw [very thick,orange] circle (\kf);
\draw [very thick,-latex,densely dashed,orange] (0,0) -- ({-sqrt(\kf^2-\ky^2)},{\ky}) node [above left] {$\mathbf{k}_i$};
\draw [line width=2pt,-latex,] (0,0) -- ({asin(-\ky/\kf)}:\vf) node [midway,sloped,below] {$\mathbf{v}_i$};
\draw (0:0.5) arc (0:{asin(-\ky/\kf)}:0.5);
\node at ({asin(-\ky/\kf)/2}:1) {$\theta_i$};
\draw [densely dashed] (-\kxmax,\ky) -- (\kxmax,\ky);
\draw [->] ({\kf/2.5},0) -- ++(0,\ky) node [midway,right] {$k_{y,i}$};
\end{scope}
\begin{scope}[xshift=5cm]
\def\kxmax{2.3cm}
\draw [->] (-\kxmax,0) -- (\kxmax,0) node [right] {$k_x$};
\draw [->] (0,-2.5) -- (0,2.5) node [left] {$k_y$};
\def\kf{2cm}
\draw [very thick,cyan] circle (\kf);
\draw [line width=2pt,-latex] (0,0) -- ({asin(\ky/\kf)}:\vf) node [midway,sloped,above] {$\mathbf{v}_o$};
\draw [very thick,-latex,densely dashed,cyan] (0,0) -- ({sqrt(\kf^2-\ky^2)},{\ky}) node [above right] {$\mathbf{k}_o$};
\draw (0:0.5) arc (0:{asin(\ky/\kf)}:0.5);
\node at ({asin(\ky/\kf)/2.5}:1) {$\theta_o$};
\draw [densely dashed] (-\kxmax,\ky) -- (\kxmax,\ky);
\draw [->] ({-\kf/2.5},0) -- ++(0,\ky) node [midway,left] {$k_{y,o}$};
\end{scope}
\end{tikzpicture}
}
\caption{Band offset profiles $V(x)$ and local energy band diagrams of (a) a unipolar nn junction, whose corresponding Fermi circles are shown in (b), and (c) a bipolar p-n junction, whose corresponding Fermi circles are shown in (d).}
\label{fig:Snell}
\end{figure*}

\subsubsection{Snell's law.}\label{sec snell}

Consider an infinitely extending graphene sheet arranged in the $x$-$y$ plane, subject to a potential profile $V(x)$ causing an interface along the $y$ direction. Let the left (right) region be the incoming (outgoing) region labeled by $i$ ($o$). The local Fermi energy, i.e., the highest filled energy relative to the Dirac point, is $E_i$ in the incoming region and $E_o$ in the outgoing region. We consider an incoming state occupying a wave vector $\mathbf{k}_i$ corresponding to $E_i$ and a positive $x$-component of the group velocity $\mathbf{v}_i$. The outgoing wave vector $\mathbf{k}_o$ corresponding to energy $E_o$ also has positive $x$-component of the group velocity $\mathbf{v}_o$, as shown \fig\ref{fig:Snell}. 

Regardless of the sign of $E_i$ and $E_o$, Snell's law for electrons in graphene is simply the conservation of the $y$-component of the wave vector,
\begin{equation}
k_{y,i} = k_{y,o}\ ,
\label{eq kyi=kyo}
\end{equation}
due to the translational invariance along $y$. Without loss of generality, we fix $E_o>0$ and consider two cases of $E_i>0$ and $E_i<0$, the former corresponding to a unipolar nn junction and the latter a bipolar p-n junction.

\paragraph{Unipolar nn junction.}

When both of $E_i$ and $E_o$ are positive, the band offset profile and the local energy bands are schematically shown in \fig\ref{fig band offset profile:nn}. Since the group velocity is parallel to the wave vector for n-type graphene, as indicated by \eq\ref{eq v graphene}, both of $\mathbf{k}_i$ and $\mathbf{k}_o$ have positive components along $k_x$, as shown in \fig\ref{fig Fermi circles for Snell:nn}. Since the angle of incident (refraction), $\theta_i$ ($\theta_o$), defined as the angle between the incoming (outgoing) velocity and the normal of the interface (i.e., the $x$- or equivalently the $k_x$-axis), is the same as the azimuthal angle of $\mathbf{k}_i$ ($\mathbf{k}_o$), the conservation of $k_y$ \eq\ref{eq kyi=kyo} reads $k_i\sin\theta_i = k_o\sin\theta_o$, where $k_i=|\mathbf{k}_i|$ and $k_o=|\mathbf{k}_o|$. Multiplying the equation by $\hbar v_F$, we have $E_i \sin\theta_i = E_o\sin\theta_o$.

We note, that for unipolar junctions it is even possible to achieve total internal reflection if the incident angle is large enough.

\paragraph{Bipolar p-n junction.}

For a p-n junction with $E_i<0$ and $E_o>0$ (\fig\ref{fig band offset profile:pn}), we consider $\mathbf{k}_i$ with a negative $k_x$-component in order to have a positive $x$-component of the group velocity $\mathbf{v}_i$, as shown in \fig\ref{fig Fermi circles for Snell:pn}. Because of \eq\ref{eq v graphene}, $\mathbf{v}_i$ is now antiparallel to $\mathbf{k}_i$, and the angle of $\mathbf{k}_i$ is not just the angle of incidence $\theta_i$, but $\theta_i+\pi$. Therefore, the conservation of $k_y$ \eq\ref{eq kyi=kyo} reads $k_i\sin(\theta_i+\pi) = k_o\sin\theta_o$. Multiplying the equation by $\hbar v_F$, we have $-\hbar v_F k_i\sin\theta_i=\hbar v_F k_o\sin\theta_o$. Since $k_i=|\mathbf{k}_i|$ and $k_o=|\mathbf{k}_o|$ are both positive, we have $E_i=-\hbar v_F k_i$, and the resulting Snell's law
\begin{equation}
E_i\sin\theta_i = E_o\sin\theta_o
\label{eq Snell's law}
\end{equation}
remains the same form as that in the previous case of the nn junction. Equation \ref{eq Snell's law} is valid for both unipolar and bipolar junctions, with none of $\theta_i,\theta_o,E_i,E_o$ restricted to positive. In the case of bipolar junctions with $E_i E_o<0$, the angles of incidence and refraction must, in view of \eq\ref{eq Snell's law}, be of opposite signs: $\theta_i\theta_o<0$, consistent with the negative refraction described in \sectionname\ref{sec negative refraction}.

\subsubsection{Transmission across graphene junctions.}\label{sec transmission across graphene pn}

The Snell's law \eq\ref{eq Snell's law} discussed above indicates that the role of refraction index in ray optics is played by Fermi energy in electron optics of graphene. The equation gives a constraint for the angles of incidence and refraction, but says nothing about the quantum-mechanical transmission yet. 

In the literature, two main approaches are adopted in solving the problem of transmission across graphene junctions: analytical Dirac equation and numerical tight-binding model. The former leads to useful formulas but is restricted to simple potential profiles. The latter is not restricted to any potential profile but provides only numerical results without any neat formulas. At low-energy, the tight-binding approach agrees with the analytical formulas obtained from solving the Dirac equation, as was shown in \reference\cite{Liu2012} for the case of sharp p-n junctions \cite{Cheianov2006}, linearly smooth p-n junctions \cite{Cheianov2006}, and sharp n-p-n junctions \cite{Katsnelson2006} for graphene. The case of gapless bilayer graphene n-p-n junctions discussed in \reference\cite{Katsnelson2006} was also reproduced by the numerical tight-binding method \cite{Liu2012} but below we briefly review only the cases of single layer graphene junctions.

\paragraph{Sharp p-n junctions.}

\begin{figure*}
\def\wtemp{0.23}
\centering
\subcaptionbox{\label{fig T(phi) sharp}}{
\includegraphics[width=\wtemp\textwidth,page=1]{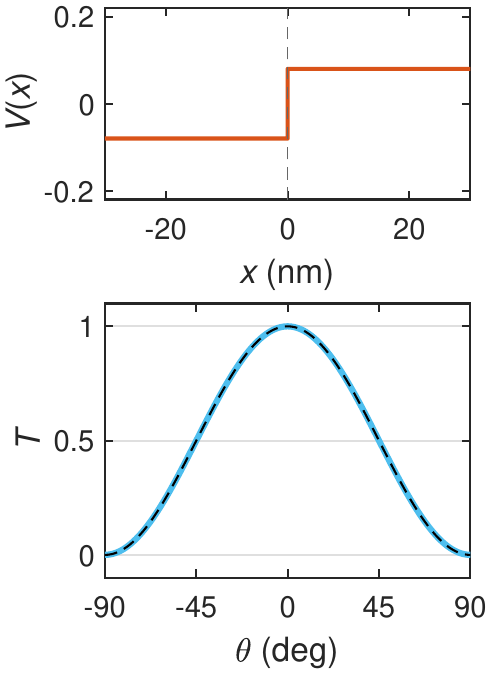}
}
\hfill
\subcaptionbox{\label{fig T(phi) smooth linear 1}}{
\includegraphics[width=\wtemp\textwidth,page=2]{fig/T_vs_phi_pn.pdf}
}
\hfill
\subcaptionbox{\label{fig T(phi) smooth linear 2}}{
\includegraphics[width=\wtemp\textwidth,page=3]{fig/T_vs_phi_pn.pdf}
}
\hfill
\subcaptionbox{\label{fig T(phi) smooth sine}}{
\includegraphics[width=\wtemp\textwidth,page=4]{fig/T_vs_phi_pn.pdf}
}
\caption{{\bf Angle-resolved transmission of symmetric graphene p-n junctions} considering (a) a sharp (abrupt) junction of potential difference 0.1~eV and (b) a linearly smooth junction of the same potential difference and smoothing thickness $d=20$~nm. Diagrams of (c) are similar to (b) with the only difference being the potential difference increased to 0.4~eV. (d) considers a nonlinear smooth p-n junction. Upper panels of each subfigure show the onsite-energy profile $V(x)$ and the lower panels show the angle-resolved transmission $T(\phi)$ based on the numerical tight-binding model (thick solid cyan curves) and analytical formulas (dashed black curves). (a)--(c) are modified and reproduced from \reference\cite{Liu2012}.}
\label{fig T(phi)}
\end{figure*}

Graphene subject to a potential profile $V(x)$ shown in \fig\ref{fig band offset profile:pn} is the case of an asymmetric sharp p-n junction. The upper panel of \fig\ref{fig T(phi) sharp} is a case of a symmetric p-n junction, across which the transmission as a function of the angle of incidence, denoted as $\theta$ for brevity, has a neat expression first derived in \reference\cite{Cheianov2006}:
\begin{equation}
T(\theta) = \cos^2\theta\ ,
\label{eq T(phi) sharp np}
\end{equation}
which is shown in the lower panel of \fig\ref{fig T(phi) sharp} by the black dashed curve. The result is independent of the potential height and can be reproduced by using the numerical tight-binding approach (thick solid cyan curve in \fig\ref{fig T(phi) sharp} \cite{Liu2012}.
The transmission at all angles is non-zero, but most strikingly, for $\theta=0$ transmission probability approaches one! This surprising effect is known as Klein tunneling~\cite{Shytov2008}.

\paragraph{Smooth p-n junctions.}\label{sec linear pn}

\begin{figure}[htbp]
	\centering
	\includegraphics[width=\columnwidth]{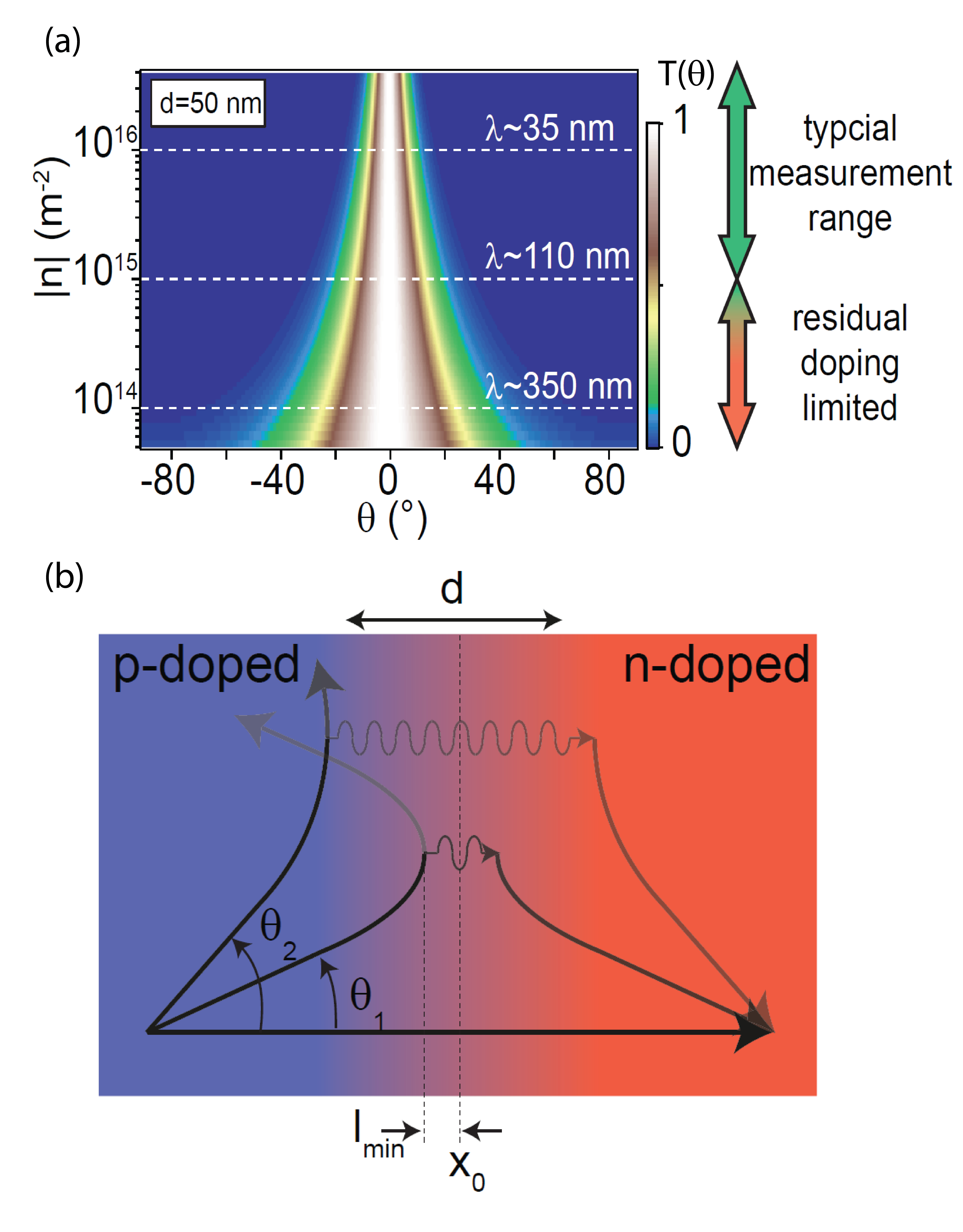}
	\caption{\textbf{Electron transmission at a p-n junction.}
       The graph in (a) shows the expected transmission probability color coded as a function of incident angle $\theta$ and carrier density $n$. Here, an equation for a smooth potential change has been used according to \reference\cite{Katsnelson2006}: $t(\theta)= \mathrm{exp}(-\pi k_F d sin(\theta)^2)$. The graph has been adapted from \reference\cite{HandschinThesis2017}. The semiclassical particle trajectories illustrate in (b) how they are repelled from the region of zero carrier density which is located at position $x_0$. The smaller the incident angle $\theta$, the closer the trajectories approach $x_0$. If the remaining classical distance $l_\textrm{min}$ gets shorter than the Fermi wavelength $\lambda_F$, the wave\-function will have an appreciable overlap on the other side leading to an increase in transmission probability. Quite remarkably, for $\theta=0$ transmission probability approaches unity. This is known as Klein tunneling.}
	\label{fig:smooth_PN}
\end{figure}

If the potential varies smoothly on the scale of the Fermi wavelength $\lambda_F$, electrons incident from the left with an angle $\theta > 0$ relative to the normal of the scattering region are smoothly bent within the scattering region. If the carrier density now truly crosses from n-type to p-type, the electrons are bent off the scattering region, never reaching the other side. This is schematically shown in \fig\ref{fig:smooth_PN}a. One would therefore predict a transmission probability of zero in case of a p-n junction with a smooth ``soft'' potential change.

As shown in \reference\cite{Cheianov2006}, the angle-resolved transmission $T(\theta)$ for a linearly smooth p-n junction such as those considered in \fig\ref{fig T(phi) smooth linear 1} or \fig\ref{fig T(phi) smooth linear 2}, can be described by
\begin{equation}
T(\theta) = \exp\left(-\pi\frac{k_F d}{2}\sin^2\theta\right)\ ,
\label{eq T(phi) linear np}
\end{equation}
when the product of the Fermi wave vector\footnote{Note that the Fermi wave vector is temporarily denoted as $k_F$ here in order to be consistent with the literature, but is mostly denoted simply as $k$ in the rest of our discussions.} $k_F$ and the smoothness $d$ fulfills $k_Fd\gg 1$. Two examples for such linearly smooth p-n junctions are shown in \fig\ref{fig T(phi) smooth linear 1} with $k_Fd\approx 1.54$ and \fig\ref{fig T(phi) smooth linear 2} with $k_Fd\approx 6.16$. 

The simulation also shows that the situation is again angle dependent. For incident electron trajectories that are nearly perpendicular to the scattering region with the line of zero density located at position $x_0$, the probability for transmission $T(\theta)$ gets appreciable.  The angle dependence shown in \fig\ref{fig:smooth_PN}a and \fig\ref{fig T(phi)}b-d can be understood in a qualitative manner as follows: appreciable transmission sets in when the classical electron trajectory approaches $x_0$ at a distance $l_{\rm min}$ with the condition that the Fermi wavelength at this position exceeds $l_{\rm min}$. In this case, there is appreciable wave\-function overlap between the two sides.

We note, that the analytical formula of \eq\ref{eq T(phi) linear np} does not match the numerical tight-binding results for the former case because $k_F d\gg 1$ is not fulfilled, while the agreement for the latter case can be seen. Figures \ref{fig T(phi) sharp}--\ref{fig T(phi) smooth linear 2} have been shown in \reference\cite{Liu2012}. For completeness, we show another p-n junction in \fig\ref{fig T(phi) smooth sine} which is the same as \fig\ref{fig T(phi) smooth linear 2} except that the junction profile is not linear but modeled by a sine function; see the upper panel of \fig\ref{fig T(phi) smooth sine}. The $T(\theta)$ is analytically difficult to solve but remains numerically straightforward. Nevertheless, because of $k_F d\gg 1$, the difference between the two approaches is not drastic, as seen in the lower panel of \fig\ref{fig T(phi) smooth sine}. We conclude that the exponential form of \eq\ref{eq T(phi) linear np} can always serve as a good approximation for smooth n-p and p-n junctions.

\paragraph{Klein tunneling.}

Whether sharp \eq\ref{eq T(phi) sharp np} or smooth \eq\ref{eq T(phi) linear np}, the transmission probability across a graphene p-n junction at normal incidence is always perfect: $T(0)=1$, which resembles the original Klein paradox in relativistic quantum electrodynamics \cite{Klein1929}, and is often referred to as the Klein tunneling \cite{Beenakker2008,CastroNeto2009,Allain2011}.

\paragraph{Interfaces at contact electrodes.}

In experiments mostly soft p-n interfaces are often reported. Suspended graphene devices have typically a soft p-n junction, while encapsulated devices can have sharper potential steps approaching the $10$~nm range, determined by the thickness of the gate insulator. However, there are inherent potential steps at the contacts themselves. This has two reasons: a) depending on the contact material, the work\-function difference between the metal and graphene results in an exchange of carriers, hence, to a region of contact doping; b), due to screening properties, the metal contacts can also be seen as an effective capacitor. Together with the gate capacitor(s), this changes the so-called gate-lever arms in the vicinity of the contact region. Again, one expects a change in carrier density in the vicinity of contacts. Since the metal electrode is in direct contact with the graphene sheet, this potential change can be rather steep. Consequently, as long as diffusive scattering can be disregarded, a normal metal contacts can also be seen as a (partially) reflecting mirror as will be seen e.g. in \fig\ref{fig:FigCS3}.

\subsection{Ballistic conductance}

For a two-dimensional material of width $W$ at energy $E$ corresponding to the magnitude of the wave vector $k$, the number of modes is given by 
\begin{equation}
M=\frac{k W}{\pi}\ ,
\label{eq M=kW/pi}
\end{equation}
which is obtained from $2k/(2\pi/W)$ and is just the number of $k_y$ points (assuming the transport direction is along $x$). This estimation arises from the assumption of applying the periodic boundary condition along the $y$ dimension, and becomes exceptionally precise whenever $W$ is sufficiently large, or in simpler terms it counts how many times the half wavelength of electrons' fits into the transport channel. Taking into account the degeneracy factor $g$, the ballistic conductance of the material in the absence of any potential, defect, and disorder, is given by
\begin{equation}
G=\frac{e^2}{h}gM\ ,
\label{eq G ballistic 2D material}
\end{equation}
according to the Landauer formula \cite{Datta1995}.

\subsubsection{Pristine graphene.}

\begin{figure*}
\def\figH{7cm}
\centering
\subcaptionbox{\label{fig G vs k analytical vs numerical}}{
\includegraphics[height=\figH]{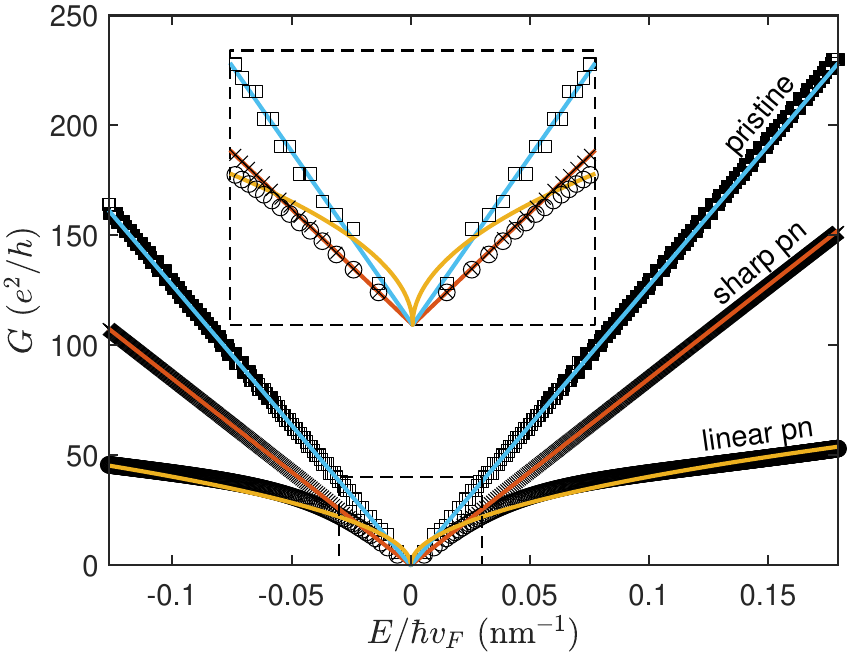}}
\subcaptionbox{\label{fig ballistic G experiment vs theory}}{
\includegraphics[height=\figH]{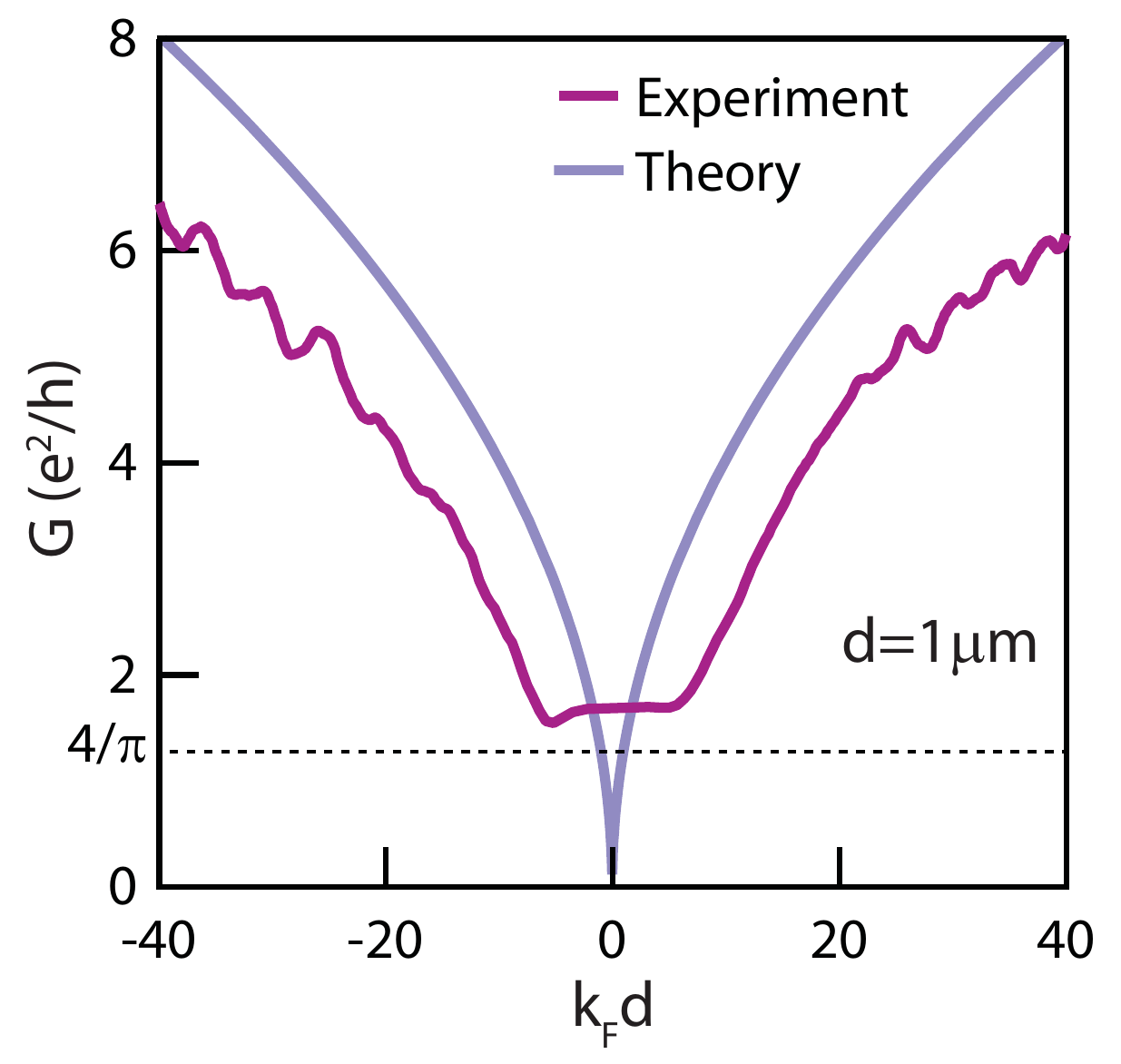}}
\caption{\textbf{Energy (and hence wave vector) dependence of graphene conductance in the clean limit.} (a) Ballistic conductance as a function of $E/\hbar v_F$ based on analytical formulas (colorful solid lines) and numerical quantum transport simulations (black markers), considering a pristine graphene sheet (cyan curve and black squares), a sharp graphene pn junction (red curve and black crosses), and a linearly smooth ($d=50$~nm) graphene pn junction (orange curve and black circles), all with width $W=1$~$\mu$m. The inset shows the range of $E/\hbar v_F\in [-0.03,0.03]$~nm$^{-1}$ and $G\in[0,40]$~$e^2/h$. (b) Ballistic conductance of a suspended graphene pn junction: experiment vs theory; taken from  \reference\cite{Rickhaus2015b}. The smoothness of the junction and the width of the graphene sample are assumed to be $d=1$~$\mu$m and $W=2$~$\mu$m, respectively. }
\end{figure*}

Using $g=4$ for graphene (\sectionname\ref{sec dos}) and substituting \eq\ref{eq M=kW/pi} into \eq\ref{eq G ballistic 2D material}, we have the ballistic conductance for pristine graphene:
\begin{equation}
G = \frac{e^2}{h}\frac{4W}{\pi}k\ ,
\label{eq G ballistic graphene}
\end{equation}
which is plotted in \fig\ref{fig G vs k analytical vs numerical} (solid cyan curve), considering an example of $W=1$~$\mu$m. Note that from the above \eq\ref{eq M=kW/pi} and \eq\ref{eq G ballistic 2D material}, the ballistic conductance of all 2D materials exhibit such a linear-in-$k$ dependence, or square-root-in-$n$ dependence when using $k=\sqrt{\pi|n|}$ (\sectionname\ref{sec E vs n}), up to a different degeneracy factor $g$. 

\subsubsection{Ballistic graphene p-n junctions.}

When the pristine graphene is subject to a potential $V(x)$ without breaking the translation symmetry along $y$, the contribution of each $k_y$ mode to the conductance is generally limited by the transmission probability $0\leq T(k_y)\leq 1$. By summing over contributions from all modes, the ballistic conductance is given by
\begin{equation}
G = \frac{e^2}{h}\frac{2W}{\pi}\int_{-k}^{k}T(k_y)dk_y\ ,
\label{eq G ballistic graphene junction general}
\end{equation}
which recovers \eq\ref{eq G ballistic graphene} when $V(x)=0$ such that $T(k_y)=1$ for all $-k\leq k_y\leq k$. For arbitrary $V(x)$, the transmission $T(k_y)$ can be numerically computed, but no neat formulas for the resulting conductance \eq\ref{eq G ballistic graphene junction general} can be obtained in general, except the two special cases reviewed in \sectionname\ref{sec transmission across graphene pn}.

For symmetric sharp p-n and n-p junctions, using \eq\ref{eq T(phi) sharp np} we have $\int_{-k}^k T(k_y)dk_y=4k/3$, and the conductance is precisely given by
\begin{equation}
G =\frac{e^2}{h}\frac{8W}{3\pi}k\ ,
\label{eq G ballistic graphene sharp pn}
\end{equation}
which differs from \eq\ref{eq G ballistic graphene} only by a factor of $2/3$, indicating that an ideally sharp graphene p-n junction is rather transparent (highly conductive); see the red solid curve in \fig\ref{fig G vs k analytical vs numerical}.

For linearly smooth p-n and n-p junctions, the $\int_{-k}^k T(k_y)dk_y$ integral is not analytically solvable even using \eq\ref{eq T(phi) linear np} when $kd\gg 1$ is fulfilled. However, as we have seen in \fig\ref{fig T(phi) smooth linear 2}, $T(\theta)$ decays to zero well before $\theta=\pm\pi/2$, which means that approximating the integral as $\int_{-\infty}^\infty T(k_y)dk_y=\int_{-\infty}^\infty\exp\left(-\pi d k_y^2/2k\right)dk_y=\sqrt{2k/d}$ is quite fine. Substituted into \eq\ref{eq G ballistic graphene junction general}, the final result is \cite{Cheianov2006}
\begin{equation}
G \approx \frac{e^2}{h}\frac{2W}{\pi}\sqrt{\frac{2k}{d}}\ ,
\label{eq G ballistic graphene smooth pn}
\end{equation}
which has a square-root-in-$k$ dependence, contrary to \eq\ref{eq G ballistic graphene} and \eq\ref{eq G ballistic graphene sharp pn}, as shown by the solid orange curve in \fig\ref{fig G vs k analytical vs numerical}.

Note that the horizontal axis of \fig\ref{fig G vs k analytical vs numerical} summarizing the $k$ dependence of the ballistic conductance for a pristine graphene sheet, \eq\ref{eq G ballistic graphene}, a sharp graphene p-n junction, \eq\ref{eq G ballistic graphene sharp pn}, and a linearly smooth graphene p-n junction, \eq\ref{eq G ballistic graphene smooth pn}, is $E/\hbar v_F$ instead of $k$, because $k$ is defined positive in our discussions while $E$ can be negative. All of the three cases well agree with the numerical results (black markers) based on quantum transport simulations, to be briefly reviewed in the following \sectionname\ref{sec quantum transport simulations}. 

It is also nice to illustrate how close experimental graphene p-n junctions come to theoretical prediction for a clean device with soft potential. \Fref{fig ballistic G experiment vs theory})  shows a comparison of an actual measurement with the theoretical prediction. The measured dependence does indeed follow a square-root dependence $G\propto \sqrt{k_F}$, proofing that the potential barrier is smooth and varies in this example on a length scale of $d=1$~$\mu$m. The comparison also reveals that there is an additional contact resistance of order $2e^2/h$ in series to the junction resistance. Finally, the conductance $G$ does not approach zero at the Dirac point, as there should be a cut-off which theory predicts to be $G_\textrm{min}=4e^2/\pi h$ ~\cite{Katsnelson2006}. That the experimentally measured minimal conductance is close to the ideal ballistic limit illustrates the cleanliness of this device~\cite{RickhausThesis2015}.

\begin{figure*}
\subcaptionbox{\label{fig simulating graphene devices}}{
\includegraphics[width=0.95\textwidth]{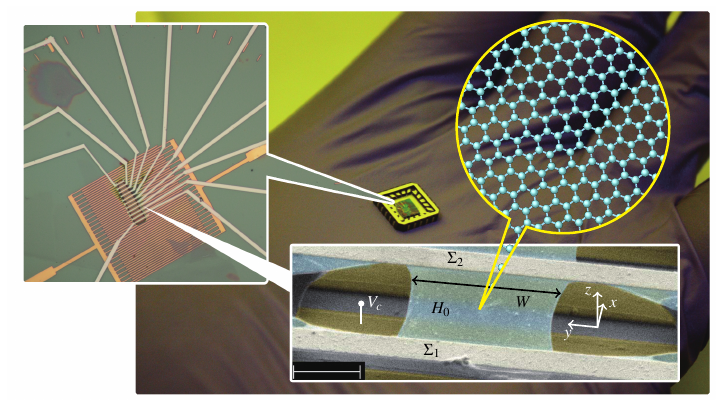}
}

\subcaptionbox{\label{fig schematic suspended device}}{
\includegraphics[scale=1.4]{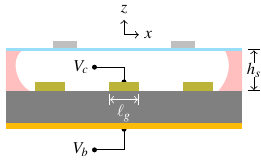}
}
\subcaptionbox{\label{fig suspended device nx}}{
\includegraphics[height=4.2cm]{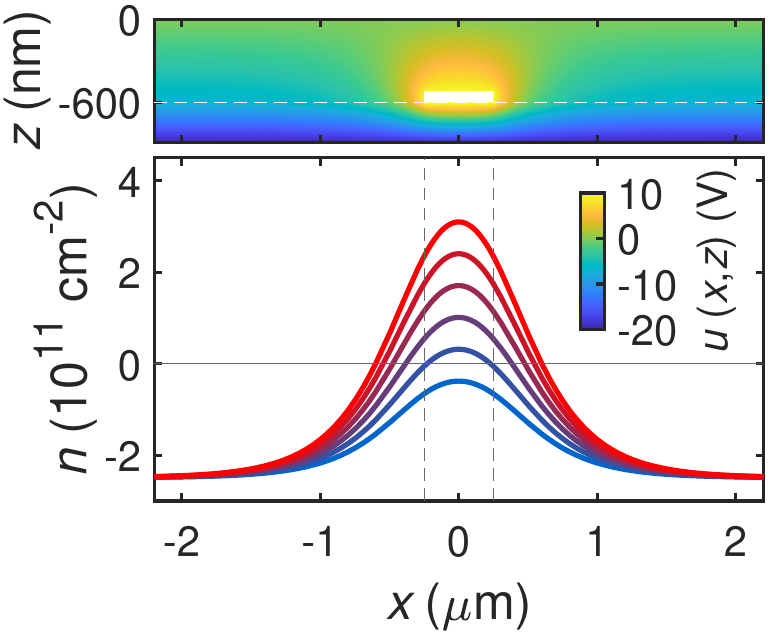}
}
\subcaptionbox{\label{fig suspended device Vx}}{
\includegraphics[height=4.2cm]{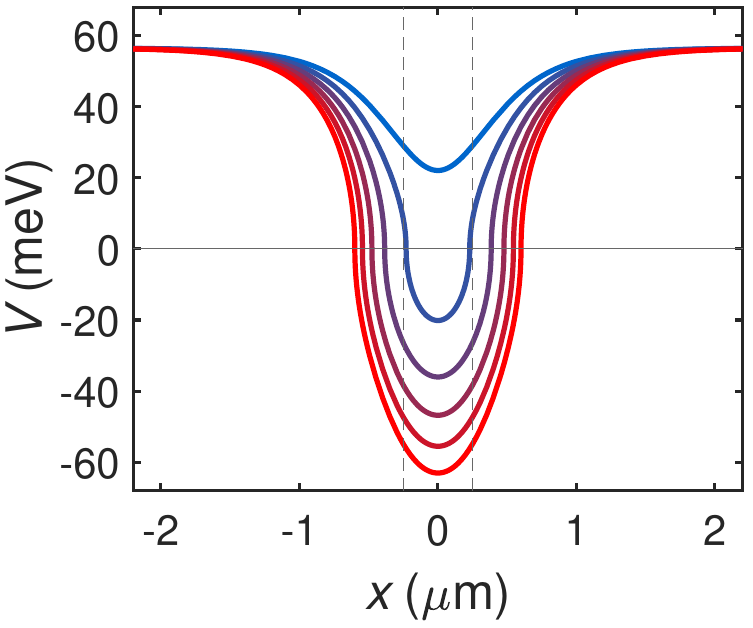}
}

\caption{(a) Optical images (background and left) of exemplary suspended graphene devices. Lower right: An SEM image of a two-terminal device made of a suspended graphene sample (described by a tight-binding Hamiltonian $H_0$) attached to two contacts (described by $\Sigma_1$ and $\Sigma_2$) and tuned by a central gate at voltage $V_c$ (marked on the figure) and a global backgate at voltage $V_b$ which is not shown in (a) but sketched in (b). (b) Schematics of the side view of the device shown in the SEM image of (a). Considering $(\ell_g,h_s) = (500,600)$~nm and $V_b=-20$~V, (c) shows carrier density profiles $n(x)$ of graphene at $V_c=0$~V (blue)$,2$~V$,\cdots,10$~V (red), and their corresponding on-site energy profiles $V(x)$ are shown in (d). The upper part of (c) shows an example of the electrostatic potential $u(x,z)$ at $(V_c,V_b)=(10,-20)$~V.}
\end{figure*}

\subsection{Quantum transport simulation for clean graphene}\label{sec quantum transport simulations}

Quantum transport in the framework of Landauer-B\"uttiker formalism \cite{Datta1995} is an exceptionally useful and powerful tool, especially for low-bias, low-temperature transport in the clean limit. To focus on clean graphene devices, let us summarize the formalism by considering an exemplary two-terminal suspended graphene device. 

\subsubsection{Real-space Green's function method.}\label{sec real-space green's function method}

As shown in \fig\ref{fig simulating graphene devices}, the system we are interested in is composed of a scattering region described by a clean Hamiltonian $H_0$ and the attaching electrical contacts described by $\Sigma_1$ and $\Sigma_2$, i.e., the so-called lead self-energies. To model and simulate electronic devices in real space, the local orbitals of the atoms composing the lattice may be chosen as the basis to represent $H_0$. Considering only the $p_z$ orbital of the carbon atoms, the real-space tight-binding Hamiltonian can be written as
\begin{equation}
H_0 = t\sum_{\langle i,j\rangle}c_i^\dag c_j + t'\sum_{\langle\!\langle i,j\rangle\!\rangle}c_i^\dag c_j\ ,
\label{eq real space tbm:H0}
\end{equation}
where $c_i^\dag$ ($c_i$) creates (annihilates) an electron on site $i$ located at position $\mathbf{r}_i$, so that $c_i^\dag c_j$ stands for an electron hopping from $\mathbf{r}_j$ to $\mathbf{r}_i$, and $\sum_{\langle i,j\rangle}$ ($\sum_{\langle\!\langle i,j\rangle\!\rangle}$) means that the sum runs over all nearest (second nearest) neighboring site pairs fulfilling $|\mathbf{r}_i-\mathbf{r}_j|=a$ ($|\mathbf{r}_i-\mathbf{r}_j|=\sqrt{3}a$); see \fig\ref{fig graphene lattice} for the definition of $a$. Despite that \eq\ref{eq real space tbm:H0} describes a finite-sized graphene without translational invariance, the hopping parameters $t$ and $t'$ are often assumed to be the same as those for the energy bands introduced in \sectionname\ref{sec BS of graphene} where translational invariance is the basic requirement. 

Apart from the clean part of the graphene Hamiltonian, $H_0$, the on-site energy term appearing as a diagonal matrix,
\begin{equation}
U = \sum_{i}V(\mathbf{r}_i)c_i^\dag c_i\ ,
\label{eq real space tbm:U}
\end{equation}
takes into account electrons' potential energy from all different sources, such as electrical gating, chemical doping, contact doping, disorder, atomic orbital energy, and so on. In the clean limit, \eq\ref{eq real space tbm:U} mainly describes the electrical gating, which will be explained in the following \sectionname\ref{sec realistic onsite}.

Together with the lead self-energies $\Sigma_1$ and $\Sigma_2$ describing the electrical contacts which serve as electron reservoirs, the effective Hamiltonian describing the contact-graphene-contact system can be written as
\begin{equation}
H(E) = H_0+U+\Sigma_1(E)+\Sigma_2(E)\ ,
\label{eq real space tbm:H}
\end{equation}
which is a function of energy $E$ because the lead self-energies depend on $E$. Commonly adopted methods for calculating the lead self-energies include eigenfunction expansion \cite{Datta1995}, eigendecomposition and Schur decomposition \cite{Wimmer2008}, and recursive Green's function \cite{Lewenkopf2013}, but are beyond the scope of this review. 

Once \eq\ref{eq real space tbm:H} is built, the retarded Green's function at energy $E$ is by definition given by
\begin{equation}
G_R (E) = \left[E-H(E)\right]^{-1}\ .
\label{eq retarded Green's function}
\end{equation}
Since all terms in \eq\ref{eq real space tbm:H} are $N_s\times N_s$ square matrices, $N_s$ being the total number of lattice sites, the above \eq\ref{eq retarded Green's function} stands for a matrix inversion, which is computationally heavy unless $N_s$ is small. For graphene, it is possible to rescale the lattice to reduce $N_s$, to be explained in the following \sectionname\ref{sec scalable tbm}. Note, however, that even $N_s$ is not too large, inverting the entire matrix $\left[E-H(E)\right]$ is not necessary, because not all elements of $G_R(E)$ are needed. Technical details are crucial at this point but are also beyond the scope of this review.

With the retarded Green's function obtained, together with the broadening matrices
\begin{equation}
\Gamma_p(E) = -2\mathrm{Im}\Sigma_p(E)
\label{eq broadening matrix}
\end{equation}
corresponding to the $p$th lead self-energy, the transmission function from lead $q$ to lead $p$ at energy $E$ can be obtained:
\begin{equation}
T_{p\leftarrow q}(E) = \mathrm{Tr}\left[\Gamma_p(E)G_R(E)\Gamma_q(E)G_A(E)\right]
\label{eq T(E) transmission function}
\end{equation}
where $G_A(E)=G_R^\dag(E)$ is the advanced Green's function matrix.

\subsubsection{Realistic on-site energy.}\label{sec realistic onsite}

When tuning the carrier density of graphene by electrical gating, what does the gate do? To a simple picture, when a positive gate voltage is applied, negative charges are induced on the surface of graphene, causing the raise of Fermi level, and vice versa. The change of the Fermi level can also be understood as the change of the entire energy bands with the Fermi level fixed. This picture is more useful when the carrier density is not uniform, corresponding to a spatially varying energy band offset, which is exactly the $V(\mathbf{r})$ in \eq\ref{eq real space tbm:U}. 

Without taking into account the correction due to quantum capacitance \cite{Luryi1988,Fang2007,Droscher2010,Liu2013}, the carrier density profile of graphene is given by the following linear superposition,
\begin{equation}
n(\mathbf{r}) = \sum_j \frac{C_j(\mathbf{r})}{e}V_j\ ,
\label{eq n(r) from C(r)}
\end{equation}
where $C_j(\mathbf{r})$ is the capacitance profile of the $j$th gate which can be either analytically described by a proper model function or numerically obtained by solving the Poisson equation \cite{Liu2013} using commercial software such as \textsc{Comsol} \cite{comsol} or finite-element-based partial differential equation (PDE) solvers such as \textsc{FEniCS} in \textsc{python} \cite{FEniCS} or \textsc{pdeModeler} (\textsc{pdetool} in older versions) of \textsc{Matlab} \cite{pde}.

To continue with the example of suspended graphene, consider the scanning electron microscopy (SEM) image of \fig\ref{fig simulating graphene devices} (lower right), whose side view is schematically shown in \fig\ref{fig schematic suspended device}. The carrier density of the graphene sample placed at $z=0$ is tuned by a central gate at voltage $V_c$ and a back gate at voltage $V_b$. Since the width of the graphene sample $W$ is sufficiently large and the geometry of the metal contacts and gates does not depend on $y$, we may consider the two-dimensional Laplace equation $\nabla^2 u = 0$ to solve for the electrostatic potential $u(x,z)$, subject to a properly assigned boundary conditions. The suspension height $h_s$ is typically several hundreds of nanometers and is roughly the smoothness of gate capacitance profiles. This means that the resulting carrier density and onsite energy profiles are completely smoothed (exhibiting no plateaus) whenever the gate length $\ell_g \lesssim h_s$, which is the case in the example shown here. 

An exemplary solution of $u(x,z)$ subject to $(V_c,V_b)=(10,-20)$~V is shown in the top panel of \fig\ref{fig suspended device nx}, considering $(\ell_g,h_s)=(500,600)$~nm. From the surface gradient of $u$ at $z=0^-$, the corresponding surface charge density (and hence the carrier density in graphene) can be obtained \cite{Liu2013}. Because the Laplace equation is linear, it is more convenient to first obtain the central gate capacitance, $C_c(x)$, by considering $(V_c,V_b) = (1,0)$~V, and back gate capacitance, $C_b(x)$, by considering $(V_c,V_b) = (0,1)$~V. For arbitrary gate voltages, the calculated $C_c(x)$ and $C_b(x)$ allow us to obtain the carrier density profile, $n(x)$, using \eq\ref{eq n(r) from C(r)}. When the correction due to quantum capacitance is considered, \eq\ref{eq n(r) from C(r)} for the carrier density in graphene needs to be modified, but the procedure remains the same \cite{Liu2013}. 

The bottom panel of \fig\ref{fig suspended device nx} shows carrier density profiles, without the quantum correction, considering various $V_c$ and fixed $V_b$ (values specified in the caption of \fig\ref{fig suspended device nx}). Since the length scale of the variation of $n(x)$ is supposed to be much larger than the atomic scale, it is legitimate to assume that the energy-carrier relation \eq\ref{eq E(n) graphene} is locally fulfilled. Therefore, considering the local energy band offset defined as
\begin{eqnarray}
V(x) = -E(n(x)) = -\mathrm{sgn}(n(x))\hbar v_F \sqrt{\pi|n(x)|}\ ,
\label{eq V(x) from n(x)}
\end{eqnarray}
which is the onsite energy term in \eq\ref{eq real space tbm:U}, the global Fermi energy for the entire graphene sample is expected to be fixed at zero. In short, using \eq\ref{eq V(x) from n(x)} in the above introduced quantum transport, the transmission \eq\ref{eq T(E) transmission function} should be evaluated at $E=0$.

\subsubsection{Scalable tight-binding model. }\label{sec scalable tbm}

From \sectionname\ref{sec real-space green's function method}, we have seen that the matrix size of \eq\ref{eq real space tbm:H0} mainly depends on the number of atoms (or the lattice sites) composing the lattice under consideration. In the present spinless case of \eq\ref{eq real space tbm:H0} with only one $p_z$ orbital per atom considered, the $H_0$ thus represented is an $N_s\times N_s$ square matrix. Using the unit cell area $|\mathbf{a}_1\times\mathbf{a}_2|=\sqrt{3}a_c^2/2$, it can be shown that the number of carbon atoms is about 38 millions per $\mu$m$^2$, which is the typical order of magnitude of the sample area used in transport experiments. Dealing with $N_s\times N_s$ matrices with $N_s\sim 10^7$ is not a simple task. When the spin degree of freedom or other orbitals are taken into account, the size of $H_0$ may be further doubled, tripled, or even more.

For graphene, luckily the lattice spacing $a$ and the nearest-neighboring hopping $t$ appears in the low energy dispersion \eq\ref{eq E(k) graphene low E} simply as a product; see \eq\ref{eq hbarvF}. Therefore, by considering a honeycomb lattice of lattice spacing $\tilde{a}=s_f a$ and nearest neighbor hopping energy $\tilde{t}=t/s_f$, its low-energy dispersion proportional to $\tilde{t}\tilde{a}=ta$ is guaranteed to be the same as that of real graphene. This scaling approach first introduced for spinless monolayer graphene in \reference\cite{Liu2015} led to the possibility of simulating micron-sized graphene samples \cite{Rickhaus2015,Rickhaus2015b,Rickhaus2015c,Rickhaus2015d,Terres2016,Xiang2016,Liu2017,Bours2017,Kolasinski2017,Makk2018,Ma2018,Brun2019,Lane2019,Kraft2020,Kang2020,Moreau2020,Moreau2021,Schrepfer2021,MrencaKolasinska2022,Brun2022,Chiu2022} and therefore made quantum transport simulations for ballistic graphene a very powerful tool. Moreover, the approach is compatible with spin-orbit coupling \cite{Zhumagulov2022} as applied in a recent work on spin-dependent transport in graphene on WSe$_2$ \cite{Rao2023}, and can also be applied to bilayer graphene as was remarked in \reference\cite{Liu2015} and applied in \reference\cite{Du2018,Chiu2022}.

\subsubsection{Periodic boundary hopping.}

Previously, in \sectionname\ref{sec transmission across graphene pn} we have mentioned that the angle-resolved transmission can be numerically computed by the tight-binding approach. This is exactly what we have described in the above \sectionname\ref{sec real-space green's function method}, except that \eq\ref{eq real space tbm:H0} for $H_0$ needs to be modified, in a way to allow periodic boundary hopping between upper and lower edge sites. Such hopping terms arise from the Bloch theorem because of the assumed translational invariance along the lateral dimension. This means, the periodic boundary hopping terms contain $k_y$, such that $H_0=H_0(k_y)$. See \reference\cite{Liu2012,Wimmer2008,Liu2012a} for details. Because of the $k_y$ dependence of the $H_0$, the lead self-energies, the effective Hamiltonian \eq\ref{eq real space tbm:H}, retarded Green's function \eq\ref{eq retarded Green's function}, the broadening matrices \eq\ref{eq broadening matrix}, and the transmission function \eq\ref{eq T(E) transmission function} all contain the $k_y$ dependence. In terms of the incidence angle, $\theta = \arcsin (k_y/k)$, the numerically computed $T(k_y)$ can be plotted as $T(\theta)$ as shown in \fig\ref{fig T(phi)}.

In addition to the angle-resolved transmission, the method of periodic boundary hopping is also very useful for quasi-one-dimensional two-terminal graphene device, such as the example of \fig\ref{fig simulating graphene devices} where $W$ is sufficiently large. In such cases, the onsite energy $V(x)$ is one-dimensional, and $T(k_y)$ can be numerically computed. Integrating over $k_y$ and using \eq\ref{eq G ballistic graphene junction general}, the conductance thus obtained is consistent with the one computed using a finite-width ribbon, but the computation is much lighter and faster. In the recent work about gate-controlled one-dimensional superlattice in graphene, the two kinds of computations were explicitly compared \cite{Kang2020}, and the results are hardly distinguishable.

In the literature, the first work applying this method to reproduce the features of the experimentally measured conductance was \cite{Liu2012a}, revisiting the experiment \cite{Young2009} showing the phase shift of Fabry-P{\'e}rot interference due to the Berry phase \cite{Shytov2008}. Subsequent applications include p-n junctions of suspended graphene \cite{Rickhaus2013}, multiple p-n junctions of graphene on substrate \cite{Drienovsky2014}, p-n-p junctions of bilayer graphene \cite{Varlet2014,Du2018}, and large-angle decoupled twisted bilayer graphene \cite{Rickhaus2020,MrencaKolasinska2022}.

\renewcommand{\arraystretch}{1.8} 
\begin{table*}[ht]
\begin{center}
\begin{tabular}{|c|c|}

\hline

Carrier density & $n$ \\ \hline

Fermi velocity & $v_F \sim 10^6 \ m/s$ \\ \hline

Wave vector & $k = \sqrt{\pi |n|}$ \\ \hline

Energy & $E = sign(n) \hbar v_F \sqrt{\pi n}$ \\ \hline

Velocity & $v = \frac{1}{\hbar} \frac{\partial E}{\partial k}$ \\ \hline

Density of states & $D = \frac{2E}{\pi \hbar v_F}$ \\ \hline

Cyclotron radius for a magnetic field $B$ & $r_C = \frac{\hbar k}{eB}$ \\ \hline

Cyclotron frequency & $\omega_C = \frac{e B v_F}{\hbar k_F}$ \\ \hline

Magnetic length & $l_B = \sqrt{\frac{h}{eB}}$ \\ \hline

Nth Landau level energy & $E_N = sign(N) \sqrt{2e\hbar v_F^2 N B}$ \\ \hline

Filling factor & $\nu = \frac{n h}{e B}$ \\ \hline

Sharp p-n junction transmission with incidence angle $\theta$ & $T(\theta) = \cos(\theta)^2$ \\ \hline

Pristine ballistic graphene conductance (width $W$) & $G = \frac{4e^2}{h} W \sqrt{\frac{|n|}{\pi}} $ \\ \hline

\end{tabular}
\caption{Main formulas summary for single layer graphene.}
\end{center}

\end{table*}
\renewcommand{\arraystretch}{1} 

\section{Fabricating clean graphene devices}
\label{sec:Fabrication}

The observation of electron optical phenomena requires high quality devices with ballistic transport where electrons can travel large distances without scattering processes. In 2DEGs buried below the surface record high mobilities were achieved, which allowed to perform electron optical experiments like magnetic focussing \cite{Mayorov2011, Taychatanapat2013, Bhandari2016, Lee2016}.  The special Dirac spectrum of graphene leads to protection against back-scattering events which promises large electron mobilities. However, in the first transport experiments rather low quality devices (at least compared to 2DEGs) have been realized with a mean free path on the order of 100 nm and mobilities on the order of few 1000 cm$^2/$Vs \cite{Novoselov2004,Zhang2005}. It was soon realised that in these devices the mobility was limited by the charge traps in the SiO$_2$ that was used as a gate dielectric on silicon wafers \cite{Chen2008} and by contamination of the graphene surface mostly originating from the fabrication process. Whereas the latter can be partially eliminated by cleaning the graphene surface with forming gas (or other gases) or by AFM cleaning \cite{Goossens2012a,Lindvall2012}, for the former, separation of the graphene flake from the surface was required. 

\subsection{Suspended graphene}

\begin{figure*}[htbp]
	\centering
	\includegraphics[width=\textwidth]{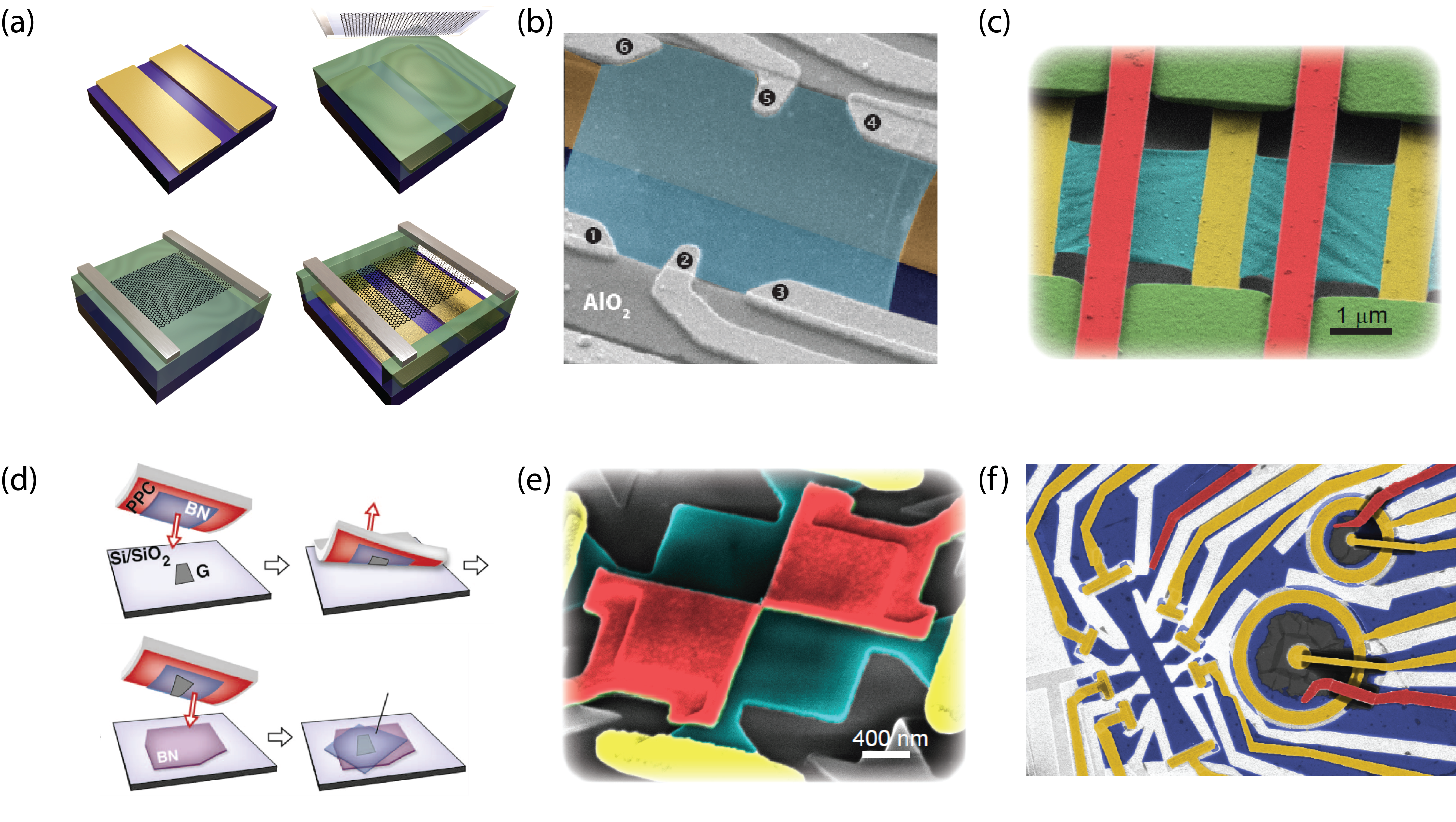}
	\caption{\textbf{High mobility graphene devices.} (a) Suspended graphene devices realised by selective removal of a sacrificial LOR polymer layer. (b)-(c) Graphene suspended devices with complicated structure were demonstrated, with several contacts and bottom gates (b) or even with suspended topgates. Images (a) and (b) are from \reference\cite{Maurand2014a}, (c) from \reference\cite{HandschinThesis2017}. (d) Encapsulating graphene with h-BN lead to high mobility devices and allowed a higher yield in fabrication  \cite{Wang2013}. Using these methods complicated gate structures and device architectures can be realized. Image (e) is from \reference\cite{HandschinThesis2017}, (f) is from C. Sch\"onenberger and coworkers.}
	\label{fig:Fig_devices}
\end{figure*}

To achieve separation from the substrate the graphene flakes were suspended above the SiO$_2$ wafer. This was in the beginning done by etching away the SiO$_2$ with buffered HF solution, followed by a critical point drying step \cite{Bolotin2008,Du2008}. First devices fabricated using this method reported mobilities on the order of 200’000$\,$cm$^2/$Vs. Later on the fabrication of top gates to suspended devices was achieved \cite{Taychatanapat2013}, which together with a doped Si substrate allowed the realisation of double gated high quality devices. Later, another method was developed by Tombros et al. \cite{Tombros2011b} and further extended by Maurand et al. \cite{Maurand2014a}, where instead of etching away the SiO$_2$ below graphene, the flakes were transferred onto a sacrificial layer that was spin-coated on top of the Si/SiO$_2$ wafer. This sacrificial layer could be locally removed by electron beam lithography and liftoff without the need for critical point drying, and the contact material was not limited by the HF etching. This method could also be extended with top-gates, but was better suited for local-bottom gated structures. The fabrication method and examples of suspended devices are shown in \fig\ref{fig:Fig_devices}a-c.

The drawback of both methods was that after fabrication the graphene flakes were covered by resist residues and other contaminants, which had to be removed at low-temperature by passing a large current through the device (current annealing). This process allowed the realization of ultra-high quality devices, however led to a substantial decrease of fabrication yield. For complicated devices with multiple contacts and graphene flakes that are shaped to a certain form, the yield became extremely low. 

\subsection{Graphene-hBN heterostructures}

The next breakthrough came when Dean et al. demonstrated that another 2D crystal, hexagonal boron nitride (hBN), could be an ideal substrate for graphene devices \cite{Dean2010}. First of all, hBN could be exfoliated similarly to graphene, and due to the 2D nature of the crystal an atomically smooth interface could be achieved between graphene and hBN, as demonstrated by TEM studies \cite{Haigh2012}. Second, hBN is an insulator with a large bandgap and can be used as a gate dielectric. The high quality hBN crystals separated the flakes from the SiO$_2$ substrates and allowed the formation of high quality devices with mobilities similar to the suspended ones. The hBN crystals grown by T. Tanaguchi and K. Watanabe led to a revolution in the field and now they are central building blocks of 2D heterostructures. In these heterostructures, where graphene was placed on top of an hBN flake, the top surface was usually cleaned with forming gas or AFM \cite{Dean2010}. In a next step it was also shown that the graphene devices can be fully encapsulated between hBN flakes \cite{Wang2013}, where the devices were made with the pick-up method shown on \fig\ref{fig:Fig_devices}d. The method relies on van der Waals interaction between 2d crystals and allows to pick up flakes from a substrate using another one, leading to the fabrication of not only hBN/Gr/hBN, but more complex heterostructures. The atomically clean interfaces were once again demonstrated by TEM measurements~\cite{Wang2013} and also signalled by the high-quality transport experiments. Since the graphene is encapsulated between hBN crystals prior to fabrication, in order to fabricate electrical contacts, an etching step was performed which exposed the edge of the device and allowed the fabrication of 1D edge contacts which worked surprisingly well, with low contact resistance. This method was further extended/altered by later works~\cite{Zomer2014a,Kretinin2014a, Pizzocchero2016,Purdie2018} and is the standard fabrication technique for most research groups. 

An important advancement in reaching high quality devices was the introduction of graphite gate electrodes. It was found that in this case potential fluctuation from the substrates was further removed in graphite/hBN/Gr/hBN/graphite devices \cite{Zibrov2017, Zeng2019,Polshyn2018}. The decrease of potential fluctuations was confirmed in scanning SET measurements as discussed in more detail e.g. in \reference\cite{Yankowitz2019b}. Finally, very recently a novel transfer method based on silicon cantilevers has been developed for the fabrication of vdW heterostructures \cite{Wang2023}. This allows polymer-free transfer and a compatibility with UHV based fabrication for extremely air-sensitive materials. This methods leads to ultra-clean devices and a very fast fabrication procedure. 

The advent of encapsulated devices gave a new boost for electron optical experiments, since the device architectures could become much more versatile compared to suspended devices, as shown by examples in \mbox{\fig\ref{fig:Fig_devices}e-f}. 

\subsection{Device characterization}

\begin{figure*}[htbp]
	\centering
	\includegraphics[width=.85\textwidth]{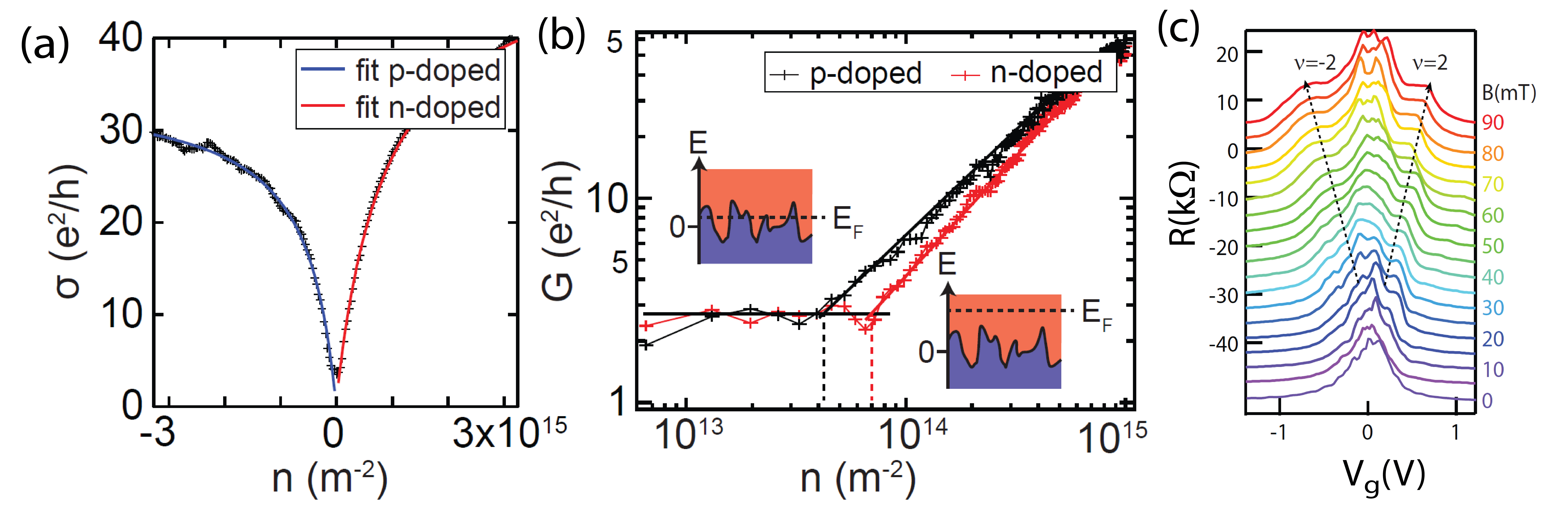}
	\caption{\textbf{Standard characterization of high mobility samples.} (a)  Conductivity versus carrier density (with fitting procedure described in the text). (b) Conductivity versus carrier density in a log log plot shows a saturation corresponding to a residual doping at charge neutrality of about $5 \times 10^9 \rm{cm}^{-2}$ (c) Resistance versus back-gate voltage at finite field showing quantum Hall plateau emerging at field as low as 50 mT which demonstrates the high quality of the device. (a) and (b) are from \reference\cite{HandschinThesis2017}, whereas panel (c) is from \reference\cite{RickhausThesis2015}.}
	\label{fig:Fig_mobility}
\end{figure*}

The quality of graphene devices is usually characterised by field effect mobility. This is extracted from conductance vs. density plots, often by dividing or by derivating the conductance by density. Other methods rely on fitting the conductance vs. density dependence with a formula taking into account contact resistance as well (for two terminal measurements), short range scattering, residual doping around the CNP and a given form of density dependent mobility (often constant). Some typical results are demonstrated on \fig\ref{fig:Fig_mobility}. The fitting procedure with the equation $\sigma^{-1}=\rho_c + (\mu e \sqrt{n_{*}^2+n^2})^{-1}$ for an encapsulated device is shown in \fig\ref{fig:Fig_mobility}(a), where a doping independent mobility $\mu$ is supposed originating from scattering either on charged impurities or from strain fluctuations. $\rho_c$ is the contact resistance and $n_{*}$ is the residual doping.  The field effect mobility can be converted into a mean free path, which determines the length-scale on which electron-optical experiments can be performed. Though this depends on the density, in high quality devices this can reach from a few micrometres up to 20-30 micrometres by now \cite{Banszerus2016, Barrier2020, Wang2023}.

For low energy experiments performed close to the CNP, another quantity, the residual doping $n_{*}$ is important. This quantifies the cut-off lowest density that can be achieved in the device, below which an inhomogenous doping profile forms with electron-hole puddles. The lowest values that can be reached were around below $10^{9}\,$cm$^{-2}$ demonstrated in suspended devices \cite{Nam2017a, RickhausThesis2015, ZihlmannThesis2015}. This is often extracted from log-log conductance-density plots, see also \fig\ref{fig:Fig_mobility}b for an example. Finally device quality can also be inferred from the magnetic field at which Shubnikov de Haas (SdH) oscillations appear. This leads to another lifetime, the quantum lifetime, which is susceptible to small angle scattering as well. This contrasts with the momentum scattering time, which is more sensitive to backscattering events involving large momentum changes. The SdH oscillations can appear in a few tens of milliTeslas, and in \fig\ref{fig:Fig_mobility}c a well-developed quantum Hall plateau is observed already at 50$\,$mT. The splitting of Landau level degeneracy also signifies high quality devices, however the strength of the electronic interactions also matter, which could also depend on the device architecture, e.g. on the distance of screening gate electrodes \cite{Coissard2022,Liu2021,Kim2020,Rickhaus2020,Stepanov2020}. In state-of-the art devices, one limiting factor for the mobility could arise from strain fluctuations, but for the best devices scattering on the edges of the device limit the mobility \cite{Wang2020a,Couto2014,Banszerus2016}. Finally, other time-scales that are usually longer than the momentum scattering time are intervalley-scattering, phase coherence or spin-relaxation time can also be important for certain set of experiments.

\section{Electron optics experimental toolbox}

\subsection{Magnetic focussing}
\label{sec:Focussing}

\begin{figure*}[htbp]
	\centering
	\includegraphics[width=\textwidth]{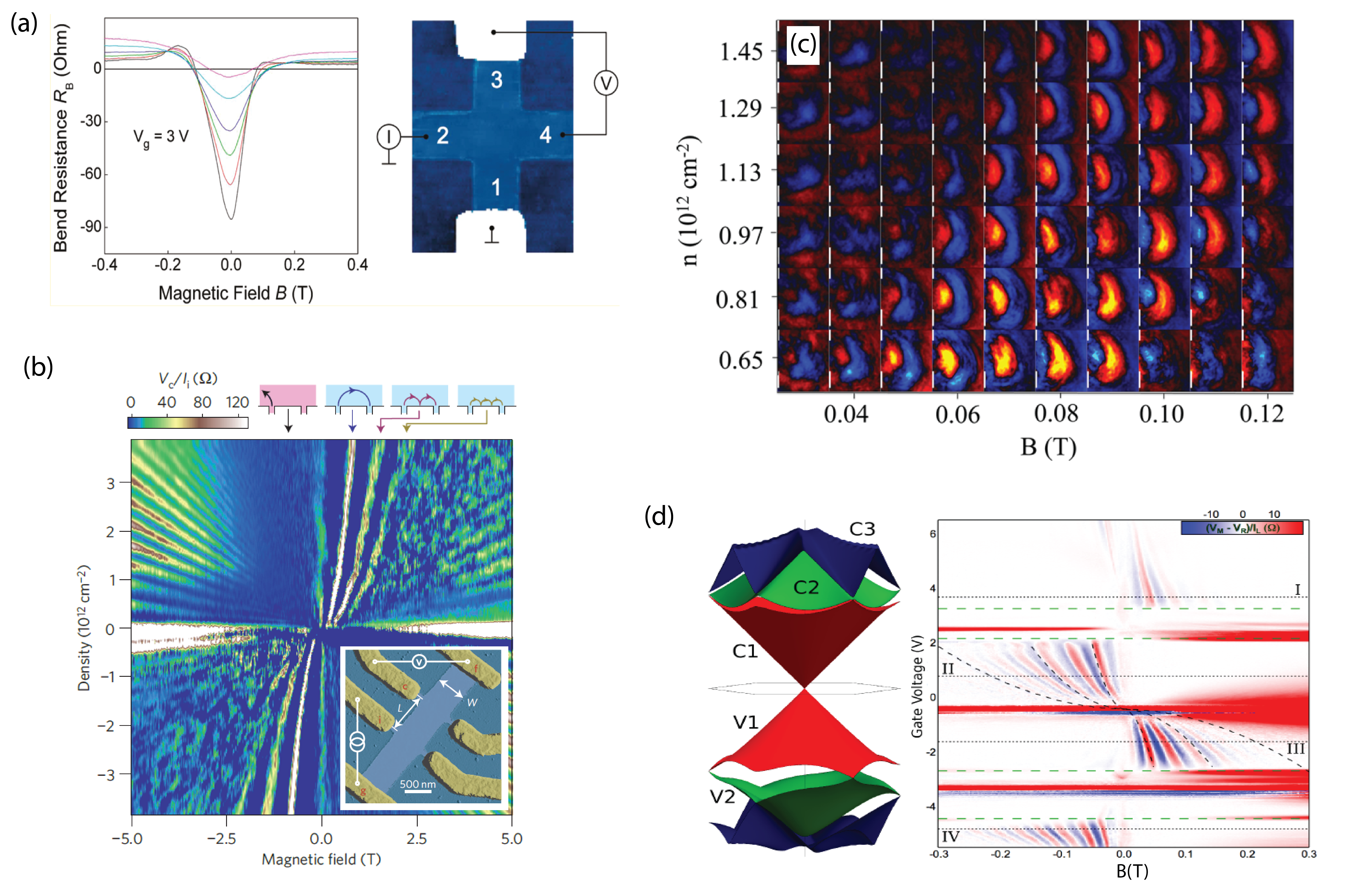}
	\caption{\textbf{Ballistic electron propagation in high mobility devices.} (a) Mayorov et.al. demonstrated in \reference\cite{Mayorov2011} negative resistance vanishing with magnetic field in a non-local measurement, consistently with a device size limited mean free path. (b) Selective focusing of electron trajectories with up to three specular reflections on the sample edge was demonstrated soon after \cite{Taychatanapat2013}. (c) The cyclotron orbits were directly imaged in \reference\cite{Bhandari2016} with a scanning gate microscope. (d) Lee and coworker were able to observe magnetic focusing originating both from the usual graphene bands (main panel) as well as the magnetic focusing of the mini bands originating from the moire potential between graphene and h-BN. The band structure modification is shown on the left \cite{Lee2016}.} 
	\label{fig:Fig_focussing}
\end{figure*}

\paragraph{Bend resistance.}
One of the first signatures of ballistic transport on the micron-scale came from Hall crosses in hBN encapsulated devices \cite{Mayorov2011}, as shown in \fig\ref{fig:Fig_focussing}a. In the experiment, a voltage is measured between terminal 3 and 4, the current is injected at contact 2, while contact 1 is grounded. The voltage-current ratio yields the bend resistance. In zero magnetic field a large negative resistance was observed originating from straight trajectories from contact 2 and 4. In perpendicular magnetic fields, however the trajectories start to bend, as described in \sectionname\ref{Sec:cyclotron} and electrons are guided towards contact 3 leading to a crossover to positive bend resistances. These measurements confirmed ballistic transport up to $3\,\mu$m. With the development of device quality similar negative bend was observed but now on the $30\,\mu$m length scale~\cite{Banszerus2016}.

\paragraph{Focussing experiments.}

Similar physics arises in Hall-bars in magnetic focusing experiments, as shown in graphene for the first time by Taychatanapat et al in \reference\cite{Taychatanapat2013}. As shown on \fig\ref{fig:Fig_focussing}b, one of the side electrodes of the Hall-bars are used as electrons injectors, whereas on a neighbouring electrode the increase of the electrochemical potential is measured as a result of electron trajectories hitting the electrode. In magnetic fields the electrons follow circular trajectories, and if twice the cyclotron radius matches the distance of the contacts, an increased voltage is observed. The cyclotron radius given in \eq\ref{eq cyclotron radius} can be rewritten as
\begin{equation}
    r_c=\hbar k/eB=\hbar\sqrt{n\pi}/eB,
    \label{Eq_Rc}
\end{equation} 
therefore both magnetic field and the electron density tunes the resonance condition. As a result these focusing resonances show up as dispersing lines in the measured voltage in gate-B-field maps, as shown in \fig\ref{fig:Fig_focussing}b for positive gate voltages and magnetic fields. The focusing peaks also show up at negative gate voltages (for holes), however for opposite magnetic fields compared to positive densities. This is the result of opposite group velocity and hence chirality in the electron and hole band for given momentum. The dashed line shows the focusing conditions based on \eq\ref{Eq_Rc}, which matches quite well the measurements. Note that for the correct peak positions the valley degeneracy has to be taken into account. In the real experiments the electrons are injected under an angular distribution, however their majority will still focus to the same position along caustics. The focusing trajectories have been imaged using scanning gate microscopy \cite{Bhandari2016,Morikawa2015,Bhandari2020} and an example is shown in \fig\ref{fig:Fig_focussing}c, which nicely demonstrates their tunability using magnetic fields and gate voltage.

\paragraph{Higher order focussing.}
A focusing condition can also be reached if the electrons bounce on the side of the sample. These account for higher focusing peaks, whenever $2r_c\times j=L$, where $j$ is an integer. The visibility of higher order peaks depends on the sample quality (mean-free path) and on the specularity of the interface on which the trajectories are scattered. A disordered edge leads to the randomization of reflection angles and to the loss of visibility for higher order focusing peaks.  This has been very recently investigated in BLG \cite{Ingla-Aynes2023}, where the edge was realised by electrostatic gating opposite to etched single layer devices, see in more detail \fig\ref{fig:TrajectoriesBLG}b in \sectionname\ref{subsec:ConfinementBLG}. It was found that in this case the smoother edge potential leads to a specular reflection demonstrated by the high visibility of higher order focusing peaks.  We note here that the magnetic focusing has also been applied in a special setting \cite{Bhandari2020},  where the edge of the graphene, on which the charge carriers bounce, was replaced by a superconducting electrode. In this case due to Andreev reflection, instead of electrons, holes can reach the detector leading to a reversed sign for the second focusing peak. 
In general, by  investigation of the temperature dependence of the focusing signal (the weight of the focusing peak) it was found that it cannot be explained by thermal broadening of the injected electrons' momenta. It was found from these experiments that at low temperature electron-electron interaction dominates the momentum scattering time with $T^{-2}$ dependence. 

\begin{figure*}[htbp]
	\centering
	\includegraphics[width=0.9\textwidth]{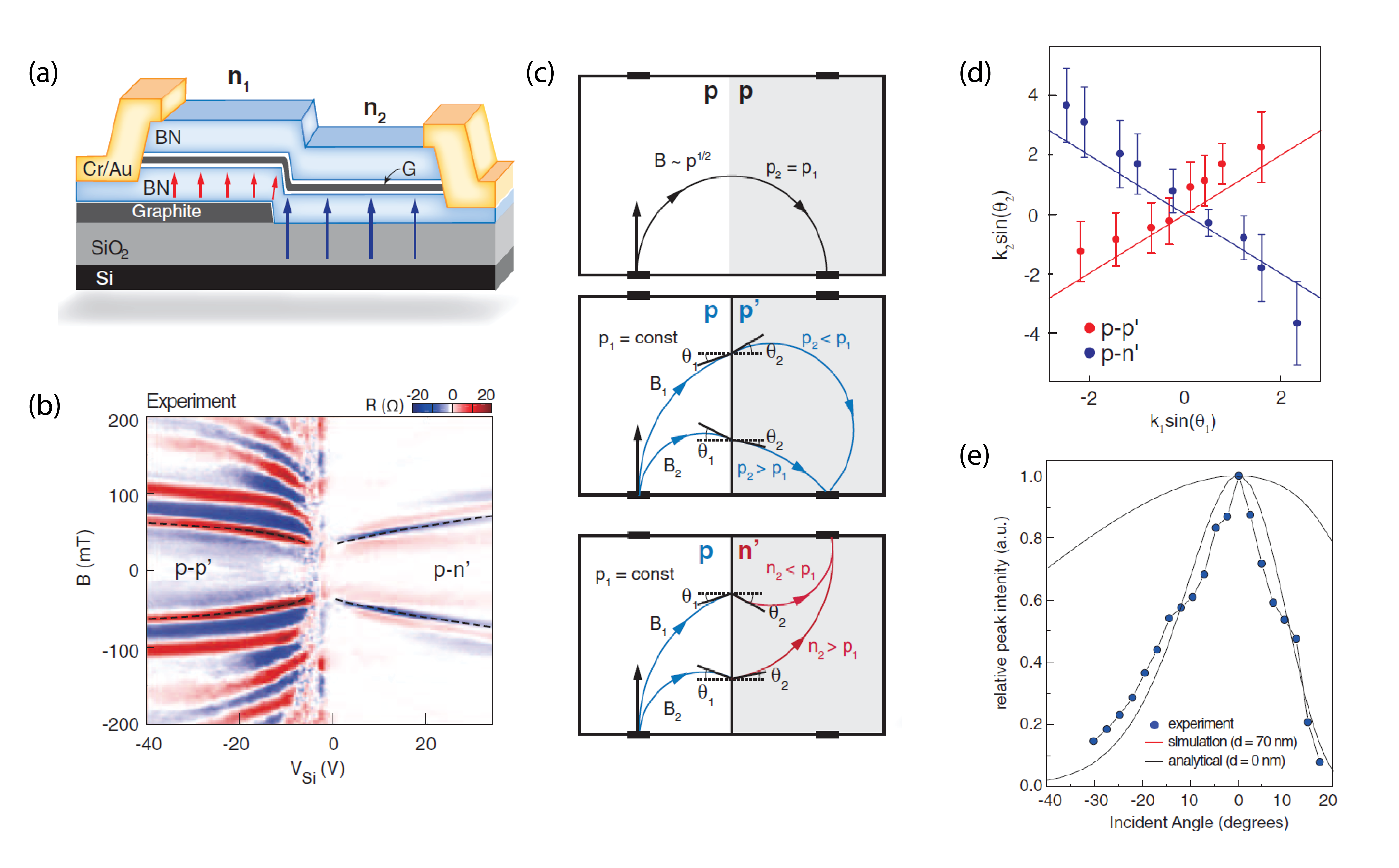}
	\caption{\textbf{Snell's law across p-n junction in ballistic graphene device.} (a) Schematic of the h-BN encapsulated graphene device: a local graphite gate below half of the device is used to create a p-n junction, the carrier density in the rest of the device is controlled by the Si backgate. (b) Non-local resistance shows focusing peak similar to the one in \fig\ref{fig:Fig_focussing}, instead of being reflected on the sample edge as in \fig\ref{fig:Fig_focussing}, the charge carriers are here deflected by the p-n junction. (c) schematic of electrons trajectories for different carrier densities. (d) The estimated outgoing angle versus incident angle (dots) follow the Snell's law (blue and red lines). (f) Focusing peak signal versus incidence angle (dots), fitted by a simulation for a 70 nm wide p-n junction. Adapted from \reference\cite{Chen2016}.}
	\label{fig:Fig_pn_focus}
\end{figure*}

\paragraph{Revealing the band structure.}

Finally, magnetic focusing is a sensitive tool to probe modification of the band structure and the Fermi surface. In a recent example the splitting of the graphene Fermi surface was realized by inducing spin-orbit interaction in graphene that is placed on a WSe$_2$ substrate \cite{Rao2023}. This led to a splitting of the Fermi surface and to signatures of splitting in the focusing peaks. Another example is coming from graphene/hBN structures, where hBN is aligned with the graphene lattice which imposes a superlattice on graphene and leads to strong modification of the band structure, as shown in \fig\ref{fig:Fig_focussing}d. As a result of the band structure modification mini-bands with secondary CNPs appear. The focusing peak position will be sensitive to the band structure and in this case several transitions between electron and hole-like carriers have been observed \cite{Lee2016}, as shown in \fig\ref{fig:Fig_focussing}f.  These measurements therefore give an important tool in the mapping of the band-structure modifications of graphene heterostructures. Recently, another work investigated the case of twisted bilayer graphene where similar transitions from electron to hole Fermi surfaces have been found \cite{Berdyugin2020}. 

\paragraph{Focussing in p-n junctions.}

Chen et al.~used a transverse focusing setting through a p-n junction to verify Snell's-law both in the unipolar (n-n') and the bipolar (p-n) regime. This is shown in \fig\ref{fig:Fig_pn_focus}a, where the doping on the left and right side of the junction can be tuned using two separate gates. These allow to generate a homogeneous doping with circular trajectories (panel c, top cartoon) or distorted trajectories by having different dopings (but with the same carrier type) on the two sides (middle cartoon), and even situations when the doping is opposite on the two sides (bottom sketch). In this case, the trajectories on the p-side bend in the opposite direction due to the opposite group velocity (see \sectionname\ref{semiclassic} or \reference\cite{Beenakker2008}), and the carriers are focused to an opposite terminal leading to a sign change in the voltage. The measurements are shown in panel b for fixed doping on the left side, as a function of right doping and magnetic field. For negative gate voltages the focusing signal can be well fitted (see dashed lines) by trajectories similar to the one shown in panel b). Moreover, for positive gate voltages the focusing through a p-n junction is observed. By tuning the magnetic field and the gate voltage on the left side the incidence angle of the carriers to the p-n junction can be set. To achieve the focusing condition the gate voltage on the right side is tuned, which also determines the angle of the electrons refracted through the junction. The refracted angle multiplied with the momentum as a function of the same quantity on the left side is plotted in \fig\ref{fig:Fig_pn_focus}d, verifying Snell's law given in \eq\ref{eq Snell's law}. Finally, the intensity of the focusing peaks can also be used to determine the angle dependent transmission through a p-n junction, which is shown in panel e. The measurements can be well reproduced with a model of a graded p-n junction with a width of $70\, \rm{nm}$.


\subsection{Electron optical elements}
\label{sec:Optical elements}

\begin{figure*}[htbp]
	\centering
	\includegraphics[width=\textwidth]{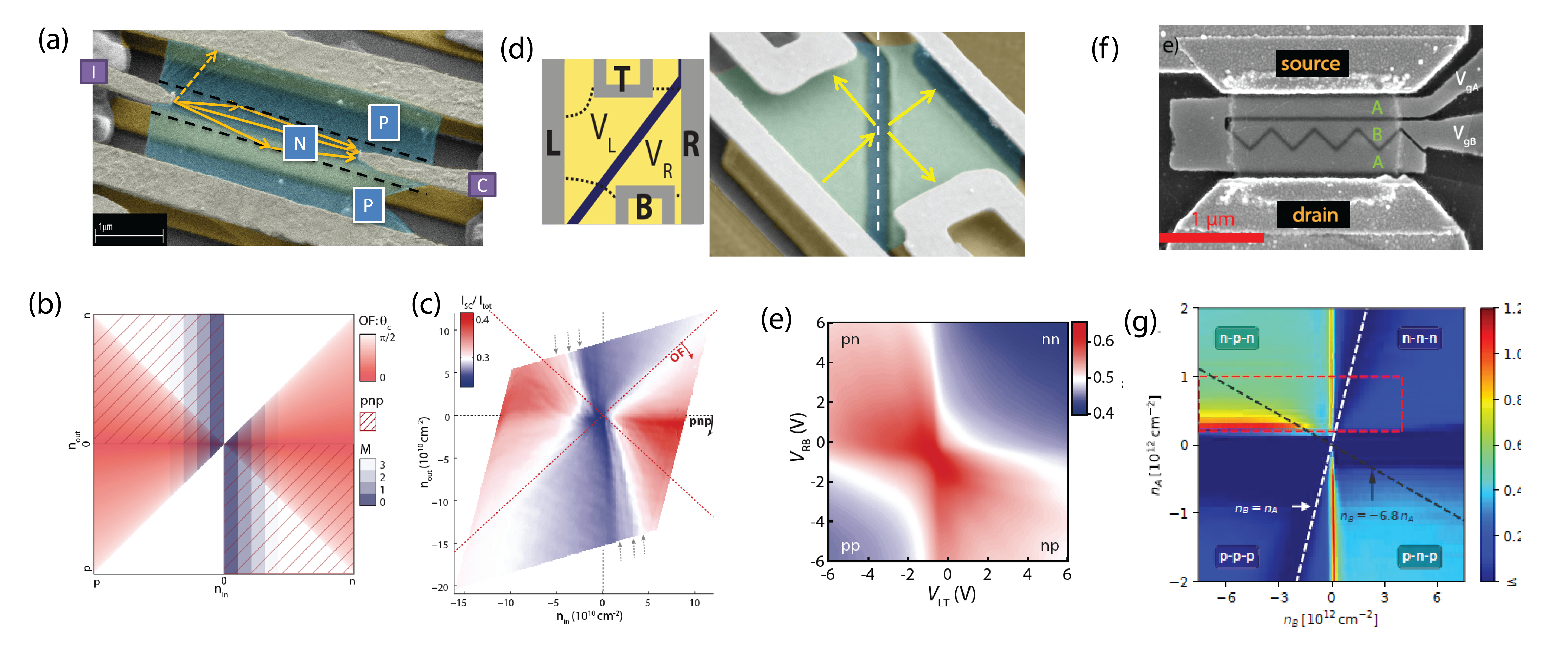}
	\caption{\textbf{Electron-optics building blocs.} a) False colored electron optical image of a suspended graphene sheet where carrier density is tuned locally such to mimic the behavior of an optical fiber. b) Different regions for device operation as function of channel and bulk carrier densities, marking regions of optical fiber guiding, p-n guiding and mode filling. c) Measured transmission reproducing well the expectations shown in panel b). d) False colored electron optical image of a suspended graphene devices with  local gates splitting the graphene sheet into two regions and effectively creating a electronic beam splitter between different contacts. e) Current reflected from the beam splitter towards terminal $T$ (normalised to the injected current) as a function of both local carrier densities. The reflection efficiency increases as the p-n junction is formed. (f) corner reflector, g) resistance of the corner reflector as a function of carrier densities in the B (horizontal axis) and A regions (vertical axis). Images in (a), (b), (c) are adapted  from \reference\cite{Rickhaus2015c} images in (d), (e) are adapted from \reference\cite{Rickhaus2015d}, images in (f) and (g) are adapted from \reference\cite{Graef2019}}.
	\label{fig:Fig_guiding}
\end{figure*}

\subsubsection{Electron guiding.}

In \sectionname\ref{sec graphene interface} we have seen that the reflection and refraction properties of n-n’ and p-n junctions depend on the densities, hence are gate tunable.  Using this with local gating different electron optical devices can be engineered. 
The critical angle of reflection is used to keep photons inside optical fibres and a similar guiding experiment has been engineered also in multiterminal suspended devices \cite{Rickhaus2015c}. A false-colored SEM image of the device is shown in \fig\ref{fig:Fig_guiding}, where the local gates below the graphene allow to tune the electron density in the outer regions and in the central channel, marked with dashed lines, separately. This allows the formation of n-n’-n, n-p-n junctions. The injectors and detectors of the channel are the suspended narrow electrodes, whereas the large electrodes are used to measure the loss. In panel b, we show by red the density regions where optical fiber guiding originating from total internal reflection is expected. Moreover, as soon as a p-n junction is formed (marked by the striped region), the guiding is expected to become more effective due to the larger reflection probability in smooth p-n junctions if the electrons do not arrive close to perpendicular to the junction. Finally for narrow channels, the appearance of mode-by-mode filling is possible, as shown by the blue regions. The experiment shown in panel c shows strong resemblance to the expectation with guiding efficiencies close to 50\%. The experiments were reproduced by tight-binding based calculations and even signatures of mode filling were observed. The guiding efficiency was limited by electrons injected perpendicularly from the side of the contact to the interface (see dashed arrow), which could then easily reach the large electrodes via Klein tunnelling. With proper shaping of the devices and the versatility of hBN encapsulated device architectures higher efficiencies could be reached now.

\subsubsection{Tunable-beam splitters and reflectors.}
The angle-dependent transmission was also used to make a gate tunable beam-splitter \cite{Rickhaus2015d}. The device architecture is shown in \fig\ref{fig:Fig_guiding}d. The electrons are injected from electrode L, the current is measured on the rest of the terminals and tilted gates are used to realise a junction tilted with respect to the injected current. The percentage of the current measured at the top contact in the bipolar region (p-n, n-p) clearly increases compared to the unipolar region (n-n or p-p) case, as demonstrated in panel e. It was found that most of the current is diverted to the top contact when a p-n junction is formed. This study was followed by studies on similar architectures, where even higher tunability was demonstrated \cite{Wang2019a,Elahi2019}. 
The geometry shown on \fig\ref{fig:Fig_guiding}f relies on similar principles \cite{Graef2019,Morikawa2017}. In the corner-reflector devices the source and the drain is connected by two gates: gate A used for collimation and gate B used as the reflector. The principle is similar to that of a prism, however here the refractive index of the inner and outer region are gate tunable. The measurements from \reference\cite{Graef2019} are shown in panel f. The reflector works in the n-p-n regime, for large densities within region B. In this case, resistance larger than in the CNP region can be reached. Limitations on the visibility come from residual scattering mechanism (here phonons), since electrons which don’t arrive perpendicularly to the interface need to traverse several times within the prism, which leads to enhanced scattering probability. 

\subsection{Collimation and lensing}
\label{sec:Collimation}

\begin{figure*}[htbp]
	\centering
	\includegraphics[width=\textwidth]{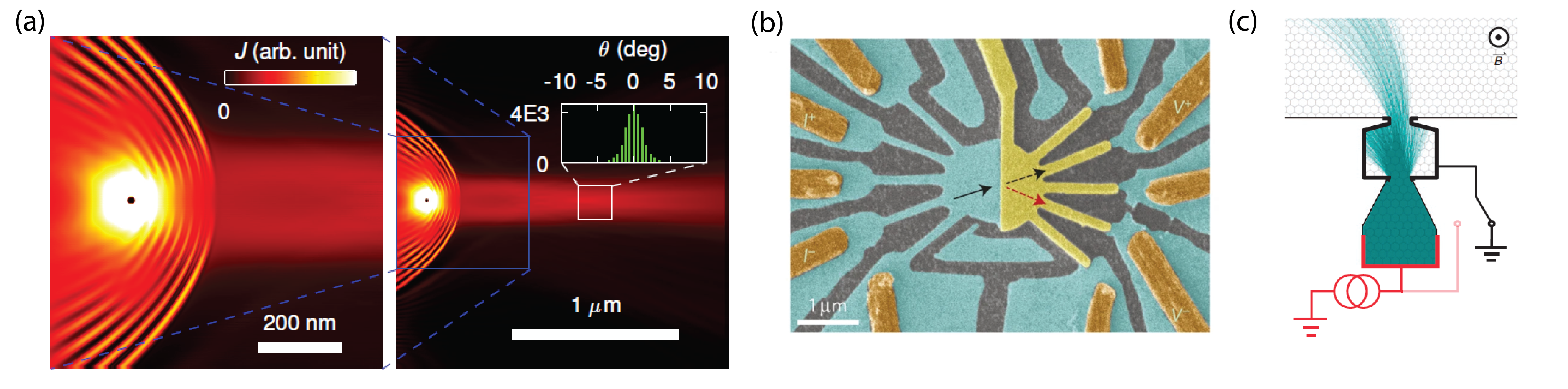}
	\caption{\textbf{Electronic flow collimation.} 
 a) Electron collimation based on a pointlike source at the focal point of a parabolic interface separating two regions with opposite charge carrier densities. Calculation shows  small scale and  large scale current distributions. In the inset this current density analyzed for the white box area.  The beam is collimated on the scale of the electronic wavelength. Adapted from Ref.~\cite{Liu2017}.
 (b) Electron collimation based on etching the graphene to specific shape \reference\cite{Lee2015}. (c) Pinhole configuration used to generate a directional beam in \reference\cite{Barnard2017}. }
	\label{fig:Fig_collimation}
\end{figure*}

\subsubsection{Collimation using p-n junctions.}
For the experiments above the collimation effect of a p-n junctions was important. As shown in \sectionname\ref{sec transmission across graphene pn} and demonstrated on \fig\ref{fig:Fig_pn_focus} a smooth p-n junction only transmits electrons under small incidence angles. Smooth p-n junctions are easy to realize in suspended samples, where the gate distance is large. However, for several experiments, this has to be combined with sharp p-n interfaces, which poses technical challenges. An improved version of p/n collimation have been suggested by Liu and coworkers in \reference\cite{Liu2017}. Specifically, by combining negative refraction and Klein collimation at a parabolic p-n interface, highly collimated, non-dispersive electron beams can be engineered, which stay focused over scales of several microns, as shown in \fig\ref{fig:Fig_collimation}a. Such "beams of electron waves" can be bent and steered by a magnetic field without
losing collimation, see \fig\ref{fig:TrajectorieshBN}.
This provides a setup for observing high-resolution angle-dependent Klein tunneling and high-fidelity transverse magnetic focusing \cite{Liu2008}.

This setting is based on feeding charge carriers through a point-like source (vertically) into the system. Indeed, such sub-$100\,$nm point contacts have
already been experimentally realized in graphene \cite{Handschin2015}. However, the intensity of the created beams suffers from the fact that only an angular segment of the isotropically emitted wave is collimated. Enclosing the point source by a cavity to collect such losses should, in principle, allow for strongly enhancing the beam intensity.

\begin{figure*}[htbp]
	\centering
	\includegraphics[width=0.8\textwidth]{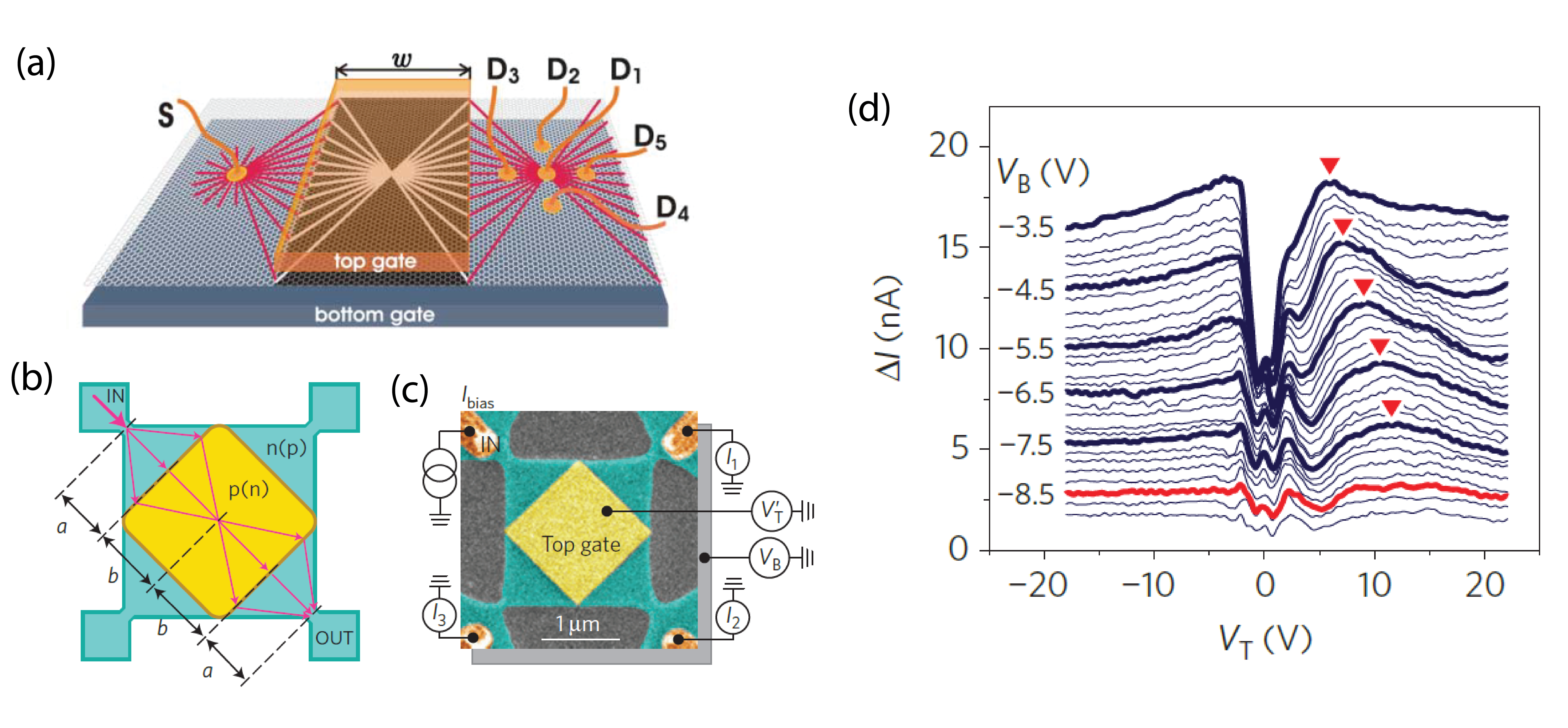}
	\caption{\textbf{Negative refraction and Veselago lensing.} (a) Theoretical concept of a graphene based Veselago lens using negative refraction \cite{Cheianov2007}.  (b) Experimental design with electron trajectories traced in red. (c) False colored scanning electron microscope of the device. (d) Constant current is injected from top left contact and the collected current at bottom left contact is measured. Background (taken at $V_b = -9.5\,$V) is substracted. Red triangles indicate current enhancement. Image adapted from \reference\cite{Lee2015}.}
	\label{fig:Fig_Veselago}
\end{figure*}

\subsubsection{Geometrical collimation.}
Other methods to generate collimated beams are shown in panels b and c of \fig\ref{fig:Fig_collimation}. 
In panel b the geometrical shaping of the devices is used to make narrow injector contacts \cite{Lee2015}. This allows a well-defined injection angle to the central region and the p-n junction that is realized at the boundary of the yellow gate. In such architecture the injector part has to be ballistic as well, which puts more serious constrains on the device quality. Moreover, the edge of the injector part of the devices often suffer from edge roughness due to the etching procedure used to define them, which leads to random scattering and hence can randomize the outgoing electron trajectories close to the injection point. 
A solution for this problem is shown in \fig\ref{fig:Fig_collimation}c, where the side of the injector contact is used to drain the electrons that are not injected under a narrow angular distribution \cite{Barnard2017}. Here the injection is done from the bottom  of the injector-collimator element (shown in red), and after the constriction the black contact can be used to drain the uncollimated electrons. The viability of this method was tested using magnetic focusing experiments \cite{Barnard2017}.

\begin{figure*}[htb]
\includegraphics[width=1\textwidth]{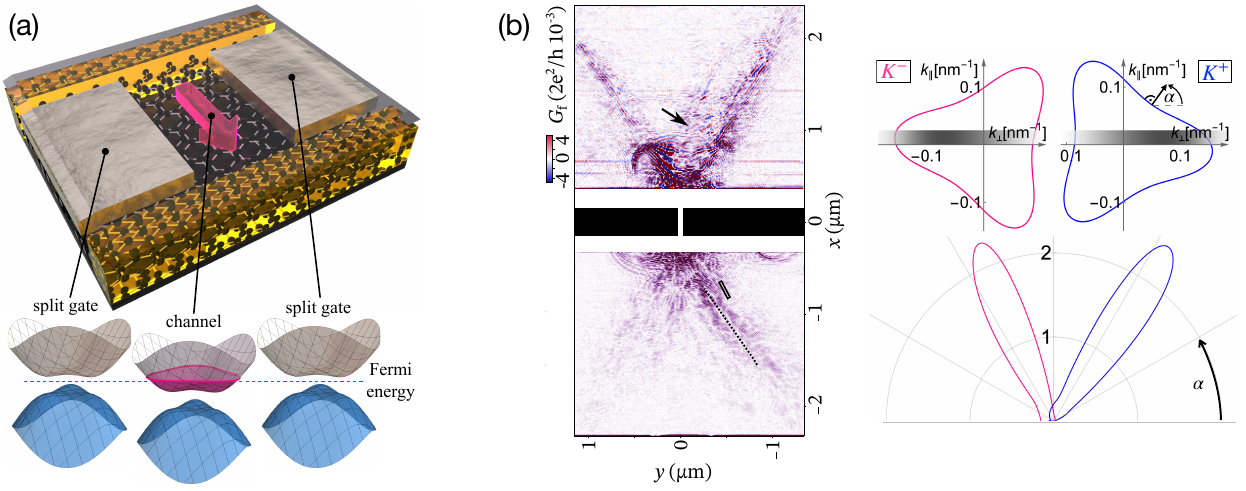}
\caption{{\bf Electron guiding and anisotropic propagation in \BLG{}.} (a) Device schematic for electrostatic soft confinement and guiding in gapped \BLG{}: Multiple gates locally modulate both the gap and the charge carrier density such that, e.g., the Fermi energy is within the band gap underneath the split gates, but in the conduction band in the centre, defining a 1D channel. (b) Collimated electron jets behind a gate-defined channel in gapped \BLG{}. Left: spatial structure of electron flow in the bilayer graphene bulk on both sides of the channel measured by \SGM{}. Right: the trigonally warped Fermi lines in \BLG{} are anisotropic in both $K^{\pm}$ valleys, entailing an anisotropic, valley dependent velocity distribution of the charge carriers.  Panel (a) © C. Schulz and A. Knothe \href{https://creativecommons.org/licenses/by/4.0/}{CC BY 4.0}, panel (b) adapted from \reference\cite{Gold2021}}
\label{fig:BLGQPCJets}
\end{figure*}

\subsubsection{The Veselago lens.}

A very peculiar negative refraction property of p-n junction can also be used to make a "perfect lens", called Veselago lens. In their early work Cheinaov and coworkers have suggested \cite{Cheianov2007} that a flat p-n surface can be used to focus the trajectories originating from a point like source to a point like detector. This is shown in \fig\ref{fig:Fig_Veselago}a where the trajectories originating from S source electrode are refocused to D1 after two reflection through the p-n junction. The experimental challenge lies in the formation of point like sources~\cite{Handschin2015} and in the realization of sharp p-n interfaces. This is important, since as described in \sectionname\ref{sec linear pn}, for smoother junctions only trajectories that are close to perpendicular to the junction are transmitted. The first signatures of the lensing was shown in \reference\cite{Lee2015} in a geometry shown in \fig\ref{fig:Fig_Veselago}b. The trajectories are injected through a constriction, which similarly to the proposal are twice refracted through a p-n junction. The SEM image of the device is shown in panel c. The geometrical parameters of the device, $a$ and $b$ determines the density ratio at which the focusing to the drain ("OUT") electrode should happen. Weak signatures of this focusing effect were observed in the measurements shown in \fig\ref{fig:Fig_Veselago}d, the position where a weak enhancement is observed is marked by red arrows. The different curves correspond to different doping in the green region of the device, and for the position of the arrows the doping in the yellow region is set to the focusing condition. Later work  have shown signatures of this lensing effect in interference experiments \cite{Zhang2022}.

\subsection{Anisotropic Fermi surfaces}
   \label{collimation}

\subsubsection{Gate-defined electron guiding in ballistic \BLG{}.}
\label{subsec:ConfinementBLG}

Early designs of quantum nanostructures in mono- and \BLG{} confined the charge carriers by physically etching the graphene sheet~\cite{bischoff2015, bischoff2014, ihn2010}. This method allowed to observe some quantum confinement effects in graphene, but introducing hard wall boundaries by etching is intrinsically flawed. Edge disorder and, consequently, randomly localised states along the sample edges are inevitable, leading to a loss of coherence and control of the charge carriers' degrees of freedom.

An alternative for confinement in bilayer 2D materials uses electrostatic gating \cite{Droscher2012, Overweg2018, Goossens2012, Allen2012}. In a bilayer lattice, an external electric field breaks the inversion symmetry and opens a band gap. Then, in the gapped device, e.g.,  two split gates may define a channel: adjusting the potentials of the gates tunes the Fermi energy into the conduction band within the channel region but into the band gap underneath the split gates, as shown in \fig\ref{fig:BLGQPCJets}a. This method of electrostatic confinement introduces smoother confinement potentials and avoids edge-induced device perturbations. Over the recent years, such gate-defined soft electrostatic potentials have developed into a formidable tool for electron confinement, steering, and control in bilayer graphene. Immense progress in sample quality and gating has enabled the demonstration of electron confinement and control in a series of  gate-defined quantum nano\-structures, including quantum wires \cite{Overweg2018a,Lee2020,Gold2021,Lane2019,Banszerus2020c,Knothe2018,Kraft2018,Ingla-Aynes2023}, quantum dots \cite{Eich2018a,Eich2018,Banszerus2018,moller2021,Tong2021,Tong2022,Knothe2020,Banszerus2020b,Kurzmann2021,Garreis2021,Banszerus2022,Banszerus2020e,Banszerus2021,Knothe2022,arXiv:2303.10201, mayerTuningConfinedStates2023}, and electron interferometers~\cite{Iwakiri2022,Fu2023,Mirzakhani2023}.

   \begin{figure*}[htb]
\includegraphics[width=1\textwidth]{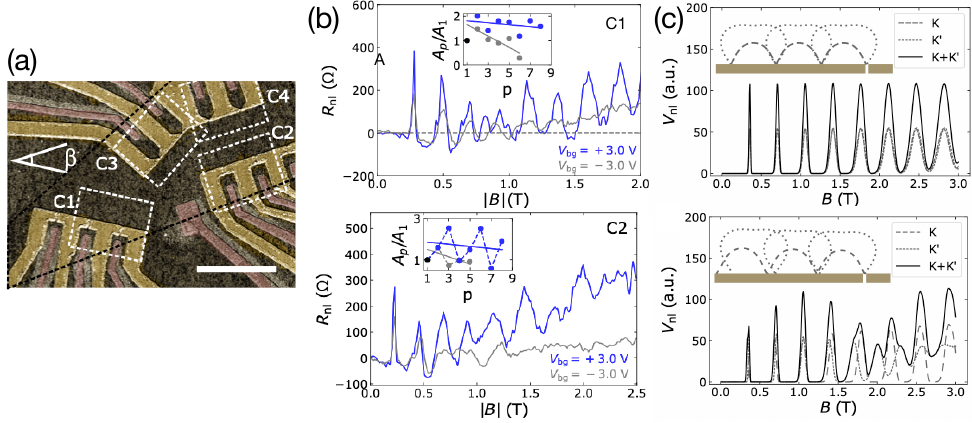}
\caption{{\bf Anisotropic magnetic trajectories in \BLG{}.} Due to the $C_{3}$ symmetry of the trigonally warped Fermi lines, transverse electron focusing in \BLG{}  depends on the orientation of emission and detection with respect to the lattice (a). (b) TEF resonances for two differently oriented devices. (c) Simulated TEF signal for perfectly aligned (top) and 3$^{\circ}$ misaligned (bottom) \QPC{}s with respect to \BLG{}'s armchair crystallographic direction taking into account the trigonally warped and valley dependent Fermi lines. Images adapted from \reference\cite{Ingla-Aynes2023}.}
\label{fig:TrajectoriesBLG}
\end{figure*}

Due to the high quality of the gate-induced electrostatic confinement and the hBN-encapsulated \BLG{} samples \cite{Banszerus2016, Lee2016, Gold2021, Berdyugin2020} charge carriers in such confined structures often propagate in a largely ballistic manner and can be guided and controlled by virtue of the external gates.

\subsubsection{Lowered rotational symmetry of the electronic structure and anisotropic charge carrier dynamics.} 
\label{subsec:Anisotropy}
  \begin{figure*}[htb]
\includegraphics[width=1\textwidth]{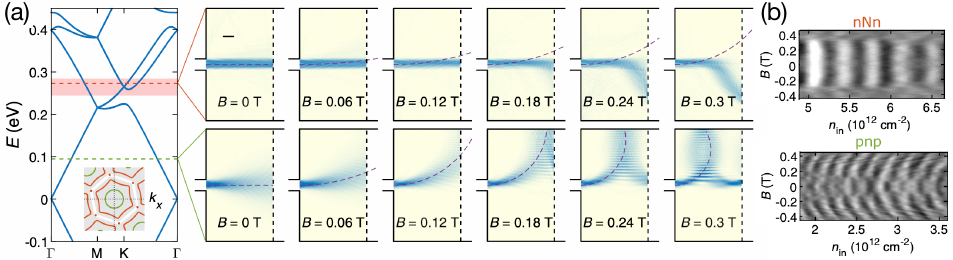}
\caption{{\bf Magnetotransport in graphene/hBN superlattices depending on the Fermi line shape.} (a) Left: the band structure of a graphene/hBN moir\'e superlattice exhibits hexagonal (red)  or circular (green) Fermi lines at the K point of the mini-Brillouin zone depending on doping. Right: charge carrier beams from simulations in the two different regimes show bending or no bending in a magnetic field depending on the possible directions prescribed by the Fermi line symmetry. (b) Bending or no bending of the ballistic charge carrier trajectories was evidenced in dispersive or nondispersive interference fringes in graphene/hBN Fabry-P{\'e}rot cavities adapted from \reference\cite{Kraft2020} (see \sectionname\ref{sec:Quantum-interference-in-zero-field} and \sectionname\ref{sec:non-zero-low-magnetic-field}  for a detailed discussion about interferences in Fabry-P{\'e}rot cavities).}
\label{fig:TrajectorieshBN}
\end{figure*}

The tunable band gap and the anisotropic, trigonally warped low-energy dispersion are the key factors for unusual ballistic electron optics in bilayer graphene (BLG) different from the standard Dirac case. The low-energy band-structure of BLG is given by \eq\ref{eqn:BLG_disp}. The presence of $v_3$ skew hopping parameter breaks rotational symmetry and induces trigonal warping, leading to triangularly deformed Fermi surfaces with opposing orientation in the  $K^{\pm}$ valleys, as can be seen in \fig\ref{fig:TrajectoriesBLG}b. Since an anisotropic dispersion entails a corresponding anisotropy in the charge carriers' velocity distribution, $\mathbf{v}=\frac{1}{\hbar}\nabla_{\mathbf{k}}E$, the trigonally warped dispersion of the bilayer graphene leads to directional and valley dependence of the ballistic electronic transport.

Anisotropic ballistic charge carrier dynamics in \BLG{} has been observed directly in scanning gate microscopy (\SGM) experiments: In \reference\cite{Gold2021}, they raster-scan the \BLG{} areas behind a gate-defined channel with an \SGM{} tip and measure the linear conductance between source and  drain as a function of the tip position. The resulting conductance map in \fig\ref{fig:BLGQPCJets}b shows two narrow jets emanating from a gate-defined \BLG{} channel predominantly at an angle of 60° with respect to each other. These collimated jets at this specific angle directly result from the reduced symmetry of \BLG{}'s trigonally warped dispersion. For charge carriers at a given Fermi energy, the triangular Fermi line gives rise to three distinct preferred directions per valley (normals to the triangle's flat legs). The electronic states emerging from the channel populate these directional states according to the distribution of the occupied channel mode. As a result, there are two jets behind the channel corresponding to one of the preferred directions per valley. For recent additional evidence for the anisotropic transport, see \reference\cite{inglaaynés2023ballistic}). 

The charge carriers' anisotropic velocity distribution is an intrinsic material property of \BLG{} and its trigonally warped dispersion impacts different aspects of ballistic propagation.

For one, the real-space anisotropy translates into the Fresnel and Snell laws for diffraction and reflection at p/n-junctions in \BLG{}. The discussion is similar to that of monolayer graphene laid out in \sectionname\ref{sec snell}, but for bilayer graphene, the non-isotropic Fermi lines and opening of a gap must be taken into account \cite{Seemann2023, Peterfalvi2012}. The unique interplay between anisotropic scattering and anisotropic ballistic propagation in bilayer graphene has prompted investigations into the potential for intraband electron focusing and valley-selective electronic Veselago lenses in \BLG{} \cite{Peterfalvi2012}. These concepts allow for manipulating and controlling electron motion based on a material's properties and symmetries. Further research has explored the motion of electrons in gate-defined bilayer graphene cavities, revealing unusual regular and chaotic trajectory dynamics due to the anisotropies and nonlinearities induced by the trigonally warped Fermi lines \cite{Schrepfer2021, Seemann2023}. {We discuss electronic cavities in detail in \sectionname\ref{subsec:EmissionCavity}.

Moreover, the deformation of the Fermi lines in bilayer graphene has consequences for the trajectories of charge carriers when exposed to weak magnetic fields. Section \ref{sec:Focussing} and \sectionname\ref{sec:non-zero-low-magnetic-field}  describe how the combined influence of the magnetic field and the charge carrier density on the radius of circular cyclotron orbits of circularly symmetric dispersions leads to transport resonances in transverse magnetic focusing experiments. Specifically, the deformation of the Fermi lines and the corresponding cyclotron orbits changes these caustics of transverse electron focusing resonances. For example, in bilayer graphene, the valley dependence and deformation of the Fermi lines change the position and shape of these caustics depending on the orientation of the device with respect to the lattice, and hence the positions where the charge carriers are focused and defocused as they move through the material. These deformations have implications for how to steer and confine bilayer graphene's charge carriers in magnetotransport, cf.~\fig\ref{fig:TrajectoriesBLG}  from \reference\cite{Ingla-Aynes2023}: Here, the authors study magnetic focussing between gate-defined \BLG{} channels along different crystallographic directions, demonstrating the effect of the non-rotationally symmetric, trigonally warped cyclotron orbits.

 The discussion above about anisotropic electron optics in ballistics \BLG{} relies on the anisotropic band structure and resulting velocity distribution. This line of argumentation is not limited to Bernal stacked bilayer graphene but applies to any material with an anisotropic dispersion. Further common examples of materials with anisotropic Fermi lines include heterostructures of graphene and hBN, twisted multilayer graphene structures, and graphene with gate-defined superlattices. In the former instance, the moir\'e superlattice potential induced by the adjacent hBN leads to reconstructed spectra with circular, trigonal, or hexagonal symmetry as a function of doping, cf.~\fig\ref{fig:TrajectorieshBN} on transport studies of anomalous cyclotron motion in hBN/graphene/hBN) \cite{Kraft2020}  and \fig\ref{fig:Fig_focussing} (\cite{Lee2016}). 

 In the left panel of \fig\ref{fig:TrajectorieshBN}a the computed modified band structure is shown, which exhibits strong electron-hole asymmetry. As visible in the left inset, the Fermi surface can assume both isotropic or hexagonal textures, depending on the energy regime. This leads to distinctly different electron propagation in magnetic field as shown for two different doping situations: The lower sequence of panels shows for the circular Fermi surface the magnetic-field dependent propagation of bended electron waves along usual cyclotron orbit segments. In contrast to that, at an energy corresponding to the hexagonal Fermi contour, electron waves stay straight and nearly $B$-field unaffected at weaker fields and follow peculiar hexagonal-shaped orbits for stronger fields \cite{Kraft2020}.
 
 In the case of graphitic multilayers, e.g., large angle twisted \BLG{} \cite{Berdyugin2020} and twisted monolayer–bilayer graphene \cite{Xu2021} show trigonally distorted Fermi lines with various shapes at different Fermi energies. Gate-defined superlattices allow inducing diverse and variable potential modulations by virtue of patterned gates \cite{Huber2020,Huber2022, Drienovsky2014,Drienovsky2018}, entailing Fermi surfaces of various shapes and symmetries \cite{MrencaKolasinska2023}. Furthermore, external perturbations such as mechanical strain and shear can lower spatial symmetries and induce anisotropies in a material's electronic dispersion \cite{Moulsdale2020, Varlet2015}. Generally, the discussion of anisotropic ballistic electron optics highlights the complex nature of electron motion in materials with anisotropic dispersions.

\subsection{Directed emission from single- and bilayer graphene cavities}
\label{subsec:EmissionCavity}
\subsubsection{Tailoring charge carrier emission from graphene disks.}

There are different ways to tailor graphene-based cavity regions. Complementary to gates, as used in several of the charge carrier guiding and steering experiments reviewed in \sectionname\ref{subsec:ConfinementBLG}, disk-like cavities have been created by employing a scanning tunneling setting, see \fig\ref{fig:Strocio-exp}a. Within such circular  p-n junctions, whispering-gallery type resonant states that are confined through the ring-shaped p-n junction have been probed experimentally~\cite{Zhao2015}, see also \reference\cite{Brun2022, Ge2021}. Such resonant states are exceptionally long-lived and stable against decay from the cavity due to Klein tunneling suppressing the tunneling of waves with grazing incidence. In \fig\ref{fig:Strocio-exp}b, such whispering-gallery modes are depicted for different angular momenta, respectively. Panels c) and d) display corresponding observed and calculated $dI/dV$ spectra of the whispering-gallery resonant states.
Subsequently, non-reciprocity of such whispering gallery modes was theoretically predicted \cite{Rodriguez-Nieva2015}.
Earlier theoretical works had addressed the influence of the classical charge carrier dynamics' character (integrable versus chaotic) on transport through open cavities~\cite{Wurm2009, Bardarson2009, Wurm2010, Schneider2011, Hein2013, Schneider2014} and spectra of closed cavities~\cite{Wurm2011, Wurm2011a, Xu2018} of different shapes.
Recently, in \reference\cite{Bercioux2023} ballistic graphene disks with spin-orbit interaction have been shown to host chiral spin channels with the spin fully in-plane and radially polarized upon tuning certain parameters.


\begin{figure}
\includegraphics[width=\columnwidth]{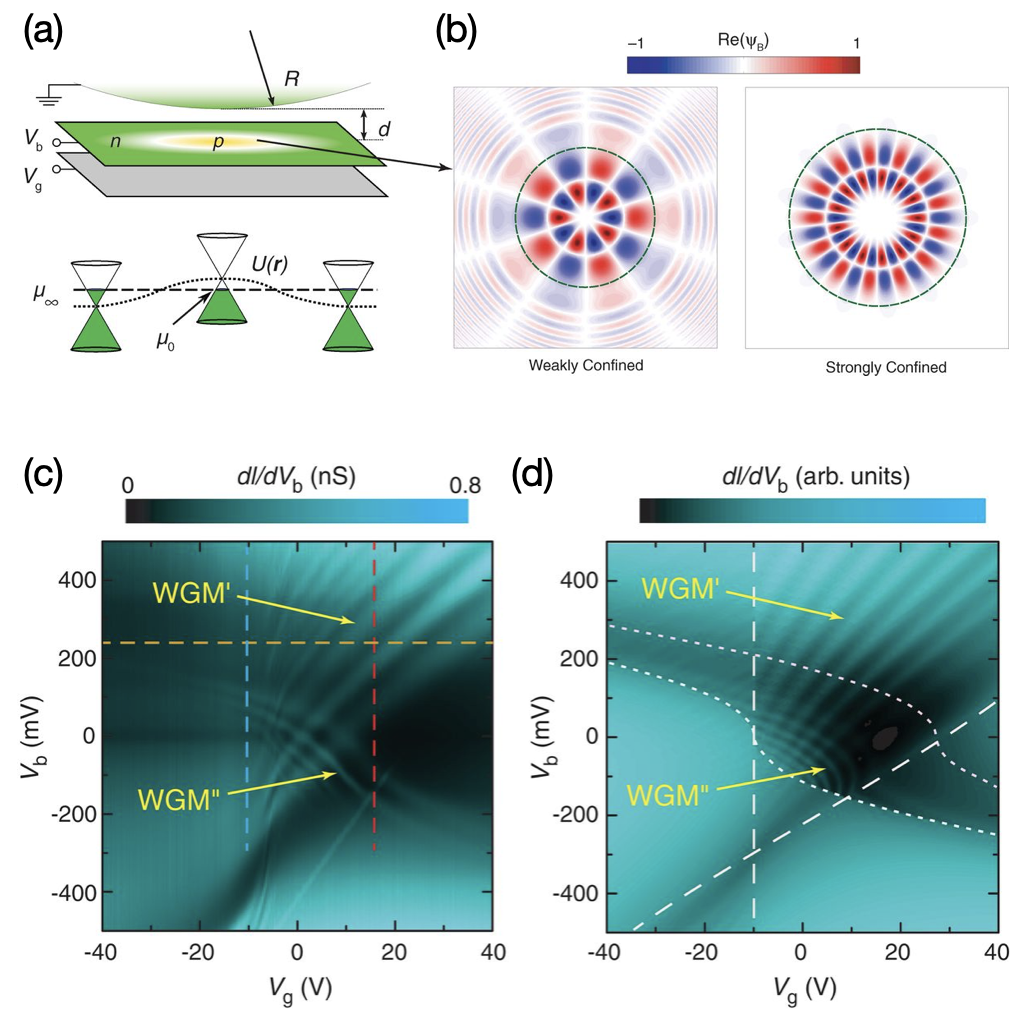}
\caption{
{\bf Confining and probing electronic states in single-layer graphene cavities.} 
(a) Disk-like microscopic cavity (hosting resonant states) based on a circular p-n junction created by the combined effect of an STM tip with voltage bias (Vb) and a back-gate voltage (Vg) inducing a ring-type p-n junction (see \reference\cite{Zhao2015}). 
It creates a sharp boundary with associated Klein scattering of charge carrier waves giving rise to confined whispering-gallery resonant states.
(b) Spatial profile of calculated whispering-gallery modes. The effective confinement is stronger for larger angular momentum with more oblique wave incidence angles. The left (right) panel shows modes with weak (strong) confinement, respectively.
(c) Differential tunneling conductance (dI/dVb) as a function of Vb and Vg. The two fans of interference features, marked WGM' and WGM'', originate from different WGM resonances. (d) Corresponding calculations based on an effective Dirac model. From \reference\cite{Zhao2015}. Reprinted with permission from AAAS.}
\label{fig:Strocio-exp}
\end{figure}



\begin{figure*}[htb]
\includegraphics[width=\textwidth]{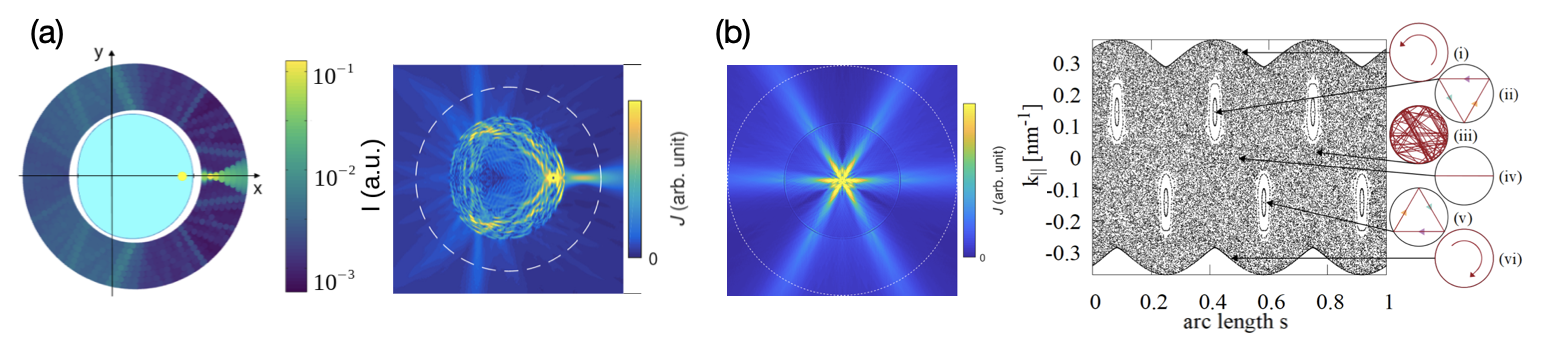}
\caption{
(a) {\bf Directed charge carrier emission from tailored graphene-disks.} Ray-wave correspondence in single-layer graphene billiards of Lima\c con shape with effective refractive index $n = -1$. The light blue area marks the cavity. Left: Ray simulations for point source injection marked as yellow dot. Right: Corresponding wave simulations for the same source position as on the left. The color scale represents the electronic wave function intensity. From \reference\cite{Schrepfer2021}. (b) {\bf Effects of $k$-space anisotropy.} Left: Local charge carrier density for charges injected from a point-like source at the center of a bilayer graphene disk (diameter 1$\mu$m), dashed line marks the midfield region ($r_m=2 \mu$m. Adapted from \reference\cite{Schrepfer2021}. Right: Poincaré surface of section revealing different types of trajectory dynamics in a gate-defined \BLG{} electron cavity: chaotic dynamics (iii), stable triangular, periodic orbits (ii, v), unstable periodic orbits along the diameter (iv), and whispering-gallery-like orbits (i, vi). Images taken from \reference\cite{Seemann2023}.
}
\label{fig:CavitiesAll}
\end{figure*}

The experimental realization of disk-like cavities has opened up several intriguing directions, arising from the fact that the physics of resonant states in single-layer graphene cavities has much in common with the field of mesoscopic optics:
 there, corresponding settings for electromagnetic radiation have been devised for controlling highly directional emission from asymmetrically shaped, lasing cavities \cite{Noeckel1997}. In these dielectric microcavities, total internal reflection partially confines light in whispering-gallery type modes~\cite{Hentschel2002}. Breaking of the rotational symmetry was found~\cite{Noeckel1997, Wiersig2008} to lead to directional light emission of decaying resonant states. These emission characteristics were understood by invoking optical ray-wave correspondence. The cavity geometry determines the phase space structure of the rays inside the cavity in the classical ray limit of optics. Controlling the ray phase space structure by deforming the cavity allowed one to steer directional emission and lasing in the optics context. 
 
Based on a corresponding ray-wave correspondence approach for electrons in graphene, such mesoscopic optics concepts have recently been transferred to specific graphene cavity setups. 
These cavities are defined by the p-n interface geometry that, in turn, is determined by the gate voltage step from the inner to the outer region, i.e., $V_{\rm in}$ to $V_{\rm out}$ where $V_{\rm in}$ is related to $V_{\rm out}$ by an effective index $n$ of refraction, $V_{\rm in} = n V_{\rm out}$, as explained in \sectionname\ref{sec snell}.
A back-gate voltage provides a tunable parameter to mimic different effective refractive indices and, thereby, the corresponding Fresnel laws at the boundaries.
The possibility of readily realizing negative refractive indices in graphene adds to the fascination of such studies.

In \reference\cite{Lai2018}, the decay features of integrable disk- and chaotic stadium-type cavities were studied based on classical ray tracing.  \reference\cite{Schrepfer2021} explores charge carrier trapping and (directed) emission for deformed leaky graphene micro-disks by considering the complete ray-wave correspondence through classical and quantum simulations. 
As depicted in \fig\ref{fig:CavitiesAll}a, the corresponding ray and wave results agree semi-quantitatively. They both exhibit a pronounced directed emission of electrons, leaving the cavity to the right for that setting.
More generally, one finds various emission characteristics depending on the position of the source where charge carriers are fed into the cavities.
Furthermore, single-layer and double-layer graphene cavities exhibit Klein- and anti-Klein tunneling at the cavity boundary, respectively, leading to distinct differences concerning dwell times and resulting emission profiles of the cavity states. Moreover, bilayer-based cavities offer the additional possibility to tune between Klein and anti-Klein tunneling by varying a respective asymmetry parameter~\cite{Mccann2013, Varlet2014, Varlet2016, Du2018, Seemann2023, Elahi2022}.
For bilayer graphene, trapping of resonant states is more efficient, and the emission characteristics depend less on the source position~\cite{Schrepfer2021}. 
Recently, in \reference\cite{Elahi2022} the trapping, respectively transmission, of charge carriers in a single- and bilayer graphene-based Corbino disk has been studied and proposed as a signature of Klein and anti-Klein tunneling, respectively. 

\subsubsection{Symmetry breaking through anisotropic Fermi surfaces.}
\label{subsec:AnisotropicFermi}

The left panel of \fig\ref{fig:CavitiesAll}b shows a typical resonant state in a bilayer-graphene-based disk. Interestingly, despite the circular cavity geometry, the wave simulation result displays distinct emission directions, which can be understood as follows.
While the current is injected isotropically from the central point source, the reflected waves (through anti-Klein tunneling) return to the point injector, then acting as a scatterer. However, the scattering is non-isotropic because the underlying Fermi contour gets non-circular for bilayer graphene at finite energies. Hexagonal contributions to the Fermi contour lead to six predominant velocity
directions~\cite{Kraft2020} that are filtered out, cf \sectionname\ref{subsec:Anisotropy}, \fig\ref{fig:TrajectorieshBN}. They determine the resonant state in \fig\ref{fig:CavitiesAll}b and thereby peculiar directional charge carrier emission. 
In \reference\cite{Seemann2023}, the emergence of non-standard fermion optics solely due to anisotropic material characteristics, i.e., $k$-space structure, has been examined in much more detail. There it is shown how the anisotropic dispersion of bilayer graphene induces chaotic and regular charge carrier dynamics depending on the gate voltage, despite the high symmetry of the circular cavity, cf.~ the right panel of \fig\ref{fig:CavitiesAll}b.

\begin{figure*}[htbp]
	\centering
	\includegraphics[width=\textwidth]{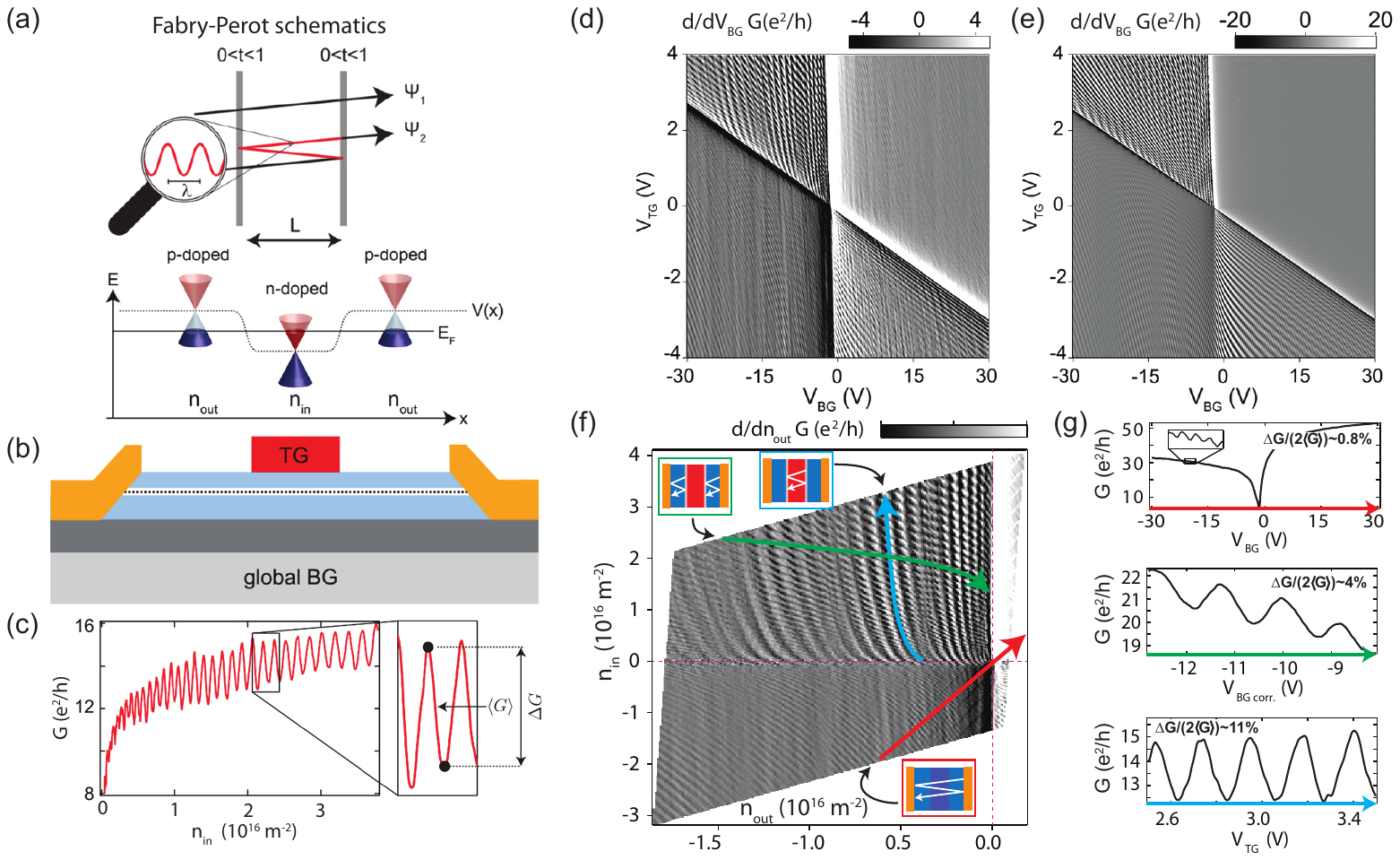}
	\caption{\textbf{FP oscillations in p-n-p or n-p-n cavities.}
       (a) Schematics of a Fabry-P{\'e}rot (FP) setup with two planar mirrors separated by the distance $L$. The mirrors have a fixed transparency of $t$. In the lower part, the band\-structure is shown for such a three-section graphene device in a p-n-p gate configuration. (b) Cross section of the encapsulated graphene device with the two outer sections controlled by the global back gate (BG) and the middle one controlled by the local top gate (TG). In (c) the measured conductance $G(n)$ as a function of carrier density $n$ is shown for a typical device. The conductance modulation corresponds to $\sim 10$\%. (d,e) The derivative $d/dV_{\textrm{BG}}$ of the two-terminal conductance is shown as a function of back-gate and top-gate voltages. The left graph (d) is the experiment and the right one (e) the simulation. (f) $G$ as a function of normalized densities in the outer, $n_{\textrm{out}}$,  and inner region, $n_{\textrm{in}}$. On the right side (g) three cuts along different directions (red, green, blue) are shown. The modulation is clearly largest along the blue cut, which is the directions along which only the carrier density of the inner region is changed. (a-e) are adapted from~\cite{HandschinThesis2017}, the calculation in (e) was performed by Ming-Hau Liu and (f,g) are adapted from~\cite{Handschin2017a}.}
	\label{fig:FigCS2}
\end{figure*}

These findings imply that directional emission can be steered by breaking the cavity geometry's symmetry and through an anisotropic dispersion. Compared to standard mesoscopic optics, graphene electron optics provides an additional, fundamentally different further mechanism for symmetry breaking and steering electron beams. Besides, contrary to electromagnetic optics, the charge carrier dynamics in graphene can be further manipulated through external magnetic fields, opening another angle of research. {We discuss various aspects of $k$-space anisotropies in ballistic graphene-based systems in \sectionname\ref{subsec:Anisotropy}}.

\section{Quantum-interference in graphene p-n junctions} 

\subsection{Quantum-interference in zero magnetic field} \label{sec:Quantum-interference-in-zero-field}

\subsubsection{Fabry-P{\'e}rot interferometer based on p-n junctions.}
\label{sec:Fabry-Perot-interferometer}

In a ballistic 2D electron system, a combination of semitransparent mirrors can result in cavity resonances when wave propagation is coherent. A very well-known example from optics is the Fabry-P{\'e}rot interferometer. It consists of two parallel semitransparent mirrors that are spaced by a distance $L_c$. An illustration is shown in \fig\ref{fig:FigCS2}a. An optical plane wave incident from left under an angle $\Theta$ relative to the normal of the mirrors enters the cavity with transmission probability $T$. At the other mirror the wave is reflected with probability $R=1-T$. This process can be continued to a finite number of partial waves limited by temporal and spatial coherence. Once summing up all transmitted partial waves $\sum_i \psi_i$ one obtains for the transmitted intensity the well-known relation $1/(1+K\sin(\gamma/2)^2)$, where $K=4R/T^2$ is proportional to the square of the so called Finesse $F=\pi \sqrt{R}/T$ and $\gamma$ the propagation phase between two successive emitted partial waves: $\gamma=4\pi n L_c \cos(\Theta)/\lambda$. Here, $n$ is the refractive index and $\lambda$ the wavelength. The finesse can be seen as the quality factor, which in a practical optical etalon or narrow-band filter can take large values up to $F\sim 400 000$. In analogy, electrons in a ballistic 2D electron systems at the Fermi energy are coherent if temperature is small enough and phase randomization due, for example, gate noise can be neglected. Since p-n junctions in graphene have the properties of a semitransparent mirror, they can serve as elements to realize an electronic Fabry-P{\'e}rot interferometer. An example is sketched in \fig\ref{fig:FigCS2}b and a typical interference pattern as measured by the two-terminal conductance $G$, which is proportional to the total transmission probability, is shown in \fig\ref{fig:FigCS2}c. Here, an encapsulated graphene device is sketched that consists of three gateable regions: the central inner one and two outer regions that are gated by the back gate \textrm{BG}, while an additional top gate \textrm{TG} is used for the inner region. Here, we can realize a n-p-n, p-n-p or a general n-n’-n device. Let us assume that we gate the device into the p-n-p regime as indicated by the energy diagram. The measured interference pattern in $G$ is shown as a function of carrier density in the middle section $n_{\textrm{in}}$. The carrier density in the inner n-region is increased starting from the CNP at $n_{\textrm{in}}=0$ to $4 \cdot 10^{16}$~m$^{-2}$. The conductance increases with carrier density as expected, but on top of this general trend conductance oscillations are seen. The visibility is of order 10\%. This is very different to optical cavities where the visibility is $\sim~100$\%. This shows that the transmission probability of the mirrors in graphene, averaged over all angles is surprisingly large. This is due to Klein tunneling which leads to a large transmission at normal incidence. If we would only consider waves travelling normal to the p-n junctions, the visibility would disappear. It is important to remember that the low visibility in graphene interferometers based on p-n junctions as semitransparent mirrors is not caused by limitations in coherence but is intrinsic and caused by Klein tunneling~\cite{Shytov2008}.


The band\-structure of graphene at the tight-binding level is rather simple. However, combining all electrodes and contacts with graphene is not straightforward to model, but it can be performed using the scaling approach introduced in detail in \sectionname\ref{sec scalable tbm}. Figure \ref{fig:FigCS2} shows a comparison between a measurement (panel d) and the respective parameter-free simulation (panel e). The correspondence between the two is remarkable, even detailed features are reproduced. Note, to enhance the features, the derivative of the conductance versus the back-gate voltage is plotted here. There are four regions, clearly separated by two pronounced lines. The vertical line at $V_{\textrm{BG}}=0$ corresponds to the CNP in the outer two regions. The line with the negative slope corresponds respectively to the CNP of the middle region. This line has a finite slope because $n_{\textrm{in}}$ is tuned by both gate voltages. Of the four quadrants, the top left corresponds to the gating situation p-n-p, the top right to n-n’-n, bottom right to n-p-n, and bottom left to p-p’-p. We see Fabry-P{\'e}rot (FP) interference effects in all four quadrants, though they are the weakest in the n-n’-n case. In contrast, they are most pronounced in the bipolar regions. Here, the main contribution originates form the expected FP resonances formed in the inner region. For the unipolar situations, the electron waves propagate all the way from the left source to the right drain contacts, where reflections can happen, too. That the oscillations are fainter in the n-n’-n case as compared to the p-p’-p case suggests that the contacts are n-doped. In this case, there are also p-n junctions present at the contacts enhancing the reflection probability. It is instructive to inspect the interference pattern even further. A zoomed-in graph is shown in \fig\ref{fig:FigCS2}f, now plotted as a function of normalized coordinates, $n_{\textrm{in}}$ and $n_{\textrm{out}}$. The modulation in $G$ that one experiences in a cut along the blue line corresponds to the proper FP oscillations of the inner cavity (see panel g for the conductance traces along these lines). The oscillation along the green cut shows oscillations that are due to FP oscillations generated in the left and right outer regions separately. And finally, the red cut corresponds to unipolar gating with the largest cavity size. The visibility is the weakest in the red cut, intermediate for the green one, and largest for the blue cut.

FP resonances have become the key signature for claiming ballistic transport in all kinds of graphene devices~\cite{Shytov2008,Zhang2009,Divari2010,Campos2012,Rickhaus2013,Oksanen2014,Varlet2014,Calado2015,Liu2015,Taychatanapat2015,BenShalom2016,Allen2017,Handschin2017a,Nanda2017,Zhu2018,Pandey2019,Veyrat2019} which is only qualitatively correct as detailed below. FP resonances not only appear in the linear-response conductance, but also when superconductors are involved, for example in the critical current of graphene-based Josephson junctions~\cite{Calado2015,BenShalom2016,Nanda2017,Zhu2018} and in Andreev reflection in normal metal-graphene-superconductor devices~\cite{Pandey2019}. They would also show up in thermoelectrical properties~\cite{Divari2010, Jung2016} and higher moments in charge transfer, for example in noise properties. FP interferences go beyond monolayer graphene. They have been observed in bilayer~\cite{Varlet2014} and trilayer graphene~\cite{Campos2012}. Additionally, in graphene superlattices the secondary Dirac points can give rise to additional (quasi-)bipolar barriers, yielding more complex interference patterns~\cite{Handschin2017a, Kraft2020, Rickhaus2019}.

While the observation of FP interferences is widely used as evidence for ballistic transport, based on the bare observation one should not claim that the scattering mean-free path is larger than the sample size: $l_{\textrm{mfp}} > L$. Even if there is appreciable disorder, e.g. close to sample edges, there is still a distribution in scattering so that there might still be enough electron trajectories that remain ballistic in the interior of the sample.  A detailed study would require modelling the visibility of the interference pattern taking also finite temperature into account~\cite{Mueller2009}. As mentioned before, p-n junctions are never abrupt, and they are rather smooth in high-quality graphene devices which can be operated at lower carrier concentrations for which $\lambda_F$ is large. To assess the effective width of the p-n junction barriers, it is helpful to deduce the effective cavity length $L_c$ from the experiment. It can be estimated if we assume hard wall potentials bounding the cavity. For waves travelling normal to the cavity mirrors, constructive interference occurs if the path difference between the adjacent partial waves is a multiple of $\lambda_F$. This leads to the condition $2 L_c = j\lambda_F$, where $j$ is an integer. Since the Fermi wavelength $\lambda_F$ depends on the carrier density as $k_F=2\pi/\lambda_F=\sqrt{\pi n}$, we obtain for the $j$-th constructive interference maximum the condition $L_c\sqrt{n_j}=j\sqrt{\pi}$. In the experiment we can note down the carrier density for two adjacent conductance maxima, $n_{j+1}$ and $n_j$, to obtain for the effective cavity length $L_c$ the equation $L_{c,j}=\sqrt{\pi}/(\sqrt{n_{j+1}}-\sqrt{n_j})$. This length depends on the index $j$. In practice it should be a constant if the barrier were indeed hard wall potentials. However, it strongly varies in graphene devices. In a p-n-p graphene cavity, the extracted $L_{c,j}$ strongly depends both on the inner and outer carrier density. $L_c$ expands (shrinks) if the inner carrier density is increased (decreased), and vice versa, $L_c$ shrinks (expands) if the outer carrier density is increased (decreased).  The change in cavity length can be large and can account for a $\sim 100$\% change~\cite{HandschinThesis2017,Handschin2017a}. Until today, Fabry-P{\'e}rot interference effects were only studied in the low bias regime, where the oscillations display a checkerboard pattern in conductance on bias-gate maps. It would be interesting to explore interference effects also in the non-linear transport regime, which is a topic of increasing interest~\cite{Ideue2021}. The temperature dependence of FP resonances has been studied in detail in \reference\cite{Mueller2009}.

\subsection{Non-zero magnetic field (low field regime)} \label{sec:non-zero-low-magnetic-field}
\subsubsection{Different magnetic-field regimes.} \label{sec:different-magnetic-field-regimes}

\begin{figure*}[htbp]
	\centering
	\includegraphics[width=0.95\textwidth]{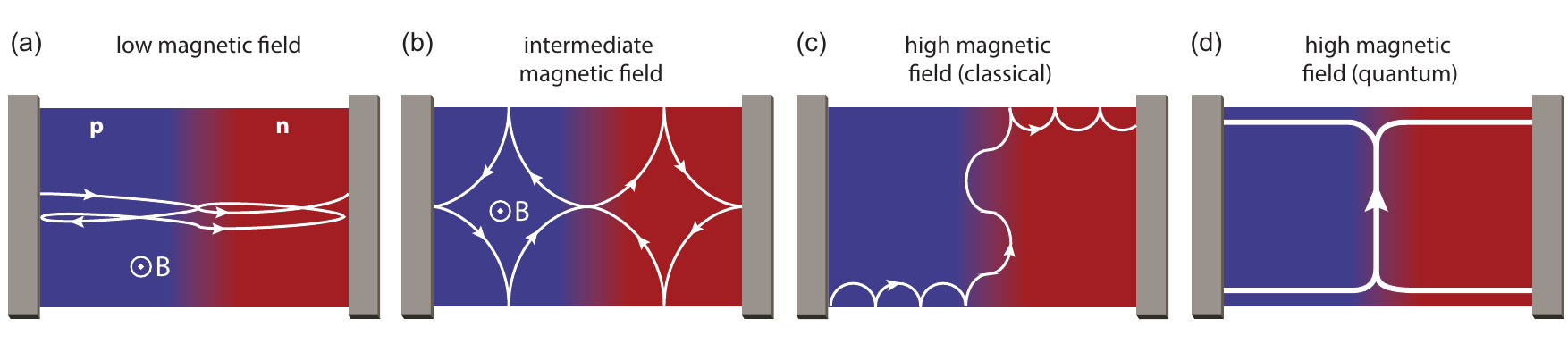}
	\caption{\textbf{Different field regimes.}
       This schematic emphasizes on the four different magnetic-field regimes. In the Fabry-P{\'e}rot regime (a) the straight classical electron trajectories are only slightly bent, while for larger fields (b) two-dimensional cavity bound states, also known as scar states, can form. If the magnetic field is as large that the cyclotron orbit is smaller than the sample dimension both in width and length, skipping orbits are formed (c). These orbits are chiral, meaning that there is a unidirectional propagation determined by the sign of magnetic field and the sign of the charge carriers (electron or holes) along each edge. At a p-n junction, and for a constant magnetic field, snake states lead to a current flow that follows the p-n junction, crossing one side of the sample to the other. If the field is even higher (d), one enters the quantum Hall regime where skipping orbits become one-dimensional edge channels. Image taken from~\cite{Rickhaus2015}.}
	\label{fig:FigCS3}
\end{figure*}

In the following, we introduce the four different magnetic field regimes for p-n junctions, depicted in \fig\ref{fig:FigCS3}. We assume phase-coherent ballistic transport and consider a single p-n junction in the center of a two-terminal graphene device. Since there are two cavities, Fabry-P{\'e}rot resonances may appear on both sides of the p-n junction, provided there is enough specular scattering of electron waves at the contacts. We have already mentioned that contact doping is a general phenomenon in graphene devices. It can be n or p-type depending on the contact material and fabrication processes involved. It is a parameter that is still today not very well understood. Let us further assume that the size of the sample is much larger than the Fermi wavelength $\lambda_F$ and that the graphene edge is ideal with a large probability for specular reflection. In this case, we can consider electron propagation in the form of wave packets that follow semiclassical electron-optical trajectories. A magnetic field $B$ applied perpendicular to the graphene plane causes a Lorentz force to act on the electrons which consequently are deflected. If the deflection measured at an angle is much smaller than one, we are in low field limit which is indicated in \fig\ref{fig:FigCS3}a. Here, The Fabry-P{\'e}rot condition for constructive interference is only slightly modified. But there is an interesting effect in the acquired phase which has been used as evidence for Klein tunneling~\cite{Young2009}. It will be discussed further below. The electrons in a homogeneous magnetic field will follow cyclotron motion, as detailed in \sectionname\ref{Sec:cyclotron}. The cyclotron radius, $r_c$ shrinks with increasing magnetic field. If $r_c$ is of order sample size (length or width), the corrections to the conventional Fabry-P{\'e}rot resonance are becoming large. New bound states can form, so called scar-states~\cite{Huang2009}. A very symmetric scar-state is indicated as an example in \fig\ref{fig:FigCS3}b. Scar states can be much more complex having intersecting electron trajectories, for example. If $r_c$ is smaller than the sample size, full cyclotron orbits fit into the two cavities. Now, so-called skipping orbits are formed along the edges of the device as seen in \fig\ref{fig:FigCS3}c. They lead to charge transport along the edges in a directional manner determined by both the sign of the magnetic field and the carrier type. The cyclic motion is opposite in a p-type materials as compared to an n-type one, as discussed in \sectionname\ref{semiclassic}. If we follow the cyclic motion along the edges, we see that propagation can be pinned to a p-n junction, due to the opposite chirality on opposite sides of the junctions. The charge motion now follows the partial cyclotron orbits along the p-n junction. This propagating state is known as snake-state~\cite{Ghosh2008,Oroszlany2008,Milovanovic2014a,Rickhaus2015,Taychatanapat2015,Kolasinski2017,Cohnitz2016,Bercioux2019}. This is a very peculiar addition to charge transport as it connects the two edges of the graphene device. Snake-states were proposed already before graphene for 2D electron systems realized in semiconducting hetero\-structures~\cite{Sim1998,Nogaret2000,Reijniers2000,Reijniers2002}. Since this electron gas is unipolar (electrons), one requires a magnetic field reversal in the two areas to obtain a snake state. This is much more difficult to realize. Hence, the ambipolar nature of graphene has given us a neat playground to study this special kind of electron state~\cite{Rickhaus2015,Taychatanapat2015}. If the magnetic field is increased further, the skipping orbits evolve into quantum Hall edge states~\cite{Halperin1982,MacDonald1984,Kane1987,Buettiker1988,Washburn1988, Makk2018}. To understand when and how one enters the quantum regime it is instructive to look at the cyclotron frequency $\omega_c = e B v_F / k_F$. To reach the quantum Hall regime, $\omega_c$ times the scattering time in the bulk should be larger than one. This ensures that there are full cycles that need to be quantized along the usual Sommerfeld-Bohr condition. Additionally, $\hbar\omega_c$ should be larger than the thermal energy. There is an additional length parameter, known as the magnetic length $l_B$ which follows from the Landau quantization: $l_B = \sqrt{h/eB}$. Now, we can formulate the two conditions required to remain in the regime where semiclassical electron orbits can be considered: $l_B \gg \lambda_F$ and $r_c \gg \lambda_F$. To remain in the semiclassical regime, the magnetic field cannot be too large. What is very interesting and peculiar is the dependence on electron density. The cyclotron radius shrinks with decreasing density. Hence, the closer one approaches the Dirac point (zero density), the less trustable is the assumption of semiclassical transport. If one crosses from an n-type to a p-type region, the semiclassical approach must break down in the center of the junction. It is therefore clear that a semiclassical approach to snake states can only deliver qualitative results and that a proper quantum treatment is required to make quantitative predictions.

Another interesting relation is given by $(l_B/\lambda_F)^2=r_c/\lambda_F=\nu/4\pi$, where $\nu$ denotes the filling factor given by $\nu=nh/eB$. Semiclassical transport, as discussed in the following two sections, requires large filling factors. To have an idea regarding absolute values, examples for typical graphene densities at a still low magnetic field of $0.1$~T are given in \tab\ref{Table QHE}:

\begin{table*}[ht]
  \centering
    
    \vspace{5mm}
    \label{tab:magnetic-field-parameters}
    \begin{tabular}{|c|c|c|c|c|c|}
      \hline \hline
      $V_g$~(V) & $n$~(cm$^{-2}$) & $\lambda_F$ & $r_c$~(at $0.1$~T) & $l_B$~(at $0.1$~T) & $\nu$~(at $0.1$~T) \\
      \hline
      $0.01$ & $10^8$ & $3$~$\mu$m & $12$~nm & $200$~nm & $0.04$ \\
      $0.1$ & $10^9$ & $1$~$\mu$m & $40$~nm & $200$~nm & $0.4$ \\
      $1.0$ & $10^{10}$ & $330$~nm & $120$~nm & $200$~nm & $4.0$ \\
      $10$ & $10^{11}$ & $100$~nm & $400$~nm & $200$~nm & $40$  \\
      \hline
    \end{tabular}
    \caption{This table illustrates the dependence of the Fermi wave\-length $\lambda_F$ and cyclotron radius $r_c$ at a relatively weak magnetic field of $0.1$~T on the gate voltage $V_g$ or the respective carrier-density $n$. The first two rows show cases where a classical description is not valid, as the filling factor $\nu$ drops below one. It is important to rationalize that while passing across a p-n junction the carrier-density must change sign. Hence, there is a region in the center of the p-n junction where a classical description in terms of electron trajectories can only provide an approximation.}
\label{Table QHE}
\end{table*}

\subsubsection{Correction to Fabry-P{\'e}rot oscillations for the lowest magnetic fields.}
\label{sec:corrections-to-FP-in-lowest-magnetic-fields}

\begin{figure*}[htbp]
	\centering
	\includegraphics[width=0.95\textwidth]{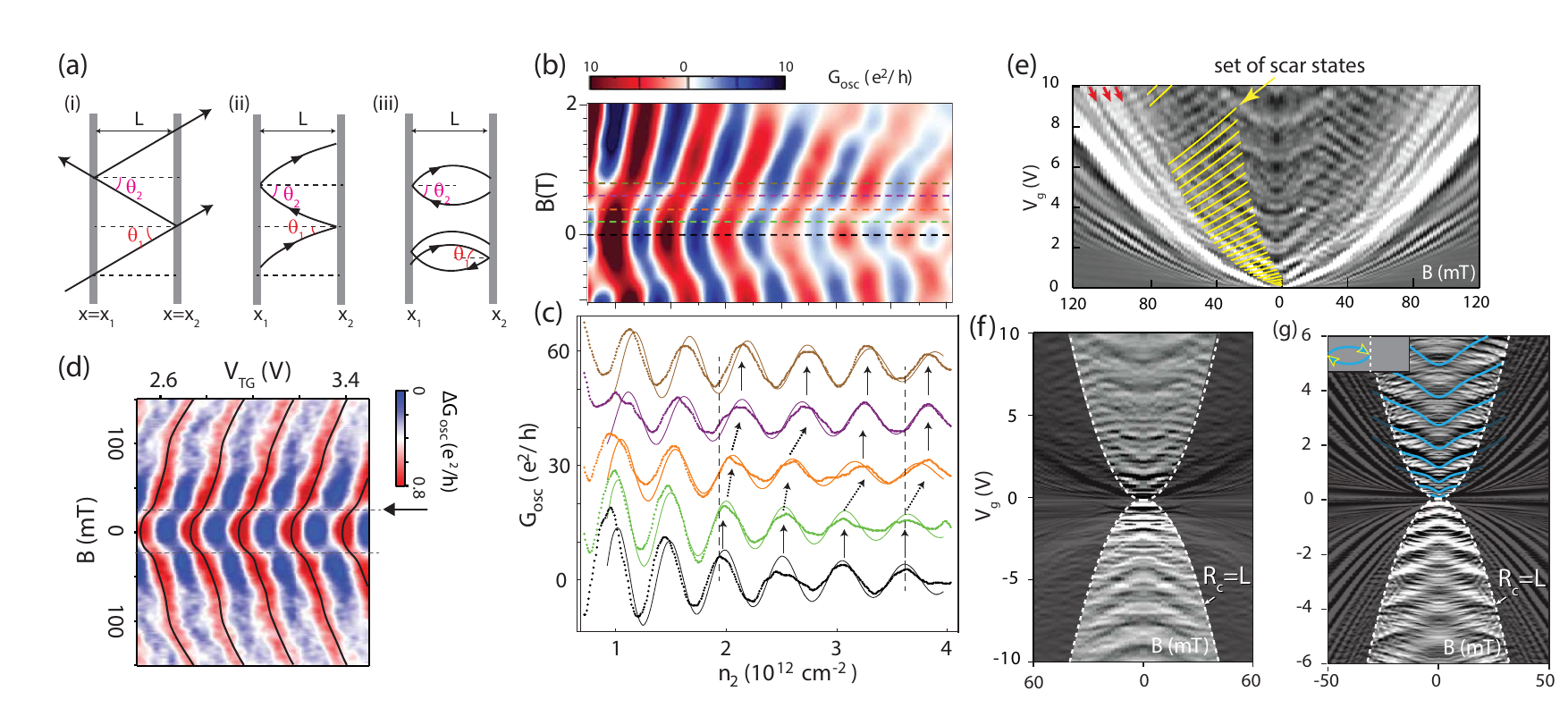}
	\caption{\textbf{Low and intermediate field regime with $\pi$-shift and scar states.}
      (a) schematically shows classical electron trajectories for a FP arrangement with two mirrors $m_1$ and $m_2$ at position $x_1$ and $x_2$ in (i) zero magnetic and (ii,iii) in a weak magnetic field. For the latter one must distinguish between two cases: (ii) when the two scattering angles have opposite or (iii) the same sign. The contribution from the symmetric loop, where the angles have the same sign, dominates for larger fields. This change in sign gives rise to a $\pi$-shift in the interference pattern. (b,c) Experimental data of the $\pi$-shift adapted from Young and Kim~\cite{Young2009}. 
      The shift appears at around $0.5$~T. (d) Similar experimental data obtained with an encapsulated graphene device, taken from \reference\cite{HandschinThesis2017}. Here, the shift already appears at a much lower field of $\sim 20$~mT. (e) shows another low-field interference pattern obtained from an ultra\-clean suspended graphene device, taken from \reference\cite{RickhausThesis2015}. The fine regular pattern, which is emphasized in yellow on the negative field side, is caused by scar states. Measurement (f) and a simulation (g) for interference in a single p-n junction as a function of magnetic field and gate axis taken along the bipolar direction. What is emphasized here, is the border between the low field and higher-field regime determined by the cavity size $L$ and the cyclotron radius $r_c$. The upwards dispersing curves emphasized in light blue belong to the small angle FP interferences. The additional interference pattern within the low field region is due to cavity states (scar states)~\cite{RickhausThesis2015}.}
	\label{fig:FigCS4}
\end{figure*}

For low magnetic fields the classical electron-optical trajectories bend very little. It is important to recall how the FP signal appears in a graphene p-n-p device without a magnetic field. Due to Klein tunneling, there are no contributions to the interference from trajectories that propagate normal to the two mirrors. These trajectories have angles $\Theta_1=\Theta_2=0$. Only trajectories with a finite angle add to the interference. Since the reflection amplitude at each mirror drops fast with angle, due to the typical soft potential steps in realistic devices, the main contribution to the interference is due to electron trajectories with finite but {\em small} angles~\cite{Young2009}. A scattering situation with finite angle in zero magnetic field is shown in \fig\ref{fig:FigCS4}a. The two angles at the two mirrors have opposite sign, $\Theta_1 = -\Theta_2$ in (i). As pointed out by Shytov~et~al.~\cite{Shytov2008} the reflection amplitude of the mirrors, $r_{1,2}(\Theta)$, must be an odd function, since $r_{1,2}$ goes through zero when $\Theta$ changes sign. This sign change is thus an additional feature of Klein tunneling. It has measurable consequences when a magnetic field bends the electron trajectories. There are bent trajectories of two kinds possible: (ii) one for which the two angles remain of opposite sign, and one kind for which the two angles have the same sign. In the former, and thus for small magnetic fields, a phase shift of $\pi$ adds to the interference contribution due to the product $r_1\cdot r_2$ acquiring a minus sign. With increasing magnetic field, the symmetric trajectories that enclose the origin start to dominate and the $\pi$ shift disappears. Hence, it has been proposed that a hallmark of Klein tunneling would be a $\pi$-shift of the FP oscillation pattern when a magnetic field is added~\cite{Shytov2008}. This has indeed been observed, first by Young and Kim~\cite{Young2009}. The magnetic field scale for this transition is determined by the inverse of the cavity area given by the length of the cavity times its width.

Figure \ref{fig:FigCS4}b shows the first experimental evidence for the $\pi$-shift due to Klein tunneling~\cite{Young2009}. One can see that the interference fringes shift a bit faster with magnetic field around a field value of $0.5$~T. The additional five cross-sections in (c) make this effect a bit clearer. Due to the generally observed dispersion of the interference fringes with magnetic field the $\pi$-shift is not as evident as one would like to have it. In higher mobility samples, the shift can show up in a more pronounced manner. This is illustrated with \fig\ref{fig:FigCS4}d where the cross-over already appears at around $25$~mT~\cite{HandschinThesis2017}. This result was obtained in an encapsulated graphene device with a much longer cavity length compared to the previous mentioned example. Figure \ref{fig:FigCS4}e shows FP resonances measured in a suspended ultra\-clean graphene device in a larger parameter range~\cite{RickhausThesis2015}. For magnetic fields $\gtrsim 15$~mT further oscillations appear. They are highlighted with yellow lines on the negative magnetic field side. These resonances are so-called scar-states. They were discovered in studies of bound states of quasi-classical trajectories of electrons in a two-dimensional electron gas with stadium boundaries~\cite{McDonald1979}. Unlike the FP resonances, which can be seen as one-dimensional bound states, scar states are cavity states that include scattering at the edges of the sample. Additionally, a measurement and a simulation of conductance oscillations in a graphene device with a single p-n junction is reproduced in \fig\ref{fig:FigCS4}f,g. Here the boundary between the ``low-field'' regime and the ``large-field'' regime is emphasized. The boundary is determined by the condition $r_c=L$, where $L$ is the cavity size and $r_c$ the cyclotron radius. Outside this boundary, states with a parabolic-like dispersion are seen. These are due to skipping orbits and will be discussed in the next chapter. The resonances emphasized in blue are FP resonances where the typical $\pi$-shift is evident too. In addition, one can clearly see many more interference features in the low-field regime caused to two-dimensional cavity states. A remarkable good agreement between simulation and experiment is found~\cite{RickhausThesis2015}.

\subsubsection{Skipping orbits and snake states at intermediate to high magnetic fields.}
\label{sec:skipping-orbits-and-sbake-states}

\begin{figure*}[htbp]
	\centering
	\includegraphics[width=0.95\textwidth]{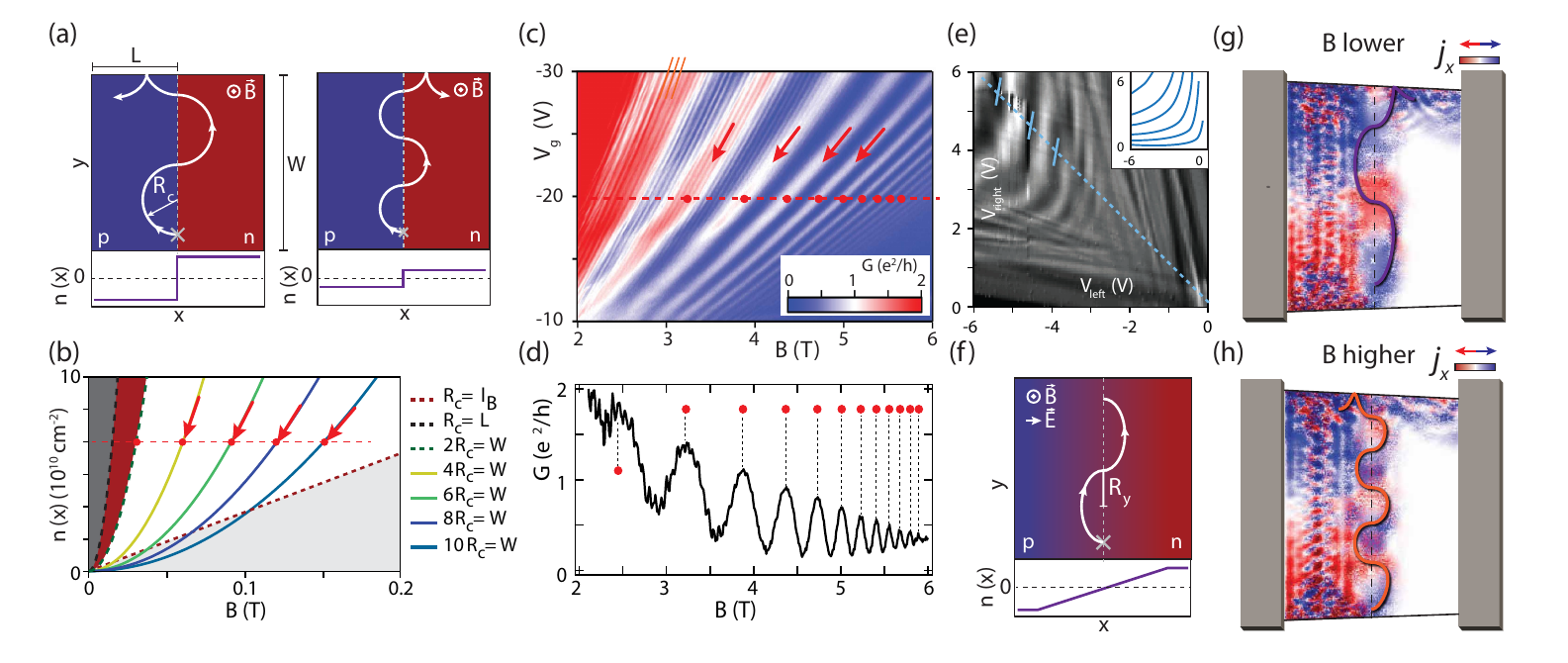}
	\caption{\textbf{Intermediate to high-field regime with quasi-classical skipping and snake-state orbits.}
       (a) Illustration of snake-states that propagate along p-n junctions. If we assume a common starting point (crosses) at the bottom, the trajectory may either end up on the left or right side, depending on the ratio between the width $W$ and the cyclotron radius $r_c$. This results in conductance oscillations. Lines for a fixed ratio $W/r_c$ are parabolic in a density $n$ versus magnetic field $B$ map, as seen in (b). For a constant density, neighboring peaks in $G(B)$ are expected to be equidistantly spaced (red dots). (c,d) show an experimental result obtained in an h-BN encapsulated graphene p-n device. The snake-state oscillation is emphasized by the red arrows and red points. It is found that the peaks are not equidistantly spaced. This is caused by a distortion of the orbits due to the gradual density change at the p-n junction, sketched in (f), see text for further explanations. (e) shows the oscillation pattern in a gate-gate map at constant magnetic field. The peaks and dips follow a hyperbolic-like pattern (inset). Finally, (g) and (f) show numerical simulations of p-n junctions for two different magnetic fields. Plotted in color is the current density along the horizontal axis. If one follows the pattern along the zero-density line (dashed), one can see that there is a periodic sign change consistent with the notion of snake-states. In addition to the snake-state, one can see other bound states residing on the left side. These simulations were performed by Ming-Hau Liu, see~\cite{RickhausThesis2015}. Note, there are further equidistant oscillations (orange lines) seen in (c). They are of Aharonov-Bohm type and are addresses in further chapters.} 
	\label{fig:FigCS5}
\end{figure*}

In the intermediate to high magnetic-field regime, the cyclotron radius $r_c$ is smaller than the length and width of the cavity: $r_c < L,W$. In agreement with \fig\ref{fig:FigCS3}c we assume here a graphene device with a single p-n junction in the middle. The p-doped region is on the left side and has a hole-carrier density of $p_{\textrm{left}} > 0$. In analogy, the n-doped region is on the right side with electron carrier density of $n_{\textrm{right}}=p_{\textrm{left}}$. We assume first, that the carrier-density jumps abruptly at the p-n junction located at coordinate $x=0$ (the $x$-axis points along the sample direction from source to drain and the $y$-axis is transverse along the p-n junction). In the classical electron optical picture and for a global constant perpendicular magnetic field $B$, the carrier trajectory would alternate between the p and n-side with skipping orbits being half circles with opposite chirality. This situation is illustrated in \fig\ref{fig:FigCS5}a. The picture also suggest why one should observe conductance oscillations due to these so-called snake-states~\cite{Davies2012,Chen2012,Milovanovic2013}. If the last half-cycle ends on the left side, the charge is reflected back to the source contact on the left. In contrast, if it ends on the right side, it will be transmitted to the drain contact on the right. Hence, the conductance should oscillate with a period given by $4r_c$, if we assume that $p_{\textrm{left}}=n_{\textrm{right}}=n>0$. The conductance modulation $\delta G$ can be written to be proportional to $\cos(\pi W/2r_c)$, leading to conductance minima and maxima whenever $W$ is an even multiple of $r_c$. Since $r_c$ is proportional to $\sqrt{n}/B$, lines of constant phase in the $n$ versus $B$ plane follow a parabolic dependence: $n\propto B^2$. This is sketched in \fig\ref{fig:FigCS5}b. In the experiment we expect the conductance modulation to follow the colored curves. The dashed red line denotes the condition $l_B=r_c$, with $l_B=\sqrt{h/eB}$ is the magnetic length. In the shaded region, where $l_B>r_c$, the quasi-classical description must break down. In this regime of large magnetic field and/or small carrier concentration, Landau quantization needs to be considered.

Figure \ref{fig:FigCS5}c shows an experimental result~\cite{RickhausThesis2015}. Note, the gate voltage $V_g$ controls both the carrier density on the p- and n-side with an equal magnitude. The stronger intensity modulation which starts with a spacing in magnetic field of $\sim 0.5$~T, which then seems to decrease, is thought to be due to snake-states~\cite{Taychatanapat2015,Rickhaus2015}. The previous reasoning for the snake-state oscillation predicts a constant period $\delta B$ in magnetic field at constant carrier density. This is not what is observed, as shown in panel d. Clearly, if we consider a horizontal cut at constant density, the spacing narrows with increasing magnetic field. In the data, there is another set of faster but weaker oscillations seen superimposed. This oscillation pattern is to a good approximation equidistant in magnetic field. It likely originates from Aharonov-Bohm oscillations due to edge-states forming along the p-n junctions at small densities~\cite{Morikawa2015,Makk2018}. This physics will be covered in \sectionname\ref{pn mach zehnder}.

One could also think that other quasi-classical electron trajectories could contribute to the current. For example, trajectories crossing the p-n junction not at normal incidence, but with a shallower or larger angle. However, one expects that these contribute less to the total conductance. The main contribution is due to trajectories that cross the interface at (or close to) normal incidence, since for those trajectories the transmission probability is maximal due to the Klein effect, while for all others the transmission probability is strongly suppressed. This argument does not really hold for a sharp step in carrier density, but only for a smooth potential changes. In real devices, however, the step varies smoothly over a length of $>20$~nm.

It is not straightforward to calculate the shape of the snake-state electron trajectory in a self-consistent manner, accounting accurately for the gradual potential change. The semicircles deform by elongating along the $y$-direction as indicated in \fig\ref{fig:FigCS5}f. However, it is possible to evaluate the length $R_y$ assuming a linear density change between the p and the n side~\cite{Makk2018}. One obtains:
\begin{equation}
  R_y=\left(\frac{\pi\hbar}{eB}\right)^2\frac{|p_{\textrm{left}}-n_{\textrm{right}}|}{2d} \textrm{.}
\end{equation}
Here, $d$ is the width over which the density changes from the p to the n-side. Interestingly, the skipping orbit length at fixed magnetic field is determined by the local electric field. Figure \ref{fig:FigCS5}e shows measurements at fixed magnetic field of $0.12$~T as a function of the two gate voltages $V_{\textrm{left}}$ and $V_{\textrm{right}}$ that control the densities in the two regions. The inset shows lines of constant electric field $E_x$ obtained numerically. The oscillation pattern appears in the gate-gate map in the form of hyperbola. The modified equation for the skipping-orbit length preserves the relation $n(B)\propto B^2$ that is followed by intensity maxima and minima as introduced with \fig\ref{fig:FigCS5}b. However, it does change the periodicity in magnetic field for constant densities. One does not expect a constant $\Delta B$ anymore, but rather $\Delta B\propto 1/B$. This fits much better to the experimental observation in \fig\ref{fig:FigCS5}c,d where it is evident that for a horizontal cut the spacing between adjacent conductance maxima decreases with increasing magnetic field. This observation can be taken as a confirmation that the potential drops gradually at the p-n junction.

Figure \ref{fig:FigCS5}f also shows 2D-simulations of current patterns. Here, the current component in $x$-direction is shown for two different scenarios: a small and a large field. If we concentrate on the region of the p-n junction, with the zero-density line indicated by a dashed line, we can recognize the alternating skipping pattern. Following along the dashed line from the bottom to the top, the current alternates. It is first blue (positive), then changes to red (negative), and so forth. Hence, the picture of commensurate snake orbits determining the conductance is appealing. However, it assumes a fixed starting point. This starting point is indicated in \fig\ref{fig:FigCS5}a with a cross. One might think that the conductance oscillation could average out if one varies the starting point. There has been simulation of this problem in which the sum of all trajectories was considered. Interestingly, caustics can show up causing repeated refocusing which in a similar manner can cause the conductance modulation~\cite{Davies2012,Lee2016}. If we look at the left half of the sample in the simulations, we see two things: (i) a fast oscillation pattern which is due to scar states, but (ii) we also see a peculiar blue region close to the bottom sample edge and an opposite red one close to the upper sample edge. These regions can be interpreted as the starting points of the quasi-classical picture with snake trajectories. The charge current is fed into the sample along the bottom edge with $j_x > 0$ (blue region). The current stream then follows the snake orbit. The part that is reflected ends on the top edge with $j_x < 0$ (red region). The part that is transmitted to the right can nicely been seen as skipping orbits along the upper edge in the right half of the sample. What exactly determines these ``starting points'' is at present not known.


Since graphene is a zero-bandgap semimetal, the Fermi energy can continuously by moved from the valence band into the conductance band. Hence, the discovery of graphene had made it possible to study snake state physics in quite some detail for the first time~\cite{Rickhaus2015,Taychatanapat2015}. Other geometries that were theoretically studied are cylindrical magnetic field patterns, which can be realized in graphene with a cylindrical gate. Such a gate can confine electrons through boundary snake-states. Many more electron-optical devices can be conceived based on confinement geometries defined by bipolar junctions. Further examples are discussed in the \sectionname\ref{sec:Optical elements} and  \sectionname\ref{sec:Collimation}.


\section{Edge state interferometers}
\label{edge}


In the previous section we saw that semiclassical arguments well describe transport in a wide range of magnetic field strengths, see \fig\ref{fig:FigCS3}a-c.  We now move to the quantum regime, when semiclassical skipping orbits and snake states break down and give way to fully formed quantum Hall edge channels  -- such channels exist wherever the electronic density changes enough for at least one Landau level to be crossed, which is for example the case both at the physical edge of the graphene flake and on each side of a p-n junction, see \fig\ref{fig_edge_states_pn}.  We thus present electron quantum optics developments in graphene setups using such edge channels as waveguides.  The first step is to outline the basics of quantum Hall transport.  Fundamental concepts, including the physics of decoherence, will be introduced from a general perspective, while the peculiarities of graphene will be highlighted when necessary.  The second step is a detailed review of different edge state interferometers realised in graphene, in particular Mach-Zehnder interferometers based on p-n junctions and Fabry-P{\'e}rot ones formed by quantum point contacts, much as in 2DEGs.
We also present some considerations, both experimental and theoretical, on decoherence effects in these experiments, and how they compare with their AlGaAs/GaAs counterpart.

\subsection{Chiral edge electronics: theory essentials}

The basics of integer quantum Hall transport can be understood within a single (quasi)particle picture via the Landauer-B\"uttiker formalism \cite{Buettiker1992,Datta1995}.  In the simplest linear-response scenario, electrons injected from a given reservoir propagate phase-coherently and independently from each other along the available 1D edge channels -- one for each filled Landau level -- and are finally absorbed by a second reservoir.  The conductances of arbitrary multi-terminal setups are obtained once the single-particle transmission amplitudes at the Fermi energy $t^{mn}_{\alpha\beta}$ are known, where $m\,(n)$ labels a given quantum channel from/into reservoir $\alpha\,(\beta)$.  Numerous extensions of the formalism were worked out, \eg to deal with non-linearities \cite{Christen1996} in various contexts \cite{Sanchez2004,Sanchez2013,Gorini2014,Texier2018} or to consider AC transport and current fluctuations via Floquet scattering theory \cite{Moskalets2002,Samuelsson2005}.  Though simple\footnote{Its simplicity is actually deceiving, as it hides numerous subtleties rooted in mesoscopic physics (non-locality of responses, role of contacts, invasiveness of probes\dots).  See \eg references~\cite{Akkermansbook,Imrybook} for some details.}, this approach provides a clear and remarkably well-working physical picture of edge transport. It is on this basis that some key concepts behind electron quantum optics eventually developed \cite{Ji2003,Samuelsson2004,Chung2005,Samuelsson2005}.  In spite of its successes, this intuitive construction has limitations coming from its two main requirements: (i) idealised, featureless 1D chiral edge channels in one-to-one correspondence with bulk Landau levels; (ii) Fermi liquid premises, \ie free quasielectrons propagating from reservoir to reservoir.  Though fair initial assumptions, neither turns out to be particularly accurate.  

Consider first the edge state problem, starting from a 2D system at ${\bf B}=0$.  The sample edges are defined by a confining potential $\Phi(\br)$, to which one can (semiclassicaly) associate a smooth local electronic density $n_0(\br)$ which decreases to zero as it approaches the sides. 
Now let ${\bf B}\neq0$.  The textbook picture of adiabatically bending the Landau level energies $E_n(k)$ as they approach the sample sides \cite{Datta1995}
\be
E_n(k) \to E_n(k) -e\Phi(\br_k),\;\br_k = (x_k,y),\;x_k=kl_B^2,
\ee 
yields sharply defined 1D channels with velocity $v_n \sim \partial_k E_n(k)$, and an associated electronic density $n_B(\br)$ which is discontinuous at each Landau level crossing.  This is shown in \fig\ref{fig_edge_states}a-c.  The smooth density profile at $B=0$, sketched in red, is massively distorted and becomes step-like when the magnetic field is switched on, $n_0(\br) \to n_B(\br)$.  It was realised \cite{Beenakker1990,Chang1990} and formalised \cite{Chklovskii1992, Chklovskii1993} early on that this cannot be accurate:  
the electrostatically defined profile $n_0(\br)$ can only be modified slightly in the presence of ${\bf B}$, since $\hbar\omega_c \ll |e\Phi_0|$.
At the same time the Landau levels do not simply adiabatically bend.  The qualitatively correct picture is shown in \fig\ref{fig_edge_states}d-f.  
\begin{figure}
	\includegraphics[width=0.49\textwidth]{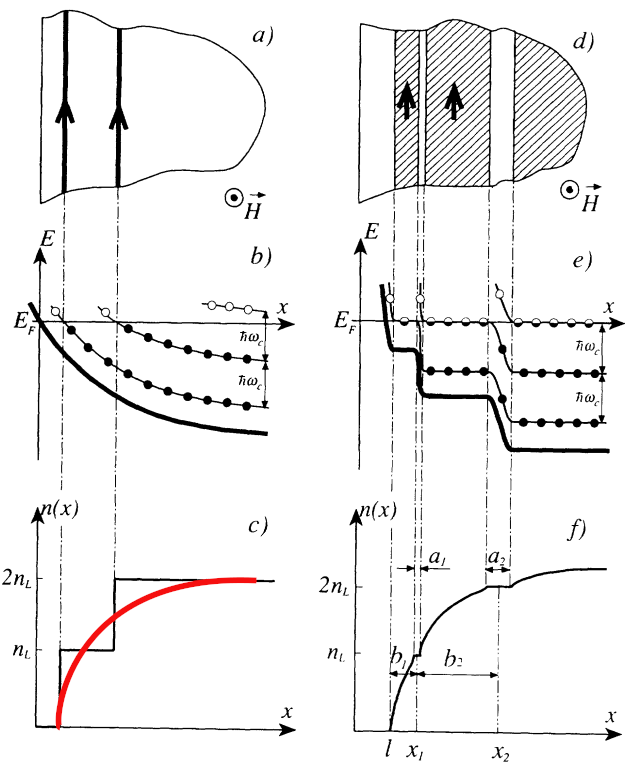}
	\caption{(a)-(c): Single-particle Landauer-B\"uttiker picture of widely-separated, exactly 1D edge channels.  Arrows indicate the propagation direction. (d)-(f): Chklovskii-Shklovskii-Glazman self-consistent (Hartree) picture.  Edge channels are compressible (shaded) strips of finite width a distance $b_i, i=1,2$ apart, separated by narrow incompressible (white) strips of width $a_i, i=1,2$.  Figure adapted from \reference\cite{Chklovskii1992}.}
\label{fig_edge_states}
\end{figure}
It implies the formation of {\it compressible} $(\partial_n \mu \neq 0)$ electronic strips -- the edge states -- separated by {\it incompressible} $(\partial_n \mu = \infty)$ regions -- the gapped Landau levels.  The problem must be solved self-consistently, since the electrostatic potential, the electronic density and the spectrum form a set of coupled non-linear equations.  The original construction by Chklovskii, Shklovskii, Glazman and Matveev \cite{Chklovskii1992, Chklovskii1993} was recently improved via self-consistent numerics \cite{Armagnat2020} and in this form applied to graphene p-n junctions \cite{Flor2022}, see \fig\ref{fig_edge_states_pn}.  It is a Hartree-level construction.
\begin{figure}
	\includegraphics[width=0.48\textwidth]{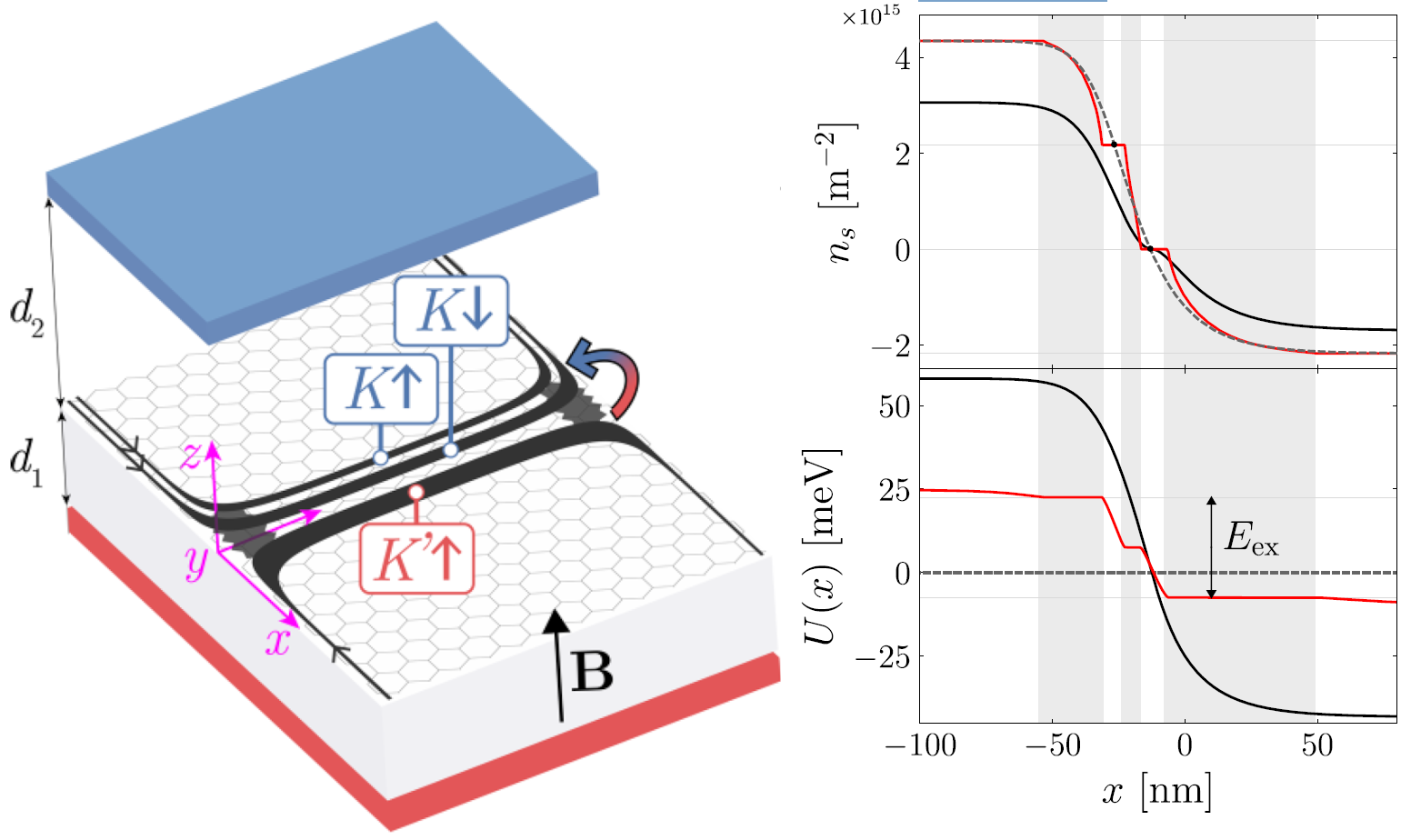}
	\caption{\textbf{Edge channel structure at a graphene p-n junction.} The top (blue) and bottom (red) gates define the junction and the filling factors $\nu$, here $\nu=2, -1$ respectively on the n and p side.  The n and p quantum Hall channels are compressible regions of finite width (shaded gray in the right panel) running along the graphene edges and following the junction profile.  They are labelled by valley $(K,K')$ and spin $(\uparrow,\downarrow)$.  The red curves in the right panel show the density and electrostatic profiles for $B=9T, d_1=d_2=20{\rm nm}$. Figure adapted from \reference\cite{Flor2022}.}
\label{fig_edge_states_pn}
\end{figure}
What if non-local exchange (Fock) is also taken into account?  Remarkably, in this case the bulk-boundary correspondence ``one edge state for each Landau level'' does not necessarily hold \cite{Chamon1994}.  This ``edge state reconstruction'' was experimentally confirmed in different systems \cite{Paradiso2012,Bhattacharyya2019}, and recently further investigated \cite{Khanna2021,Khanna2022}.  One concludes that the edge state properties of a quantum Hall droplet are less universal than those of the bulk topological phase.  In particular, the edges have in general a non-trivial internal structure possibly hosting multiple co- and counter-propagating modes of varying width -- which may also be more or less strongly coupled to one another, see below.  This is a general conclusion, affecting both integer and fractional quantum Hall phases.  In fact, since transport experiments probe edge state excitations, it is not always obvious how to relate these with excitations of the topological bulk, \eg the fractionally charged quasiparticles of a fractional quantum Hall phase \cite{Beenakker1990}.  Note that full edge reconstruction with counter-propagating modes is expected to take place for smoothly confined quantum Hall droplets \cite{Chamon1994,Khanna2022}, but the situation in graphene is varied, as it depends on how edges are experimentally realised.  Pristine edges obtained by exfoliation can be sharp and clean, showing no sign of reconstruction \cite{li2013, Coissard2023}, while in etched samples the formation of incompressible strips is important \cite{Cui2016}, and counter-propagating modes were also observed  \cite{seredinski2019}. Edges obtained by gating are clean but smooth on the $l_B$ scale at higher fields \cite{Armagnat2020, Flor2022}, so that full reconstruction cannot always be excluded \textit{a priori}. 

Let us now reconsider the Fermi liquid assumption.  This is somewhat questionable for narrow edge states, since in 1D the Fermi liquid picture breaks down and is substituted by the Luttinger liquid one \cite{Vignalebook}.  The breakdown is brought about by electron-electron interactions, whose critical importance in 1D can be formalised by the bosonisation procedure \cite{Giamarchibook}.  It turns out that the (almost) free quasiparticles of a 1D fermionic many-body state are not the quasielectrons of Fermi liquids, but collective bosonic modes, the simplest being charge density waves.  
Ideal 1D quantum Hall edge states are actually realisations of {\it chiral} Luttinger liquids, \ie, Luttinger liquids hosting either only left- or only right-propagating modes.  Explicitly, considering spinless electrons for simplicity's sake, the chiral (right) Luttinger model may be brought into the following basic form \footnote{See \eg \reference\cite{Vignalebook} for details.}  
\be
\label{eq_H_cLL}
H_{cLL} = \sum_{\bq>0} \hbar v_\bq q \, \hb^\dagger_\bq \hb_\bq,
\ee 
with $\hb^\dagger_\bq, \hb_\bq$ bosonic operators creating/annihilating collective density fluctuations propagating with velocity $v_\bq \geq 0$, the latter depending on details of the electron-electron interaction.  

An in-depth discussion of the Tomonaga-Luttinger low-energy model does not belong here \cite{Giamarchibook,Vignalebook,Haldane1981,Vondelft1998,Imambekov2012,Levchenko2021}.  It suffices to say that it is a standard starting point to study (time-dependent) quantum Hall edge transport \cite{Ferraro2014,Fujisawa2022}, though not always necessary \cite{Foerster2005,Marquardt2005,Chung2005,Lunde2010}.  
Indeed, the internal structure of the compressible edge strips (varying width, lack of perfect homogeneity) and the coexistence of different channels in close vicinity, see figures \ref{fig_edge_states} and \ref{fig_edge_states_pn}, remind us of the approximate nature of \eq\ref{eq_H_cLL}, and that strips are not ideal 1D objects.  Their width should notably influence their character (Fermi vs. Luttinger) and the velocity of excitations propagating through them \cite{Chalker2007,Armagnat2020}, as well as posing fine practical problems in the definition of a surface they may enclose \cite{Feldman2022}.  On the other hand disorder of different origins may cause phase transitions within the (quasi) 1D state, as well as enhancing inter-channel coupling \cite{Fujisawa2022,Vignalebook}.  The latter is a fundamental point, as inter-channel coupling is a major source of decoherence in quantum Hall setups \cite{Levkivskyi2008, Neder2006, Roulleau2007, Roulleau2008, Altimiras2010, LeSueur2010}, together with intrinsic non-linearities of the 1D electron liquid beyond the ideal Luttinger construction \cite{Haldane1981,Imambekov2012,Levchenko2021}.

Decoherence indicates the loss of phase memory of a quantum state.  It is intimately related with {\it irreversible} loss of (quantum) information, which takes place whenever the quantum state interacts with additional entities -- a heat bath, a fluctuating electromagnetic environment and so on -- whose dynamics is beyond our control \cite{Zurek2003,Akkermansbook}.  The quantum state here is an electronic excitation propagating along an edge channel, be it of Fermi or Luttinger nature.  Its phase-memory loss comes \eg from interactions with a bath \cite{Safi2004}, which may explicitly be classical \cite{Foerster2005} or quantum \cite{Marquardt2005,Degiovanni2009}, or from intra-channel \cite{Chalker2007,Neuenhahn2009} and inter-channel coupling \cite{Komiyama1989, Levkivskyi2008, Neder2006, Roulleau2007, Roulleau2008, LeSueur2010}.   

To be definite, consider the case of inter-channel coupling in an ideal Luttinger liquid scenario.  The propagating signal is an eigenmode of the Hamiltonian \eq\ref{eq_H_cLL}, thus by definition it never decays.  In presence of a second chiral Luttinger strip coupled to the first by $e$-$e$ interactions \cite{Chalker2007,Levkivskyi2008,Chirolli2013} the Hamiltonian becomes
\be
H = H_{cLL}^1 + H_{cLL}^2 + \delta H^{12}_{ee}.
\ee
Consider an excitation propagating through the first Luttinger strip.  If the second channel sits unbiased nearby it will act as a bath, which can be traced out breaking the unitarity of time evolution within the first channel.  The dynamics induced by mutual coupling is however richer in general, since excitations become coherent objects spreading across two channels \cite{Levkivskyi2008,Lunde2010,Chirolli2013}.  The latter may lose phase coherence by coupling to additional environment entities, \eg other channels, the bulk, nearby gates.  Notice that qualitatively similar conclusions would be reached if the strips were Fermi liquid in character.  In fact, decoherence processes can be modelled quite effectively via phenomenological ``B\"uttiker probes", \ie fictitious floating voltage probes which spoil phase memory without affecting the overall charge transfer \cite{Buettiker1986,Chung2005}.  This widely employed approach is successful also in graphene \cite{Ma2018}, but obviously cannot provide substantial microscopic insight.   

We have thus seen that the theory basis for electron quantum optics in the quantum Hall regime is the competition between electron-electron interactions (electrostatics and beyond) and the applied strong magnetic field.
Specifically concerning graphene as a platform, some of its characteristics set is aside from traditional semiconductor systems.  Besides its ``relativistic'' Landau level spectrum, see \sectionname\ref{Sec:QHE}, two are of central importance: (i) the valley and sublattice internal degrees of freedom (isospin) and their locking to momentum; (ii) the finer control one has over electrostatics and specifically screening, since metallic gates can be very close to the transport sample.   
The latter is in particular a great advantage, since it allows to realise compact p-n interferometers and to effectively screen edge channels from the environment and each other, largely increasing the coherence length within each \cite{jo2022scaling}.


\subsection{P-n junction based Mach-Zehnder interferometers}
\label{pn mach zehnder}
\vspace{-1em}
\begin{center}
\end{center}
\nopagebreak
\vspace{1em}



\begin{figure}[ht]
	\includegraphics[width=\columnwidth]{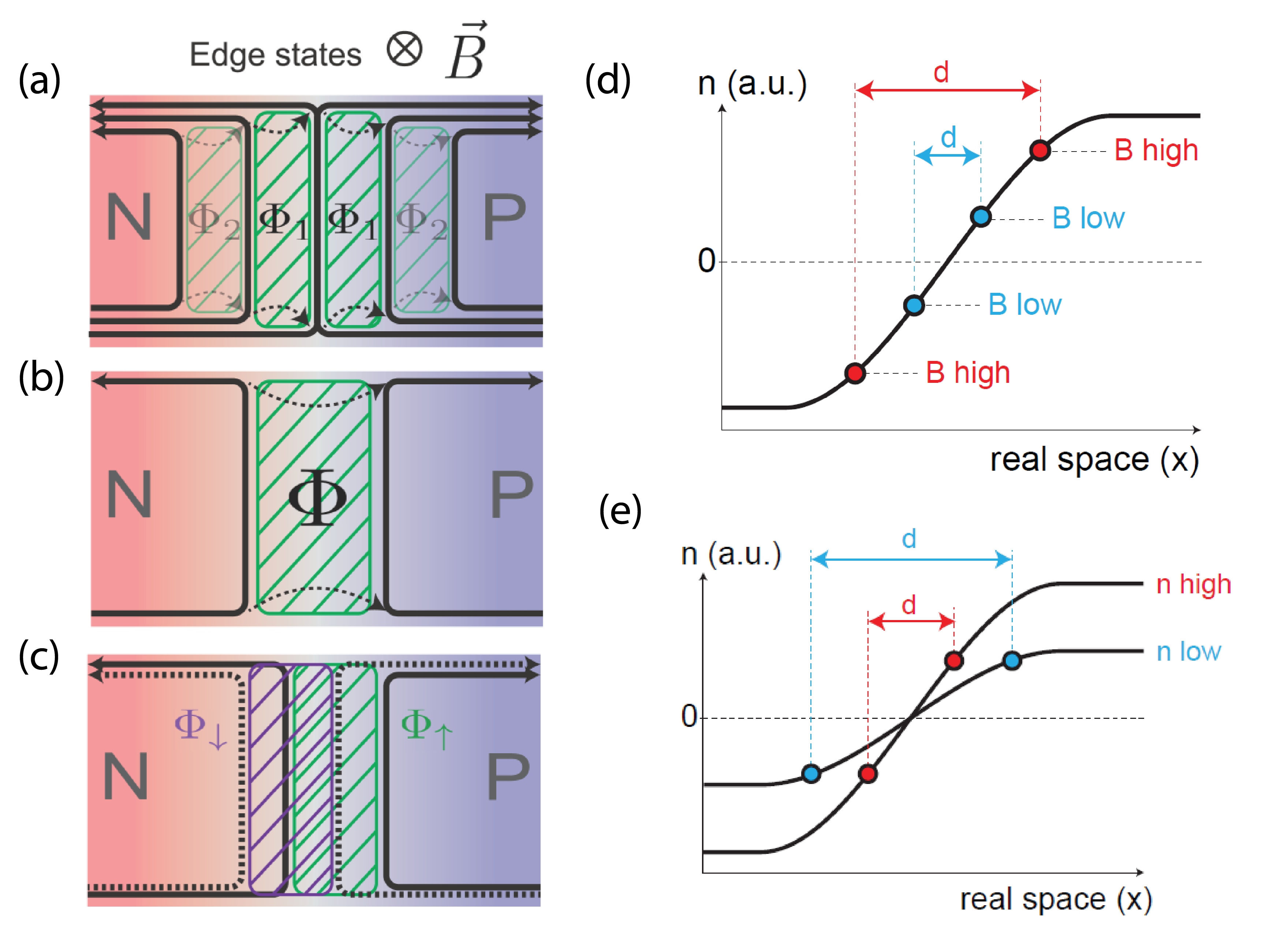}
	\caption{\small{\textbf{The formation of MZ interferometers in graphene p-n junctions.} (a) Edge states along a p-n junction, with the an electron-hole like edge states propagating along the p-n junction. Scattering between the channels is possible at the edges. Green dashed line mark possible MZ interformeter areas. (b) For larger fields the lowest LL is split up and the gapped region between electron and hole trajectories gives the interferometer's area. (c) If spin spilitting is present mixing is only possible between states with the same spin. (d) The positions of the edge states, that which reside at a given filling factor along the p-n junction are marked with small circles with at two different magnetic fields. (e) Position of the edge states is shown for two different dopings. Figure is reproduced from Ref.~\cite{HandschinThesis2017}}}
	\label{Fig_MZ_Intro1}
\end{figure}

The first observation of a graphene p-n junction based Mach-Zehnder (MZ) interferometer was reported by Morikawa and coworkers in \reference\cite{morikawa2015edge}. They have realized a p-n-p junction using two gate electrodes. At high magnetic fields edge states form which propagate along the sample edges, and the outermost electron and hole edge states, corresponding to filling factor 2 and -2, merge at the p-n interface where they co-propagate at $\nu=0$, as shown in \fig\ref{Fig_MZ_Intro1}a. However, if as written in \sectionname\ref{Sec:QHE}, the interactions split the lowest Landau level up, at lower field or moderate interaction strength, into a doublet ($K$ and $K'$) than the edge states at the p-n interface will move away from the interface to regions where the doping corresponds to $\nu=\pm1$ (separating gapped bulk regions of $\nu=0$ and $\nu=\pm2$). This is shown in panel b of Fig.\ref{Fig_MZ_Intro1}. As a result, co-propagating edge states are formed at the p-n junction. If there is no coupling of these edge states, this would lead to an insulating behaviour in transport measurements. As long as these edge states are close (smaller magnetic field) the coupling along the full length of the p-n interface will result in an oscillating motion of electrons originating from the source electrode, which is the quantum description of the quasi-classical snake states. Edge states with larger filling factor on both sides can participate in the formation of quasi-classical trajectories \cite{Makk2018}. As the magnetic field is increased, the edge states become separated in the bulk, but as was suggested by Morikawa and coworkers they can be coupled at the bottom and top edge of the sample, as shown in Figure~\ref{Fig_MZ_Intro1}b, and hence an interferometer loop can be formed. This interference loop results in an oscillating conductance, where the conductance maxima is separated by a change of flux quantum within the loop. Therefore one expects that the conductance maxima follows lines on gate-gate or gate-magnetic field maps, where the flux within the interference loop is constant. 

For simplicity, let us assume, that a symmetric, smooth p-n junction is formed. In this case, by increasing the doping on the two sides, the density gradient at the p-n junction increases, therefore the edge states which reside at fix filling factor move inward, as shown in panel f of \fig\ref{Fig_MZ_Intro1}. In order to keep the flux constant, the magnetic field has to be increased. Without a detailed derivation, this leads to parabolic-like oscillation maxima in the conductance in gate-magnetic field maps, as can be in \fig\ref{Fig_MZ_Intro2}a.
Using similar arguments, it can be shown, that for constant magnetic field if one of the gate voltages is changed, to keep the area constant the other gate needs to be used for compensation, leading to hyperbolic lines in the conductance in gate-gate maps as shown in \fig\ref{Fig_MZ_Intro2}b-d. 

\begin{figure*}[ht]
	\vspace{-10pt}
	\begin{center} 
		\includegraphics[width=\textwidth]{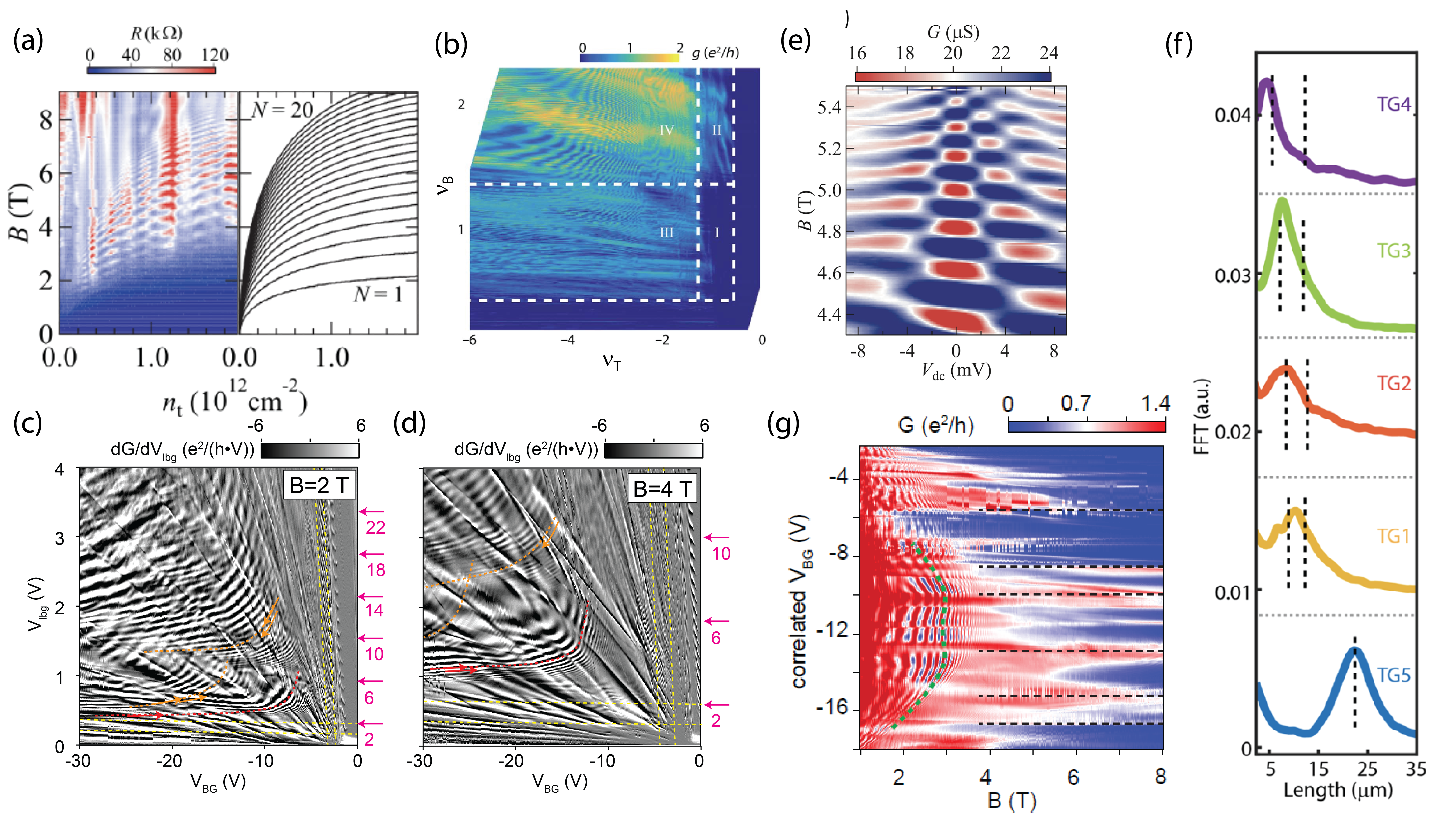}
	\end{center}
	\caption{\small{\textbf{First graphene based MZ interferometers.} (a) Magneto-resistance oscillations dispersing as a function of B-field  and doping. Simple calculations reproducing the findings based on the edge state positions is shown in the right \cite{morikawa2015edge}. b) Magneto-conductance oscillation as a function of the filling factors on the two sides of the p-n junction \cite{Wei2017}. (c)-(d) Similar features shown from Ref.~\cite{Makk2018} in 2 and 4T magnetic field, respectively, as a function of the two gate voltages. The derivative of the conductance is shown to highlight the oscillations. Different oscillations marked with different colors are attributed to different origin. (e) Bias dependence of the magneto conductance oscillation as function of magnetic field demonstrating a checkerboard pattern \cite{morikawa2015edge}.  (f) Oscillation periodicity for MZ oscillations is extracted for different samples, and the from that length of the interferometer is calculated, matching nicely the dependence on p-n junction length \cite{Wei2017}. (g) Magnetoconductance oscillations attributed to valley-isospin physics shown in gate-B-field maps with horizontal lines. For this figure the gate voltage is tuned such that only the p-n junction position is changed, whereas the potential profile remains the same \cite{Handschin2017b}.}}
	\label{Fig_MZ_Intro2}
    \vspace{20pt}
\end{figure*}

If the magnetic field or the interaction strength is further increased then the lowest Landau level splits further. In this case, spin and valley split edge states co-propagate along the p-n interface (\fig\ref{Fig_MZ_Intro1}c). Since spin-scattering is prohibited even on the edge of the sample, only edge states with the same spin can mix. This was first observed by in \reference\cite{Wei2017} (\fig\ref{Fig_MZ_Intro2}b) and later in \reference\cite{Makk2018} (panel c-d). The Mach-Zehnder interpretation was also corroborated in \reference\cite{Wei2017} by investigating interferometers with different p-n junction lengths leading to different oscillation periodicity (\fig\ref{Fig_MZ_Intro2}f). Finally, the bias dependence of these oscillations was also studied, for which an example is shown in \fig\ref{Fig_MZ_Intro2}e, more details can be found in references~\cite{morikawa2015edge, Wei2017, Makk2018}.  
We note that even in the non-split Landau level case (\fig\ref{Fig_MZ_Intro1}a) Aharonov-Bohm oscillations are possible between, e.g., the lowest and higher lying Landau levels. Since all these different MZ oscillations and the snake states give similar signatures in both magnetic field-gate and gate-gate maps, see, e.g., all the oscillations in \fig\ref{Fig_MZ_Intro2}c-d with different colors, their identification is possible only based on further bias or temperature dependent measurements. Whereas snake states can be observed up to $100\,$K, MZ oscillations disappear at few Kelvins or below.
Finally we would draw the attention to another set of very prominent oscillations which stem radially from the charge neutrality point (marked by black dashed lines in \fig\ref{Fig_MZ_Intro2}g. As seen in panel (g) they seem to be magnetic field dependent. Along such lines and the position of the p-n junction is fixed and these oscillations have been attributed to valley-isospin oscillations probing the microscopic character of the edges \cite{Handschin2017b}. This is discussed in the next section.

\subsection{Tunable Mach Zehnder inteferometers}
\vspace{-1em}
\nopagebreak
\vspace{1em}

A controlled approach to achieve Mach-Zehnder interferences was demonstrated in a recent study 
\cite{jo2021quantum} where electronic beam splitters were utilized, leveraging the valley degree of freedom in graphene. The concept of valley beam splitters builds upon theoretical work by \cite{Tworzydlo,Trifunovic} and earlier experimental work of Refs. \cite{Handschin2017b, Rehmann2019}, where the crystalline structure at the corner of a graphene p-n junction enables electron scattering between p-n interface channels with opposite valley polarizations of quantum Hall edge channels.
In the experiment, the researchers employed small electrostatic side gates to tune the mixing point of the edge channels along the edge of the graphene flake, thereby controlling the scattering process. This allowed for the reliable modulation of electronic transmission through the valley beam splitters, ranging from zero to near unity. Notably, this work demonstrated the complete tunability of Aharonov-Bohm (AB) interference by adjusting the side gate voltage and magnetic field. The resulting AB oscillations exhibited stability and reproducibility.\\

The sample schematic is depicted in \fig\ref{fig4}a, illustrating an encapsulated graphene in a bipolar quantum Hall state. In the N region, the Landau-level filling factor is $\nu_\mathrm{N}=2$, resulting in two counterclockwise circulating channels with opposite spins ($\uparrow$, $\downarrow$) along the boundary. On the other hand, the p region has a filling factor of $\nu_\mathrm{P}=-1$, featuring only one clockwise circulating spin-down channel.\\
When an injected current of $I_0/2$ carried by spin-down carriers is introduced, it can interact with the edge current flowing from the p region. Consequently, this interaction leads to a contribution to the transmitted current $I_\mathrm{T}$. The flow of the spin-down current is regulated by splitting it into p-n interface channels that possess opposite valley isospins \cite{Wei2018}.
After the physical top edge of graphene intersects with the electrostatically defined p-n interface, the current proceeds along either the p -side or the n -side of the interface. The transmission probability along the P -side is denoted as $T_1 = |t_1|^2$, while the reflection probability on the N side is given by $|r_1|^2 = 1 - T_1$.
In the presence of a strong perpendicular magnetic field, the valley degeneracy is lifted, as discussed in \sectionname\ref{Sec:QHE}. As a result, the currents on the n -side and p -side exhibit opposite valley isospins, represented as $\pm \vec{w}$. The probability $T_1$ reflects the degree of valley-channel splitting, which can be described by a quantum-mechanical superposition.

\begin{equation}
|\Psi_\mathrm{initial} \rangle =r_1|\uparrow,\vec{w}\rangle+t_1|\uparrow,-\vec{w}\rangle 
\label{initialstate}
\end{equation}
At the n-side interface, the spin-up state is represented as $|\uparrow,\vec{w}\rangle$, while at the p-side interface, it is denoted as $|\uparrow, -\vec{w}\rangle$. The valley-isospin undergoes a change from that of the top edge channel to $\pm \vec{w}$, resulting in a significant momentum shift. This change is attributed to the atomic structure at the intersection \cite{Trifunovic}.\\

By applying voltages to the side gates, they could modify the electrostatic potential profile at both ends of the p-n interface, as depicted in the schematic diagram \fig\ref{Fig_MZ_Intro1}. In a recent investigation employing the Chklovski-Shklovskii-Glazman formalism, the precise positioning of edge states in a graphene p-n junction was determined through rigorous quantitative calculations \cite{Flor2022}.  
When the filling factor below a side gate was set to $\nu \leq -1$, the p-n junction intersected the physical edge, creating a sharp potential change at the atomic distance scale. This sharp potential change facilitated the mixing of the valley channels. On the other hand, by setting the filling factor to $\nu = 0$, the p-n junction intersected an electrically defined edge where the potential landscape was smooth, resulting in no change in valley isospin.
Through the manipulation of these side gates and the associated filling factors, the researchers could control the extent of valley-channel splitting and restrict the mixing of valleys. This allowed for precise control over the transmission probability and the preservation of valley isospin within the system.\\

\begin{figure*}[ht]
	\begin{center}
		\includegraphics[width=\textwidth]{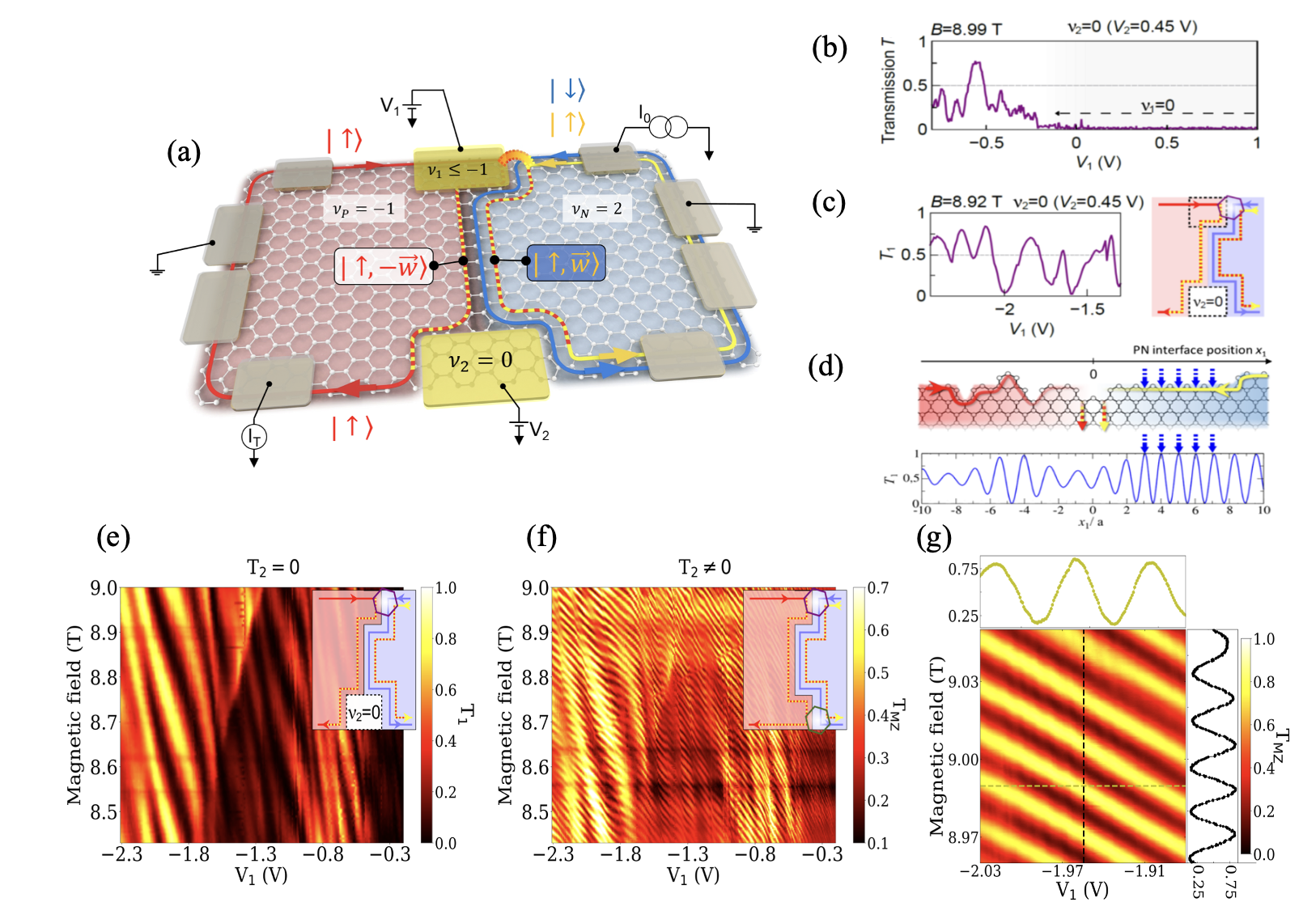}
	\end{center}
	\caption{\small{(a) Schematic of a fully tunable Mach-Zehnder interferometer using valley degrees of freedom of graphene \cite{jo2021quantum}. (b) Measured transmission $T_1$ of the top valley splitter as a function of the side gate voltage i.e controlling the filling fraction below it. (c) Oscillation of transmission of a valley splitter. Right panel: The edge state configuration in this condiction. (d) KWANT simulation of transmission ($T_1$) as a function of the position along the p-n interface. (e) Measured $T_1$ as a function of gate voltage and magnetic field. (f) Transmission in the MZ interferometer configuration ($T_{MZ}$) when both (top and bottom) the splitters are allowed to have valley mixing. (g) $T_{MZ}$ as a function of magnetic field and gate voltage.   }}
	\label{fig4}
\end{figure*}

The authors of this study first demonstrated the ability to tune the transmission probability $T_1$, defined as the ratio of transmitted current $I_\mathrm{T}$ to half of the injected current $I_\mathrm{0}/2$. This tuning was achieved by adjusting the voltage $V_1$ applied to the top side gate, as shown in \fig\ref{fig4}b.
To achieve valley-channel splitting at the top intersection while suppressing it at the bottom, they set the filling factor $\nu_1$ below the top side gate to $\nu_1 \leq -1$, and $\nu_2$ below the bottom side gate to $\nu_2 = 0$. When a positive non-zero voltage was applied to $V_1$, it resulted in $\nu_1 = 0$, ensuring that the edge channels only intersected at electrostatically defined edges without valley-channel splitting, leading to a vanishing transmission.
Conversely, for $V_1 < 0$, they ensured $\nu_1 \leq -1$, causing the p-n junction to intersect the top physical edge. This enabled valley-channel splitting and resulted in a finite transmission. Subsequently, the authors demonstrated the full tunability of the transmission probability $T_1$ from zero to nearly unity, as depicted in \fig\ref{fig4}c, by varying the voltage $V_1$. Importantly, they also showed that $T_1$ could be tuned by changing the magnetic field, as illustrated in \fig\ref{fig4}e.
The period of the dependence on $V_1$ was estimated to be around $\Delta V_1 \sim 100$ meV on average. It was calculated that this change in voltage caused the p-n interface to shift by approximately $\sim 1$ nm below the top side gate.
The period of the magnetic field ($B$) dependence was found to be $\Delta B_1 \sim 300$ mT, corresponding to a change of approximately $\sim 0.2$ nm in the magnetic length at $B = 9.2$ T. These length scales are comparable to the interatomic distance of pristine graphene (approximately $0.142$ nm) and the period of atomic edge structures (e.g., $0.246$ nm for the zigzag edge), but significantly shorter than the spatial variation of the electrostatic potential induced by gate voltage. This strongly suggests that the transmission probability $T_1$ can be controlled by the atomic structure at the top intersection. The shift of the p-n interface, estimated from experimental data, was made possible by independent control of the top and bottom side gates, which was not achievable in previous works \cite{Wei2017,Handschin2017b}. Similar experiments were conducted with the bottom side gate, yielding a comparable physical scenario and trend in the results. In the transmission results, some irregular but reproducible oscillations were observed. These oscillations were attributed to the roughness in the upper physical edge of the graphene. The theoretical simulation using KWANT simulations, as depicted in \fig\ref{fig4}d, supported this explanation. Overall, the experiments and simulations provided strong evidence that the transmission probability $T_1$ in the system could be controlled by the atomic structure at the top intersection, and the observed irregularities in the results were attributed to edge roughness effects.\\

Finally, the investigated the Mach-Zehnder interferometer by utilizing both the top and bottom valley splitters. The edge channels from the two sides of the p-n junction acted as the arms of the interferometer, while the valley splitters served as the beam splitters, as illustrated in \fig\ref{fig4}(f). Notably, smaller-scale Aharonov-Bohm oscillations were observed with a magnetic field period of approximately $25$ mT. The area of the interferometer was calculated to be $0.15 \ \mu m^2$, indicating a separation of around $110$ nm between the edge channels. This separation is attributed to electron-electron interactions. When both valley splitters were set to a half-transmission configuration, the interferometer exhibited regular oscillations, as depicted in \fig\ref{fig4}g. The visibility of the Mach-Zehnder interferometer, defined as $(T_{\mathrm{max}}-T_{\mathrm{min}})/(T_{\mathrm{max}}+T_{\mathrm{min}})$, was approximately $60\% $.\\

\begin{figure*}[t]
	\begin{center}
		\includegraphics[width=\textwidth]{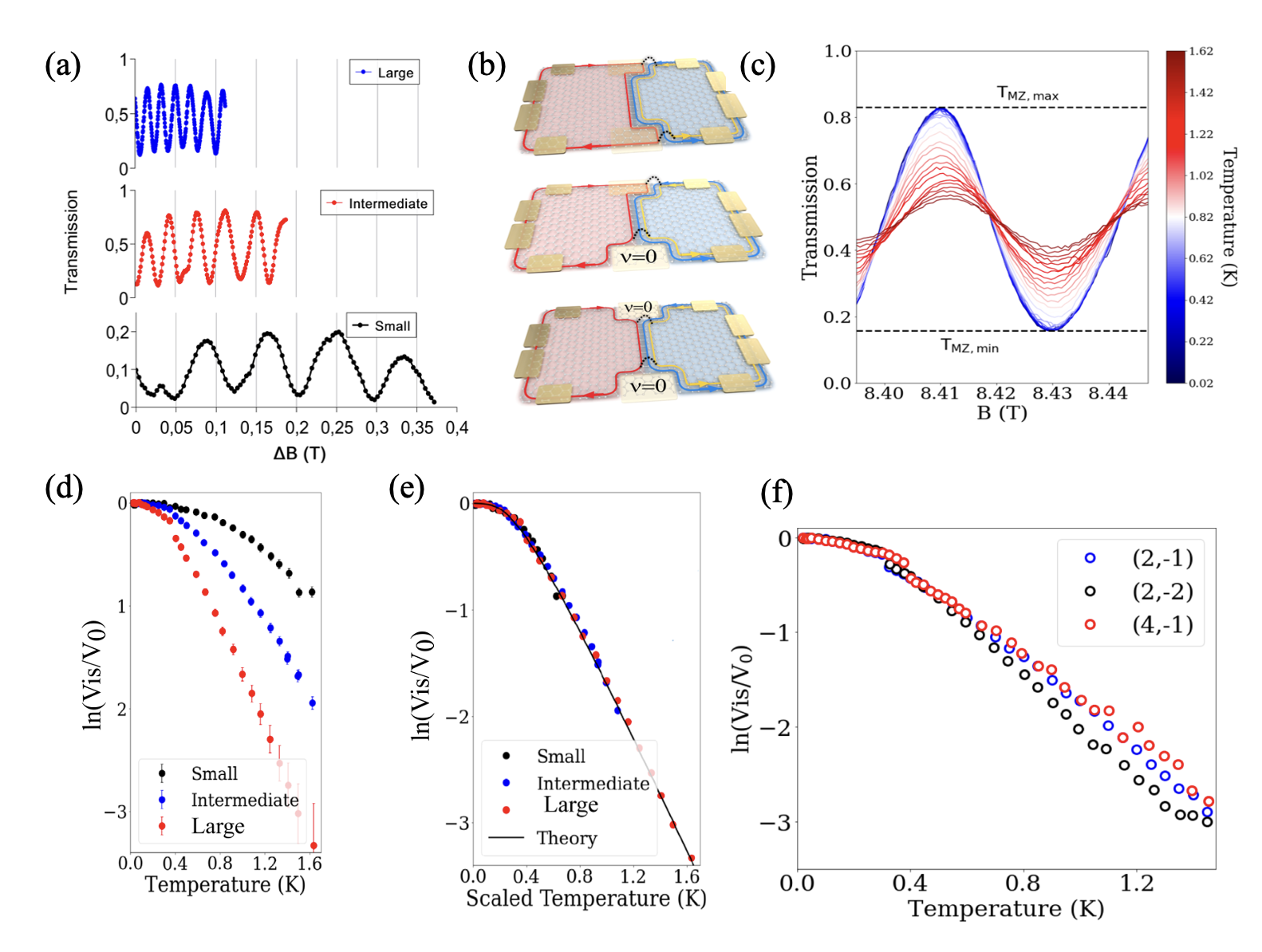}
	\end{center}
	\caption{\small{(a) AB oscillation in three MZ interferometer of different lengths. (b) The schematic of channel mixing points in the interferometer. (c)Temperature dependence of the transmission for the large interferometer. (d) Visibility decay of with temperature for all the interferometer. (e) Universal scaling behaviour of visibility.  (f) Visibility decay profile in presence of different number of  edge channels. \cite{jo2022scaling}.  }}
	\label{fig5H}
\end{figure*}

The paper also discussed the coherence properties of the valley-split state in relation to the energy of the transported electrons. A lobe pattern was observed in the transmission probability $T_{MZ}$ as a function of the bias voltage, which is a typical behavior in Mach-Zehnder interferometers (MZIs) fabricated in conventional GaAs heterostructures. The microscopic origin of this lobe pattern is now well understood and discussed in detail in the \reference\cite{jo2022scaling}. 
Moreover for the graphene MZI studied in these works, the value of $V_{lob}$ was found to be $210$ $\mu$eV, which is relatively large compared to the reported value of $20 \ \mu$eV for MZIs fabricated in GaAs/AlGaAs heterostructures. This suggests that the graphene MZI exhibits robust phase coherence over a wider energy range. \\
This robustness of phase coherence of graphene MZI has been harnessed recently to demonstrate coherent phase manipulation of periodically injected single electronic state\cite{assouline2023emission}. This development opens up further avenues for exploring electronic experiment analogous to optics in graphene platform.

\subsection{Decoherence and relaxation in quantum Hall Mach-Zehnder interferometers.}
\begin{center}
\end{center}

Single electron coherence in quantum edge channels is notably highlighted by Mach–Zehnder interferometry experiments. But it also digs up a number of questions on the electron decoherence mechanism in these systems. A large majority of experiments dealing with quantum Hall interferometers in conventional semiconductors suffers from decoherence, which can come from different sources like edge reconstruction due to the presence of impurity around \cite{Chamon1994}, inter-edge interaction \cite{seelig2001charge, youn2008nonequilibrium} and intra-edge Coulomb interaction \cite{roulleau2008noise, Levkivskyi2008, bocquillon2013separation}. Most of the time, those interactions are intertwined which makes it hard to address them separately. Within the last decade, the majority of experimental and theoretical works addressed the issue of inter-edge interaction \cite{Neder2006, Roulleau2007, Sukhorukov2007,Litvin2008, Roulleau2008, roulleau2008noise, Levkivskyi2008, Degiovanni2009, Altimiras2009, LeSueur2010,Degiovanni2010,Huynh2012, Bocquillon2013,Ferraro2014, Freulon2015, Lunde2016, Marguerite2016, Gurman2016, Tewari2016,Marguerite2017,Itoh2018, Cabart2018, Duprez2019,Rodriguez2020,Fujisawa2022} but there is still an ongoing debate about the observed results. One possible way out is demonstrated in \cite{jo2022scaling} by using a fully tunable graphene MZ interferometer utilizing p-n junctions. In the experiment, three interferometers of different lengths (\fig\ref{fig5H}a-c) were studied, showing a persistence of the interferences up to $1.6$ K, relatively high compared to the operating temperature of GaAs interferometers. The visibility of the interferences, as shown in \fig\ref{fig5H}d, was shown to have two distinct regimes in temperature. Visibility decay is found to be algebraic instead of exponential below 1K which signifies the suppression of thermal route of decoherence. This ‘new-found’ algebraic decay regime was not observed before in conventional semiconductor system, and it does not depend on the different configurations (e.g. different edge configurations of p and n side) of the interferometer. Electron heating giving rise to this kind of scaling behavior was ruled out by careful measurement of thermal noise at each temperature confirming the electrons to be well thermalized. Interestingly, the decay profile for all the three interferometers lies on a single curve (\fig\ref{fig5H}e) if plotted against a  scaled temperature $LT/L_{0}$ where $T$ is temperature, $L$ is the interferometer length, and $L_{0}$ is the length of the large interferometer. This scaling behaviour is in good agreement with an intra-channel interaction model. To access the effect of the presence of adjacent edge channels, the temperature dependence of interference visibility was monitored by changing the number of edge channels which is shown in \fig\ref{fig5H}f. The short-range inter-channel interaction fractionalizes the electron flow in fast and slow modes which causes decoherence. No significant change in decay profile was observed, indicating the absence of influence of inter-edge channel interaction mechanisms. In a van-der-Waals architecture, the possibility to position the electrostatic gates very close (vertical distance~ $30$~nm) to the 2DEG  provides screening between nearest edge channels.  Therefore, the decoherence mechanism due to inter-edge channel interaction can be efficiently suppressed and one can only talk about the intra-edge channel interaction.  

Moreover, at high bias, magnons can be emitted which are also a source of decoherence \cite{wei2018electrical, assouline2021excitonic}.


\subsection{Graphene quantum point contacts in the QH regime}
\vspace{-1em}
\nopagebreak
\vspace{1em}

\begin{figure*}[ht]
	\begin{center}
		\includegraphics[width=\textwidth]{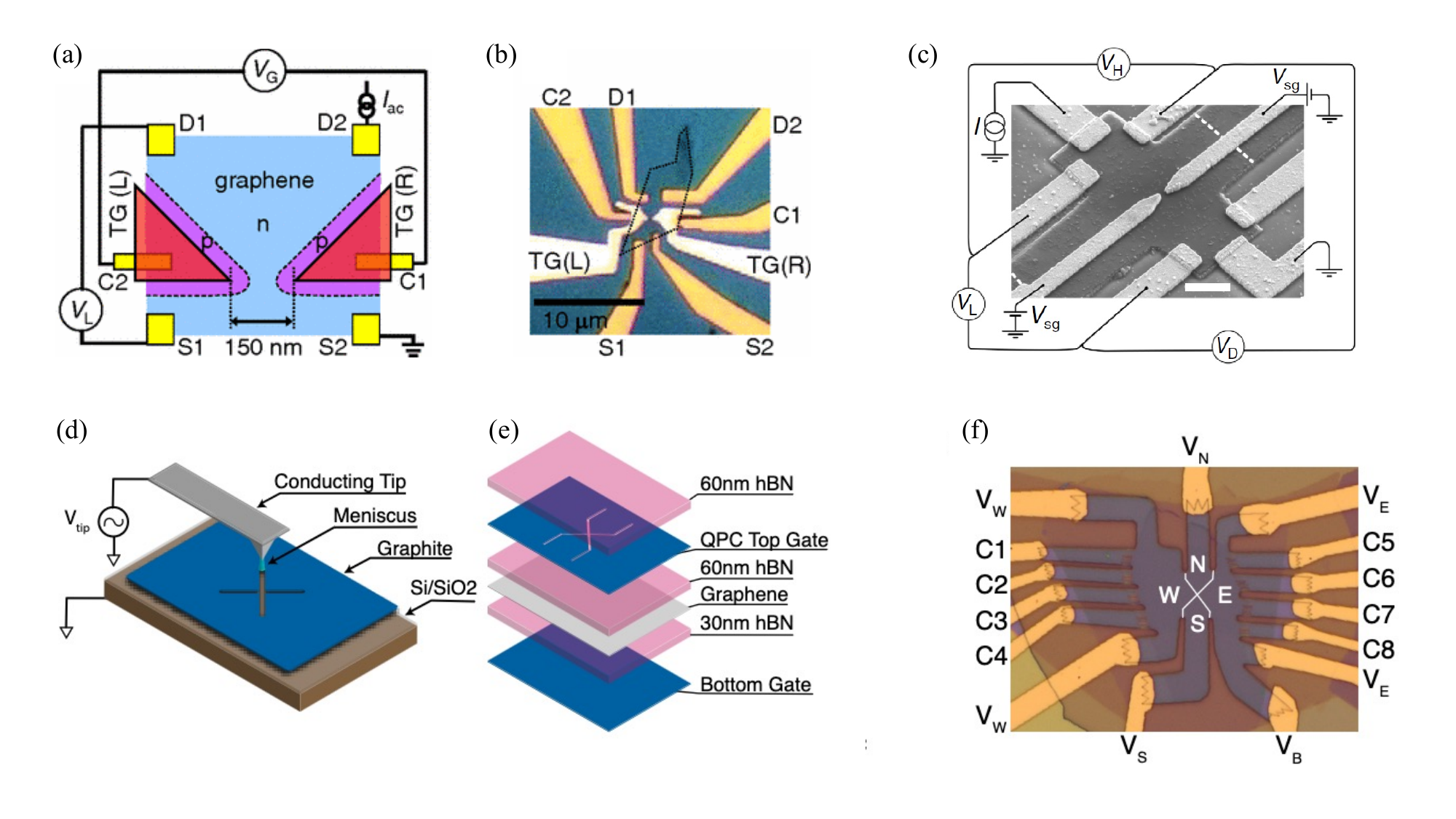}
	\end{center}
	\caption{{\bf Graphene quantum Hall QPC devices.} (a) and (b), First generation reported in \cite{Nakaharai2011}, where a thin layer of aluminium oxide is used to separate the gates from the graphene flake deposited on a SiO2 substrate. (a), device schematics and (b) optical micrograph of the sample. (c) Scanning electron micrograph of a QPC realized in hBN-encapsulated graphene \cite{Zimmermann2017}, with metallic QPC gates deposited on the surface of the hBN/graphene/hBN stack. (d), (e) and (f) high-mobility, dual graphite gates devices reported in \cite{Cohen2022}, where the top graphite gate is divided in 4 regions by anodic oxidation (d). (e) stack structure and (f) optical micrograph of the sample. The regions denoted \textit{N} and \textit{S} correspond to the QPC gates.
 }
 	\label{figQPCdevices}
\end{figure*}

Quantum points contacts realized in conventional 2DEGs rely on the ability to deplete the electron gas locally using electrostatic gates. This is not possible in single layer graphene at zero magnetic field since in this condition graphene is gapless. Under a high magnetic field, one can rely on the gap between different Landau levels to locally confine the edge channels with gates until backscattering occurs between the two counter propagating edges. An order of magnitude of these gaps is given by recent measurement conducted at 4 T and 1.4 K, estimating broken symmetry state gaps to be about 100 meV \cite{Liu2022}. This QPC technique under strong magnetic field has been implemented in several recent experiments both in the integer \cite{Nakaharai2011, Xiang2016, Ahmad2019} and in the fractional QH regime \cite{Zimmermann2017,Ronen2021, Cohen2022,Cohen2022b}. Typical geometries are depicted in \fig\ref{figQPCdevices}. QPC split gates, made of either metal or thin graphite flakes, are fabricated on top of an hBN-encapsulated graphene flake, and are biased with a dc voltage such that the quantum Hall states below them is set to a filling factor smaller than the bulk filling factor, thereby expelling the edge channel from beneath. The filling factor of the bulk is tuned using a global back gate (usually in graphite) combined with additional graphite top gates in \cite{Ronen2021,Cohen2022,Cohen2022b}. The distance between the split gates is typically about $100~$nm. In \cite{Cohen2022,Cohen2022b,Ronen2021}, the QPC and bulk top gates are realized from a single top graphite flake divided into the local gates by either reactive ion etching as shown in \fig\ref{fig:5_1}c \cite{Ronen2021} or by local anodic oxidation using an AFM tip as shown in \fig\ref{figQPCdevices}d-f \cite{Li2018, Cohen2022} .

Typical QPC characterization, shown in \fig\ref{figQPCresults}, consists in maps of the electric conductance across the QPC measured as a function of both QPC gate voltage and back/top gate voltage. They show regions of quantized conductance, corresponding to an integer number of channels perfectly transmitted across the QPC. When applying larger negative voltage on the QPC, a depleted region is first created below the gate. At higher voltages this region become populated with negatively charged carriers leading to an effective p-n-p barrier. In earlier devices \cite{Nakaharai2011, Xiang2016, Ahmad2019}, equilibration among channels in p and n regions led to an effective short cut of the split gates who could then not work as a QPC. These devices thus required the filling factor below the top gate to be fixed at $\nu=0$, the gap of which prevents equilibration across the gate. Because of the finite density range over which $\nu=0$ is defined, the QPC gate voltage could only be tuned in a limited range (typically, $\Delta V_\mathrm{QPC}\sim100~$mV at 5 T, for an approx.~50~nm thick BN). The global back gate and additional top gates were thus tuned in combination with the QPC gates, such that the electrostatic potential at the saddle point is raised or lowered while the filling factor below the QPC gates, and in the bulk of the sample, are fixed. This leads to configurations typically depicted in \fig\ref{figQPCresults}g-i, where an integer number of edge channels can be ballistically transmitted across the QPC in a controlled fashion, leading to the conductance plateaus shown, {e.g.}, in \fig\ref{figQPCresults}f. The condition for the split gates to operate correctly as a QPC in these early devices depending on the carrier concentration below the gate and a critical magnetic field has been extensively studied for example in \cite{Veyrat2019}. Most recent devices with higher mobility and operating at higher field do not present signs of equilibration \cite{Zimmermann2017, Deprez2021, Ronen2021, Cohen2022}. In particular, they show extended regions of zero conductance, demonstrating the ability to effectively pinch those devices thanks to the different quantum Hall gaps and even in the absence of an intrinsic bandgap. 

The ideal point contact signature has been the subject of debates in the GaAs community recently \cite{biswas2022, nakamura2023half, schiller2022}. While these discussions focused on the use of QPCs for shot noise measurement, similar questions arise for the use of QPCs for interferometry, and it is not clear whether a QPC which is ideal for all measurements exists (or is even possible). Notably, a recent article showed for the first time tunneling measurements across a QPC between edge channels at filling factors $\nu=1$ and $\nu=1/3$, and  observed the scaling laws for the bias and temperature dependence of the tunneling conductance predicted by the Tomonaga-Luttinger liquid theory \cite{Cohen2022b}.

\begin{figure*}[ht]
	\begin{center}
		\includegraphics[width=\textwidth]{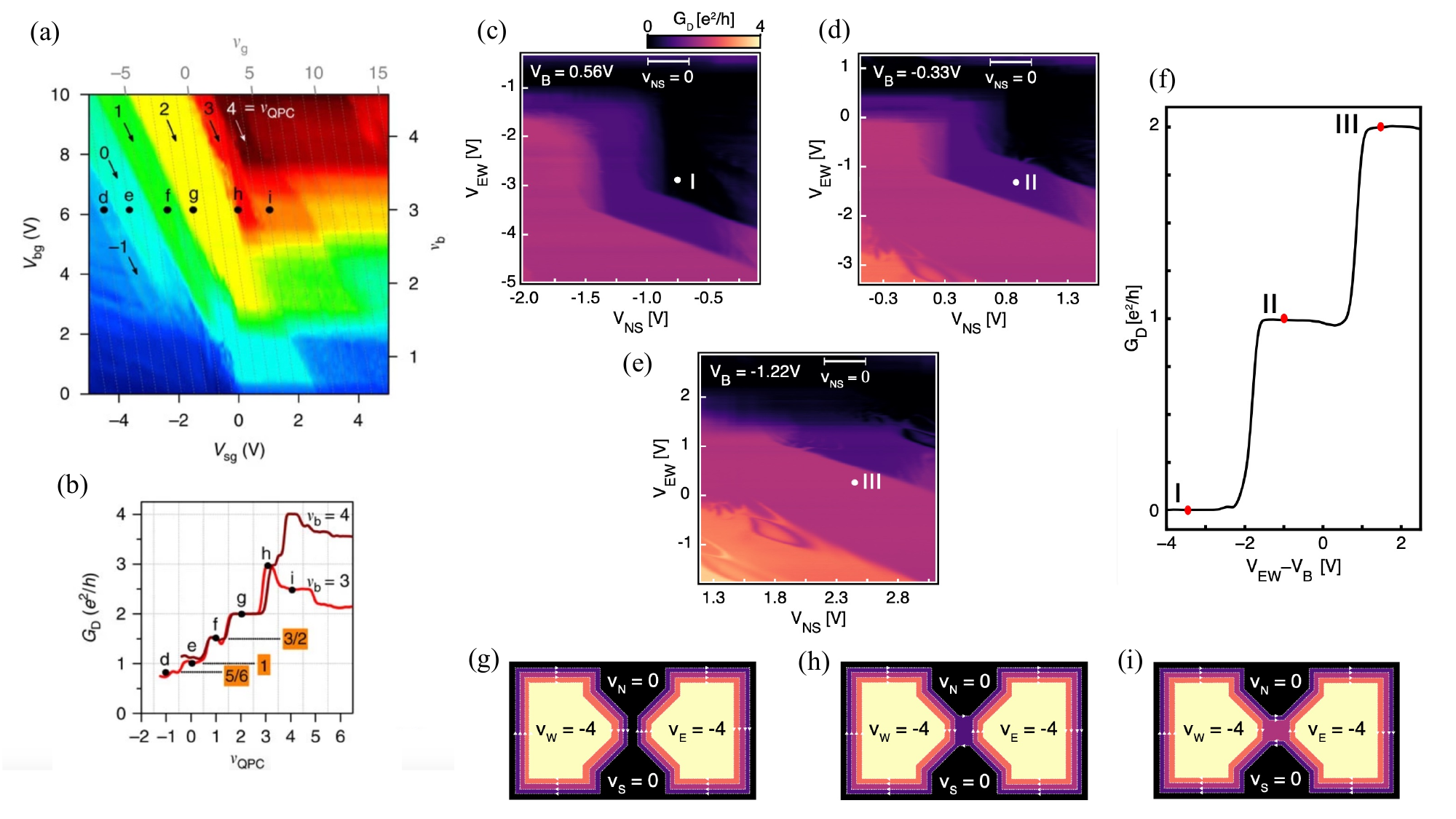}
	\end{center}
\caption{{\bf Conductance measurements in graphene QPCs.}  (a) conductance map of the QPC realized in \reference\cite{Zimmermann2017}, as a function of the back gate ($V_\mathrm{bg}$) and QPC gate ($V_\mathrm{sg}$) voltages. (b) line cuts of the data plotted in (a) at fixed back gate voltage, showing the quantized conductance plateaus. (c), (d) and (e) conductance map of the QPC realized in \reference\cite{Cohen2022} as a function of the bulk top gates (($V_\mathrm{EW}$) and QPC gates ($V_\mathrm{NS}$) voltages, for 3 different values of the bulk back gate ($V_\mathrm{B}$) voltage. The white bar indicates the QPC gate voltage range over which the filling factor below the QPC gates is $\nu=0$. (e) line cut of the conductance data as a function of a combination of the bulk top and back gates, showing the quantized conductance plateaus.  (g), (h) and (i) schematic representation of the filling factors in the vicinity of the QPC corresponding to the points marked I, II and III in (c-f).}

 	\label{figQPCresults}
\end{figure*}

\subsection{Quantum Hall Fabry-P{\'e}rot interferometers}
\vspace{-1em}
\nopagebreak
\vspace{1em}

\begin{figure*}
    \centering
   \includegraphics[width=\textwidth]{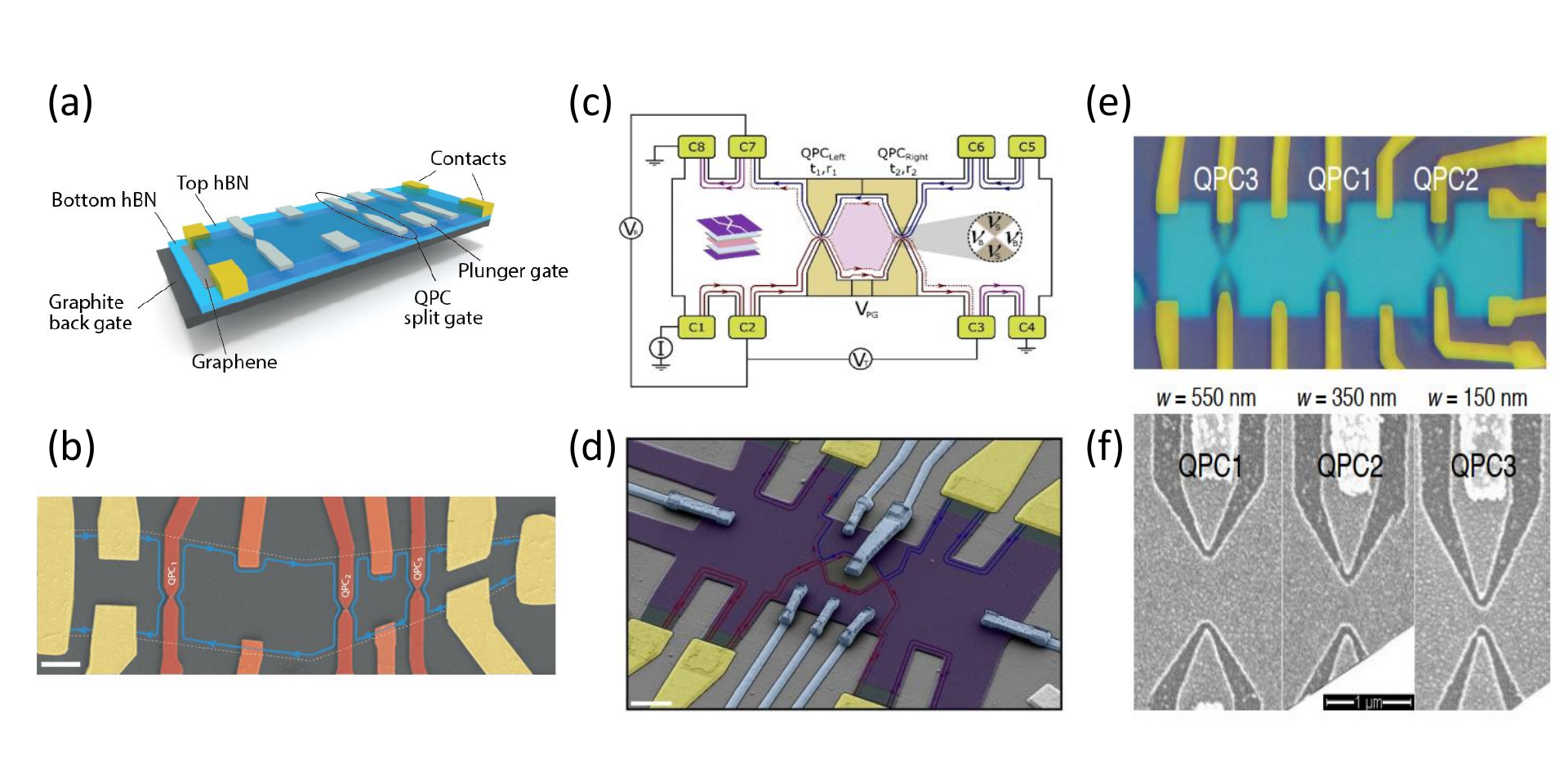}
    \caption{{ \bf Quantum Hall Fabry-P{\'e}rot interferometers.} (a) Schematic and (b) SEM image of the device studied in \reference\cite{Deprez2021}: 1D Ohmic contact made by Cr/Au, QPC split gates and plunger gates are made by depositing Pd electrodes on the hBN top flake. (c) Schematic and (d) SEM image of the device studied in \reference\cite{Ronen2021}. QPC and plunger gates are made by selectively etching the uppermost graphite flake. (e) Optical image and (f) SEM image of the device studied in \reference\cite{zhao2022}. QPC and plunger gates are made by etching both the top h-BN and the graphene flake.}
    \label{fig:5_1}
\end{figure*}

\subsubsection{Principle of the FPI experiment.}
\hfill\\
The quantum Hall Fabry-P{\'e}rot interferometer (FPI) is a pivotal tool for accessing the exchange statistic of exotic quasiparticles and realizing anyonic braiding \cite{feldman2021,Halperin2011}. In analogy with optical FPI, where semi-transparent mirrors reflect the incident light back and forth and enable photon interference, quantum point contacts are utilized as electron beam splitters to backscatter the chiral edge channels of the quantum Hall states. The electronic QH FPI can be built with two QPCs in series in a two-dimensional electron gas in semiconducting heterostructures or 2D materials in the QH regime, where selective partitioning of the edge channels leads to interferences. In this configuration, interference of electrons propagating along the periphery of the cavity can be controlled by the Aharonov-Bohm phase $\varphi_{AB}$=$2\pi AB/\phi_0$, where $A$ and $B$ are the enclosed area and magnetic field, and $\phi_0$=h/e is the magnetic flux quantum.

Historically, this type of interferometer was first realized in a 2D electron gas embedded in GaAs/AlGaAs quantum wells \cite{van1989,ofek2010,mcclure2012,sivan2018,nakamura2019}. However, the presence of charging effects between the edge modes and the compressible bulk has long hindered the measurement of the Aharonov-Bohm phase, not to mention the exploration of braiding statistics \cite{sivan2018,nakamura2019,choi2015, Zhang2009}. Recently, new GaAs heterostructures, purposely designed to mitigate charging effects by incorporating additional quantum wells serving as screening layers, have enabled the long-awaited observation of Aharonov-Bohm interference of a 1/3 fractional QH edge \cite{nakamura2019,Nakamura2020} and subsequently at filling factor 2/5 \cite{Nakamura2023}. Nevertheless, the quest for smaller interferometers with negligible charging energy is fervently pursued for practical applications. 

Graphene-based van-der-Waals heterostructures offer a promising alternative platform for realizing QH FPI due to their intrinsically advantageous dielectric environment. The presence of a graphite backgate, typically positioned in close proximity ($\sim 20-60~\rm{nm}$) to the graphene, naturally provides electrostatic screening, which effectively reduces the device's charging energy. Moreover, crystallographic edges create a hard-wall potential, limiting possible edge reconstructions \cite{Coissard2023} that might otherwise generate undesired additional integer, fractional and even neutral modes. These neutral modes are known to be detrimental to coherence \cite{Bhattacharyya2019}. 

Recently, monolayer and bilayer graphene-based FPIs have been successfully fabricated. We review here the various strategies employed in the design of interferometers with minimal charging energy, as well as the observation of Aharonov-Bohm conductance oscillations in the integer quantum Hall regime \cite{Deprez2021, Ronen2021, Fu2023}.

\subsubsection{Monolayer graphene based FPI}
\hfill\\

\begin{figure*}
    \centering
   \includegraphics[width=.8\textwidth]{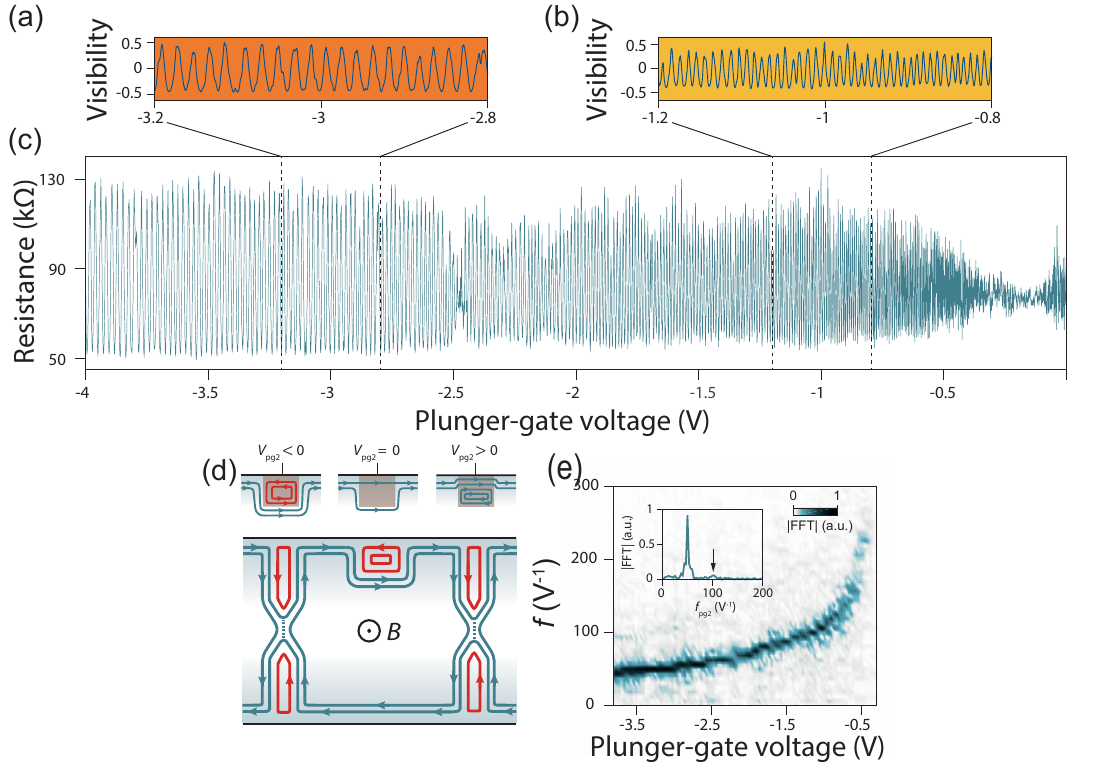}
    \caption{ \textbf{Plunger gate dependant oscillations.} Resistance oscillations of outer channel of $\nu = 1.5$ at $B = 14 \ \rm{T}$.Plunger gate voltage sweeps from -3.2 V to -2.8 V in (a), from -1.2 V to -0.8 V in (b) and from -4 V to 0 V in (c). (d) Schematic of the edge channel position at different plunger gate voltages. (e) Amplitude of sliding Fourier transform  of the resistance oscillations shown in (c) as a function of plunger gate voltage and plunger gate frequency. Adapted from \cite{Deprez2021}.}
    \label{fig:5_2}
\end{figure*}

Figure~\ref{fig:5_1} illustrates representative graphene-based devices. These heterostructures consist of hexagonal boron nitride (h-BN) encapsulated graphene deposited on top of a graphite backgate as detailed in reference~\cite{Deprez2021, Ronen2021}. Two distinct technical approaches have been employed to define the FPI cavity for quantum Hall edge channels. In the first approach, the physical edge of graphene, in conjunction with QPCs and plunger gates made from Pd deposited atop the uppermost h-BN, were used by D\'{e}prez and coworkers \cite{Deprez2021} as shown in figures~\ref{fig:5_1}a-b. Plunger gates placed between the QPCs manipulate the electron trajectory, enabling the modulation of the interference loop area. Multiple FPIs can be constructed by incorporating additional QPCs, as shown in figures~\ref{fig:5_1}a-b, where three QPCs define three cavities of different sizes. 
A second approach, developed by Ronen and coworkers \cite{Ronen2021}, involves an additional graphite layer on top of the hBN heterostructure. This layer is locally etched to establish QPC and plunger gates (as depicted in figures~\ref{fig:5_1}c-d). The interferometer cavity is thus entirely determined electrostatically through these top-gates. These gates offer the flexibility to tune the smoothness of the electrostatic edge potential, thereby enabling a modulation of the edge channel velocity. Last, Zhao and coworkers \cite{zhao2022} implemented a similar device where the QPCs are constructed by etching both graphene and top h-BN layer.



\begin{figure*}
    \centering
   \includegraphics[width=\textwidth]{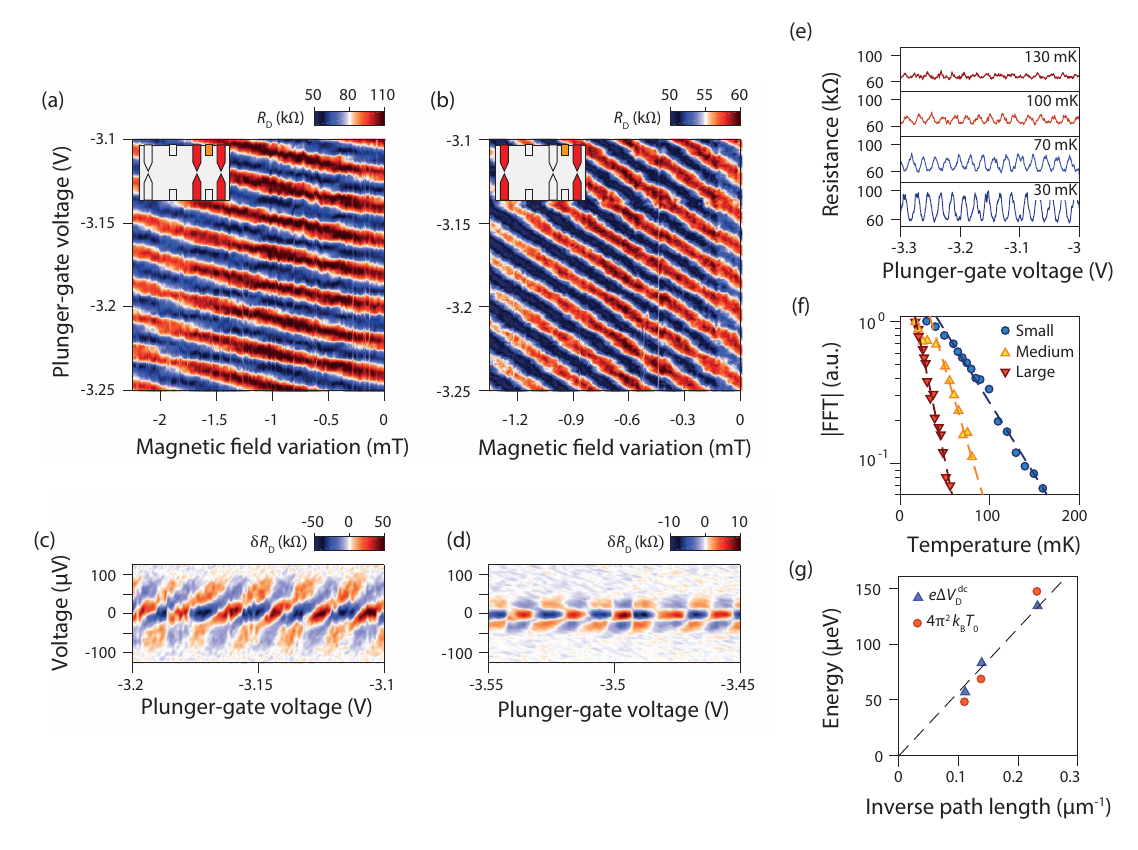}
    \caption{ { \bf Resistance oscillations with magnetic field and plunger gate voltage.} (a) in a $3~\mu\rm{m^2}$ interferometer and (b) in a $15~\mu\rm{m^2}$ interferometer (b). Checkerboard pattern obtained from the $3~\mu\rm{m^2}$ interferometer in (c) and from the $15~\mu\rm{m^2}$ interferometer in (d). (e) Resistance oscillations with plunger gate voltage at different temperatures. (f) Oscillation amplitudes extracted from a Fourier transform as a function of temperature for three interferometers ($A = 3.1~\mu\rm{m^2}$, $A = 10.7~\mu\rm{m^2}$  and $A = 14.7~\mu\rm{m^2}$). (g) Energy scale extracted from the checkerboard patterns in (c) and (d) and the temperature dependence in (f) as a function of the inverse path length. Adapted from \cite{Deprez2021}}
    \label{fig:5_3}
\end{figure*}

A standard measurement setup, as shown in \fig\ref{fig:5_1}c, is commonly used in most experiments \cite{Deprez2021, Ronen2021}. The measurement of the longitudinal or diagonal resistance of the FPI are performed using lock-in amplifiers. 

The new interesting aspect of the graphene platform lies in the extensive tunability provided by the plunger gate. In contrast to GaAs heterostructures, where the plunger gate tunability is constrained by the depletion of electrons beneath the gate, graphene's gapless band structure allows for a broad gate sweep covering multiple quantum Hall states and filling factors. Figure~\ref{fig:5_2} illustrates characteristic resistance oscillations with high visibility across a wide voltage range. Applying a negative voltage to the plunger gate depletes the electron gas and eventually accumulates holes beneath the gate. Consequently, the electron trajectory is continuously pushed toward the device interior with decreasing plunger gate voltage (see \fig\ref{fig:5_2}d).  

\begin{figure*}
    \centering
   \includegraphics[width=1.1\textwidth]{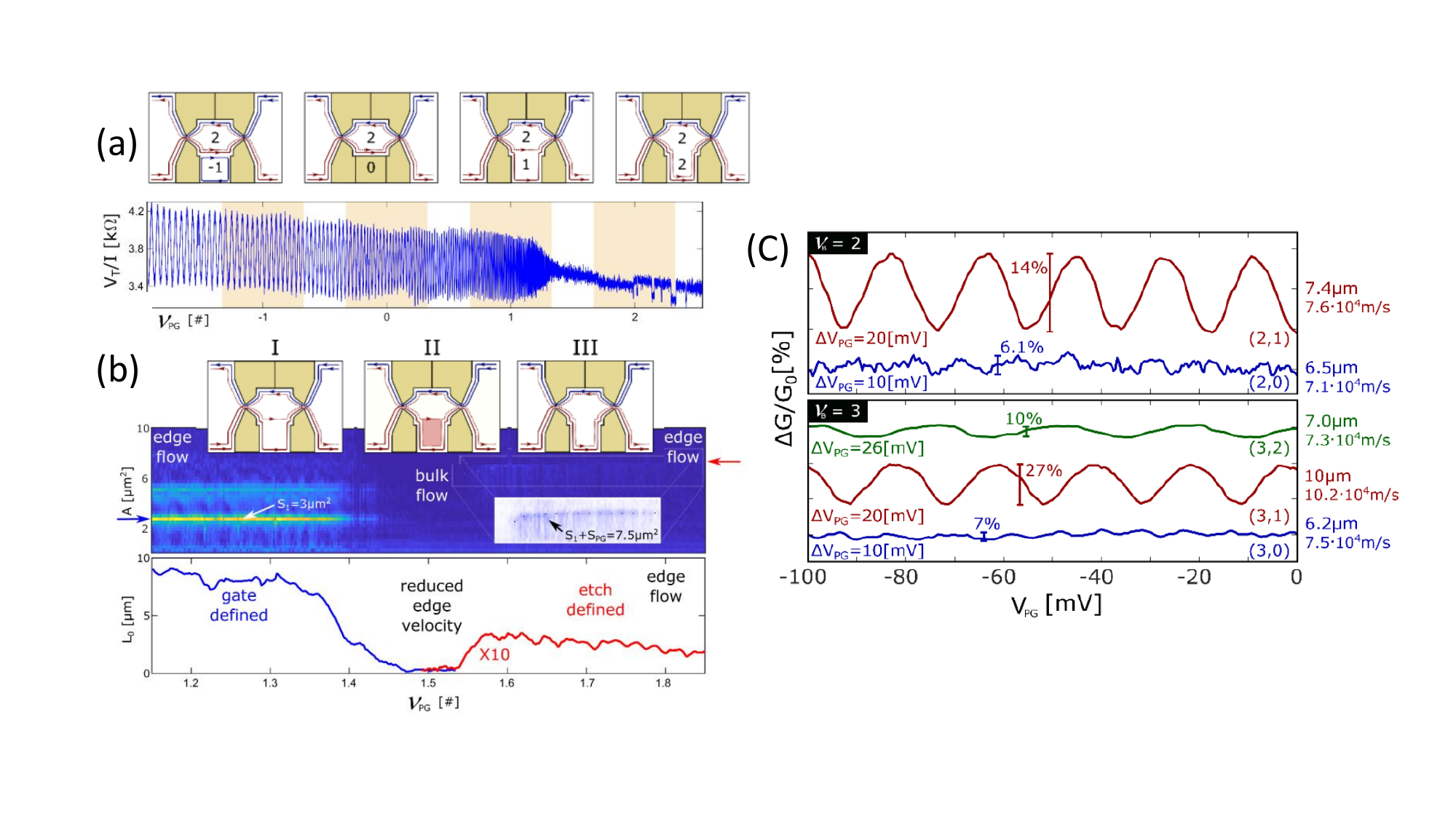}
    \caption{(a) Resistance oscillations with plunger gate voltages from -2 V to 3 V. (b) Interfering area extracted from oscillation periodicity in (a) from 1 V to 2 V (middle panel), top schematics indicate the quantum Hall edges path for each oscillation regime. Bottom panel is the phase coherence length extracted from the oscillations' visibility. (c) Oscillation visibility at bulk filling factor 2 (up) and 3 (bottom) for different edge channel interference. Adapted from \cite{Ronen2021}}
    \label{fig:5_4}
\end{figure*}


Graphene-based FPIs exhibit less charging effect compared to GaAs-based ones of similar sizes. Figures~\ref{fig:5_3}a-b depict 2D plots of the resistance $R_{\rm{D}}$ as a function of both magnetic field and plunger gate voltage for the outer channel interference of two interferometers of different sizes ($A = 3.1~\mu\rm{m^2}$ and $A = 14.7~\mu\rm{m^2}$). Effective interferometer areas ($A$) calculated from magnetic field periodicities ($\Delta B$) using the formula $A = \phi_0/\Delta B$, are found to be in excellent agreement with  area defined by lithography. The periodic stripes in the plot correspond to lines of constant phase. The direction of these lines is used to distinguish whether the interference is dominated by Aharonov-Bohm effect or Coulomb interactions \cite{Halperin2011}. A constant Aharonov-Bohm phase results in stripes with a negative slope due to the diminished area being compensated by an enhanced magnetic field, as expressed by the equation $\Delta \varphi_{\rm{AB}} = 2 \pi / \phi_0 (B\Delta A + A \Delta B) = 0$. 
The observed negative slopes in figures~\ref{fig:5_3}a-b, along with the accurate estimation of interference area, provide clear evidence that these interferometers are operating in the Aharonov-Bohm regime. It is noteworthy that GaAs-based interferometers of similar size were reported to be dominated by Coulomb interactions \cite{Zhang2009}. 
 

Edge velocity can be probed through the oscillation dependence on both the DC bias voltage and the temperature. When a source-drain DC bias voltage is applied, electrons experience a  dynamical phase shift given by $\varphi_{\rm{dyn}} = 2\pi e V_{\rm{DC}}2L/(hv) = 4 \pi e V_{\rm{DC}}/E_{\rm{Th}}$, where $L$ is the length of edge propagation between the two QPCs, $v$ is the edge velocity, and $E_{\rm{Th}} =  hv/L$ is the Thouless energy. 
Depending on the energy relaxation processes consecutive to the current flow, and on the electrostatic coupling between the cavity, the back gate, the source and the drain, the electrochemical potential in the cavity will adjust itself at a value intermediate between that of the source and that of the drain. The resulting conductance oscillations depending on the potential drop across the interferometer have been calculated in \cite{Deprez2021} by extending the theory of \cite{DeC.Chamon1997}. It reads
\begin{align}
  \label{eq-single-freq}
   G = g_{osc} \Bigg[\beta \cos \left( 2 \pi \frac{\varphi}{\phi_0} - \frac{2L}{\hbar v} eV\beta \right) \\  + \overline{\beta}\cos \left( 2 \pi \frac{\varphi}{\phi_0} + \frac{2L}{\hbar v} eV\beta \right) \Bigg],
\end{align}
 where $\beta$ and $\overline{\beta}$ are asymmetry parameters describing how symmetric is the voltage drop on the two side of the interferometer (see Supplementary Information in \cite{Deprez2021} for definition), and $g_{osc}=\frac{e^2}{h}2\sqrt{R_1R_2}$ with $R_1$ and $R_2$ the reflection coefficients of each QPCs.
Equation (\ref{eq-single-freq}) reduces to $\Delta G \sim \cos{(2 \pi \phi/\phi_0-4\pi e V_{\rm{DC}}/E_{\rm{Th}})}$ for a fully asymmetric potential drop across the interferometer and  $\Delta G \sim \cos{(2 \pi \phi/\phi_0)} \cos{(2\pi eV_{\rm{DC}}/E_{\rm{Th}})}$ for a fully symmetric potential drop. Figures~\ref{fig:5_3}c and d illustrates the resulting conductance oscillations as a function of plunger gate voltage and voltage bias for the asymmetric (small interferometer) and symmetric (large interferometer) cases, respectively. The edge velocity $v$ determined from the bias periodicity via $v = \Delta V_{\rm{DC}}  L/ \!h$, is found to be approximately $1.4 \times 10^5  \ \rm{m}/\rm{s}$ (at 14 T). Another method to estimate the Thouless energy stems from the temperature dependence of the oscillation amplitude, described by $\exp{(-T/T_0)}$, as exemplified in figures~\ref{fig:5_3}e-f Figure \ref{fig:5_3}f \cite{Deprez2021}. The fits of the Fourier peaks are used to estimate $T_0$ and the Thouless energy $E_{\rm{Th}} = 4\pi^2 k_{\rm{B}} T_0 L$. Figure \ref{fig:5_3}g demonstrates the excellent agreement between the Thouless energies estimated using these two methods.

The interferometer geometry in \reference\cite{Ronen2021} shown in \fig\ref{fig:5_1}d, allows to study the phase coherence of interfering edges defined either by gating or by etching. Figure \ref{fig:5_4}a shows plunger gate dependent oscillations of the inner channel at filling factor 2, where interfering channels undergo distinct potential confinements, as schematized on top of the figure. Figure \ref{fig:5_4}b shows the edge channel positions (upper panel), interfering area (middle panel) and coherence length (lower panel) in these different configurations.
In regime I ($1 \ \rm{V} < V_{\rm{pg}} < 1.4 \ \rm{V}$), oscillations' visibility remains relatively constant and the extracted interfering area corresponds to the inner channel traveling along the plunger gate. In regime III ($V_{\rm{pg}} > 1.6 \ \rm{V}$), the oscillations' visibility is dramatically reduced and the extracted area now includes the plunger gate delimited area, suggesting that both channels are propagating along the etched graphene edge. Note that the coherence lengths are about $7 \ \mu \rm{m}$ for gate defined channels and $400 \ \rm{nm}$ for etched defined  channels. The detrimental effect on the coherence observed in etched-edge defined FPIs can be attributed to edge disorder and charge accumulation at the physical edge. Interestingly graphene crystal edges that have not undergone etching, as in the experiment conducted by D\'{e}prez and coworkers \cite{Deprez2021}, led to a coherence length assessment of $10\,\mu$m, implying that pristine edges are equally good for coherence.



\begin{figure*}[ht]
	\begin{center}
		\includegraphics[width=\textwidth]{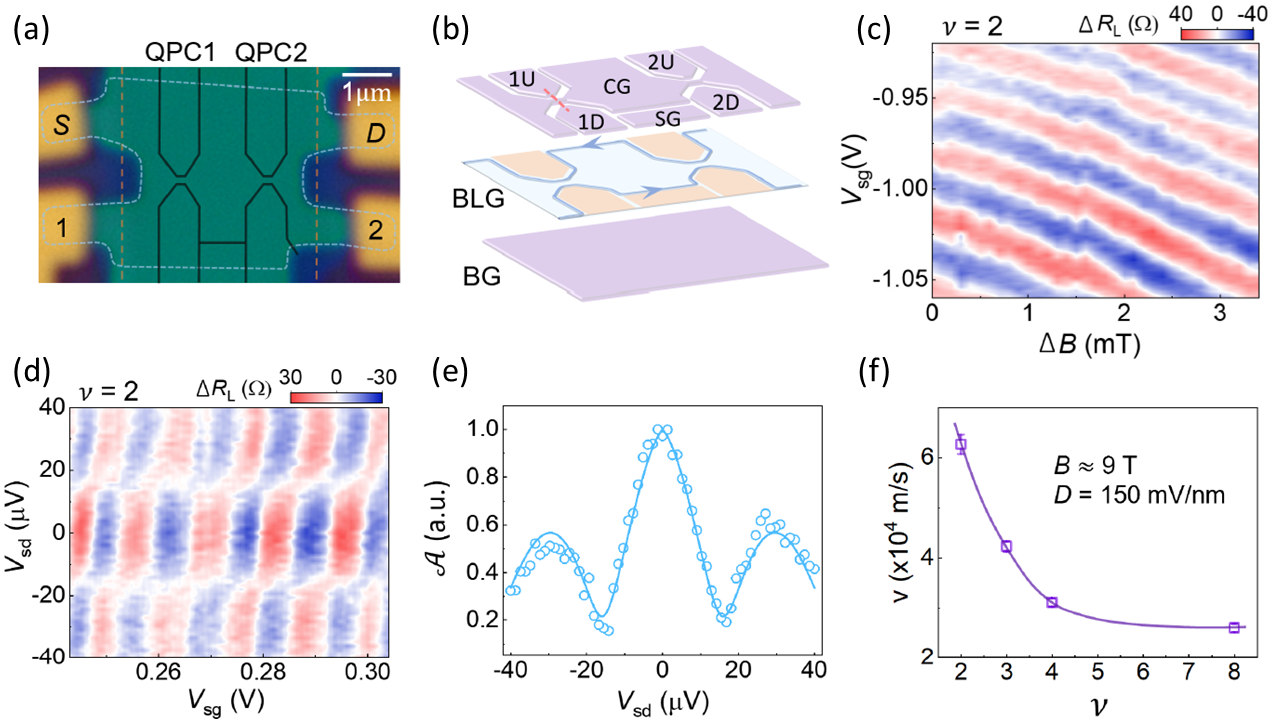}
	\end{center}
	\caption{ {\bf Bilayer graphene based Fabry-P{\'e}rot interferometer.} a) Optical image of the device. b) Schematic of the device in an exploded view. c) Resistance oscillations showing an Aharonov Bohm type pyjama at filling factor 2 for a field about 9 T. d)  The resistance oscillations as a function of side gate voltage and source drain voltage display a typical checkerboard pattern (B = 9T, filling factor 2, 20 mK). e) The oscillation amplitude at the oscillation frequency shows a lobe structure as a function of source drain voltage. The author extract both a Thouless energy and an edge velocity by fitting the lobe structure with the usual theory \cite{Deprez2021,Ronen2021} f) Edge channel velocity (extracted from lobe structure as in e) versus filling factor. The decrease in velocity for increasing filling factor can be understood in a non-interacting edge channel picture as the channels are further and further from the edge as the filling factor increase.\cite{Fu2023}. }
	\label{fig7}
\end{figure*}

 The authors of reference~\cite{Ronen2021} systematically investigated the interference of various channels at filling factors 2 and 3. Figure \ref{fig:5_4}c summarizes the plunger gate periodicity, oscillation visibility, edge velocity, and phase coherence length for these channels. The colors blue, red, and green are used to distinguish the innermost, middle, and outermost edges, respectively. The coherence lengths are of the order of tens of microns. The highest velocity and coherence length are obtained in the middle channel of $\nu_{\rm{B}} = 3$ and are accounted for by interaction screening by adjacent edges.

\subsubsection{Bilayer graphene based FPI. }
\hfill\\

Bilayer graphene also hosts numerous fractional QH states, particularly those with even denominator fractions are believed to involve non-Abelian quasi-particles \cite{li2017}. BLG is thus considered as a highly promising platform for performing non-Abelian braiding through interferometry. In the study reported by Fu and coworkers \cite{Fu2023}, the first Fabry-P{'e}rot quantum Hall interferometer in bilayer graphene was introduced. In this investigation, bilayer graphene is encapsulated by h-BN flakes, with a global graphite gate at the bottom and several split gates on the top (see Figure~\ref{fig7}a,b). The devices operate in the Aharonov-Bohm regime as demonstrated in \fig\ref{fig7}c. The checkerboard pattern (\fig\ref{fig7}d) is analysed through the same framework as in \reference\cite{Deprez2021,Ronen2021}, and the edge channel velocity is extracted from the lobe structure (\fig\ref{fig7}e) at different filling factors (\fig\ref{fig7}f). A careful analysis show that this velocity decreases as the distance of the edge channel to the edge increases, as expected. This work opens the door for further studies that might benefit from recent technical improvement, such as the use of air bridges \cite{Ronen2021} or atomic force microscopy etching \cite{Cohen2022}. 

\begin{table*}[ht]
  \begin{center}
    
    \vspace{5mm}
    \label{tab:acronym}
    \begin{tabular}{|c|c|c|c|c|}
\hline
 1(2)D & one(two) dimensional  \\ 
 \hline
 2DEG & two dimensional electron gas  \\ 
 \hline
BZ & Brillouin zone  \\  
 \hline
 SQUID & superconducting quantum interference device  \\  
 \hline
ODE & ordinary differential equation  \\  
 \hline
 PDE & partial differential equation   \\ 
 \hline
SEM & scanning electron microscope   \\ 
 \hline
 AFM & atomic force microscope   \\ 
 \hline
TEM & transmission electron microscope   \\ 
 \hline
STM & scanning tunnelling microscope   \\ 
 \hline
SET & single electron transistor   \\ 
 \hline 
SGM & scanning gate microscope   \\ 
 \hline 
TEF & transverse electron focusing   \\ 
 \hline
h-BN & hexagonal boron nitride   \\ 
 \hline
Gr & graphene (usually monolayer)   \\ 
 \hline
BLG & bilayer graphene    \\ 
 \hline
LOR & lift off resist   \\ 
 \hline
PDMS & Polydiméthylsiloxane   \\ 
 \hline
 PC & polycarbonate (polymer)   \\ 
 \hline
 PPC & Polypropylene carbonate (polymer)   \\ 
 \hline
CNP & charge neutrality point of graphene   \\ 
 \hline
SdH & Shubnikov de Haas   \\ 
 \hline
WGM & whispering gallery mode   \\ 
 \hline
TG & top gate   \\ 
 \hline
 BG & back gate   \\ 
  \hline
p-n junction & interface between positively (p) and negatively (n) charged graphene regions    \\ 
 \hline
 FP(I) & Fabry-P{\'e}rot (interferometer)  \\ 
 \hline
  MZ(I) & Mach Zehnder (interferometer)  \\ 
 \hline
QPC & quantum point contact  \\ 
 \hline
  AB & Aharonov Bohm  \\ 
 \hline
  QH & quantum Hall  \\ 
   \hline
  FT & Fourier transform  \\ 
 \hline
    \end{tabular}
    \caption{List of acronyms}
  \end{center}
\end{table*}

In conclusion, the robust oscillations observed in single and bilayer graphene-based quantum Hall FPIs with various designs, characterized by strongly suppressed charging energy and an extended phase coherence length, offer compelling reasons to advance and deepen these studies into fractional quantum Hall interferometry. Concurrently, some correlated phenomena in the integer quantum Hall regime remain to be understood. Very recent experiments have presented evidence for electron pairing at bulk filling factors $\nu\geq 2$ \cite{werkmeister2023,yang2023}, reproducing the phenomenology of GaAs FPIs \cite{choi2015,Biswas2023}, and even tripling of electrons at $\nu=3$ \cite{yang2023}. These phenomena are characterized by an anomalous Aharonov-Bohm period of $h/2e$ and $h/3e$, respectively. Interpretations suggest that these intriguing phenomena result from attracting pairing emerging via the exchange of neutral modes \cite{Frigeri2020} or inter-edge states charging effects \cite{werkmeister2023}. They underscore the genuinely complex nature of these FPI devices, challenging initial expectations based on an apparently simple, non-interacting theory \cite{DeC.Chamon1997}. Exploring the physics of anyons in the fractional quantum Hall regime will undoubtedly require careful and systematic experiments, given our current, still partial, understanding of integer quantum Hall FPIs.

\section{Conclusion and perspectives}

We have seen throughout this review that graphene has emerged as an excellent and unique platform for the realization of electron optical devices. This stems, among other things, from the ultra-high mobility of graphene structures and the ability of forming gapless p-n interfaces, where the doping changes continuously from electron to hole doping. This allowed, as detailed before, observing magnetic focusing and snake states and realizing interferometers based on quantum-Hall edge channels. The presence of the lattice and valley degree of freedom (or the layer in case of bilayer graphene) leads to a richer and much more fascinating behaviour than in conventional 2DEGs. Moreoever, due to the semimetallic nature of graphene, it can be contacted with ferromagnetic and superconducting electrodes. This yields ballistic spintronic devices that can show chiral properties, on the one hand, and ballistic interferometers that are governed by unconventional Andreev physics at p-n interfaces, on the other hand. The ability to combine graphene with other 2D materials and to stack different materials together with arbitrary twist angles opens a large parameter space, allowing to tailor the bandstructure and to design materials with yet unknown phenomenology. Correspondingly, these developments, including emergent phenomena such as unconventional superconductivity, also open up new challenges to theory.
Properties, like superconducting pairing, spin-orbit interaction and magnetic exchange can also be introduced in monolayer and bilayer graphene through proximity to TMDCs or van der Waals superconductors and ferromagnets. This is a largely unexplored area where the combination with ballistic carriers opens up a new playground for quantum electron optics.

\section*{Acknowledgments}
This common work was supported by various funding organizations.

PM acknowledges support from the Multi-Spin and 2DSOTECH FlagERA networks, the Twistrain ERC grant.  This research was supported by the Ministry of Culture and Innovation and the National Research, Development and Innovation Office within the Quantum Information National Laboratory of Hungary (Grant No. 2022-2.1.1-NL-2022-00004). 

MHL acknowledges National Science and Technology Council (NSTC) of Taiwan (grant number NSTC 112-2112-M-006-019-MY3) for financial support. He further received funding through the {\it  2023 International Presidential Visiting  Scholarship} of University of Regensburg.

KR acknowledges support from the Deutsche Forschungsgemeinschaft (DFG, German Research Foundation) within Project-ID 314695032 -- SFB 1277 (project A07).

AK acknowledges support from the  Deutsche Forschungsgemeinschaft (DFG, German Research Foundation) within DFG Individual grant KN 1383/4.

CS acknowledges support from the European Research Council (ERC) under the European Union’s Horizon 2020 research and innovation program: grant agreement No 787414, ERC-Adv TopSupra, the Swiss NCCR Quantum Science and Technology (QSIT), the Swiss Nanoscience Institute (SNI) and the University of Basel.

FDP acknowledges support from the ERC (ERC-2018-STG QUAHQ), the “Investissements d’Avenir” LabEx PALM (ANR-10-LABX-0039-PALM), and the Major Interest Domain (DIM) SIRTEQ and QUANTIP funded by the Ile-de-France Region.

CG acknowledges support from the Agence Nationale de la Recherche (ANR, French National Research Agency; project DADI), and thanks the STherQO members for pointing out certain relevant publications.

DP acknowledges support from the RobustSuperQ program. WY acknowledges support from the Quantera program.

PR and HC acknowledges support from the Agence Nationale de la Recherche (ANR, French National Research Agency; project DADI).

\section*{References}

\bibliographystyle{ieeetr}

\normalem
\bibliography{Review_biblio_alpha_omega}

\begin{thebibliography}{100}

\bibitem{Novoselov2004}
K.~S. Novoselov, A.~K. Geim, S.~V. Morozov, D.~Jiang, Y.~Zhang, S.~V. Dubonos,
  I.~V. Grigorieva, and A.~A. Firsov, ``{Electric field in atomically thin
  carbon films},'' {\em Science}, vol.~306, pp.~666--669, oct 2004.

\bibitem{Cao2018}
Y.~Cao, V.~Fatemi, S.~Fang, K.~Watanabe, T.~Taniguchi, E.~Kaxiras, and
  P.~Jarillo-Herrero, ``{Unconventional superconductivity in magic-angle
  graphene superlattices},'' {\em Nature}, vol.~556, pp.~43--50, Apr. 2018.

\bibitem{Hanbury-Brown1954}
R.~Hanbury~Brown and R.~Q. Twiss, ``A new type of interferometer for use in
  radio astronomy,'' {\em Philosophical Magazine}, vol.~45, no.~366,
  pp.~663--682, 1954.

\bibitem{Hanbury-Brown1956}
R.~Hanbury~Brown and R.~Q. Twiss, ``Correlation between photons in two coherent
  beams of light,'' {\em Nature}, vol.~177, no.~4497, pp.~27--29, 1956.

\bibitem{huard2007transport}
B.~Huard, J.~Sulpizio, N.~Stander, K.~Todd, B.~Yang, and D.~Goldhaber-Gordon,
  ``Transport measurements across a tunable potential barrier in graphene,''
  {\em Physical review letters}, vol.~98, no.~23, p.~236803, 2007.

\bibitem{lemme2007graphene}
M.~C. Lemme, T.~J. Echtermeyer, M.~Baus, and H.~Kurz, ``A graphene field-effect
  device,'' {\em IEEE Electron Device Letters}, vol.~28, no.~4, pp.~282--284,
  2007.

\bibitem{williams2007quantum}
J.~Williams, L.~DiCarlo, and C.~Marcus, ``Quantum hall effect in a
  gate-controlled pn junction of graphene,'' {\em Science}, vol.~317, no.~5838,
  pp.~638--641, 2007.

\bibitem{abanin2007quantized}
D.~Abanin and L.~Levitov, ``Quantized transport in graphene pn junctions in a
  magnetic field,'' {\em Science}, vol.~317, no.~5838, pp.~641--643, 2007.

\bibitem{Katsnelson2006}
M.~I. Katsnelson, K.~S. Novoselov, and A.~K. Geim, ``Chiral tunnelling and the
  {K}lein paradox in graphene,'' {\em Nature Phsics}, vol.~2, pp.~620--625, sep
  2006.

\bibitem{Cheianov2007}
V.~V. Cheianov, V.~Fal'ko, and B.~L. Altshuler, ``The focusing of electron flow
  and a veselago lens in graphene p-n junctions,'' {\em Science}, vol.~315,
  pp.~1252--1255, mar 2007.

\bibitem{Cheianov2006}
V.~V. Cheianov and V.~I. Fal'ko, ``Selective transmission of {D}irac electrons
  and ballistic magnetoresistance of np junctions in graphene,'' {\em Physical
  Review B}, vol.~74, p.~041403, jul 2006.

\bibitem{silvestrov2007quantum}
P.~Silvestrov and K.~Efetov, ``Quantum dots in graphene,'' {\em Physical Review
  Letters}, vol.~98, no.~1, p.~016802, 2007.

\bibitem{Dingle1978}
R.~Dingle, H.~L. St{\"{o}}rmer, A.~C. Gossard, and W.~Wiegmann, ``{Electron
  mobilities in modulation‐doped semiconductor heterojunction
  superlattices},'' {\em Applied Physics Letters}, vol.~33, pp.~665--667, oct
  1978.

\bibitem{Ketterson1985}
A.~Ketterson, F.~Ponse, T.~Henderson, J.~Klem, and H.~Morko{\c{c}},
  ``{Extremely low contact resistances for AlGaAs/GaAs modulation‐doped
  field‐effect transistor structures},'' {\em Journal of Applied Physics},
  vol.~57, pp.~2305--2307, mar 1985.

\bibitem{Chung2022}
J.~Y. Chung, A.~Gupta, K.~W. Baldwin, K.~W. West, M.~Shayegan, and L.~N.
  Pfeiffer, ``Understanding limits to mobility in ultrahigh-mobility gaas
  two-dimensional electron systems: $100$ million cm$^2$/vs and beyond,'' {\em
  Phys. Rev. B}, vol.~106, p.~075134, aug 2022.

\bibitem{VanWees1988}
B.~J. van Wees, H.~van Houten, C.~W.~J. Beenakker, J.~G. Williamson, L.~P.
  Kouwenhoven, D.~van~der Marel, and C.~T. Foxon, ``{Quantized conductance of
  point contacts in a two-dimensional electron gas},'' {\em Physical Review
  Letters}, vol.~60, pp.~848--850, feb 1988.

\bibitem{Bolotin2008}
K.~I. Bolotin, K.~J. Sikes, Z.~Jiang, M.~Klima, G.~Fudenberg, J.~Hone, P.~Kim,
  and H.~L. Stormer, ``{Ultrahigh electron mobility in suspended graphene},''
  {\em Solid State Communications}, vol.~146, pp.~351--355, jun 2008.

\bibitem{Du2008}
X.~Du, I.~Skachko, A.~Barker, and E.~Y. Andrei, ``{Approaching ballistic
  transport in suspended graphene},'' {\em Nature Nanotechnology 2008 3:8},
  vol.~3, pp.~491--495, jul 2008.

\bibitem{Dean2010}
C.~R. Dean, A.~F. Young, I.~Meric, C.~Lee, L.~Wang, S.~Sorgenfrei, K.~Watanabe,
  T.~Taniguchi, P.~Kim, K.~L. Shepard, and J.~Hone, ``{Boron nitride substrates
  for high-quality graphene electronics},'' {\em Nature Nanotechnology},
  vol.~5, pp.~722--726, oct 2010.

\bibitem{Wang2012}
L.~Wang, Z.~Chen, C.~R. Dean, T.~Taniguchi, K.~Watanabe, L.~E. Brus, and
  J.~Hone, ``Negligible environmental sensitivity of graphene in a hexagonal
  boron nitride/graphene/h-bn sandwich structure,'' {\em ACS Nano}, vol.~6,
  pp.~9314--9319, oct 2012.

\bibitem{Wang2013}
L.~Wang, I.~Meric, P.~Y. Huang, Q.~Gao, Y.~Gao, H.~Tran, T.~Taniguchi,
  K.~Watanabe, L.~M. Campos, D.~A. Muller, J.~Guo, P.~Kim, J.~Hone, K.~L.
  Shepard, and C.~R. Dean, ``One-dimensional electrical contact to a
  two-dimensional material,'' {\em Science}, vol.~342, pp.~614--617, Nov. 2013.

\bibitem{Young2009}
A.~F. Young and P.~Kim, ``Quantum interference and {K}lein tunnelling in
  graphene heterojunctions,'' {\em Nature Physics}, vol.~5, pp.~222--226, mar
  2009.

\bibitem{Rickhaus2013}
P.~Rickhaus, R.~Maurand, M.-H. Liu, M.~Weiss, K.~Richter, and
  C.~Sch{\ifmmode\ddot{o}\else\"{o}\fi}nenberger, ``{Ballistic interferences in
  suspended graphene},'' {\em Nat Commun}, vol.~4, p.~2342, Aug. 2013.

\bibitem{Banszerus2016}
L.~Banszerus, M.~Schmitz, S.~Engels, M.~Goldsche, K.~Watanabe, T.~Taniguchi,
  B.~Beschoten, and C.~Stampfer, ``Ballistic {{Transport Exceeding}} 28
  {$\mu$}m in {{CVD Grown Graphene}},'' {\em Nano Letters}, vol.~16,
  pp.~1387--1391, Feb. 2016.

\bibitem{Shytov2008}
A.~V. Shytov, M.~S. Rudner, and L.~S. Levitov, ``Klein backscattering and
  {Fabry-P\'erot} interference in graphene heterojunctions,'' {\em Phys. Rev.
  Lett.}, vol.~101, p.~156804, Oct 2008.

\bibitem{Williams2017}
J.~R. Williams, ``Electron optics with graphene p-n junctions,'' in {\em 2D
  Materials: Properties and Devices}, pp.~141--158, Cambridge University Press,
  2017.

\bibitem{Rickhaus2015}
P.~Rickhaus, P.~Makk, M.-H. Liu, E.~T{\'{o}}v{\'{a}}ri, M.~Weiss, R.~Maurand,
  K.~Richter, and C.~Sch{\"{o}}nenberger, ``{Snake trajectories in ultraclean
  graphene p–n junctions},'' {\em Nature Communications 2015 6:1}, vol.~6,
  pp.~1--6, mar 2015.

\bibitem{Novoselov2005}
K.~S. Novoselov, A.~K. Geim, S.~V. Morozov, D.~Jiang, M.~I. Katsnelson, I.~V.
  Grigorieva, S.~V. Dubonos, and A.~A. Firsov, ``Two-dimensional gas of
  massless {Dirac} fermions in graphene,'' {\em Nature}, vol.~438,
  pp.~197--200, nov 2005.

\bibitem{Zhang2005}
Y.~B. Zhang, Y.~W. Tan, H.~L. Stormer, and P.~Kim, ``Experimental observation
  of the quantum {Hall} effect and {Berry's} phase in graphene,'' {\em Nature},
  vol.~438, pp.~201--204, nov 2005.

\bibitem{Zhang2009}
Y.~Zhang, D.~T. McClure, E.~M. Levenson-Falk, C.~M. Marcus, L.~N. Pfeiffer, and
  K.~W. West, ``{Distinct signatures for Coulomb blockade and Aharonov-Bohm
  interference in electronic Fabry-Perot interferometers},'' {\em Physical
  Review B}, vol.~79, no.~24, p.~241304, 2009.

\bibitem{Ji2003}
Y.~Ji, Y.~Chung, D.~Sprinzak, M.~Heiblum, D.~Mahalu, and H.~Shtrikman, ``{An
  electronic Mach{\textendash}Zehnder interferometer},'' {\em Nature},
  vol.~422, pp.~415--418, Mar. 2003.

\bibitem{Feve2007}
G.~F\`eve, A.~Mahe', J.-M. Berroir, T.~Kontos, B.~Pla{\c{c}}ais, D.~C. Glattli,
  A.~Cavanna, B.~Etienne, and Y.~Jin, ``An on-demand coherent single-electron
  source,'' {\em Science}, vol.~316, pp.~1169--1172, May 2007.

\bibitem{Dubois2013}
J.~Dubois, T.~Jullien, F.~Portier, P.~Roche, A.~Cavanna, Y.~Jin,
  W.~Wegscheider, P.~Roulleau, and D.~C. Glattli, ``{Minimal-excitation states
  for electron quantum optics using levitons},'' {\em Nature}, vol.~502,
  pp.~659--663, Oct. 2013.

\bibitem{Henny1999}
M.~Henny, S.~Oberholzer, C.~Strunk, T.~Heinzel, K.~Ensslin, M.~Holland, and
  C.~Sch{\ifmmode\ddot{o}\else\"{o}\fi}nenberger, ``The fermionic {Hanbury
  Brown and Twiss} experiment,'' {\em Science}, vol.~284, pp.~296--298, apr
  1999.

\bibitem{Oliver1999}
W.~D. Oliver, J.~Kim, R.~C. Liu, and Y.~Yamamoto, ``{Hanbury Brown and
  Twiss}-type experiment with electrons,'' {\em Science}, vol.~284,
  pp.~299--301, apr 1999.

\bibitem{Bocquillon2012}
E.~Bocquillon, F.~D. Parmentier, C.~Grenier, J.-M. Berroir, P.~Degiovanni,
  D.~C. Glattli, B.~Pla{\c{c}}ais, A.~Cavanna, Y.~Jin, and G.~F{\`{e}}ve,
  ``Electron quantum optics: Partitioning electrons one by one,'' {\em Physical
  Review Letters}, vol.~108, p.~196803, may 2012.

\bibitem{Bocquillon2013}
E.~Bocquillon, V.~Freulon, J.-M. Berroir, P.~Degiovanni,
  B.~Pla{\ifmmode\mbox{\c{c}}\else\c{c}\fi}ais, A.~Cavanna, Y.~Jin, and
  G.~F{\ifmmode\grave{e}\else\`{e}\fi}ve, ``Coherence and indistinguishability
  of single electrons emitted by independent sources,'' {\em Science},
  vol.~339, pp.~1054--1057, Mar. 2013.

\bibitem{Nakamura2020}
J.~Nakamura, S.~Liang, G.~C. Gardner, and M.~J. Manfra, ``{Direct observation
  of anyonic braiding statistics},'' {\em Nat Phys}, vol.~16, pp.~931--936,
  Sept. 2020.

\bibitem{Bartolomei2020}
H.~Bartolomei, M.~Kumar, R.~Bisognin, A.~Marguerite, J.-M. Berroir,
  E.~Bocquillon, B.~Pla{\ifmmode\mbox{\c{c}}\else\c{c}\fi}ais, A.~Cavanna,
  Q.~Dong, U.~Gennser, Y.~Jin, and G.~F{\ifmmode\grave{e}\else\`{e}\fi}ve,
  ``{Fractional statistics in anyon collisions},'' {\em Science}, vol.~368,
  pp.~173--177, Apr. 2020.

\bibitem{Nayak2008}
C.~Nayak, S.~H. Simon, A.~Stern, M.~Freedman, and S.~Das~Sarma, ``{Non-Abelian
  anyons and topological quantum computation},'' {\em Rev. Mod. Phys.},
  vol.~80, pp.~1083--1159, Sept. 2008.

\bibitem{Goerbig2011}
M.~O. Goerbig, ``{Electronic properties of graphene in a strong magnetic
  field},'' {\em Reviews of Modern Physics}, vol.~83, pp.~1193--1243, nov 2011.

\bibitem{Young2012}
A.~F. Young, C.~R. Dean, L.~Wang, H.~Ren, P.~Cadden-Zimansky, K.~Watanabe,
  T.~Taniguchi, J.~Hone, K.~L. Shepard, and P.~Kim, ``{Spin and valley quantum
  Hall ferromagnetism in graphene},'' {\em Nature Physics}, vol.~8,
  pp.~550--556, July 2012.

\bibitem{Kharitonov2012ML}
M.~Kharitonov, ``Phase diagram for the $\ensuremath{\nu}=0$ quantum hall state
  in monolayer graphene,'' {\em Phys. Rev. B}, vol.~85, p.~155439, Apr 2012.

\bibitem{Kharitonov2012BL}
M.~Kharitonov, ``Canted antiferromagnetic phase of the
  $\ensuremath{\nu}\mathbf{=}0$ quantum hall state in bilayer graphene,'' {\em
  Phys. Rev. Lett.}, vol.~109, p.~046803, Jul 2012.

\bibitem{Knothe2016}
A.~Knothe and T.~Jolicoeur, ``Phase diagram of a graphene bilayer in the
  zero-energy landau level,'' {\em Phys. Rev. B}, vol.~94, p.~235149, Dec 2016.

\bibitem{Allen2012}
M.~T. Allen, J.~Martin, and A.~Yacoby, ``Gate-defined quantum confinement in
  suspended bilayer graphene,'' {\em Nature Communications}, vol.~3, p.~934,
  July 2012.

\bibitem{Droscher2012}
S.~Dr{\"o}scher, C.~Barraud, K.~Watanabe, T.~Taniguchi, T.~Ihn, and K.~Ensslin,
  ``Electron flow in split-gated bilayer graphene,'' {\em New Journal of
  Physics}, vol.~14, no.~10, p.~103007, 2012.

\bibitem{Goossens2012}
A.~S.~M. Goossens, S.~C.~M. Driessen, T.~A. Baart, K.~Watanabe, T.~Taniguchi,
  and L.~M.~K. Vandersypen, ``Gate-defined confinement in bilayer
  graphene-hexagonal boron nitride hybrid devices,'' {\em Nano Letters},
  vol.~12, pp.~4656--4660, Sept. 2012.

\bibitem{Overweg2018}
H.~Overweg, H.~Eggimann, X.~Chen, S.~Slizovskiy, M.~Eich, R.~Pisoni, Y.~Lee,
  P.~Rickhaus, K.~Watanabe, T.~Taniguchi, V.~Fal'ko, T.~Ihn, and K.~Ensslin,
  ``Electrostatically induced quantum point contacts in bilayer graphene,''
  {\em Nano Letters}, vol.~18, pp.~553--559, Jan. 2018.

\bibitem{Guinea2010}
F.~Guinea, A.~K. Geim, M.~I. Katsnelson, and K.~S. Novoselov, ``Generating
  quantizing pseudomagnetic fields by bending graphene ribbons,'' {\em Phys.
  Rev. B}, vol.~81, p.~035408, Jan 2010.

\bibitem{Choi2010}
S.-M. Choi, S.-H. Jhi, and Y.-W. Son, ``Effects of strain on electronic
  properties of graphene,'' {\em Physical Review B}, vol.~81, feb 2010.

\bibitem{Grassano2020}
D.~Grassano, M.~D'Alessandro, O.~Pulci, S.~G. Sharapov, V.~P. Gusynin, and
  A.~A. Varlamov, ``Work function, deformation potential, and collapse of
  landau levels in strained graphene and silicene,'' {\em Phys. Rev. B},
  vol.~101, p.~245115, Jun 2020.

\bibitem{Wang2021}
L.~Wang, A.~Baumgartner, P.~Makk, S.~Zihlmann, B.~S. Varghese, D.~I. Indolese,
  K.~Watanabe, T.~Taniguchi, and C.~Sch\"{o}nenberger, ``Global strain-induced
  scalar potential in graphene devices,'' {\em Communications Physics}, vol.~4,
  jun 2021.

\bibitem{Young2014}
A.~F. Young, J.~D. Sanchez-Yamagishi, B.~Hunt, S.~H. Choi, K.~Watanabe,
  T.~Taniguchi, R.~C. Ashoori, and P.~Jarillo-Herrero, ``{Tunable symmetry
  breaking and helical edge transport in a graphene quantum spin Hall state},''
  {\em Nature}, vol.~505, pp.~528--532, jan 2014.

\bibitem{Veyrat2019}
L.~Veyrat, A.~Jordan, K.~Zimmermann, F.~Gay, K.~Watanabe, T.~Taniguchi,
  H.~Sellier, and B.~Sac{\'{e}}p{\'{e}}, ``Low-magnetic-field regime of a
  gate-defined constriction in high-mobility graphene,'' {\em Nano Letters},
  vol.~19, pp.~635--642, feb 2019.

\bibitem{Knothe2015}
A.~Knothe and T.~Jolicoeur, ``Edge structure of graphene monolayers in the
  {$N$}=0 quantum {{Hall}} state,'' {\em Phys.~Rev.~B}, vol.~92, p.~165110,
  2015.

\bibitem{Mccann2006}
E.~McCann and V.~I. Fal'ko, ``Landau-level degeneracy and quantum {{Hall}}
  effect in a graphite bilayer,'' {\em Physical Review Letters}, vol.~96,
  p.~086805, Mar. 2006.

\bibitem{Mccann2007}
E.~McCann, D.~S. Abergel, and V.~I. Fal'ko, ``The low energy electronic band
  structure of bilayer graphene,'' {\em The European Physical Journal Special
  Topics}, vol.~148, pp.~91--103, Sept. 2007.

\bibitem{Mccann2013}
E.~McCann and M.~Koshino, ``The electronic properties of bilayer graphene,''
  {\em Reports on Progress in Physics}, vol.~76, no.~5, p.~056503, 2013.

\bibitem{Chang1995}
M.-C. Chang and Q.~Niu, ``Berry phase, hyperorbits, and the {Hofstadter}
  spectrum,'' {\em Phys. Rev. Lett.}, vol.~75, pp.~1348--1351, Aug 1995.

\bibitem{Ashcroft1976}
N.~W. Ashcroft and N.~D. Mermin, {\em Solid State Physics}.
\newblock New York: Holt, Rinehart and Winston, 1976.

\bibitem{Liu2012}
M.-H. Liu, J.~Bundesmann, and K.~Richter, ``{Spin-dependent \uppercase{K}lein
  tunneling in graphene: Role of \uppercase{R}ashba spin-orbit coupling},''
  {\em Phys. Rev. B}, vol.~85, p.~085406, Feb 2012.

\bibitem{HandschinThesis2017}
C.~Handschin, {\em Quantum-Transport in Encapsulated Graphene P-N junctions}.
\newblock PhD thesis, Department of Physics, University of Basel, Basel, Mar
  2017.
\newblock Magna Cum Laude.

\bibitem{Klein1929}
O.~{Klein}, ``{Die Reflexion von Elektronen an einem Potentialsprung nach der
  relativistischen Dynamik von Dirac},'' {\em Zeitschrift f{\"u}r Physik},
  vol.~53, pp.~157--165, Mar. 1929.

\bibitem{Beenakker2008}
C.~W.~J. Beenakker, ``Colloquium: Andreev reflection and {Klein} tunneling in
  graphene,'' {\em Reviews Of Modern Physics}, vol.~80, pp.~1337--1354, oct
  2008.

\bibitem{CastroNeto2009}
A.~H. Castro~Neto, F.~Guinea, N.~M.~R. Peres, K.~S. Novoselov, and A.~K. Geim,
  ``{The electronic properties of graphene},'' {\em Rev. Mod. Phys.}, vol.~81,
  p.~109, 2009.

\bibitem{Allain2011}
P.~Allain and J.~Fuchs, ``{Klein tunneling in graphene: optics with massless
  electrons},'' {\em Eur. Phys. J. B}, vol.~83, pp.~301--317, 2011.

\bibitem{Datta1995}
S.~Datta, {\em Electronic Transport in Mesoscopic Systems}.
\newblock Cambridge University Press, 1995.

\bibitem{Rickhaus2015b}
P.~Rickhaus, {\em Electron optics in ballistic graphene}.
\newblock PhD thesis, University of Basel, 2015.

\bibitem{RickhausThesis2015}
P.~Rickhaus, {\em Electron Optics in Ballistic Graphene}.
\newblock PhD thesis, Department of Physics, University of Basel, Basel, Sep
  2015.

\bibitem{Wimmer2008}
M.~Wimmer, {\em Quantum transport in nanostructures: From computational
  concepts to spintronics in graphene and magnetic tunnel junctions}.
\newblock PhD thesis, Universit{\"a}t Regensburg, 2008.

\bibitem{Lewenkopf2013}
C.~H. Lewenkopf and E.~R. Mucciolo, ``The recursive {Green's} function method
  for graphene,'' {\em Journal of Computational Electronics}, vol.~12,
  pp.~203--231, may 2013.

\bibitem{Luryi1988}
S.~Luryi, ``{Quantum capacitance devices},'' {\em Appl. Phys. Lett.}, vol.~52,
  no.~6, pp.~501--503, 1988.

\bibitem{Fang2007}
T.~Fang, A.~Konar, H.~Xing, and D.~Jena, ``{Carrier statistics and quantum
  capacitance of graphene sheets and ribbons},'' {\em Appl. Phys. Lett.},
  vol.~91, no.~9, p.~092109, 2007.

\bibitem{Droscher2010}
S.~Dr{\"o}scher, P.~Roulleau, F.~Molitor, P.~Studerus, C.~Stampfer, K.~Ensslin,
  and T.~Ihn, ``{Quantum capacitance and density of states of graphene},'' {\em
  Applied Physics Letters}, vol.~96, no.~15, p.~152104, 2010.

\bibitem{Liu2013}
M.-H. Liu, ``{Theory of carrier density in multigated doped graphene sheets
  with quantum correction},'' {\em Phys. Rev. B}, vol.~87, p.~125427, Mar 2013.

\bibitem{comsol}
C.~Multiphysics, ``Introduction to comsol multiphysics{\textregistered},'' {\em
  COMSOL Multiphysics, Burlington, MA, accessed Feb}, vol.~9, p.~2018, 1998.

\bibitem{FEniCS}
A.~Logg, K.-A. Mardal, G.~N. Wells, {\em et~al.}, {\em Automated Solution of
  Differential Equations by the Finite Element Method}.
\newblock Springer, 2012.

\bibitem{pde}
The MathWorks, Inc., {\em Partial Differential Equation Toolbox$^{TM}$ User's
  Guide}, matlab 2023a~ed., 2023.

\bibitem{Liu2015}
M.-H. Liu, P.~Rickhaus, P.~Makk, E.~T\'ov\'ari, R.~Maurand, F.~Tkatschenko,
  M.~Weiss, C.~Sch\"onenberger, and K.~Richter, ``Scalable tight-binding model
  for graphene,'' {\em Physical Review Letters}, vol.~114, p.~036601, Jan 2015.

\bibitem{Rickhaus2015c}
P.~Rickhaus, M.-H. Liu, P.~Makk, R.~Maurand, S.~Hess, S.~Zihlmann, M.~Weiss,
  K.~Richter, and C.~Sch{\"{o}}nenberger, ``Guiding of electrons in a few-mode
  ballistic graphene channel,'' {\em Nano Letters}, vol.~15, pp.~5819--5825,
  sep 2015.

\bibitem{Rickhaus2015d}
P.~Rickhaus, P.~Makk, M.-H. Liu, K.~Richter, and C.~Sch{\"{o}}nenberger,
  ``{Gate tuneable beamsplitter in ballistic graphene},'' {\em Applied Physics
  Letters}, vol.~107, dec 2015.

\bibitem{Terres2016}
B.~Terres, L.~A. Chizhova, F.~Libisch, J.~Peiro, D.~Joerger, S.~Engels,
  A.~Girschik, K.~Watanabe, T.~Taniguchi, S.~V. Rotkin, J.~Burgdoerfer, and
  C.~Stampfer, ``Size quantization of dirac fermions in graphene
  constrictions,'' {\em Nature Communications}, vol.~7, p.~11528, may 2016.

\bibitem{Xiang2016}
S.~Xiang, A.~Mre{\'n}ca-Kolasi{\'n}ska, V.~Miseikis, S.~Guiducci,
  K.~Kolasi{\'n}ski, C.~Coletti, B.~Szafran, F.~Beltram, S.~Roddaro, and
  S.~Heun, ``Interedge backscattering in buried split-gate-defined graphene
  quantum point contacts,'' {\em Physical Review B}, vol.~94, no.~15,
  p.~155446, 2016.

\bibitem{Liu2017}
M.-H. Liu, C.~Gorini, and K.~Richter, ``Creating and steering highly
  directional electron beams in graphene,'' {\em Phys. Rev. Lett.}, vol.~118,
  p.~066801, Feb 2017.

\bibitem{Bours2017}
L.~Bours, S.~Guiducci, A.~Mre{\'{n}}ca-Kolasi{\'{n}}ska, B.~Szafran, J.~C.
  Maan, and S.~Heun, ``{Manipulating quantum Hall edge channels in graphene
  through scanning gate microscopy},'' {\em Physical Review B}, vol.~96,
  p.~195423, nov 2017.

\bibitem{Kolasinski2017}
K.~Kolasinski, A.~Mrenca-Kolasinska, and B.~Szafran, ``Imaging snake orbits at
  graphene $n\ensuremath{-}p$ junctions,'' {\em Phys. Rev. B}, vol.~95,
  p.~045304, jan 2017.

\bibitem{Makk2018}
P.~Makk, C.~Handschin, E.~T{\'{o}}v{\'{a}}ri, K.~Watanabe, T.~Taniguchi,
  K.~Richter, M.-H. Liu, and C.~Sch{\"{o}}nenberger, ``{Coexistence of
  classical snake states and Aharonov-Bohm oscillations along graphene p n
  junctions},'' {\em Physical Review B}, vol.~98, p.~035413, jul 2018.

\bibitem{Ma2018}
Q.~Ma, F.~D. Parmentier, P.~Roulleau, and G.~Fleury, ``{Graphene
  $n\ensuremath{-}p$ junctions in the quantum Hall regime: Numerical study of
  incoherent scattering effects},'' {\em Phys Rev B}, vol.~97, p.~205445, May
  2018.

\bibitem{Brun2019}
B.~Brun, N.~Moreau, S.~Somanchi, V.-H. Nguyen, K.~Watanabe, T.~Taniguchi, J.-C.
  Charlier, C.~Stampfer, and B.~Hackens, ``Imaging dirac fermions flow through
  a circular veselago lens,'' {\em Phys. Rev. B}, vol.~100, p.~041401, Jul
  2019.

\bibitem{Lane2019}
T.~L.~M. Lane, A.~Knothe, and V.~I. Fal'ko, ``Semimetallic features in quantum
  transport through a gate-defined point contact in bilayer graphene,'' {\em
  Physical Review B}, vol.~100, p.~115427, Sept. 2019.

\bibitem{Kraft2020}
R.~Kraft, M.-H. Liu, P.~B. Selvasundaram, S.-C. Chen, R.~Krupke, K.~Richter,
  and R.~Danneau, ``Anomalous cyclotron motion in graphene superlattice
  cavities,'' {\em Physical Review Letters}, vol.~125, p.~217701, Nov. 2020.

\bibitem{Kang2020}
W.-H. Kang, S.-C. Chen, and M.-H. Liu, ``Cloning of zero modes in
  one-dimensional graphene superlattices,'' {\em Phys. Rev. B}, vol.~102,
  p.~195432, Nov 2020.

\bibitem{Moreau2020}
N.~Moreau, B.~Brun, S.~Somanchi, K.~Watanabe, T.~Taniguchi, C.~Stampfer, and
  B.~Hackens, ``{Upstream modes and antidots poison graphene quantum Hall
  effect},'' {\em Nat Commun}, vol.~12, pp.~1--7, July 2021.

\bibitem{Moreau2021}
N.~Moreau, B.~Brun, S.~Somanchi, K.~Watanabe, T.~Taniguchi, C.~Stampfer, and
  B.~Hackens, ``{Contacts and upstream modes explain the electron-hole
  asymmetry in the graphene quantum Hall regime},'' {\em Phys Rev B}, vol.~104,
  p.~L201406, Nov. 2021.

\bibitem{Schrepfer2021}
J.-K. Schrepfer, S.-C. Chen, M.-H. Liu, K.~Richter, and M.~Hentschel, ``Dirac
  fermion optics and directed emission from single- and bilayer graphene
  cavities,'' {\em Phys. Rev. B}, vol.~104, p.~155436, Oct 2021.

\bibitem{MrencaKolasinska2022}
A.~Mre{\'{n}}ca-Kolasi{\'{n}}ska, P.~Rickhaus, G.~Zheng, K.~Richter, T.~Ihn,
  K.~Ensslin, and M.-H. Liu, ``Quantum capacitive coupling between large-angle
  twisted graphene layers,'' {\em 2D Materials}, vol.~9, p.~025013, Feb 2022.

\bibitem{Brun2022}
B.~Brun, V.-H. Nguyen, N.~Moreau, S.~Somanchi, K.~Watanabe, T.~Taniguchi, J.-C.
  Charlier, C.~Stampfer, and B.~Hackens, ``Graphene whisperitronics:
  Transducing whispering gallery modes into electronic transport,'' {\em Nano
  Letters}, vol.~22, pp.~128--134, Jan. 2022.

\bibitem{Chiu2022}
S.-B. Chiu, A.~Mre\'{n}ca-Kolasi\'{n}ska, K.~L. Lei, C.-H. Chiu, W.-H. Kang,
  S.-C. Chen, and M.-H. Liu, ``Manipulating electron waves in graphene using
  carbon nanotube gating,'' {\em Phys. Rev. B}, vol.~105, p.~195416, May 2022.

\bibitem{Zhumagulov2022}
Y.~Zhumagulov, T.~Frank, and J.~Fabian, ``Edge states in proximitized graphene
  ribbons and flakes in a perpendicular magnetic field: Emergence of lone
  pseudohelical pairs and pure spin-current states,'' {\em Phys. Rev. B},
  vol.~105, may 2022.

\bibitem{Rao2023}
Q.~Rao, W.-H. Kang, H.~Xue, Z.~Ye, X.~Feng, K.~Watanabe, T.~Taniguchi, N.~Wang,
  M.-H. Liu, and D.-K. Ki, ``{Ballistic transport spectroscopy of
  spin-orbit-coupled bands in monolayer graphene on WSe$_2$},'' {\em Nature
  Communications}, vol.~14, p.~6124, mar 2023.

\bibitem{Du2018}
R.~Du, M.-H. Liu, J.~Mohrmann, F.~Wu, R.~Krupke, H.~{von L{\"o}hneysen},
  K.~Richter, and R.~Danneau, ``Tuning anti-{{Klein}} to {{Klein}} tunnelingin
  bilayer graphene,'' {\em Physical Review Letters}, vol.~121, p.~127706, Sept.
  2018.

\bibitem{Liu2012a}
M.-H. Liu and K.~Richter, ``{Efficient quantum transport simulation for bulk
  graphene heterojunctions},'' {\em Phys. Rev. B}, vol.~86, p.~115455, Sep
  2012.

\bibitem{Drienovsky2014}
M.~Drienovsky, F.-X. Schrettenbrunner, A.~Sandner, D.~Weiss, J.~Eroms, M.-H.
  Liu, F.~Tkatschenko, and K.~Richter, ``Towards superlattices: {{Lateral}}
  bipolar multibarriers in graphene,'' {\em Physical Review B}, vol.~89,
  p.~115421, Mar. 2014.

\bibitem{Varlet2014}
A.~Varlet, M.-H. Liu, V.~Krueckl, D.~Bischoff, P.~Simonet, K.~Watanabe,
  T.~Taniguchi, K.~Richter, K.~Ensslin, and T.~Ihn, ``{Fabry-P\'erot}
  interference in gapped bilayer graphene with broken anti-{Klein} tunneling,''
  {\em Phys. Rev. Lett.}, vol.~113, p.~116601, Sep 2014.

\bibitem{Rickhaus2020}
P.~Rickhaus, M.-H. Liu, M.~Kurpas, A.~Kurzmann, Y.~Lee, H.~Overweg, M.~Eich,
  R.~Pisoni, T.~Taniguchi, K.~Watanabe, K.~Richter, K.~Ensslin, and T.~Ihn,
  ``The electronic thickness of graphene,'' {\em Science Advances}, 2020.

\bibitem{Mayorov2011}
A.~S. Mayorov, R.~V. Gorbachev, S.~V. Morozov, L.~Britnell, R.~Jalil, L.~A.
  Ponomarenko, P.~Blake, K.~S. Novoselov, K.~Watanabe, T.~Taniguchi, and A.~K.
  Geim, ``{Micrometer-scale ballistic transport in encapsulated graphene at
  room temperature},'' {\em Nano Letters}, vol.~11, pp.~2396--2399, jun 2011.

\bibitem{Taychatanapat2013}
T.~Taychatanapat, K.~Watanabe, T.~Taniguchi, and P.~Jarillo-Herrero,
  ``{Electrically tunable transverse magnetic focusing in graphene},'' {\em
  Nature Physics 2013 9:4}, vol.~9, pp.~225--229, feb 2013.

\bibitem{Bhandari2016}
S.~Bhandari, G.-H. Lee, A.~Klales, K.~Watanabe, T.~Taniguchi, E.~Heller,
  P.~Kim, and R.~M. Westervelt, ``Imaging cyclotron orbits of electrons in
  graphene,'' {\em Nano letters}, 2016.

\bibitem{Lee2016}
M.~Lee, J.~R. Wallbank, P.~Gallagher, K.~Watanabe, T.~Taniguchi, V.~I. Fal'ko,
  and D.~{Goldhaber-Gordon}, ``Ballistic miniband conduction in a graphene
  superlattice,'' {\em Science}, vol.~353, pp.~1526--1529, Sept. 2016.

\bibitem{Chen2008}
J.~H. Chen, C.~Jang, S.~Xiao, M.~Ishigami, and M.~S. Fuhrer, ``{Intrinsic and
  extrinsic performance limits of graphene devices on SiO2},'' {\em Nature
  Nanotechnology 2008 3:4}, vol.~3, pp.~206--209, mar 2008.

\bibitem{Goossens2012a}
A.~M. Goossens, V.~E. Calado, A.~Barreiro, K.~Watanabe, T.~Taniguchi, and L.~M.
  Vandersypen, ``{Mechanical cleaning of graphene},'' {\em Applied Physics
  Letters}, vol.~100, p.~73110, feb 2012.

\bibitem{Lindvall2012}
N.~Lindvall, A.~Kalabukhov, and A.~Yurgens, ``{Cleaning graphene using atomic
  force microscope},'' {\em Journal of Applied Physics}, vol.~111, mar 2012.

\bibitem{Tombros2011b}
N.~Tombros, A.~Veligura, J.~Junesch, J.~{Jasper Van Den Berg}, P.~J. Zomer,
  M.~Wojtaszek, I.~J. {Vera Marun}, H.~T. Jonkman, and B.~J. {Van Wees},
  ``{Large yield production of high mobility freely suspended graphene
  electronic devices on a polydimethylglutarimide based organic polymer},''
  {\em Journal of Applied Physics}, vol.~109, may 2011.

\bibitem{Maurand2014a}
R.~Maurand, P.~Rickhaus, P.~Makk, S.~Hess, E.~T{\'{o}}v{\'{a}}ri, C.~Handschin,
  M.~Weiss, and C.~Sch{\"{o}}nenberger, ``{Fabrication of ballistic suspended
  graphene with local-gating},'' {\em Carbon}, vol.~79, pp.~486--492, nov 2014.

\bibitem{Haigh2012}
S.~J. Haigh, A.~Gholinia, R.~Jalil, S.~Romani, L.~Britnell, D.~C. Elias, K.~S.
  Novoselov, L.~A. Ponomarenko, A.~K. Geim, and R.~Gorbachev,
  ``{Cross-sectional imaging of individual layers and buried interfaces of
  graphene-based heterostructures and superlattices},'' {\em Nature Materials
  2012 11:9}, vol.~11, pp.~764--767, jul 2012.

\bibitem{Zomer2014a}
P.~J. Zomer, M.~H. Guimar{\~{a}}es, J.~C. Brant, N.~Tombros, and B.~J. {Van
  Wees}, ``{Fast pick up technique for high quality heterostructures of bilayer
  graphene and hexagonal boron nitride},'' {\em Applied Physics Letters},
  vol.~105, jul 2014.

\bibitem{Kretinin2014a}
A.~V. Kretinin, Y.~Cao, J.~S. Tu, G.~L. Yu, R.~Jalil, K.~S. Novoselov, S.~J.
  Haigh, A.~Gholinia, A.~Mishchenko, M.~Lozada, T.~Georgiou, C.~R. Woods,
  F.~Withers, P.~Blake, G.~Eda, A.~Wirsig, C.~Hucho, K.~Watanabe, T.~Taniguchi,
  A.~K. Geim, and R.~V. Gorbachev, ``{Electronic properties of graphene
  encapsulated with different two-dimensional atomic crystals},'' {\em Nano
  Letters}, vol.~14, pp.~3270--3276, jun 2014.

\bibitem{Pizzocchero2016}
F.~Pizzocchero, L.~Gammelgaard, B.~S. Jessen, J.~M. Caridad, L.~Wang, J.~Hone,
  P.~B{\o}ggild, and T.~J. Booth, ``{The hot pick-up technique for batch
  assembly of van der Waals heterostructures},'' {\em Nature Communications
  2016 7:1}, vol.~7, pp.~1--10, jun 2016.

\bibitem{Purdie2018}
D.~G. Purdie, N.~M. Pugno, T.~Taniguchi, K.~Watanabe, A.~C. Ferrari, and
  A.~Lombardo, ``{Cleaning interfaces in layered materials heterostructures},''
  {\em Nature Communications 2018 9:1}, vol.~9, pp.~1--12, dec 2018.

\bibitem{Zibrov2017}
A.~A. Zibrov, C.~Kometter, H.~Zhou, E.~M. Spanton, T.~Taniguchi, K.~Watanabe,
  M.~P. Zaletel, and A.~F. Young, ``{Tunable interacting composite fermion
  phases in a half-filled bilayer-graphene Landau level},'' {\em Nature 2017
  549:7672}, vol.~549, pp.~360--364, sep 2017.

\bibitem{Zeng2019}
Y.~Zeng, J.~I. Li, S.~A. Dietrich, O.~M. Ghosh, K.~Watanabe, T.~Taniguchi,
  J.~Hone, and C.~R. Dean, ``High-quality magnetotransport in graphene using
  the edge-free corbino geometry,'' {\em Physical Review Letters}, vol.~122,
  p.~137701, apr 2019.

\bibitem{Polshyn2018}
H.~Polshyn, H.~Zhou, E.~M. Spanton, T.~Taniguchi, K.~Watanabe, and A.~F. Young,
  ``Quantitative transport measurements of fractional quantum {Hall} energy
  gaps in edgeless graphene devices,'' {\em Physical Review Letters}, vol.~121,
  p.~226801, nov 2018.

\bibitem{Yankowitz2019b}
M.~Yankowitz, Q.~Ma, P.~Jarillo-Herrero, and B.~J. LeRoy, ``{van der Waals
  heterostructures combining graphene and hexagonal boron nitride},'' {\em
  Nature Reviews Physics 2018 1:2}, vol.~1, pp.~112--125, jan 2019.

\bibitem{Wang2023}
W.~Wang, N.~Clark, M.~Hamer, A.~Carl, E.~Tovari, S.~Sullivan-Allsop,
  E.~Tillotson, Y.~Gao, H.~de~Latour, F.~Selles, J.~Howarth, E.~G. Castanon,
  M.~Zhou, H.~Bai, X.~Li, A.~Weston, K.~Watanabe, T.~Taniguchi, C.~Mattevi,
  T.~H. Bointon, P.~V. Wiper, A.~J. Strudwick, L.~A. Ponomarenko, A.~Kretinin,
  S.~J. Haigh, A.~Summerfield, and R.~Gorbachev, ``{Ultra-clean assembly of van
  der Waals heterostructures},'' {\em Nature Electronics}, pp.~1--10, aug 2023.

\bibitem{Barrier2020}
J.~Barrier, P.~Kumaravadivel, R.~{Krishna Kumar}, L.~A. Ponomarenko, N.~Xin,
  M.~Holwill, C.~Mullan, M.~Kim, R.~V. Gorbachev, M.~D. Thompson, J.~R. Prance,
  T.~Taniguchi, K.~Watanabe, I.~V. Grigorieva, K.~S. Novoselov, A.~Mishchenko,
  V.~I. Fal'ko, A.~K. Geim, and A.~I. Berdyugin, ``{Long-range ballistic
  transport of Brown-Zak fermions in graphene superlattices},'' {\em Nature
  Communications 2020 11:1}, vol.~11, pp.~1--7, nov 2020.

\bibitem{Nam2017a}
Y.~Nam, D.~K. Ki, D.~Soler-Delgado, and A.~F. Morpurgo, ``{Electron–hole
  collision limited transport in charge-neutral bilayer graphene},'' {\em
  Nature Physics 2017 13:12}, vol.~13, pp.~1207--1214, aug 2017.

\bibitem{ZihlmannThesis2015}
S.~Zihlmann, {\em Spin and charge relaxation in graphene}.
\newblock PhD thesis, Department of Physics, University of Basel, Basel, Mar
  2018.

\bibitem{Coissard2022}
A.~Coissard, D.~Wander, H.~Vignaud, A.~G. Grushin, C.~Repellin, K.~Watanabe,
  T.~Taniguchi, F.~Gay, C.~B. Winkelmann, H.~Courtois, H.~Sellier, and
  B.~Sac{\'{e}}p{\'{e}}, ``{Imaging tunable quantum Hall broken-symmetry orders
  in graphene},'' {\em Nature 2022 605:7908}, vol.~605, pp.~51--56, may 2022.

\bibitem{Liu2021}
X.~Liu, Z.~Wang, K.~Watanabe, T.~Taniguchi, O.~Vafek, and J.~I. Li, ``{Tuning
  electron correlation in magic-angle twisted bilayer graphene using Coulomb
  screening},'' {\em Science}, vol.~371, pp.~1261--1265, mar 2021.

\bibitem{Kim2020}
M.~Kim, S.~G. Xu, A.~I. Berdyugin, A.~Principi, S.~Slizovskiy, N.~Xin,
  P.~Kumaravadivel, W.~Kuang, M.~Hamer, R.~Krishna~Kumar, R.~V. Gorbachev,
  K.~Watanabe, T.~Taniguchi, I.~V. Grigorieva, V.~I. Fal{'}ko, M.~Polini, and
  A.~K. Geim, ``{Control of electron-electron interaction in graphene by
  proximity screening},'' {\em Nat Commun}, vol.~11, pp.~1--6, May 2020.

\bibitem{Stepanov2020}
P.~Stepanov, I.~Das, X.~Lu, A.~Fahimniya, K.~Watanabe, T.~Taniguchi, F.~H.
  Koppens, J.~Lischner, L.~Levitov, and D.~K. Efetov, ``{Untying the insulating
  and superconducting orders in magic-angle graphene},'' {\em Nature 2020
  583:7816}, vol.~583, pp.~375--378, jul 2020.

\bibitem{Wang2020a}
L.~Wang, P.~Makk, S.~Zihlmann, A.~Baumgartner, D.~I. Indolese, K.~Watanabe,
  T.~Taniguchi, and C.~Sch{\"{o}}nenberger, ``Mobility enhancement in graphene
  by in situ reduction of random strain fluctuations,'' {\em Physical Review
  Letters}, vol.~124, p.~157701, apr 2020.

\bibitem{Couto2014}
N.~J. Couto, D.~Costanzo, S.~Engels, D.~K. Ki, K.~Watanabe, T.~Taniguchi,
  C.~Stampfer, F.~Guinea, and A.~F. Morpurgo, ``{Random strain fluctuations as
  dominant disorder source for high-quality on-substrate graphene devices},''
  {\em Physical Review X}, vol.~4, p.~041019, oct 2014.

\bibitem{Morikawa2015}
S.~Morikawa, Z.~Dou, S.~W. Wang, C.~G. Smith, K.~Watanabe, T.~Taniguchi,
  S.~Masubuchi, T.~Machida, and M.~R. Connolly, ``{Imaging ballistic carrier
  trajectories in graphene using scanning gate microscopy},'' {\em Applied
  Physics Letters}, vol.~107, dec 2015.

\bibitem{Bhandari2020}
S.~Bhandari, G.~H. Lee, K.~Watanabe, T.~Taniguchi, P.~Kim, and R.~M.
  Westervelt, ``Imaging {Andreev} reflection in graphene,'' {\em Nano Letters},
  vol.~20, pp.~4890--4894, jul 2020.

\bibitem{Ingla-Aynes2023}
J.~Ingla-Aynés, A.~L.~R. Manesco, T.~S. Ghiasi, S.~Volosheniuk, K.~Watanabe,
  T.~Taniguchi, and H.~S. van~der Zant, ``Specular electron focusing between
  gate-defined quantum point contacts in bilayer graphene,'' {\em Nano
  Letters}, vol.~23, p.~5453–5459, jun 2023.

\bibitem{Chen2016}
S.~Chen, Z.~Han, M.~M. Elahi, K.~M.~M. Habib, L.~Wang, B.~Wen, Y.~Gao,
  T.~Taniguchi, K.~Watanabe, J.~Hone, A.~W. Ghosh, and C.~R. Dean, ``Electron
  optics with p-n junctions in ballistic graphene,'' {\em Science}, vol.~353,
  pp.~1522--1525, sep 2016.

\bibitem{Berdyugin2020}
A.~I. Berdyugin, B.~Tsim, P.~Kumaravadivel, S.~G. Xu, A.~Ceferino, A.~Knothe,
  R.~K. Kumar, T.~Taniguchi, K.~Watanabe, A.~K. Geim, I.~V. Grigorieva, and
  V.~I. Fal'ko, ``Minibands in twisted bilayer graphene probed by magnetic
  focusing,'' {\em Science Advances}, vol.~6, p.~eaay7838, Apr. 2020.

\bibitem{Graef2019}
H.~Graef, Q.~Wilmart, M.~Rosticher, D.~Mele, L.~Banszerus, C.~Stampfer,
  T.~Taniguchi, K.~Watanabe, J.~M. Berroir, E.~Bocquillon, G.~F{\`{e}}ve, E.~H.
  Teo, and B.~Pla{\c{c}}ais, ``{A corner reflector of graphene Dirac fermions
  as a phonon-scattering sensor},'' {\em Nature Communications 2019 10:1},
  vol.~10, pp.~1--9, jun 2019.

\bibitem{Wang2019a}
K.~Wang, M.~M. Elahi, L.~Wang, K.~M. Habib, T.~Taniguchi, K.~Watanabe, J.~Hone,
  A.~W. Ghosh, G.~H. Lee, and P.~Kim, ``{Graphene transistor based on tunable
  Dirac fermion optics},'' {\em Proceedings of the National Academy of Sciences
  of the United States of America}, vol.~116, pp.~6575--6579, apr 2019.

\bibitem{Elahi2019}
M.~M. Elahi, K.~M. {Masum Habib}, K.~Wang, G.~H. Lee, P.~Kim, and A.~W. Ghosh,
  ``{Impact of geometry and non-idealities on electron "optics" based graphene
  p-n junction devices},'' {\em Applied Physics Letters}, vol.~114, jan 2019.

\bibitem{Morikawa2017}
S.~Morikawa, Q.~Wilmart, S.~Masubuchi, K.~Watanabe, T.~Taniguchi,
  B.~Pla{\c{c}}ais, and T.~Machida, ``{Dirac fermion reflector by ballistic
  graphene sawtooth-shaped npn junctions},'' {\em Semiconductor Science and
  Technology}, vol.~32, p.~045010, mar 2017.

\bibitem{Lee2015}
G.~H. Lee, G.~H. Park, and H.~J. Lee, ``{Observation of negative refraction of
  Dirac fermions in graphene},'' {\em Nature Physics 2015 11:11}, vol.~11,
  pp.~925--929, sep 2015.

\bibitem{Barnard2017}
A.~W. Barnard, A.~Hughes, A.~L. Sharpe, K.~Watanabe, T.~Taniguchi, and
  D.~Goldhaber-Gordon, ``{Absorptive pinhole collimators for ballistic Dirac
  fermions in graphene},'' {\em Nature Communications 2017 8:1}, vol.~8,
  pp.~1--6, may 2017.

\bibitem{Liu2008}
G.~Liu, J.~Velasco, Jr., W.~Bao, and C.~N. Lau, ``Fabrication of graphene p-n-p
  junctions with contactless top gates,'' {\em Applied Physics Letters},
  vol.~92, may 2008.

\bibitem{Handschin2015}
C.~Handschin, B.~Fülöp, P.~Makk, S.~Blanter, M.~Weiss, K.~Watanabe,
  T.~Taniguchi, S.~Csonka, and C.~Schönenberger, ``Point contacts in
  encapsulated graphene,'' {\em Applied Physics Letters}, vol.~107, p.~183108,
  Nov 2015.

\bibitem{Gold2021}
C.~Gold, A.~Knothe, A.~Kurzmann, A.~{Garcia-Ruiz}, K.~Watanabe, T.~Taniguchi,
  V.~Fal'ko, K.~Ensslin, and T.~Ihn, ``Coherent jetting from a gate-defined
  channel in bilayer graphene,'' {\em Physical Review Letters}, vol.~127,
  p.~046801, July 2021.

\bibitem{Zhang2022}
X.~Zhang, W.~Ren, E.~Bell, Z.~Zhu, K.~T. Tsai, Y.~Luo, K.~Watanabe,
  T.~Taniguchi, E.~Kaxiras, M.~Luskin, and K.~Wang, ``{Gate-tunable Veselago
  interference in a bipolar graphene microcavity},'' {\em Nature Communications
  2022 13:1}, vol.~13, pp.~1--6, nov 2022.

\bibitem{bischoff2015}
{Bischoff Dominik}, {Simonet Pauline}, {Varlet Anastasia}, {Overweg Hiske C.},
  {Eich Marius}, {Ihn Thomas}, and {Ensslin Klaus}, ``The importance of edges
  in reactive ion etched graphene nanodevices,'' {\em physica status solidi
  (RRL) \textendash{} Rapid Research Letters}, vol.~10, pp.~68--74, Aug. 2015.

\bibitem{bischoff2014}
D.~Bischoff, F.~Libisch, J.~Burgd{\"o}rfer, T.~Ihn, and K.~Ensslin,
  ``Characterizing wave functions in graphene nanodevices: {{Electronic}}
  transport through ultrashort graphene constrictions on a boron nitride
  substrate,'' {\em Physical Review B}, vol.~90, p.~115405, Sept. 2014.

\bibitem{ihn2010}
T.~Ihn, J.~G{\"u}ttinger, F.~Molitor, S.~Schnez, E.~Schurtenberger,
  A.~Jacobsen, S.~Hellm{\"u}ller, T.~Frey, S.~Dr{\"o}scher, C.~Stampfer, and
  K.~Ensslin, ``Graphene single-electron transistors,'' {\em Materials Today},
  vol.~13, pp.~44--50, Mar. 2010.

\bibitem{Overweg2018a}
H.~Overweg, A.~Knothe, T.~Fabian, L.~Linhart, P.~Rickhaus, L.~Wernli,
  K.~Watanabe, T.~Taniguchi, D.~S{\'a}nchez, J.~Burgd{\"o}rfer, F.~Libisch,
  V.~I. Fal'ko, K.~Ensslin, and T.~Ihn, ``Topologically nontrivial valley
  states in bilayer graphene quantum point contacts,'' {\em Physical Review
  Letters}, vol.~121, p.~257702, Dec. 2018.

\bibitem{Lee2020}
Y.~Lee, A.~Knothe, H.~Overweg, M.~Eich, C.~Gold, A.~Kurzmann, V.~Klasovika,
  T.~Taniguchi, K.~Wantanabe, V.~Fal'ko, T.~Ihn, K.~Ensslin, and P.~Rickhaus,
  ``Tunable {{Valley Splitting}} due to {{Topological Orbital Magnetic Moment}}
  in {{Bilayer Graphene Quantum Point Contacts}},'' {\em Physical Review
  Letters}, vol.~124, p.~126802, Mar. 2020.

\bibitem{Banszerus2020c}
L.~Banszerus, B.~Frohn, T.~Fabian, S.~Somanchi, A.~Epping, M.~M{\"u}ller,
  D.~Neumaier, K.~Watanabe, T.~Taniguchi, F.~Libisch, B.~Beschoten, F.~Hassler,
  and C.~Stampfer, ``Observation of the spin-orbit gap in bilayer graphene by
  one-dimensional ballistic transport,'' {\em Physical Review Letters},
  vol.~124, p.~177701, May 2020.

\bibitem{Knothe2018}
A.~Knothe and V.~Fal'ko, ``Influence of minivalleys and {{Berry}} curvature on
  electrostatically induced quantum wires in gapped bilayer graphene,'' {\em
  Physical Review B}, vol.~98, p.~155435, Oct. 2018.

\bibitem{Kraft2018}
R.~Kraft, I.~V. Krainov, V.~Gall, A.~P. Dmitriev, R.~Krupke, I.~V. Gornyi, and
  R.~Danneau, ``Valley subband splitting in bilayer graphene quantum point
  contacts,'' {\em Physical Review Letters}, vol.~121, p.~257703, Dec. 2018.

\bibitem{Eich2018a}
M.~Eich, R.~Pisoni, H.~Overweg, A.~Kurzmann, Y.~Lee, P.~Rickhaus, T.~Ihn,
  K.~Ensslin, F.~Herman, M.~Sigrist, K.~Watanabe, and T.~Taniguchi, ``Spin and
  valley states in gate-defined bilayer graphene quantum dots,'' {\em Physical
  Review X}, vol.~8, p.~031023, July 2018.

\bibitem{Eich2018}
M.~Eich, R.~Pisoni, A.~Pally, H.~Overweg, A.~Kurzmann, Y.~Lee, P.~Rickhaus,
  K.~Watanabe, T.~Taniguchi, K.~Ensslin, and T.~Ihn, ``Coupled quantum dots in
  bilayer graphene,'' {\em Nano Letters}, vol.~18, pp.~5042--5048, Aug. 2018.

\bibitem{Banszerus2018}
L.~Banszerus, B.~Frohn, A.~Epping, D.~Neumaier, K.~Watanabe, T.~Taniguchi, and
  C.~Stampfer, ``Gate-defined electron hole double dots in bilayer graphene,''
  {\em Nano Letters}, vol.~18, pp.~4785--4790, Aug. 2018.

\bibitem{moller2021}
S.~M{\"o}ller, L.~Banszerus, A.~Knothe, C.~Steiner, E.~Icking, S.~Trellenkamp,
  F.~Lentz, K.~Watanabe, T.~Taniguchi, L.~I. Glazman, V.~I. Fal'ko, C.~Volk,
  and C.~Stampfer, ``Probing {{Two-Electron Multiplets}} in {{Bilayer Graphene
  Quantum Dots}},'' {\em Physical Review Letters}, vol.~127, p.~256802, Dec.
  2021.

\bibitem{Tong2021}
C.~Tong, R.~Garreis, A.~Knothe, M.~Eich, A.~Sacchi, K.~Watanabe, T.~Taniguchi,
  V.~Fal'ko, T.~Ihn, K.~Ensslin, and A.~Kurzmann, ``Tunable valley splitting
  and bipolar operation in graphene quantum dots,'' {\em Nano Letters},
  vol.~21, pp.~1068--1073, Jan. 2021.

\bibitem{Tong2022}
C.~Tong, A.~Kurzmann, R.~Garreis, W.~W. Huang, S.~Jele, M.~Eich, L.~Ginzburg,
  C.~Mittag, K.~Watanabe, T.~Taniguchi, K.~Ensslin, and T.~Ihn, ``Pauli
  blockade of tunable two-electron spin and valley states in graphene quantum
  dots,'' {\em Physical Review Letters}, vol.~128, p.~067702, Feb. 2022.

\bibitem{Knothe2020}
A.~Knothe and V.~Fal'ko, ``Quartet states in two-electron quantum dots in
  bilayer graphene,'' {\em Physical Review B}, vol.~101, p.~235423, June 2020.

\bibitem{Banszerus2020b}
L.~Banszerus, A.~Rothstein, T.~Fabian, S.~M{\"o}ller, E.~Icking,
  S.~Trellenkamp, F.~Lentz, D.~Neumaier, K.~Watanabe, T.~Taniguchi, F.~Libisch,
  C.~Volk, and C.~Stampfer, ``Electron hole crossover in gate-controlled
  bilayer graphene quantum dots,'' {\em Nano Letters}, vol.~20, pp.~7709--7715,
  Oct. 2020.

\bibitem{Kurzmann2021}
A.~Kurzmann, Y.~Kleeorin, C.~Tong, R.~Garreis, A.~Knothe, M.~Eich, C.~Mittag,
  C.~Gold, F.~K. de~Vries, K.~Watanabe, T.~Takashi, F.~Vladimir, M.~Yigal,
  I.~Thomas, and E.~Klaus, ``Kondo effect and spin--orbit coupling in graphene
  quantum dots,'' {\em Nature communications}, vol.~12, no.~1, p.~6004, 2021.

\bibitem{Garreis2021}
R.~Garreis, A.~Knothe, C.~Tong, M.~Eich, C.~Gold, K.~Watanabe, T.~Taniguchi,
  V.~Fal'ko, T.~Ihn, K.~Ensslin, and A.~Kurzmann, ``Shell filling and trigonal
  warping in graphene quantum dots,'' {\em Physical Review Letters}, vol.~126,
  p.~147703, Apr. 2021.

\bibitem{Banszerus2022}
L.~Banszerus, K.~Hecker, S.~M{\"o}ller, E.~Icking, K.~Watanabe, T.~Taniguchi,
  C.~Volk, and C.~Stampfer, ``Spin relaxation in a single-electron graphene
  quantum dot,'' {\em Nature Communications}, vol.~13, p.~3637, June 2022.

\bibitem{Banszerus2020e}
L.~Banszerus, S.~M{\"o}ller, E.~Icking, K.~Watanabe, T.~Taniguchi, C.~Volk, and
  C.~Stampfer, ``Single-electron double quantum dots in bilayer graphene,''
  {\em Nano Letters}, vol.~20, pp.~2005--2011, Mar. 2020.

\bibitem{Banszerus2021}
L.~Banszerus, S.~M{\"o}ller, C.~Steiner, E.~Icking, S.~Trellenkamp, F.~Lentz,
  K.~Watanabe, T.~Taniguchi, C.~Volk, and C.~Stampfer, ``Spin-valley coupling
  in single-electron bilayer graphene quantum dots,'' {\em Nature
  communications}, vol.~12, no.~1, p.~5250, 2021.

\bibitem{Knothe2022}
A.~Knothe, L.~I. Glazman, and V.~I. Fal'ko, ``Tunneling theory for a bilayer
  graphene quantum dot's single- and two-electron states,'' {\em New Journal of
  Physics}, vol.~24, p.~043003, Apr. 2022.

\bibitem{arXiv:2303.10201}
L.~Banszerus, S.~Möller, K.~Hecker, E.~Icking, K.~Watanabe, T.~T. ..., and
  C.~Stampfer, ``{Particle–hole symmetry protects spin-valley blockade in
  graphene quantum dots},'' {\em Nature}, pp.~1--6, 2023.

\bibitem{mayerTuningConfinedStates2023}
D.~Mayer and A.~Knothe, ``Tuning‐{Confined} {States} and {Valley}
  {G}‐{Factors} by {Quantum} {Dot} {Design} in {Bilayer} {Graphene},'' {\em
  physica status solidi (b)}, p.~2300395, Nov. 2023.

\bibitem{Iwakiri2022}
S.~Iwakiri, F.~K. {de Vries}, E.~Portol{\'e}s, G.~Zheng, T.~Taniguchi,
  K.~Watanabe, T.~Ihn, and K.~Ensslin, ``Gate-defined electron interferometer
  in bilayer graphene,'' {\em Nano Letters}, vol.~22, pp.~6292--6297, Aug.
  2022.

\bibitem{Fu2023}
H.~L. Fu, K.~Huang, K.~Watanabe, T.~Taniguchi, M.~Kayyalha, and J.~Zhu,
  ``Aharonov-bohm oscillations in bilayer graphene quantum hall edge state
  fabry-perot interferometers,'' {\em Nano Letters}, vol.~23, pp.~718--725, jan
  2023.

\bibitem{Mirzakhani2023}
M.~Mirzakhani, N.~Myoung, F.~M. Peeters, and H.~C. Park, ``Electronic
  mach-zehnder interference in a bipolar hybrid monolayer-bilayer graphene
  junction,'' {\em Carbon}, vol.~201, pp.~734--744, oct 2023.

\bibitem{inglaaynés2023ballistic}
J.~Ingla-Aynés, A.~L.~R. Manesco, T.~S. Ghiasi, K.~Watanabe, T.~Taniguchi, and
  H.~S.~J. van~der Zant, ``A ballistic electron source with
  magnetically-controlled valley polarization in bilayer graphene,'' {\em
  arXiv}, 2023.

\bibitem{Seemann2023}
L.~Seemann, A.~Knothe, and M.~Hentschel, ``Gate-tunable regular and chaotic
  electron dynamics in ballistic bilayer graphene cavities,'' {\em Physical
  Review B}, vol.~107, p.~205404, May 2023.

\bibitem{Peterfalvi2012}
C.~G. P{\'e}terfalvi, L.~Oroszl{\'a}ny, C.~J. Lambert, and J.~Cserti,
  ``Intraband electron focusing in bilayer graphene,'' {\em New Journal of
  Physics}, vol.~14, p.~063028, June 2012.

\bibitem{Xu2021}
S.~Xu, M.~M. Al~Ezzi, N.~Balakrishnan, A.~{Garcia-Ruiz}, B.~Tsim, C.~Mullan,
  J.~Barrier, N.~Xin, B.~A. Piot, T.~Taniguchi, K.~Watanabe, A.~Carvalho,
  A.~Mishchenko, A.~K. Geim, V.~I. Fal'ko, S.~Adam, A.~H.~C. Neto, K.~S.
  Novoselov, and Y.~Shi, ``Tunable van {{Hove}} singularities and correlated
  states in twisted monolayer\textendash bilayer graphene,'' {\em Nature
  Physics}, vol.~17, pp.~619--626, May 2021.

\bibitem{Huber2020}
R.~Huber, M.-H. Liu, S.-C. Chen, M.~Drienovsky, A.~Sandner, K.~Watanabe,
  T.~Taniguchi, K.~Richter, D.~Weiss, and J.~Eroms, ``Band conductivity
  oscillations in a gate-tunable graphene superlattice,'' {\em Nano Lett.},
  vol.~20, no.~11, pp.~8046--8052, 2020.

\bibitem{Huber2022}
R.~Huber, M.-N. Steffen, M.~Drienovsky, A.~Sandner, K.~Watanabe, T.~Taniguchi,
  D.~Pfannkuche, D.~Weiss, and J.~Eroms, ``Band conductivity oscillations in a
  gate-tunable graphene superlattice,'' {\em Nature Communications}, vol.~13,
  p.~2856, May 2022.

\bibitem{Drienovsky2018}
M.~Drienovsky, J.~Joachimsmeyer, A.~Sandner, M.-H. Liu, T.~Taniguchi,
  K.~Watanabe, K.~Richter, D.~Weiss, and J.~Eroms, ``Commensurability
  oscillations in one-dimensional graphene superlattices,'' {\em Phys. Rev.
  Lett.}, vol.~121, p.~026806, Jul 2018.

\bibitem{MrencaKolasinska2023}
A.~Mre{\'{n}}ca-Kolasi{\'{n}}ska, S.-C. Chen, and M.-H. Liu, ``Probing miniband
  structure and hofstadter butterfly in gated graphene superlattices via
  magnetotransport,'' {\em npj 2D Materials and Applications}, vol.~7,
  p.~Article number: 64, sep 2023.

\bibitem{Moulsdale2020}
C.~Moulsdale, A.~Knothe, and V.~Fal'ko, ``Engineering of the topological
  magnetic moment of electrons in bilayer graphene using strain and electrical
  bias,'' {\em Physical Review B}, vol.~101, p.~085118, Feb. 2020.

\bibitem{Varlet2015}
A.~Varlet, M.~{Mucha-Kruczy{\'n}ski}, D.~Bischoff, P.~Simonet, T.~Taniguchi,
  K.~Watanabe, V.~Fal'ko, T.~Ihn, and K.~Ensslin, ``Tunable {{Fermi}} surface
  topology and {{Lifshitz}} transition in bilayer graphene,'' {\em Synthetic
  Metals}, vol.~210, pp.~19--31, Dec. 2015.

\bibitem{Zhao2015}
Y.~Zhao, J.~Wyrick, F.~D. Natterer, J.~F. Rodriguez-Nieva, C.~Lewandowski,
  K.~Watanabe, T.~Taniguchi, L.~S. Levitov, N.~B. Zhitenev, and J.~A. Stroscio,
  ``Creating and probing electron whispering-gallery modes in graphene,'' {\em
  Science}, vol.~348, no.~6235, pp.~672--675, 2015.

\bibitem{Ge2021}
Z.~Ge, D.~Wong, J.~Lee, F.~Joucken, E.~A. {Quezada-Lopez}, S.~Kahn, H.-Z. Tsai,
  T.~Taniguchi, K.~Watanabe, F.~Wang, A.~Zettl, M.~F. Crommie, and J.~J.
  Velasco, ``Imaging quantum interference in stadium-shaped monolayer and
  bilayer graphene quantum dots,'' {\em Nano Letters}, vol.~21, pp.~8993--8998,
  Nov. 2021.

\bibitem{Rodriguez-Nieva2015}
J.~F. Rodriguez-Nieva and L.~S. Levitov, ``Berry phase jumps and giant
  nonreciprocity in dirac quantum dots,'' {\em Physical Review B}, vol.~94, Dec
  2016.

\bibitem{Wurm2009}
J.~Wurm, A.~Rycerz, I.~Adagideli, M.~Wimmer, K.~Richter, and H.~U. Baranger,
  ``Symmetry classes in graphene quantum dots: Universal spectral statistics,
  weak localization, and conductance fluctuations,'' {\em Physical Review
  Letters}, vol.~102, FEB 6 2009.

\bibitem{Bardarson2009}
J.~H. Bardarson, M.~Titov, and P.~W. Brouwer, ``Electrostatic confinement of
  electrons in an integrable graphene quantum dot,'' {\em Phys. Rev. Lett.},
  vol.~102, p.~226803, Jun 2009.

\bibitem{Wurm2010}
J.~Wurm, M.~Wimmer, H.~U. Baranger, and K.~Richter, ``Graphene rings in
  magnetic fields: {Aharonov-Bohm} effect and valley splitting,'' {\em
  Semiconductor Science and technology}, vol.~25, MAR 4 2010.

\bibitem{Schneider2011}
M.~Schneider and P.~W. Brouwer, ``Resonant scattering in graphene with a
  gate-defined chaotic quantum dot,'' {\em Physical Review B}, vol.~84,
  p.~115440, Sept. 2011.

\bibitem{Hein2013}
J.~Heinl, M.~Schneider, and P.~W. Brouwer, ``Interplay of {{Aharonov-Bohm}} and
  {{Berry}} phases in gate-defined graphene quantum dots,'' {\em Physical
  Review B}, vol.~87, p.~245426, June 2013.

\bibitem{Schneider2014}
M.~Schneider and P.~W. Brouwer, ``Density of states as a probe of electrostatic
  confinement in graphene,'' {\em Physical Review B}, vol.~89, p.~205437, May
  2014.

\bibitem{Wurm2011}
J.~Wurm, K.~Richter, and i.~d.~I. Adagideli, ``Edge effects in graphene
  nanostructures: From multiple reflection expansion to density of states,''
  {\em Phys. Rev. B}, vol.~84, p.~075468, Aug 2011.

\bibitem{Wurm2011a}
J.~Wurm, K.~Richter, and I.~Adagideli, ``Edge effects in graphene
  nanostructures: Semiclassical theory of spectral fluctuations and quantum
  transport,'' {\em Physical review B}, vol.~84, NOV 14 2011.

\bibitem{Xu2018}
H.-Y. Xu, G.-L. Wang, L.~Huang, and Y.-C. Lai, ``Chaos in {Dirac} electron
  optics: Emergence of a relativistic quantum chimera,'' {\em Phys. Rev.
  Lett.}, vol.~120, p.~124101, Mar 2018.

\bibitem{Bercioux2023}
D.~Bercioux, D.~Frustaglia, and A.~D. Martino, ``Chiral spin channels in curved
  graphene $pn$ junctions,'' {\em Phys. Rev. B}, vol.~108, p.~115140, Sep 2023.

\bibitem{Noeckel1997}
J.~U. Nöckel and A.~D. Stone, ``Ray and wave chaos in asymmetric resonant
  optical cavities,'' {\em Nature}, vol.~385, p.~45–47, Jan 1997.

\bibitem{Hentschel2002}
M.~Hentschel and K.~Richter, ``{Quantum chaos in optical systems: The annular
  billiard},'' {\em {Physical Review E}}, vol.~66, no.~5, 2002.

\bibitem{Wiersig2008}
J.~Wiersig and M.~Hentschel, ``Combining directional light output and ultralow
  loss in deformed microdisks,'' {\em Phys. Rev. Lett.}, vol.~100, p.~033901,
  Jan 2008.

\bibitem{Lai2018}
C.-D. Han, C.-Z. Wang, H.-Y. Xu, D.~Huang, and Y.-C. Lai, ``Decay of
  semiclassical massless {Dirac} fermions from integrable and chaotic
  cavities,'' {\em Phys. Rev. B}, vol.~98, p.~104308, Sep 2018.

\bibitem{Varlet2016}
A.~Varlet, M.-H. Liu, D.~Bischoff, P.~Simonet, T.~Taniguchi, K.~Watanabe,
  K.~Richter, T.~Ihn, and K.~Ensslin, ``Band gap and broken chirality in
  single-layer and bilayer graphene,'' {\em Physics statis solidi rapid
  research letters}, vol.~10, pp.~46--57, JAN 2016.

\bibitem{Elahi2022}
M.~M. Elahi, Y.~Zeng, C.~R. Dean, and A.~W. Ghosh, ``Direct evidence of
  {{Klein-antiKlein}} tunneling of graphitic electrons in a {{Corbino}}
  geometry,'' {\em arXiv}, Oct. 2022.

\bibitem{Handschin2017a}
C.~Handschin, P.~Makk, P.~Rickhaus, M.-H. Liu, K.~Watanabe, T.~Taniguchi,
  K.~Richter, and C.~Sch{\"o}nenberger, ``{Fabry-Perot} resonances in a
  graphene/hbn moire superlattice,'' {\em Nano Letters}, vol.~17, pp.~328--333,
  jan 2017.

\bibitem{Divari2010}
P.~Divari and G.~Kliros, ``Modeling the thermopower of ballistic graphene
  ribbons,'' {\em Physica E: Low-dimensional Systems and Nanostructures},
  vol.~42, no.~9, pp.~2431--2435, 2010.

\bibitem{Campos2012}
L.~C. Campos, A.~F. Young, K.~Surakitbovorn, K.~Watanabe, T.~Taniguchi, and
  P.~Jarillo-Herrero, ``Quantum and classical confinement of resonant states in
  a trilayer graphene {Fabry-Perot} interferometer,'' {\em Nature
  Communications}, vol.~3, p.~1239, DEC 2012.

\bibitem{Oksanen2014}
M.~Oksanen, A.~Uppstu, A.~Laitinen, D.~J. Cox, M.~F. Craciun, S.~Russo,
  A.~Harju, and P.~Hakonen, ``Single-mode and multimode {Fabry-P\'erot}
  interference in suspended graphene,'' {\em Physical Review B}, vol.~89,
  p.~121414, Mar 2014.

\bibitem{Calado2015}
V.~E. Calado, S.~Goswami, G.~Nanda, M.~Diez, A.~R. Akhmerov, K.~Watanabe,
  T.~Taniguchi, T.~M. Klapwijk, and L.~M.~K. Vandersypen, ``Ballistic josephson
  junctions in edge-contacted graphene,'' {\em Nature Nanotechnology}, vol.~10,
  p.~761, sep 2015.

\bibitem{Taychatanapat2015}
T.~Taychatanapat, J.~Y. Tan, Y.~Yeo, K.~Watanabe, T.~Taniguchi, and
  B.~Oezyilmaz, ``Conductance oscillations induced by ballistic snake states in
  a graphene heterojunction,'' {\em Nature Communications}, vol.~6, p.~6093,
  feb 2015.

\bibitem{BenShalom2016}
M.~Ben~Shalom, M.~J. Zhu, V.~I. Fal'ko, A.~Mishchenko, A.~V. Kretinin, K.~S.
  Novoselov, C.~R. Woods, K.~Watanabe, T.~Taniguchi, A.~K. Geim, and J.~R.
  Prance, ``Quantum oscillations of the critical current and high-field
  superconducting proximity in ballistic graphene,'' {\em Nature Physics},
  vol.~12, pp.~318--322, APR 2016.

\bibitem{Allen2017}
M.~T. Allen, O.~Shtanko, I.~C. Fulga, J.~I.~J. Wang, D.~Nurgaliev, K.~Watanabe,
  T.~Taniguchi, A.~R. Akhmerov, P.~Jarillo-Herrero, L.~S. Leyitov, and
  A.~Yacoby, ``Observation of electron coherence and {Fabry-Perot} standing
  waves at a graphene edge,'' {\em Nano Letters}, vol.~17, no.~12,
  pp.~7380--7386, 2017.

\bibitem{Nanda2017}
G.~Nanda, J.~L. Aguilera-Servin, P.~Rakyta, A.~Kormanyos, R.~Kleiner,
  D.~Koelle, K.~Watanabe, T.~Taniguchi, L.~M.~K. Vandersypen, and S.~Goswami,
  ``Current-phase relation of ballistic graphene {Josephson} junctions,'' {\em
  Nano Letters}, vol.~17, pp.~3396--3401, jun 2017.

\bibitem{Zhu2018}
M.~Zhu, M.~Ben~Shalom, A.~Mishchsenko, V.~I. Fal'ko, K.~Novoselov, and A.~Geim,
  ``Supercurrent and multiple {Andreev} reflections in micrometer-long
  ballistic graphene josephson junctions,'' {\em Nanoscale}, vol.~10,
  pp.~3020--3025, feb 2018.

\bibitem{Pandey2019}
P.~Pandey, R.~Kraft, R.~Krupke, D.~Beckmann, and R.~Danneau, ``Andreev
  reflection in ballistic normal metal/graphene/superconductor junctions,''
  {\em Physical Review B}, vol.~100, p.~165416, cct 2019.

\bibitem{Jung2016}
M.~Jung, P.~Rickhaus, S.~Zihlmann, P.~Makk, and C.~Schönenberger, ``Microwave
  photodetection in an ultraclean suspended bilayer graphene p–n junction,''
  {\em Nano Letters}, vol.~16, no.~11, pp.~6988--6993, 2016.

\bibitem{Rickhaus2019}
P.~Rickhaus, G.~Zheng, J.~L. Lado, Y.~Lee, A.~Kurzmann, M.~Eich, R.~Pisoni,
  C.~Tong, R.~Garreis, C.~Gold, M.~Masseroni, T.~Taniguchi, K.~Wantanabe,
  T.~Ihn, and K.~Ensslin, ``Gap opening in twisted double bilayer graphene by
  crystal fields,'' {\em Nano Letters}, vol.~19, no.~12, pp.~8821--8828, 2019.

\bibitem{Mueller2009}
M.~Mueller, M.~Braeuninger, and B.~Trauzettel, ``Temperature dependence of the
  conductivity of ballistic graphene,'' {\em Physical Review Letters},
  vol.~103, p.~196801, Nov 6 2009.

\bibitem{Ideue2021}
T.~Ideue and Y.~Iwasa, ``Symmetry breaking and nonlinear electric transport in
  van der waals nanostructures,'' in {\em Annual Review Of Condensed Matter
  Physics, vol 12, 2021} (A.~P. Mackenzie and M.~C. Marchetti, eds.), vol.~12
  of {\em Annual Review of Condensed Matter Physics}, pp.~201--223, Thomson
  Reuters, 2021.

\bibitem{Huang2009}
L.~Huang, Y.-C. Lai, D.~K. Ferry, R.~Akis, and S.~M. Goodnick, ``Transmission
  and scarring in graphene quantum dots,'' {\em Journal of Physics-Condensed
  Matter}, vol.~21, aug 2009.

\bibitem{Ghosh2008}
T.~K. Ghosh, A.~De~Martino, W.~H\"ausler, L.~Dell'Anna, and R.~Egger,
  ``Conductance quantization and snake states in graphene magnetic
  waveguides,'' {\em Phys. Rev. B}, vol.~77, p.~081404, Feb 2008.

\bibitem{Oroszlany2008}
L.~Oroszl\'any, P.~Rakyta, A.~Korm\'anyos, C.~J. Lambert, and J.~Cserti,
  ``Theory of snake states in graphene,'' {\em Phys. Rev. B}, vol.~77,
  p.~081403, Feb 2008.

\bibitem{Milovanovic2014a}
S.~P. Milovanovic, M.~R. Masir, and F.~M. Peeters, ``Interplay between snake
  and quantum edge states in a graphene hall bar with a pn-junction,'' {\em
  Applied Physics Letters}, vol.~105, p.~123507, sep 2014.

\bibitem{Cohnitz2016}
L.~Cohnitz, A.~De~Martino, W.~H\"ausler, and R.~Egger, ``Chiral interface
  states in graphene $p\text{\ensuremath{-}}n$ junctions,'' {\em Physical
  Review B}, vol.~94, p.~165443, Oct 2016.

\bibitem{Bercioux2019}
D.~Bercioux and A.~De~Martino, ``Spin-orbit interaction and snake states in a
  graphene p-n junction,'' {\em Phys. Rev. B}, vol.~100, p.~115407, Sep 2019.

\bibitem{Sim1998}
H.-S. Sim, K.-H. Ahn, K.~J. Chang, G.~Ihm, N.~Kim, and S.~J. Lee, ``Magnetic
  edge states in a magnetic quantum dot,'' {\em Phys. Rev. Lett.}, vol.~80,
  pp.~1501--1504, Feb 1998.

\bibitem{Nogaret2000}
A.~Nogaret, S.~J. Bending, and M.~Henini, ``Resistance resonance effects
  through magnetic edge states,'' {\em Phys. Rev. Lett.}, vol.~84,
  pp.~2231--2234, Mar 2000.

\bibitem{Reijniers2000}
J.~Reijniers and F.~M. Peeters, ``Snake orbits and related magnetic edge
  states,'' {\em Journal of Physics: Condensed Matter}, vol.~12, p.~9771, Nov
  2000.

\bibitem{Reijniers2002}
J.~Reijniers, A.~Matulis, K.~Chang, F.~M. Peeters, and P.~Vasilopoulos,
  ``Confined magnetic guiding orbit states,'' {\em Europhysics Letters},
  vol.~59, p.~749, sep 2002.

\bibitem{Halperin1982}
B.~I. Halperin, ``Quantized {Hall} conductance, current-carrying edge states,
  and the existence of extended states in a two-dimensional disordered
  potential,'' {\em Phys. Rev. B}, vol.~25, pp.~2185--2190, Feb 1982.

\bibitem{MacDonald1984}
A.~H. MacDonald and P.~Streda, ``Quantized {Hall} effect and edge currents,''
  {\em Phys. Rev. B}, vol.~29, pp.~1616--1619, Feb 1984.

\bibitem{Kane1987}
B.~E. Kane, D.~C. Tsui, and G.~Weimann, ``Evidence for edge currents in the
  integral quantum {Hall} effect,'' {\em Phys. Rev. Lett.}, vol.~59,
  pp.~1353--1356, Sep 1987.

\bibitem{Buettiker1988}
M.~B\"uttiker, ``Absence of backscattering in the quantum {Hall} effect in
  multiprobe conductors,'' {\em Phys. Rev. B}, vol.~38, pp.~9375--9389, Nov
  1988.

\bibitem{Washburn1988}
S.~Washburn, A.~B. Fowler, H.~Schmid, and D.~Kern, ``Quantized {Hall} effect in
  the presence of backscattering,'' {\em Phys. Rev. Lett.}, vol.~61,
  pp.~2801--2804, Dec 1988.

\bibitem{McDonald1979}
S.~W. McDonald and A.~N. Kaufman, ``Spectrum and eigenfunctions for a
  hamiltonian with stochastic trajectories,'' {\em Phys. Rev. Lett.}, vol.~42,
  pp.~1189--1191, Apr 1979.

\bibitem{Davies2012}
N.~Davies, A.~A. Patel, A.~Cortijo, V.~Cheianov, F.~Guinea, and V.~I. Fal'ko,
  ``Skipping and snake orbits of electrons: Singularities and catastrophes,''
  {\em Phys. Rev. B}, vol.~85, p.~155433, apr 2012.

\bibitem{Chen2012}
J.-C. Chen, X.~C. Xie, and Q.-F. Sun, ``Current oscillation of snake states in
  graphene $p$-$n$ junction,'' {\em Phys. Rev. B}, vol.~86, p.~035429, Jul
  2012.

\bibitem{Milovanovic2013}
S.~P. Milovanovic, M.~R. Masir, and F.~M. Peeters, ``Spectroscopy of snake
  states using a graphene {Hall} bar,'' {\em Applied Physics Letters},
  vol.~103, p.~233502, dec 2013.

\bibitem{Buettiker1992}
M.~Büttiker, ``Chapter 4: The quantum hall effect in open conductors,'' in
  {\em Semiconductors and Semimetals} (M.~Reed, ed.), vol.~35 of {\em
  Semiconductors and Semimetals}, pp.~191--277, Elsevier, 1992.

\bibitem{Christen1996}
T.~Christen and M.~B\"uttiker, ``Low-frequency admittance of quantized {Hall}
  conductors,'' {\em Phys. Rev. B}, vol.~53, pp.~2064--2072, Jan 1996.

\bibitem{Sanchez2004}
D.~S\'anchez and M.~B\"uttiker, ``Magnetic-field asymmetry of nonlinear
  mesoscopic transport,'' {\em Phys. Rev. Lett.}, vol.~93, p.~106802, Sep 2004.

\bibitem{Sanchez2013}
D.~S\'anchez and R.~L\'opez, ``Scattering theory of nonlinear thermoelectric
  transport,'' {\em Phys. Rev. Lett.}, vol.~110, p.~026804, Jan 2013.

\bibitem{Gorini2014}
C.~Gorini, D.~Weinmann, and R.~A. Jalabert, ``Scanning-gate-induced effects in
  nonlinear transport through nanostructures,'' {\em Phys. Rev. B}, vol.~89,
  p.~115414, Mar 2014.

\bibitem{Texier2018}
C.~Texier and J.~Mitscherling, ``Nonlinear conductance in weakly disordered
  mesoscopic wires: Interaction and magnetic field asymmetry,'' {\em Phys. Rev.
  B}, vol.~97, p.~075306, Feb 2018.

\bibitem{Moskalets2002}
M.~Moskalets and M.~B\"uttiker, ``Floquet scattering theory of quantum pumps,''
  {\em Phys. Rev. B}, vol.~66, p.~205320, Nov 2002.

\bibitem{Samuelsson2005}
P.~Samuelsson and M.~B\"uttiker, ``Dynamic generation of orbital quasiparticle
  entanglement in mesoscopic conductors,'' {\em Phys. Rev. B}, vol.~71,
  p.~245317, Jun 2005.

\bibitem{Akkermansbook}
A.~E. and M.~G., {\em Mesoscopic physics of electrons and photons}.
\newblock Cambridge University Press, 2007.

\bibitem{Imrybook}
I.~Y., {\em Introduction to Mesoscopic Physics}.
\newblock Oxford University Press, 2008.

\bibitem{Samuelsson2004}
P.~Samuelsson, E.~V. Sukhorukov, and M.~B\"uttiker, ``Two-particle
  {Aharonov-Bohm} effect and entanglement in the electronic {Hanbury
  Brown--Twiss} setup,'' {\em Phys. Rev. Lett.}, vol.~92, p.~026805, Jan 2004.

\bibitem{Chung2005}
V.~S.-W. Chung, P.~Samuelsson, and M.~B\"uttiker, ``Visibility of current and
  shot noise in electrical {Mach-Zehnder and Hanbury Brown Twiss}
  interferometers,'' {\em Phys. Rev. B}, vol.~72, p.~125320, Sep 2005.

\bibitem{Beenakker1990}
C.~W.~J. Beenakker, ``Edge channels for the fractional quantum {Hall} effect,''
  {\em Phys. Rev. Lett.}, vol.~64, pp.~216--219, Jan 1990.

\bibitem{Chang1990}
A.~Chang, ``A unified transport theory for the integral and fractional quantum
  {Hall} effects: Phase boundaries, edge currents, and transmission/reflection
  probabilities,'' {\em Solid State Communications}, vol.~74, no.~9,
  pp.~871--876, 1990.

\bibitem{Chklovskii1992}
D.~B. Chklovskii, B.~I. Shklovskii, and L.~I. Glazman, ``Electrostatics of edge
  channels,'' {\em Phys. Rev. B}, vol.~46, pp.~4026--4034, Aug 1992.

\bibitem{Chklovskii1993}
D.~B. Chklovskii, K.~A. Matveev, and B.~I. Shklovskii, ``Ballistic conductance
  of interacting electrons in the quantum {Hall} regime,'' {\em Phys. Rev. B},
  vol.~47, p.~12605, 1993.

\bibitem{Armagnat2020}
P.~Armagnat and X.~Waintal, ``Reconciling edge states with compressible stripes
  in a ballistic mesoscopic conductor,'' {\em Journal of Physics: Materials},
  vol.~3, p.~02LT01, mar 2020.

\bibitem{Flor2022}
I.~M. Fl{\'{o}}r, A.~Lacerda-Santos, G.~Fleury, P.~Roulleau, and X.~Waintal,
  ``{Positioning of edge states in a quantum Hall graphene p n junction},''
  {\em Physical Review B}, vol.~105, p.~L241409, jun 2022.

\bibitem{Chamon1994}
C.~d.~C. Chamon and X.~G. Wen, ``Sharp and smooth boundaries of quantum {Hall}
  liquids,'' {\em Phys. Rev. B}, vol.~49, pp.~8227--8241, Mar 1994.

\bibitem{Paradiso2012}
N.~Paradiso, S.~Heun, S.~Roddaro, L.~Sorba, F.~Beltram, G.~Biasiol, L.~N.
  Pfeiffer, and K.~W. West, ``Imaging fractional incompressible stripes in
  integer quantum {Hall} systems,'' {\em Phys. Rev. Lett.}, vol.~108,
  p.~246801, Jun 2012.

\bibitem{Bhattacharyya2019}
R.~Bhattacharyya, M.~Banerjee, M.~Heiblum, D.~Mahalu, and V.~Umansky, ``Melting
  of interference in the fractional quantum {Hall} effect: Appearance of
  neutral modes,'' {\em Phys. Rev. Lett.}, vol.~122, p.~246801, Jun 2019.

\bibitem{Khanna2021}
U.~Khanna, M.~Goldstein, and Y.~Gefen, ``Fractional edge reconstruction in
  integer quantum {Hall} phases,'' {\em Phys. Rev. B}, vol.~103, p.~L121302,
  Mar 2021.

\bibitem{Khanna2022}
U.~Khanna, M.~Goldstein, and Y.~Gefen, ``Emergence of neutral modes in
  {Laughlin}-like fractional quantum {Hall} phases,'' {\em Phys. Rev. Lett.},
  vol.~129, p.~146801, Sep 2022.

\bibitem{li2013}
G.~Li, A.~Luican-Mayer, D.~Abanin, L.~Levitov, and E.~Y. Andrei, ``Evolution of
  {Landau} levels into edge states in graphene,'' {\em Nature communications},
  vol.~4, no.~1, p.~1744, 2013.

\bibitem{Coissard2023}
A.~Coissard, A.~G. Grushin, C.~Repellin, L.~Veyrat, K.~Watanabe, T.~Taniguchi,
  F.~Gay, H.~Courtois, H.~Sellier, and B.~Sacépé, ``Absence of edge
  reconstruction for quantum {Hall} edge channels in graphene devices,'' {\em
  Science Advances}, vol.~9, p.~eadf7220, Sep 2023.

\bibitem{Cui2016}
Y.-T. Cui, B.~Wen, E.~Y. Ma, G.~Diankov, Z.~Han, F.~Amet, T.~Taniguchi,
  K.~Watanabe, D.~Goldhaber-Gordon, C.~R. Dean, and Z.-X. Shen,
  ``Unconventional correlation between quantum hall transport quantization and
  bulk state filling in gated graphene devices,'' {\em Phys. Rev. Lett.},
  vol.~117, p.~186601, Oct 2016.

\bibitem{seredinski2019}
A.~Seredinski, A.~W. Draelos, E.~G. Arnault, M.-T. Wei, H.~Li, T.~Fleming,
  K.~Watanabe, T.~Taniguchi, F.~Amet, and G.~Finkelstein, ``Quantum hall--based
  superconducting interference device,'' {\em Science advances}, vol.~5, no.~9,
  p.~eaaw8693, 2019.

\bibitem{Vignalebook}
C.~Giuliani and G.~Vignale, {\em Quantum Theory of the Electron Liquid}.
\newblock Cambridge, England, UK: Cambridge University Press, 2005.

\bibitem{Giamarchibook}
T.~Giamarchi, {\em Quantum Physics in One Dimension}.
\newblock Oxford University Press, 2003.

\bibitem{Haldane1981}
F.~D.~M. Haldane, ``Luttinger liquid theory of one-dimensional quantum fluids.
  i. properties of the {Luttinger} model and their extension to the general 1d
  interacting spinless fermi gas,'' {\em Journal of Physics C: Solid State
  Physics}, vol.~14, p.~2585, jul 1981.

\bibitem{Vondelft1998}
J.~von Delft and H.~Schoeller, ``Bosonization for beginners —
  refermionization for experts,'' {\em Annalen der Physik}, vol.~510, no.~4,
  pp.~225--305, 1998.

\bibitem{Imambekov2012}
A.~Imambekov, T.~L. Schmidt, and L.~I. Glazman, ``One-dimensional quantum
  liquids: Beyond the {Luttinger} liquid paradigm,'' {\em Rev. Mod. Phys.},
  vol.~84, pp.~1253--1306, Sep 2012.

\bibitem{Levchenko2021}
A.~Levchenko and T.~Micklitz, ``Kinetic processes in {Fermi–Luttinger}
  liquids,'' {\em JETP}, vol.~132, p.~675, 2021.

\bibitem{Ferraro2014}
D.~Ferraro, B.~Roussel, C.~Cabart, E.~Thibierge, G.~F{\`{e}}ve, C.~Grenier, and
  P.~Degiovanni, ``{Real-Time Decoherence of Landau and Levitov Quasiparticles
  in Quantum Hall Edge Channels},'' {\em Physical Review Letters}, vol.~113,
  p.~166403, oct 2014.

\bibitem{Fujisawa2022}
T.~Fujisawa, ``Nonequilibrium charge dynamics of {Tomonaga–Luttinger} liquids
  in quantum hall edge channels,'' {\em Annalen der Physik}, vol.~534, no.~4,
  p.~2100354, 2022.

\bibitem{Foerster2005}
H.~F\"orster, S.~Pilgram, and M.~B\"uttiker, ``Decoherence and full counting
  statistics in a {Mach-Zehnder} interferometer,'' {\em Phys. Rev. B}, vol.~72,
  p.~075301, Aug 2005.

\bibitem{Marquardt2005}
F.~Marquardt, ``Fermionic {Mach-Zehnder} interferometer subject to a quantum
  bath,'' {\em Europhysics Letters}, vol.~72, p.~788, nov 2005.

\bibitem{Lunde2010}
A.~M. Lunde, S.~E. Nigg, and M.~B\"uttiker, ``Interaction-induced edge channel
  equilibration,'' {\em Phys. Rev. B}, vol.~81, p.~041311, Jan 2010.

\bibitem{Chalker2007}
J.~T. Chalker, Y.~Gefen, and M.~Y. Veillette, ``Decoherence and interactions in
  an electronic {Mach-Zehnder} interferometer,'' {\em Phys. Rev. B}, vol.~76,
  p.~085320, Aug 2007.

\bibitem{Feldman2022}
D.~E. Feldman and B.~I. Halperin, ``Robustness of quantum {Hall}
  interferometry,'' {\em Phys. Rev. B}, vol.~105, p.~165310, Apr 2022.

\bibitem{Levkivskyi2008}
I.~P. Levkivskyi and E.~V. Sukhorukov, ``{Dephasing in the electronic
  Mach-Zehnder interferometer at filling factor $\ensuremath{\nu}=2$},'' {\em
  Phys Rev B}, vol.~78, p.~045322, July 2008.

\bibitem{Neder2006}
I.~Neder, M.~Heiblum, Y.~Levinson, D.~Mahalu, and V.~Umansky, ``{Unexpected
  behavior in a two-path electron interferometer},'' {\em Physical Review
  Letters}, vol.~9, no.~1, p.~016804, 2023.

\bibitem{Roulleau2007}
P.~Roulleau, F.~Portier, D.~Glattli, P.~Roche, A.~Cavanna, G.~Faini,
  U.~Gennser, and D.~Mailly, ``{Finite bias visibility of the electronic
  Mach-Zehnder interferometer},'' {\em Physical Review B}, vol.~76, p.~161309,
  oct 2007.

\bibitem{Roulleau2008}
P.~Roulleau, F.~Portier, P.~Roche, A.~Cavanna, G.~Faini, U.~Gennser, and
  D.~Mailly, ``Direct measurement of the coherence length of edge states in the
  integer quantum {Hall} regime,'' {\em Phys Rev Lett}, vol.~100, p.~126802,
  Mar. 2008.

\bibitem{Altimiras2010}
C.~Altimiras, H.~le~Sueur, U.~Gennser, A.~Cavanna, D.~Mailly, and F.~Pierre,
  ``Tuning energy relaxation along quantum hall channels,'' {\em Phys. Rev.
  Lett.}, vol.~105, p.~226804, Nov 2010.

\bibitem{LeSueur2010}
H.~le~Sueur, C.~Altimiras, U.~Gennser, A.~Cavanna, D.~Mailly, and F.~Pierre,
  ``{Energy Relaxation in the Integer Quantum Hall Regime},'' {\em Physical
  Review Letters}, vol.~105, p.~056803, jul 2010.

\bibitem{Zurek2003}
W.~H. Zurek, ``Decoherence, einselection, and the quantum origins of the
  classical,'' {\em Rev. Mod. Phys.}, vol.~75, pp.~715--775, May 2003.

\bibitem{Safi2004}
I.~Safi and H.~Saleur, ``One-channel conductor in an ohmic environment: Mapping
  to a {Tomonaga-Luttinger} liquid and full counting statistics,'' {\em Phys.
  Rev. Lett.}, vol.~93, p.~126602, Sep 2004.

\bibitem{Degiovanni2009}
P.~Degiovanni, C.~Grenier, and G.~F{\`{e}}ve, ``{Decoherence and relaxation of
  single-electron excitations in quantum Hall edge channels},'' {\em Physical
  Review B}, vol.~80, p.~241307, dec 2009.

\bibitem{Neuenhahn2009}
C.~Neuenhahn and F.~Marquardt, ``Universal dephasing in a chiral 1d interacting
  fermion system,'' {\em Phys. Rev. Lett.}, vol.~102, p.~046806, Jan 2009.

\bibitem{Komiyama1989}
S.~Komiyama, H.~Hirai, S.~Sasa, and S.~Hiyamizu, ``Violation of the integral
  quantum hall effect: Influence of backscattering and the role of voltage
  contacts,'' {\em Phys. Rev. B}, vol.~40, pp.~12566--12569, Dec 1989.

\bibitem{Chirolli2013}
L.~Chirolli, F.~Taddei, R.~Fazio, and V.~Giovannetti, ``Interactions in
  electronic {Mach-Zehnder} interferometers with copropagating edge channels,''
  {\em Phys. Rev. Lett.}, vol.~111, p.~036801, Jul 2013.

\bibitem{Buettiker1986}
M.~B\"uttiker, ``Role of quantum coherence in series resistors,'' {\em Phys.
  Rev. B}, vol.~33, pp.~3020--3026, Mar 1986.

\bibitem{jo2022scaling}
M.~Jo, J.-Y.~M. Lee, A.~Assouline, P.~Brasseur, K.~Watanabe, T.~Taniguchi,
  P.~Roche, D.~Glattli, N.~Kumada, F.~Parmentier, {\em et~al.}, ``Scaling
  behavior of electron decoherence in a graphene {Mach-Zehnder}
  interferometer,'' {\em Nature Communications}, vol.~13, no.~1, p.~5473, 2022.

\bibitem{morikawa2015edge}
S.~Morikawa, S.~Masubuchi, R.~Moriya, K.~Watanabe, T.~Taniguchi, and
  T.~Machida, ``Edge-channel interferometer at the graphene quantum {Hall} pn
  junction,'' {\em Applied Physics Letters}, vol.~106, no.~18, p.~183101, 2015.

\bibitem{Wei2017}
D.~S. Wei, T.~van~der Sar, J.~D. Sanchez-Yamagishi, K.~Watanabe, T.~Taniguchi,
  P.~Jarillo-Herrero, B.~I. Halperin, and A.~Yacoby, ``{Mach-Zehnder}
  interferometry using spin- and valley-polarized quantum {Hall} edge states in
  graphene,'' {\em Sci Adv}, vol.~3, p.~e1700600, Aug. 2017.

\bibitem{Handschin2017b}
C.~Handschin, P.~Makk, P.~Rickhaus, R.~Maurand, K.~Watanabe, T.~Taniguchi,
  K.~Richter, M.-H. Liu, and C.~Sch{\"o}nenberger, ``Giant valley-isospin
  conductance oscillations in ballistic graphene,'' {\em Nano Letters},
  vol.~17, pp.~5389--5393, sep 2017.

\bibitem{jo2021quantum}
M.~Jo, P.~Brasseur, A.~Assouline, G.~Fleury, H.-S. Sim, K.~Watanabe,
  T.~Taniguchi, W.~Dumnernpanich, P.~Roche, D.~Glattli, {\em et~al.}, ``Quantum
  {Hall} valley splitters and a tunable {Mach-Zehnder} interferometer in
  graphene,'' {\em Physical Review Letters}, vol.~126, no.~14, p.~146803, 2021.

\bibitem{Tworzydlo}
J.~Tworzyd\l{}o, I.~Snyman, A.~R. Akhmerov, and C.~W.~J. Beenakker,
  ``Valley-isospin dependence of the quantum {Hall} effect in a graphene
  $p$-$n$ junction,'' {\em Phys. Rev. B}, vol.~76, p.~035411, Jul 2007.

\bibitem{Trifunovic}
L.~Trifunovic and P.~W. Brouwer, ``Valley isospin of interface states in a
  graphene $pn$ junction in the quantum {Hall} regime,'' {\em Phys. Rev. B},
  vol.~99, p.~205431, May 2019.

\bibitem{Rehmann2019}
M.~K. Rehmann, Y.~B. Kalyoncu, M.~Kisiel, N.~Pascher, F.~J. Giessibl,
  F.~Müller, K.~Watanabe, T.~Taniguchi, E.~Meyer, M.-H. Liu, and D.~M.
  Zumbühl, ``Characterization of hydrogen plasma defined graphene edges,''
  {\em Carbon}, vol.~150, pp.~417--424, 2019.

\bibitem{Wei2018}
D.~S. Wei, T.~van~der Sar, S.~H. Lee, K.~Watanabe, T.~Taniguchi, B.~I.
  Halperin, and A.~Yacoby, ``{Electrical generation and detection of spin waves
  in a quantum Hall ferromagnet},'' {\em Science}, vol.~362, pp.~229--233, Oct.
  2018.

\bibitem{assouline2023emission}
A.~Assouline, L.~Pugliese, H.~Chakraborti, S.~Lee, L.~Bernabeu, M.~Jo,
  K.~Watanabe, T.~Taniguchi, D.~Glattli, N.~Kumada, {\em et~al.}, ``Emission
  and coherent control of levitons in graphene,'' {\em Science}, vol.~382,
  no.~6676, pp.~1260--1264, 2023.

\bibitem{seelig2001charge}
G.~Seelig and M.~B{\"u}ttiker, ``Charge-fluctuation-induced dephasing in a
  gated mesoscopic interferometer,'' {\em Physical Review B}, vol.~64, no.~24,
  p.~245313, 2001.

\bibitem{youn2008nonequilibrium}
S.-C. Youn, H.-W. Lee, and H.-S. Sim, ``Nonequilibrium dephasing in an
  electronic {Mach-Zehnder} interferometer,'' {\em Physical review letters},
  vol.~100, no.~19, p.~196807, 2008.

\bibitem{roulleau2008noise}
P.~Roulleau, F.~Portier, P.~Roche, A.~Cavanna, G.~Faini, U.~Gennser, and
  D.~Mailly, ``Noise dephasing in edge states of the integer quantum {Hall}
  regime,'' {\em Physical Review Letters}, vol.~101, no.~18, p.~186803, 2008.

\bibitem{bocquillon2013separation}
E.~Bocquillon, V.~Freulon, J.-M. Berroir, P.~Degiovanni, B.~Pla{\c{c}}ais,
  A.~Cavanna, Y.~Jin, and G.~F{\`e}ve, ``Separation of neutral and charge modes
  in one-dimensional chiral edge channels,'' {\em Nature communications},
  vol.~4, no.~1, p.~1839, 2013.

\bibitem{Sukhorukov2007}
E.~Sukhorukov and V.~Cheianov, ``{Resonant Dephasing in the Electronic
  Mach-Zehnder Interferometer},'' {\em Physical Review Letters}, vol.~99,
  p.~156801, oct 2007.

\bibitem{Litvin2008}
L.~V. Litvin, A.~Helzel, H.-P. Tranitz, W.~Wegscheider, and C.~Strunk,
  ``{Edge-channel interference controlled by Landau level filling},'' {\em
  Physical Review B}, vol.~78, p.~075303, aug 2008.

\bibitem{Altimiras2009}
C.~Altimiras, H.~le~Sueur, U.~Gennser, A.~Cavanna, D.~Mailly, and F.~Pierre,
  ``{Non-equilibrium edge-channel spectroscopy in the integer quantum Hall
  regime},'' {\em Nature Physics}, vol.~6, pp.~34--39, oct 2009.

\bibitem{Degiovanni2010}
P.~Degiovanni, C.~Grenier, G.~F{\`{e}}ve, C.~Altimiras, H.~le~Sueur, and
  F.~Pierre, ``{Plasmon scattering approach to energy exchange and
  high-frequency noise in $\nu$ = 2 quantum Hall edge channels},'' {\em
  Physical Review B}, vol.~81, p.~121302, mar 2010.

\bibitem{Huynh2012}
P.-A. Huynh, F.~Portier, H.~le~Sueur, G.~Faini, U.~Gennser, D.~Mailly,
  F.~Pierre, W.~Wegscheider, and P.~Roche, ``{Quantum Coherence Engineering in
  the Integer Quantum Hall Regime},'' {\em Physical Review Letters}, vol.~108,
  p.~256802, jun 2012.

\bibitem{Freulon2015}
V.~Freulon, A.~Marguerite, J.-M. Berroir,
  B.~Pla{\ifmmode\mbox{\c{c}}\else\c{c}\fi}ais, A.~Cavanna, Y.~Jin, and
  G.~F{\ifmmode\grave{e}\else\`{e}\fi}ve, ``{Hong-Ou-Mandel experiment for
  temporal investigation of single-electron fractionalization},'' {\em Nat.
  Commun.}, vol.~6, pp.~1--6, Apr 2015.

\bibitem{Lunde2016}
A.~M. Lunde and S.~E. Nigg, ``{Statistical theory of relaxation of high-energy
  electrons in quantum Hall edge states},'' {\em Physical Review B}, vol.~94,
  p.~045409, jul 2016.

\bibitem{Marguerite2016}
A.~Marguerite, C.~Cabart, C.~Wahl, B.~Roussel, V.~Freulon, D.~Ferraro,
  C.~Grenier, J.-M. Berroir, B.~Pla{\c{c}}ais, T.~Jonckheere, J.~Rech,
  T.~Martin, P.~Degiovanni, A.~Cavanna, Y.~Jin, and G.~F{\`{e}}ve,
  ``{Decoherence and relaxation of a single electron in a one-dimensional
  conductor},'' {\em Physical Review B}, vol.~94, p.~115311, sep 2016.

\bibitem{Gurman2016}
I.~Gurman, R.~Sabo, M.~Heiblum, V.~Umansky, and D.~Mahalu, ``{Dephasing of an
  electronic two-path interferometer},'' {\em Physical Review B}, vol.~93,
  p.~121412, mar 2016.

\bibitem{Tewari2016}
S.~Tewari, P.~Roulleau, C.~Grenier, F.~Portier, A.~Cavanna, U.~Gennser,
  D.~Mailly, and P.~Roche, ``{Robust quantum coherence above the Fermi sea},''
  {\em Physical Review B}, vol.~93, p.~035420, jan 2016.

\bibitem{Marguerite2017}
A.~Marguerite, E.~Bocquillon, J.-M. Berroir, B.~Pla{\c{c}}ais, A.~Cavanna,
  Y.~Jin, P.~Degiovanni, and G.~F{\`{e}}ve, ``{Two-particle interferometry in
  quantum Hall edge channels},'' {\em Physica Status Solidi (b)}, vol.~254,
  p.~1600618, mar 2017.

\bibitem{Itoh2018}
K.~Itoh, R.~Nakazawa, T.~Ota, M.~Hashisaka, K.~Muraki, and T.~Fujisawa,
  ``{Signatures of a Nonthermal Metastable State in Copropagating Quantum Hall
  Edge Channels},'' {\em Physical Review Letters}, vol.~120, p.~197701, may
  2018.

\bibitem{Cabart2018}
C.~Cabart, B.~Roussel, G.~F{\`{e}}ve, and P.~Degiovanni, ``{Taming electronic
  decoherence in one-dimensional chiral ballistic quantum conductors},'' {\em
  Physical Review B}, vol.~98, p.~155302, oct 2018.

\bibitem{Duprez2019}
H.~Duprez, E.~Sivre, A.~Anthore, A.~Aassime, A.~Cavanna, A.~Ouerghi,
  U.~Gennser, and F.~Pierre, ``{Macroscopic Electron Quantum Coherence in a
  Solid-State Circuit},'' {\em Phys Rev X}, vol.~9, p.~021030, May 2019.

\bibitem{Rodriguez2020}
R.~H. Rodriguez, F.~D. Parmentier, D.~Ferraro, P.~Roulleau, U.~Gennser,
  A.~Cavanna, M.~Sassetti, F.~Portier, D.~Mailly, and P.~Roche, ``{Relaxation
  and revival of quasiparticles injected in an interacting quantum Hall
  liquid},'' {\em Nat. Commun.}, vol.~11, pp.~1--8, May 2020.

\bibitem{wei2018electrical}
D.~S. Wei, T.~Van Der~Sar, S.~H. Lee, K.~Watanabe, T.~Taniguchi, B.~I.
  Halperin, and A.~Yacoby, ``Electrical generation and detection of spin waves
  in a quantum hall ferromagnet,'' {\em Science}, vol.~362, no.~6411,
  pp.~229--233, 2018.

\bibitem{assouline2021excitonic}
A.~Assouline, M.~Jo, P.~Brasseur, K.~Watanabe, T.~Taniguchi, T.~Jolicoeur,
  D.~Glattli, N.~Kumada, P.~Roche, F.~Parmentier, {\em et~al.}, ``Excitonic
  nature of magnons in a quantum hall ferromagnet,'' {\em Nature Physics},
  vol.~17, no.~12, pp.~1369--1374, 2021.

\bibitem{Nakaharai2011}
S.~Nakaharai, J.~R. Williams, and C.~M. Marcus, ``Gate-defined graphene quantum
  point contact in the quantum {Hall} regime,'' {\em Phys. Rev. Lett.},
  vol.~107, p.~036602, July 2011.

\bibitem{Zimmermann2017}
K.~Zimmermann, A.~Jordan, F.~Gay, K.~Watanabe, T.~Taniguchi, Z.~Han,
  V.~Bouchiat, H.~Sellier, and
  B.~Sac{\ifmmode\acute{e}\else\'{e}\fi}p{\ifmmode\acute{e}\else\'{e}\fi},
  ``{Tunable transmission of quantum Hall edge channels with full degeneracy
  lifting in split-gated graphene devices},'' {\em Nat. Commun.}, vol.~8,
  pp.~1--7, Apr. 2017.

\bibitem{Cohen2022}
L.~A. Cohen, N.~L. Samuelson, T.~Wang, K.~Klocke, C.~C. Reeves, T.~Taniguchi,
  K.~Watanabe, S.~Vijay, M.~P. Zaletel, and A.~F. Young, ``Nanoscale
  electrostatic control in ultraclean van der {Waals} heterostructures by local
  anodic oxidation of graphite gates,'' {\em Nature Physics}, vol.~--, no.~--,
  pp.~--, 2023.

\bibitem{Liu2022}
X.~Liu, G.~Farahi, C.~L. Chiu, Z.~Papic, K.~Watanabe, T.~. Taniguchi, and
  A.~Yazdani, ``{Visualizing broken symmetry and topological defects in a
  quantum Hall ferromagnet},'' {\em Science}, vol.~375, no.~6578, pp.~321--326,
  2022.

\bibitem{Ahmad2019}
N.~F. Ahmad, T.~Iwasaki, K.~Komatsu, K.~Watanabe, T.~Taniguchi, H.~Mizuta,
  Y.~Wakayama, A.~M. Hashim, Y.~Morita, S.~Moriyama, and S.~Nakaharai,
  ``{Effect of gap width on electron transport through quantum point contact in
  hBN/graphene/hBN in the quantum Hall regime},'' {\em Applied Physics
  Letters}, vol.~114, p.~023101, jan 2019.

\bibitem{Ronen2021}
Y.~Ronen, T.~Werkmeister, D.~Haie~Najafabadi, A.~T. Pierce, L.~E. Anderson,
  Y.~J. Shin, S.~Y. Lee, Y.~H. Lee, B.~Johnson, K.~Watanabe, {\em et~al.},
  ``Aharonov--bohm effect in graphene-based {Fabry--P{\'e}rot} quantum {Hall}
  interferometers,'' {\em Nature nanotechnology}, vol.~16, no.~5, pp.~563--569,
  2021.

\bibitem{Cohen2022b}
L.~A. Cohen, N.~L. Samuelson, T.~Wang, T.~Taniguchi, K.~Watanabe, M.~P.
  Zaletel, and A.~F. Young, ``{Universal chiral Luttinger liquid behavior in a
  graphene fractional quantum Hall point contact},'' {\em Science}, vol.~382,
  pp.~542--547, Dec. 2023.

\bibitem{Li2018}
J.~Li, H.~Wen, K.~Watanabe, T.~Taniguchi, and J.~Zhu, ``Gate-controlled
  transmission of quantum {Hall} edge states in bilayer graphene,'' {\em
  Physical Review Letters}, vol.~120, p.~057701, jan 2018.

\bibitem{Deprez2021}
C.~D{\ifmmode\acute{e}\else\'{e}\fi}prez, L.~Veyrat, H.~Vignaud, G.~Nayak,
  K.~Watanabe, T.~Taniguchi, F.~Gay, H.~Sellier, and
  B.~Sac{\ifmmode\acute{e}\else\'{e}\fi}p{\ifmmode\acute{e}\else\'{e}\fi}, ``{A
  tunable Fabry{\textendash}P{\ifmmode\acute{e}\else\'{e}\fi}rot quantum Hall
  interferometer in graphene},'' {\em Nat Nanotechnol}, vol.~16, pp.~555--562,
  May 2021.

\bibitem{biswas2022}
S.~Biswas, R.~Bhattacharyya, H.~K. Kundu, A.~Das, M.~Heiblum, V.~Umansky,
  M.~Goldstein, and Y.~Gefen, ``Shot noise does not always provide the
  quasiparticle charge,'' {\em Nature physics}, vol.~18, no.~12,
  pp.~1476--1481, 2022.

\bibitem{nakamura2023half}
J.~Nakamura, S.~Liang, G.~C. Gardner, and M.~J. Manfra, ``Half-integer
  conductance plateau at the $\nu$= 2/3 fractional quantum {Hall} state in a
  quantum point contact,'' {\em Physical Review Letters}, vol.~130, no.~7,
  p.~076205, 2023.

\bibitem{schiller2022}
N.~Schiller, Y.~Oreg, and K.~Snizhko, ``Extracting the scaling dimension of
  quantum {Hall} quasiparticles from current correlations,'' {\em Physical
  Review B}, vol.~105, no.~16, p.~165150, 2022.

\bibitem{zhao2022}
L.~Zhao, E.~G. Arnault, T.~F. Larson, Z.~Iftikhar, A.~Seredinski, T.~Fleming,
  K.~Watanabe, T.~Taniguchi, F.~Amet, and G.~Finkelstein, ``Graphene-based
  quantum hall interferometer with self-aligned side gates,'' {\em Nano
  Letters}, vol.~22, no.~23, pp.~9645--9651, 2022.

\bibitem{feldman2021}
D.~E. Feldman and B.~I. Halperin, ``Fractional charge and fractional statistics
  in the quantum {Hall} effects,'' {\em Reports on Progress in Physics},
  vol.~84, no.~7, p.~076501, 2021.

\bibitem{Halperin2011}
B.~I. Halperin, A.~Stern, I.~Neder, and B.~Rosenow, ``Theory of the
  {Fabry-Perot quantum Hall} interferometer,'' {\em Physical Review B},
  vol.~83, no.~15, p.~155440, 2011.

\bibitem{van1989}
B.~Van~Wees, L.~P. Kouwenhoven, C.~Harmans, J.~Williamson, C.~Timmering,
  M.~Broekaart, C.~Foxon, and J.~Harris, ``Observation of zero-dimensional
  states in a one-dimensional electron interferometer,'' {\em Physical Review
  Letters}, vol.~62, no.~21, p.~2523, 1989.

\bibitem{ofek2010}
N.~Ofek, A.~Bid, M.~Heiblum, A.~Stern, V.~Umansky, and D.~Mahalu, ``Role of
  interactions in an electronic {Fabry--Perot} interferometer operating in the
  quantum {Hall} effect regime,'' {\em Proceedings of the National Academy of
  Sciences}, vol.~107, no.~12, pp.~5276--5281, 2010.

\bibitem{mcclure2012}
D.~McClure, W.~Chang, C.~M. Marcus, L.~Pfeiffer, and K.~West, ``{Fabry-Perot}
  interferometry with fractional charges,'' {\em Physical Review Letters},
  vol.~108, no.~25, p.~256804, 2012.

\bibitem{sivan2018}
I.~Sivan, R.~Bhattacharyya, H.~Choi, M.~Heiblum, D.~Feldman, D.~Mahalu, and
  V.~Umansky, ``Interaction-induced interference in the integer quantum {Hall}
  effect,'' {\em Physical Review B}, vol.~97, no.~12, p.~125405, 2018.

\bibitem{nakamura2019}
J.~Nakamura, S.~Fallahi, H.~Sahasrabudhe, R.~Rahman, S.~Liang, G.~C. Gardner,
  and M.~J. Manfra, ``{Aharonov--Bohm interference of fractional quantum Hall
  edge modes},'' {\em Nature Physics}, vol.~15, no.~6, pp.~563--569, 2019.

\bibitem{choi2015}
H.~Choi, I.~Sivan, A.~Rosenblatt, M.~Heiblum, V.~Umansky, and D.~Mahalu,
  ``Robust electron pairing in the integer quantum {Hall} effect regime,'' {\em
  Nature communications}, vol.~6, no.~1, p.~7435, 2015.

\bibitem{Nakamura2023}
J.~Nakamura, S.~Liang, G.~C. Gardner, and M.~J. Manfra, ``{Fabry-Perot
  interferometry at the $\nu $= 2/5 fractional quantum Hall state},'' {\em
  Phys. Rev. X}, vol.~13, p.~041012, Oct 2023.

\bibitem{DeC.Chamon1997}
C.~{de C. Chamon}, D.~E. Freed, S.~A. Kivelson, S.~L. Sondhi, and X.~G. Wen,
  ``{Two point-contact interferometer for quantum Hall systems},'' {\em
  Physical Review B}, vol.~55, pp.~2331--2343, jan 1997.

\bibitem{li2017}
J.~Li, C.~Tan, S.~Chen, Y.~Zeng, T.~Taniguchi, K.~Watanabe, J.~Hone, and
  C.~Dean, ``Even-denominator fractional quantum {Hall} states in bilayer
  graphene,'' {\em Science}, vol.~358, no.~6363, pp.~648--652, 2017.

\bibitem{werkmeister2023}
T.~Werkmeister, J.~R. Ehrets, Y.~Ronen, M.~E. Wesson, D.~Najafabadi, Z.~Wei,
  K.~Watanabe, T.~Taniguchi, D.~E. Feldman, B.~I. Halperin, A.~Yacoby, and
  P.~Kim, ``{Strongly coupled edge states in a graphene quantum Hall
  interferometer},'' {\em arXiv:2312.03150}, 2023.

\bibitem{yang2023}
W.~Yang, D.~Perconte, C.~Déprez, K.~Watanabe, T.~Taniguchi, S.~Dumont,
  E.~Wagner, F.~Gay, I.~Safi, H.~Sellier, and B.~Sacépé, ``Evidence for
  correlated electron pairs and triplets in quantum hall interferometers,''
  {\em arXiv:2312.14767}, 2023.

\bibitem{Biswas2023}
S.~Biswas, H.~K. Kundu, V.~Umansky, and M.~Heiblum, ``{Electron Pairing of
  Interfering Interface-Based Edge Modes},'' {\em Physical Review Letters},
  vol.~131, 2023.

\bibitem{Frigeri2020}
G.~A. Frigeri and B.~Rosenow, ``{Electron pairing in the quantum Hall regime
  due to neutralon exchange},'' {\em Phys. Rev. Res.}, vol.~2, p.~043396, Dec
  2020.

\end{thebibliography}

\end{document}